\documentclass[11 pt, a4paper]{article}
\usepackage{array,epsfig}
\usepackage{amsthm}
\usepackage{amsmath}
\usepackage{color}
\usepackage{enumitem}
\usepackage{booktabs}
\usepackage{sidecap}
\usepackage{setspace}
\usepackage{comment}
\usepackage{url}
\usepackage{float}
\usepackage{parskip}
\usepackage{standalone}
\usepackage{tabu}
\usepackage{longtable}
\usepackage{xltabular}
\usepackage{tabularx}
\usepackage{colortbl}
\usepackage{longtable, booktabs}
\usepackage[font=footnotesize,labelfont=bf]{caption}
\usepackage{bm, bbm, mdframed, mathtools, multirow, tikz, multirow, tcolorbox, soul, caption}
\usepackage[left=2.5cm, right=2.5cm, top=3cm, bottom=3cm]{geometry}
\usepackage[english]{babel}
\usepackage[mathscr]{euscript}
\newcommand{\E}{\mathrm{E}}

\numberwithin{equation}{section} 
\tcbuselibrary{breakable}
\usepackage{mathdots}
\usepackage{lscape}
\usepackage{pdflscape}

\usepackage{subcaption} 
\usepackage{lscape}
\usepackage{apacite} 
\usepackage[longnamesfirst]{natbib}
\usepackage[autostyle]{csquotes}  
\usepackage{hyperref}
\hypersetup{%
  colorlinks = true,
  citecolor = black,
  filecolor=black,
  urlcolor=black,
  linkcolor = black
}
\usepackage{glossaries}
\usepackage{appendix}
\usepackage{verbatim}
\usepackage{tcolorbox}
\usepackage{xcolor}
\usepackage{listings}
\usepackage{rotating}
\usepackage{comment}
\usepackage{graphicx}
\usepackage{dsfont}
\usepackage{adjustbox}
\usepackage[para,online,flushleft]{threeparttable}
\usepackage{ragged2e}
\usepackage{amsmath}
\usepackage{amsfonts}
\usepackage{amssymb}
\usepackage{multirow}
\usepackage{makecell}

\title{\textbf{From Average Effects to Targeted Assignment:\\ A Causal Machine Learning Analysis of Swiss Active Labor Market Policies}}

\author{
  Federica Mascolo\thanks{University of St.Gallen, Rosenbergstrasse 20-22, 9000 St.Gallen, CH, E-mail:\texttt{federica.mascolo@unisg.ch}, \texttt{nora.bearth@unisg.ch}, \texttt{fabian.muny@unisg.ch}, \texttt{michael.lechner@unisg.ch}, \texttt{jana.mareckova@unisg.ch}\\
   }
    , Nora Bearth\textsuperscript{\footnotemark[1]},
    Fabian Muny\textsuperscript{\footnotemark[1]},
    Michael Lechner\textsuperscript{\footnotemark[1]}
\textsuperscript{\footnotemark[2]}\thanks{Michael Lechner is also affiliated with CEPR, London, CESIfo, Munich, IAB, Nuremberg and IZA, Bonn. \\
Financial support from the Swiss National Science Foundation (SNSF) is gratefully acknowledged. The study is part of the project ``Chances and risks of data-driven decision making for labour market policy'' (grant number SNSF 407740\_187301) of the Swiss National Research programme ``Digital Transformation'' (NRP 77). We thank GPT-3.5 and Grammarly for editorial help.}
\addtocounter{footnote}{1},
Jana Mareckova\textsuperscript{\footnotemark[1]}
}

\date{\today}

\begin{document}
\begingroup
\let\newpage\relax
\maketitle
\endgroup

 \vspace{0.3cm}
 \begin{center}
  \textbf{Abstract} \end{center}
\vspace{0.3cm}
\begin{minipage}{\textwidth}
    \small
Active labor market policies are widely used by the Swiss government, enrolling over half of all unemployed individuals. 
This paper evaluates the effectiveness of Swiss programs in improving employment and earnings outcomes using causal machine learning and rich administrative data on unemployed individuals in 2014 and 2015, including detailed labor market histories and other covariates. The findings for Swiss citizens and immigrants with permanent residency indicate a small positive average effect of a Temporary Wage Subsidy program on employment and earnings in the third year after program start. In contrast, Basic Courses, such as job application training, exhibit negative effects on both outcomes over the same period. No significant impacts are found for Employment Programs conducted outside the regular labor market or for Training Courses such as language or computer classes. The programs are most effective for individuals with a non-EU migration background, while Temporary Wage Subsidies also benefit those with lower educational attainment. Finally, shallow policy trees provide practical guidance for improving the targeting of program assignments.
\newline
\newline

\end{minipage}

\thispagestyle{empty}

\newpage

\setcounter{page}{1}

\section{Introduction}\label{Intro}

Active labour market policies (ALMPs) are widely adopted by governments worldwide to improve the reintegration of the unemployed into the labour market. According to the OECD, the Swiss government annually spends around 0.5\% of GDP on such measures to support unemployed individuals to find a new job \citep{OECD:2024}, placing its expenditure slightly above the OECD average. However, the empirical evidence on the effectiveness of these programmes is mixed \cite*[e.g.,][]{10.1093/jeea/jvx028,  gerfin2002microeconometric, lalive2008impact}, with some even expected to have detrimental effects on re-employment probabilities and future earnings. While certain programmes may show overall ineffectiveness, they could still offer substantial benefits to specific subgroups of the unemployed. Therefore, knowing which groups profit most from a specific programme is crucial for optimising the assignment decisions. 

Despite numerous assessments of the Swiss ALMPs in the past, the current understanding of the heterogeneous effects of these measures remains limited, as noted e.g.~in \citet*{Seco:2024a}. Therefore, this paper aims to address this issue. In particular, it investigates the effects of participating in ALMPs on individual future employment and earnings by leveraging an extensive Swiss administrative dataset covering all unemployed between 2004 to 2018 and their labour market histories. By incorporating recent methodological advancements in causal machine learning (CML) and the optimal policy learning literature, the analysis offers granular estimates of effect heterogeneities among subgroups. Furthermore, it provides recommendations for improving individual programme assignments. The methodology aligns with \cite*{cockx2023priority}, utilising an unconfoundedness identification strategy and employing the Modified Causal Forest (MCF), a causal machine learning estimator introduced by \citet*{lechner2022modified} for estimation.\footnote{In a comparison of the MCF to Double Machine Learning and the Generalized Random Forest, \cite{Lechner:2024} document the MCF's reliability and robustness.}

The paper contributes to the literature in different ways. Firstly, employing flexible CML methods enhances the evaluation of programmes by yielding estimates that are robust to errors from model misspecification. Secondly, the CML methods allow us to uncover heterogeneities to a much greater extent than was previously achievable \citep{Baiardi:2024}. Beyond merely describing such heterogeneities, the paper focuses on detecting the drivers behind the effect variations, aiming to uncover the underlying mechanisms contributing to these disparities. Thirdly, the allocation of the unemployed is analysed in a data-driven way, considering factors such as interpretability and capacity constraints. This allows us to make recommendations to policymakers on whom to assign to which programme. Lastly, the paper reassesses the Swiss ALMPs to address their more recent relevance, potentially affected by changes in Switzerland's economic and labour environments due to globalisation, technological progress, and demographic shifts. This reassessment is vital to ascertain the continued effectiveness of these programmes.

The analysed programmes include Temporary Wage Subsidies, Basic and Training Courses, and an Employment Programme. The Temporary Wage Subsidy compensates individuals who accept temporary jobs with lower earnings than their unemployment benefits. Basic Courses consist of short preparatory activities, such as job application trainings. Training Courses cover skills like computer or language proficiency, while Employment Programmes place participants in unpaid positions outside the regular job market. The findings for the main sample of Swiss residents and immigrants with permanent residenceshow that all these programs have strong negative impacts on earnings and employment immediately after their initiation. However, the Temporary Wage Subsidy demonstrates positive effects in the medium-term, observed three years after programme start. In contrast, Basic Courses exhibit statistically significant negative effects in the medium-term, while Training Courses and Employment Programmes show no statistically significant effects during this period. Notably, when examining specific subgroups, the outcomes indicate that the effectiveness of the programmes is more pronounced for individuals with a place of origin outside of the European Union or Switzerland. Utilising policy learning methods constrained by observed programme shares does not improve the allocation of individuals to the programmes. However, increasing the share of individuals participating in the Temporary Wage Subsidy programme can enhance the average employment duration by up to half a month in the third year following programme initiation. Therefore, the primary policy recommendation is to expand the Temporary Wage Subsidy programme, while ensuring it remains within a range that does not introduce market distortions.

The remainder of the paper is organised as follows: The following subsection gives an overview of the related literature. Section \ref{institutional_setting} provides background information on the Swiss unemployment system. Section \ref{data} describes the data and explains how the population of interest and the different programmes have been selected. Section \ref{econometrics} outlines the econometric framework, followed by Section \ref{results}, which presents the main results. Section \ref{sensitivity_analysis} demonstrates the robustness of the results in different settings and the placebo analysis. Section \ref{discussion} discusses the implications of the results and concludes. 
Lastly, Appendix \ref{Appendix_data} contains descriptive statistics of the data, Appendix \ref{Appendix_estimator} provides further details about the estimation procedure, and Appendix \ref{Appendix_results} presents extended treatment effect and optimal policy results.

\subsection{Related Literature}\label{Literature}
The impact of ALMPs on labour market outcomes has been extensively studied in the literature. Evaluating the effectiveness of public expenditure is of interest across countries and over time. Typical outcomes of interest include post-programme earnings and employment status in the short-, medium-, and long-term. The existing literature reveals mixed effects of these policies and significant variation across countries and groups of unemployed. Recent meta-analyses \citep*{10.1093/jeea/jvx028,       vooren2019effectiveness} provide insights from various studies such as  \citet*{c56cf7f9-3aeb-3d5d-9780-3a616776b70c}, \citet*{heckman1999economics} and \citet{RePEc:hhs:ifauwp:2001_014}, indicating that Job Search Assistance and forms of Wage Subsidies have positive effects on employment probabilities. In contrast, the evidence for the positive effects of Training Courses is weaker, and the overall effectiveness of ALMPs seems limited, particularly in the short-term. 

Among the first to evaluate ALMPs in Switzerland are \citet {gerfin2002microeconometric} and \citet{lalive2008impact}. Using matching estimators, they find positive effects for Temporary Wage Subsidies, negative effects for Employment Programmes, and mixed results for Training Courses. This work extends their contributions by presenting results obtained through novel estimation techniques. In particular, a flexible CML estimator by \citet{lechner2022modified} is leveraged, building upon more recent research lines \citep[e.g.,][]{wager2018estimation, athey2019generalized}, that provides robust estimates that do not rely on restrictive functional form assumptions while enabling to explore the heterogeneity in causal effects.

Several recent contributions employ CML, specifically causal forest estimators, to investigate the heterogeneous impacts of ALMPs. \cite{cockx2023priority} focus on Belgian unemployed eligible for short- or long-term vocational training. The authors find positive effects on the medium-term outcomes. They also find substantial heterogeneity with higher positive effects for recent immigrants with low proficiency in the local language. \citet{burlat2024everybody} assesses training courses in Eastern France. She finds positive medium-term effects for technical qualification programmes and heterogeneity by education. \cite*{goller2021active} focus on ALMPs targeted at long-term unemployed in Germany. They find positive average effects for all programmes and heterogeneities by gender and region for this particular subgroup of the unemployed population. 

In two other closely related papers, \cite{knaus2022heterogeneous} and \cite{knaus2022double} exploit Swiss administrative data, including unemployed individuals in 2003, to study the effects of ALMPs using different CML approaches. In these papers, the primary focus lies in demonstrating the usefulness of the new CML methods, as indicated by utilising an established but rather old data set. Both papers find negative average effects for job-search programmes while \citet{knaus2022double} finds positive average effects of Training Courses. The heterogeneity analysis reveals differences in the effects based on gender, nationality, previous labour market success, and qualifications. While closely related to those contributions, our paper advances by employing an advanced estimator, utilising a more recent dataset and analysing various programmes.

Besides estimating average and heterogeneous effects, there has been interest in improving the programme allocation process. In an early attempt based on Swiss data, \citet{lechner2007value} find that statistical treatment rules outperform caseworker assignments. \citet{knaus2022double} and \cite{cockx2023priority} leverage optimal policy learning for optimal targeting of unemployed individuals to programmes. In particular, they follow a policy rule based on \citet*{zhou2023offline}. \cite{cockx2023priority} find that the allocations generated by shallow policy trees increase average outcomes compared to observed and random allocations. In this work, similar lines are followed to develop interpretable data-driven assignment rules that adhere to capacity constraints for the Swiss setting.

\section{Institutional Setting and Programmes} \label{institutional_setting}

Switzerland is a federalist state with an unemployment rate that fluctuated around $4.8\%$ between 2014 and 2018.\footnote{\href{https://www.bfs.admin.ch/bfs/en/home/statistics/work-income/unemployment-underemployment/ilo-unemployed.html}{Federal Statistical Office unemployment rate based on the International Labour Organisation definition of unemployed (permanent residents aged 15 or over who do not have a job, who are looking for work and who can start working within a short period of time)}. Data was retrieved on 18.08.2023.}  
Over the last decades, ALMPs gained importance in supporting the unemployed in finding new jobs. While these policies are regulated at the national level, they are implemented by regional employment offices (REO). These agencies serve as the initial point of contact for the unemployed. Upon registration, individuals who have been employed and paid social security contributions for at least 12 months in the last two years are eligible for income maintenance. The amount of unemployment benefit depends on the individual's past salary. The maximum benefit entitlement period is 24 months.

In the Swiss system, actively searching for a job is an additional condition to receive unemployment benefits. For instance, individuals must apply for a certain number of job openings within a specific time frame. They may be assigned to various ALMPs during their period of unemployment. These ALMPs vary from employment programmes to training courses such as language or computer courses. Caseworkers at the regional offices determine the allocation of the unemployed to specific ALMPs based on the information about the unemployed person, such as socio-demographic characteristics, labour-market history, and job preferences (see Section \ref{out_cov} for details). While there are general principles for allocation at the federal level, the caseworkers typically make the final decision based on the information retrieved at the moment of the unemployed's registration and during the consultation session. Additionally, caseworkers have the authority to sanction individuals if they decline to participate in programmes. Regardless of the programme, individuals must continue searching for a job and terminate the programme upon finding employment.

The REOs offer a wide array of diverse programmes, which 
are categorised into the following groups for the purpose of our programme evaluation:\footnote{The programme categorisation follows the REOs' official definitions and is comparable to previous evaluations. \citet{gerfin2002microeconometric} use the same categories, though at a more granular level. \citet{lalive2008impact} group computer and language courses together with basic courses but use the same categories otherwise.} \textit{Basic Courses (BC)} are non-training courses related to acquiring basic qualifications or competencies related to personality and job search (e.g., job application strategies, career development workshops). \textit{Training Courses (TC)} include language courses from introductory to advanced level for foreigners and Swiss citizens, computer courses that provide mostly basic IT literacy and other types of courses targeted to particular sectors (e.g., technical or commercial training, training for health and social services sectors). The \textit{Employment Programme (EP)} offers unpaid employment outside the regular labour market, with some positions offered by public institutions (e.g., cantons, municipalities) and others by private firms or NGOs. The jobs offered with the EP should resemble regular jobs but not compete with other companies. \textit{Temporary Wage Subsidies (WS)} are paid to incentivise individuals to accept jobs with lower wages than their unemployment benefits within the standard labour market.\footnote{WS are considered in the analysis despite not being officially categorised as ALMP by the REOs since they are used as policy instruments similarly to the other programmes.} This is accomplished by providing compensation through the unemployment insurance, ensuring that the total salary and compensation exceed the original unemployment benefit. From the perspective of the unemployment insurance, this option is also appealing as the compensation is less than the full unemployment benefit.\footnote{However, it is initially unclear whether the government saves money through WS, as participation in the programme extends the entitlement period for compensations. In addition, if individuals can find unsubsidised employment faster without the programme, avoiding WS may be more cost-effective.} 
To address concerns of market distortions, allowing subsidised firms to exploit cheaper labour and create unfair competition with non-subsidised companies, the REOs apply location- and occupation-specific reference salaries that eliminate any financial incentive for accepting temporary employment below these levels.
Besides the mentioned programmes, the REOs offer additional measures such as training and introduction grants, vocational work placements or internships, which are categorised as \textit{Other Programmes (OP)}.

\section{Data} \label{data}

\subsection{Study Population}
The analysis utilises a comprehensive dataset derived from diverse Swiss administrative records, offering a monthly snapshot of individuals' employment status and earnings. The population of interest includes all unemployed who became eligible for benefits between January 2014 and June 2015, following a minimum of three months of prior unsubsidised employment. This relatively short time frame is selected to ensure that individuals in the sample are unaffected by major institutional changes. In addition, since data availability is from 2004 to 2018, it permits using ten years of pre-treatment covariates and assessing medium-term outcomes for up to three years. Finally, it ensures that years affected by the aftermath of the global financial crisis are excluded. Within the specified time frame, the focus is on individuals aged between 25 and 55 at the onset of their unemployment spell. Younger individuals are excluded to avoid having to model educational choices, and older individuals are excluded to sidestep (early) retirement decisions, both of which may affect their availability for the labour market. Individuals who reported receiving disability pension in the three months before the start of the unemployment spell are also excluded, as they may be subject to distinct rules and treatments.

The resulting sample consists of Swiss nationals and immigrants with a temporary residence permit (B) or a permanent settlement permit (C).\footnote{Other immigrants are not eligible for unemployment benefits in Switzerland.} Individuals with type B permits are recent residents of Switzerland. They have shorter earnings and employment histories compared to immigrants with permanent settlement permits, who exhibit a labour market history that is more similar to Swiss-born citizens. The shorter labour market history for the first group poses a challenge for identification on a whole population.\footnote{Only 25\% of workers with temporary work permits have earning records before 2006. 50\% of this subsample shows records starting from 2010.} Therefore, the same analysis is carried out on two distinct samples:

Swiss citizens and individuals with settlement permits, denoted as \textit{Permanent Residents} (PERM), and individuals with temporary residence permits, denoted as \textit{Temporary Residents} (TEMP). The first sample includes 113,329 individuals, whereas the second comprises 34,229.

\subsection{Participation in Programmes}

The analysis is conducted for the first programme to which the unemployed is assigned.\footnote{All programmes starting within a 6-month window after the start of the unemployment spell and lasting for at least five business days are considered. Thus, very short programmes such as assessment tests conducted before allocating to a specific programme are excluded.} Focusing on the first programme seems reasonable as 70\% of individuals are not assigned to a second programme within the first six months of their unemployment spell. The group of unemployed who are not enrolled in any programme within six months of the unemployment spell constitute the control group. Finally, if an individual is initially assigned to an OP, they are excluded from the analysis.\footnote{OPs are not included in the main analysis due to the small number of participants. In addition, as these programmes are designed for particular subgroups of the unemployed, their participants are not easily comparable to those in other programmes.} 

Table \ref{table:programme_combined} reports how the populations of interest are assigned to the considered programmes. Slightly over 50\% of individuals are assigned to a programme within the first six months, with WS being the largest programme, encompassing approximately 25\% of individuals. There are no substantial differences in the programme shares between the PERM and TEMP samples.

\begin{table}[H]
\begin{adjustbox}{width=0.7\textwidth,  center}
\begin{threeparttable}
  \captionsetup{font=large}  
    \caption{Absolute and relative frequencies of individuals by programme and sample}
    \centering
    \begin{tabular}{@{}lrrrrr@{}}
    \toprule
    \multirow{2}{*}{\textbf{Programme}} & \multicolumn{2}{c}{\textbf{Permanent Residents}} & \multicolumn{2}{c}{\textbf{Temporary Residents}} &\multirow{2}{*}{\textbf{Ø Duration}}\\
    &  Individuals & Share & Individuals & Share &  \\ \midrule
            No programme & 51,020 & 45.0\%  & 14,624 & 42.7\% & \\
            Temporary Wage Subsidies & 28,584 & 25.2\%  & 9,020 & 26.4\% & 50 days\\
            Basic courses & 20,944 & 18.5\%  & 5,046 & 14.7\% & 29 days\\
            Employment programme & 6,803 & 6.0\%  & 3,565 & 10.4\% & 71 days\\
            Training courses & 5,978 & 5.3\%  & 1,974 & 5.8\% & 41 days\\
            \midrule
            Total& 113,329& & 34,229 &\\
         \bottomrule
    \end{tabular}
\begin{tablenotes}
\textit{Note:} This table shows the absolute and relative number of individuals enrolled in each programme for the samples of permanent and temporary residents.
\end{tablenotes}
\label{table:programme_combined}
\end{threeparttable}
\end{adjustbox}
\end{table}

A concern related to comparability between the control and treatment groups stems from the dynamic nature of programme assignment, where unemployed individuals can be assigned to a programme at any point during the first six months of the unemployment spell. In this context, the control group may include individuals who had already exited unemployment before the potential start of their programme, causing a potential bias.
To deal with this issue, pseudo programme start dates are assigned to the individuals in the control group \citep*[as introduced by][]{Lechner:1999}. The observation is deleted from the sample if the individual is no longer unemployed at the assigned pseudo date.\footnote{Following this procedure, 2.9\% and 4.6\% of observations in the permanent and temporary residents sample are deleted, respectively.} Hence, individuals who tend to find a job before being assigned to a programme are disregarded, given their observed characteristics. To compute the pseudo starting dates, a random sample of 20\% of the observations is drawn to train a Random Forest and predict the duration (months) from the start of the unemployment spell to the pseudo programme start. The Random Forest is trained with the same specification used in the main analysis on the sub-sample of only treated among the random 20\%. By definition, the predicted dates range between one and six months. Admittedly, this solution is imperfect as individuals in the control group may still be more likely to leave unemployment even after the pseudo programme starts, leading to an expected potential bias in the effects \citep[see the discussion in][]{cockx2023priority}. An alternative approach to address this bias is suggested in \citet{van2022long}, though it necessitates additional identification assumptions and is not (yet) implemented within a CML framework.  

\subsection{Outcome and Covariates} \label{out_cov}

As the main goal of ALMPs is to combat existing unemployment \citep{Seco:2024}, effects of programme participation on monthly employment status, monthly earnings, and various aggregates thereof during the 36 months following the programme's start are examined.
The outcomes are aggregated in three ways. Initially, the first year after the programme start is analysed, estimating the lock-in effect. During programme participation, individuals usually decrease job search efforts, resulting in a ``locked-in'' phase where their likelihood of finding a new job is lower compared to non-participation. Next, the entire 36-month period is considered, allowing for an evaluation of the programme's impact over a medium-term horizon while considering the lock-in effect costs. However, three years may not suffice to offset the lock-in effect. Therefore, the outcomes in the third year after the programme start are examined, disregarding the associated lock-in costs. This last outcome is a medium-term outcome that seems most policy-relevant, distant enough from potential sequences of programmes where individuals may enrol. Due to data limitations, long-term outcomes beyond 36 months cannot be assessed.

Panel A of Table \ref{table:desc_stat_full} presents the means and standardised differences\footnote{The standardised difference is defined as $\Delta=\frac{\left|\bar{X}^k-\bar{X}^j\right|}{\sqrt{1 / 2\left(\operatorname{Var}\left(\bar{X}^k\right)+\operatorname{Var}\left(\bar{X}^j\right)\right)}} \cdot 100$ where $\bar{X}^{k}$ and $\bar{X}^j$ indicate the sample mean of the programmes and the control group, respectively. Following \cite{rosenbaum1985constructing}, a standardised difference higher than 20 is considered `large'.} of programme participants compared to non-participants of the discussed outcome measures. The table provides a first indication of the later results before correcting for selection bias. The non-participants perform relatively well in both samples, demonstrating the highest number of months employed in the first year (no lock-in effect) and the highest cumulative earnings in the third year. Considering employment in the third year, WS recipients in the PERM sample and all programme participants in the TEMP sample tend to perform slightly better on average than the control group.

Besides outcomes, the dataset consists of an extensive set of covariates. These include socio-demographic characteristics, including age, gender, civil status, nationality, language proficiency, and residence permit type. It also incorporates information on prior job positions, encompassing function, sector, and experience. Additionally, the dataset incorporates the placement officer's assessment of the client's job search efforts, details on any sanctions imposed by the caseworker, sick days during prior unemployment periods, and whether individuals received contributions from the social security system over the last two years. Lastly, the data provides the REO identifiers. Additionally, macroeconomic indicators related to local labour market characteristics measured at the start of the unemployment spell are linked to the administrative data.

Panel B of Table \ref{table:desc_stat_full} presents the means and standardised differences for permanent and temporary residents regarding their key characteristics and labour market history. For a more complete set of descriptive statistics, including all covariates, see Appendix \ref{Descriptive}. Firstly, temporary residents are generally younger, more often male, less educated, and less proficient in the local language than permanent residents. Additionally, they appear in the administrative records about ten years later, leading to fewer months of previous employment and lower cumulative earnings histories. Regarding the programme groups, the characteristics of WS and EP participants are similar, primarily attracting individuals with low levels of education and low previous earnings. The TCs have a higher proportion of individuals with low language proficiency, reflecting the inclusion of language courses within the programme. Additionally, a larger share of the older generation participates in TCs for permanent residents, likely attributable to the basic IT courses offered. Finally, BC participants exhibit similarities to the control group, characterised by higher education levels and earnings.

\begin{table}[H]
\renewcommand{\arraystretch}{1.0} 
\begin{adjustbox}{width=0.85\columnwidth, center}
\captionsetup{font=large}  
\begin{threeparttable}
\captionsetup{font=large}
    \caption{Descriptive statistics of selected variables by programme for the PERM and TEMP samples.}
    \centering
    \begin{tabular}{@{}lcccccccccc@{}}
    \toprule
    \textbf{Variable} & \multicolumn{5}{c}{\textbf{Permanent Residents}} & \multicolumn{5}{c}{\textbf{Temporary Residents}} \\
    \toprule
     & NP & WS & BC & TC & EP & NP & WS & BC & TC & EP\\
    \midrule   
        \textit{Panel A:} Outcomes &    &   &   &   & \\
    \midrule
    Sum of months in emp.  & 5.82 & 4.50 & 3.21 & 2.97 & 3.17 &  5.38 & 4.37 & 2.96 & 2.93 & 3.22\vspace{-0.5em}\\
    first year & & (32.62) & (65.94) & (72.80) & (67.41) && (25.55) & (62.08) & (63.33) & (55.61)\\
    
    Sum of months in emp.  & 9.09 & 9.45 & 8.79 & 8.73 & 8.52 & 7.49 & 8.50 & 7.98 & 7.95 & 7.86\vspace{-0.5em}\\
    third year & & (8.42) & (6.66) & (7.96) & (12.67) && (21.65) & (9.90) & (9.35) & (7.57)\\
   
    Sum of months in emp. & 23.15 & 21.82 & 19.21 & 18.73 & 18.75 & 19.86 & 20.05 & 17.66 & 17.59 & 18.01\vspace{-0.5em}\\
    first to third year & & (12.93) & (36.94) & (41.49) & (41.66) && (1.70) & (19.75) & (20.50) & (16.79)\\
    
    Sum of earnings  & 32,729 & 21,265 & 18,270 & 16,979 & 15,210 & 27,887 & 19,069 & 15,057 & 14,958 & 14,277\vspace{-0.5em}\\
    first year & & (41.13) & (49.79) & (54.41) & (64.24) && (35.59) & (49.58) & (50.65) & (55.60)\\
    
    Sum of earnings  & 51,893 & 44,595 & 47,208 & 46,689 & 38,481 & 40,191 & 36,879 & 38,166 & 39,542 & 33,510 \vspace{-0.5em}\\
    third year & & (18.87) & (10.99) & (12.31) & (34.52) && (9.63) & (5.31) & (1.69) & (19.22)\\
    
    Sum of earnings & 132,153 & 103,751 & 105,431 & 102,108 & 86,156 & 105,824 & 87,538 & 86,062 & 87,750 & 77,498\vspace{-0.5em}\\
    first to third year & & (30.64) & (26.43) & (29.95) & (49.75) && (22.39) & (22.19) & (20.27) & (34.37)\\
    \midrule
    \multicolumn{5}{@{}l}{\textit{Panel B:} Socio-demographics variables}   &  & & & & \\
    \midrule
    Age & 38.50 & 39.72 & 40.03 & 40.62 & 40.02 & 36.36 & 36.96 & 36.56 & 36.34 & 36.39\vspace{-0.5em}\\
     & & (13.64) & (17.20) & (24.39) & (16.91) && (7.58) & (2.51) & (0.35) & (0.34)\\
    Female & 0.46 & 0.50 & 0.46 & 0.50 & 0.47 & 0.36 & 0.37 & 0.42 & 0.36 & 0.36\vspace{-0.5em}\\
    & & (8.90) & (0.26) & (8.74) & (2.63) &&  (2.66) & (12.84) & (0.14) & (1.26)\\
    Education  & 2.47 & 2.20 & 2.32 & 2.39 & 2.15 &2.33 & 1.83 & 2.32 & 2.27 & 1.88\vspace{-0.5em}\\
    (1:compulsory to 5:post-grad.)  & & (23.24) & (12.75) & (6.06) & (27.78)  && (37.82) & (0.67) & (4.23) & (32.85)\\
    Proficiency local language & 6.24 & 5.90 & 6.16 & 5.67 & 5.90  & 4.46 & 3.91 & 4.62 & 2.92 & 3.94\vspace{-0.5em}\\
     (1:basic to 7:mother tongue) & & (20.80) & (4.98) & (31.42) & (20.41) && (23.57) & (6.73) & (66.04) & (22.62)\\
    Citizenship: & 0.69 & 0.65 & 0.66 & 0.64 & 0.64  &&&&\vspace{-0.5em}\\
    Swiss & & (9.48) & (7.46) & (11.56) & (11.26)  &&&&\\
    Citizenship: &   0.12 & 0.11 & 0.12 & 0.12 & 0.10  & 0.39 & 0.29 & 0.38 & 0.25 & 0.27\vspace{-0.5em}\\
    Neighbouring countries& & (3.50) & (2.13) & (0.36) & (6.13) && (21.22) & (0.56) & (30.23) & (24.98)\\
    Citizenship: & 0.09 & 0.10 & 0.07 & 0.10 & 0.10 & 0.39 & 0.42 & 0.32 & 0.40 & 0.38\vspace{-0.5em}\\
    Rest of European Union & & (2.59) & (8.53) & (1.66) & (2.05) && (19.03) & (17.48) & (26.99) & (32.11)\\
    Citizenship: & 0.04 & 0.06 & 0.06 & 0.04 & 0.06 & 0.06 & 0.09 & 0.09 & 0.09 & 0.09\vspace{-0.5em}\\
    Rest of Europe & & (10.34) & (11.37) & (2.49) & (11.54) &&  (10.05) & (13.13) & (12.84) & (11.52)\\
    Citizenship: & 0.06 & 0.09 & 0.09 & 0.11 & 0.10 & 0.17 & 0.21 & 0.20 & 0.25 & 0.26\vspace{-0.5em}\\
    Rest of the world & & (9.82) & (10.17) & (16.71) & (14.99) && (10.80) & (9.02) & (21.74) & (22.60)\\
    Sum of months in emp.& 50.81 & 49.99 & 52.56 & 51.93 & 50.05 &  36.95 & 35.76 & 37.03 & 34.13 & 36.00\vspace{-0.5em}\\
    five years before UE spell & & (7.17) & (15.93) & (10.05) & (6.66) && (8.27) & (0.59) & (18.99) & (6.67)\\
    Sum of months in emp.  & 94.67 & 93.61 & 97.14 & 94.89 & 93.09 & 48.20 & 47.08 & 46.79 & 40.85 & 47.32\vspace{-0.5em}\\
    ten years before UE spell & & (4.42) & (10.29) & (0.88) & (6.48) && (4.14) & (5.17) & (28.05) & (3.27)\\
    Sum of earnings  & 291,634 & 236,765 & 306,595 & 299,634 & 239,902 & 198,733 & 145,381 & 184,143 & 172,006 & 147,058\vspace{-0.5em}\\
    five years before UE spell & & (27.97) & (6.62) & (3.56) & (26.48) && (31.17) & (7.07) & (13.46) & (29.75)\\
    Sum of earnings  & 509,002 & 418,462 & 539,520 & 518,918 & 425,508 &  246,858 & 186,130 & 224,138 & 201,238 & 188,317\vspace{-0.5em}\\
    ten years before UE spell & & (25.28) & (7.28) & (2.43) & (23.42) && (27.52) & (8.92) & (18.18) & (25.92)\\ 
    First year registered & 1996 & 1995 & 1995 & 1995 & 1995 & 2008 & 2007 & 2008 & 2009 & 2007 \vspace{-0.5em}\\
    & &(6.97) & (12.92) & (7.99) & (10.44) &&  (1.96) & (6.60) & (25.75) & (1.93)\\ \midrule
   
   Number of observations & 38,464 & 22,761 & 16,690 & 5,508 & 4,761 & 10,445 & 7,177 & 4,037 & 2,850 & 1,597 \\
    \bottomrule
    \end{tabular}
\begin{tablenotes}
\textit{Note:} The table shows the unconditional means and standardised differences (in brackets); relative to non-participation for selected variables by programme. NP: Non-Participation, WS: Wage Subsidy, BC: Basic Course, TC: Training Course, EP: Employment Programme. Local language knowledge: from 1: basic to 7: mother tongue. Education: 1: Compulsory school, 2: Secondary school, 3: Diploma vocational training, 4: Bachelor degree, 5: Post-graduate: Master degree, PhD. The feature ``First year registered'' states the year an individual first made social security contributions in Switzerland. 
The full list of variables is available in Appendix \ref{Descriptive}. 
\end{tablenotes}
\label{table:desc_stat_full}
\end{threeparttable}
\end{adjustbox}
\end{table}

\section{Econometrics} \label{econometrics}
\subsection{Causal Framework}
The potential outcome framework \citep{rubin1974estimating} is used to describe a multiple treatment model in an unconfoundedness setting. A causal effect is defined as the difference between two potential outcomes. However, only the potential outcome of the assigned programme for each individual can be observed. Hence, credible counterfactuals must be found to estimate the causal effect.

Let the observed sample consist of $n$ observations considered to be i.i.d.~draws from the random variables ($\tilde{X}$, $Z$, $D$, $Y$).\footnote{Capital letters denote random variables and lower case letters denote their realisations.} The variables $\tilde{X}$ are confounders needed to correct for selection bias, and $Z$ is a covariate that identifies sub-groups of the population for which the group-specific effect heterogeneity is estimated. $\tilde{X}$ and $Z$ may or may not overlap, when combined, they are denoted as $X$. There is a treatment $D$ and an observed outcome $Y$, which is the potential outcome $Y^d$ under the observed treatment $D=d$. Since only one potential outcome can be observed for each individual, the individual treatment effects, defined as the differences between the individual potential outcomes ($Y^d - Y^{d'}$ with $ d, d' \in  \{0,\dots,4\}$), are never identified from the data. Therefore, the focus is on estimating the individualised average treatment effect (IATE) $\tau_{d, d'}(x) := \mathbb{E}[ \,  Y^d - Y^{d'} \, | \, X =x]$, which is an average of effects for individuals with the same covariates $X=x$. This represents the most granular effect that is  examined. Furthermore, less granular effects are analysed, namely, the group average treatment effect (GATE) $\tau_{d, d'}(z):=  \mathbb{E}\left[ \,  Y^d - Y^{d'} \,  | \, Z = z \right]$ and the balanced group average treatment effect (BGATE) $\tau_{d, d'}^B(z) :=  \mathbb{E}\left[\mathbb{E}[ \,  Y^d - Y^{d'} \, | \,  Z = z, W]\right]$. Both GATE and BGATE capture the average impact of pre-defined subgroups, identified by $Z = z$, but the BGATE additionally ensures that the distribution of some other covariates $W$, a subgroup of variables from $X$, is the same for all the realisations of the variable of interest $Z$ \citep{bearth2025causal}. Balancing for the covariates $W$ reduces the risk of confounded heterogeneity and helps to explain which variables potentially drive the heterogeneity in the treatment effect. Finally, the average impact on the overall population (ATE) $\tau_{d, d'} :=  \mathbb{E}\left[Y^d - Y^{d'} \right]$ and the average treatment effect on the treated (ATET)  $\tau_{d, d'}(d):=  \mathbb{E}\left[ Y^d - Y^{d'} | \, D=d \right]$ are of interest. In the following sections, we will report results comparing program participation $d \in \{1, ..., 4\}$ to non-participation $d'=0$.

The learned knowledge about effect heterogeneity is subsequently utilised to find assignment rules that improve the overall outcomes of the policy. The ideas for a multiple treatment setting for optimal interpretable allocations are implemented similar to \citet*{zhou2023offline}. Let $\pi(V)$ be a policy function that maps from policy variables $V$ (subgroup of variables from $X$) to a particular programme $d$. The value function (or welfare) of the policy is then defined by $Q(\pi) =  \mathbb{E} [Y^{\pi(V)}] = \mathbb{E} [\sum_d \mathbbm{1}_{\{\pi(V) = d\}}Y^{d}]$. The object of interest is the optimal policy rule $\pi^*$ that maximises the value function over the set of candidate policy rules $\Pi$, i.e. $\pi^*=\arg\max_{\pi \in \Pi} Q(\pi)$. 

\subsection{Identification}

To identify the causal effects of interest, four identifying assumptions need to hold \citep{imbens2000role, lechner2001identification}. Most importantly, the conditional independence assumption asserts to control for all variables jointly influencing programme assignment and potential outcomes. As explained in Section \ref{institutional_setting}, caseworkers decide which individuals to assign to which programme. Therefore, controlling for all variables that influence the caseworker's decision and are linked to the outcome variable is crucial. This study incorporates most of the information available to caseworkers during programme assignments, enabling to control for the underlying information guiding the assignment mechanism. For example, available variables include various socio-demographic characteristics, past earnings, and labour market history, including information about past employment status and details about previous occupations. In addition, factors such as the number of sickness days, the caseworker's evaluation of job search efforts during prior unemployment spells, the individual's dependence on other social insurances, and employment preferences are included. After controlling for all those variables, the remaining biases should be small according to \cite{lechner2013sensitivity}. However, compared to their paper, this study lacks detailed information about the last employer (e.g., firm size and age, workforce composition, turnover).
Furthermore, health information is limited to the degree of disability and health issues leading to sickness during previous unemployment spells. Valid concerns may also arise about unobserved factors such as the personality of the unemployed, social networks, or attitude, which are not controlled for in this study. However, \citet*{caliendo2017unobservable} suggest that these factors should not significantly affect the programme's impact on employment and earnings when considering the employment history, as is the case in this paper. Additionally, the placebo study in Section \ref{sensitivity_analysis} shows no sign of violations of conditional independence. 

The common support assumption requires that a unit with any given value of the confounding variables can be observed in any treatment state. To ensure this assumption holds, propensity scores $\hat{p}_d(x)=\hat P(D = d|X=x)$ are estimated using Random Forests, and observations outside the support are excluded, implementing the min-max trimming procedure as in \citet{lechner2019practical}. This process results in losing 5\% of the observations in the PERM sample and 12\% in the TEMP sample. Table \ref{sd_main_dropped} in the Appendix shows the standardised differences in key covariates between the kept and dropped observations. There are no substantial differences for both samples, indicating that the population of our analysis remains comparable after trimming. Appendix \ref{Common Support} provides further details on the trimming procedure.

The Stable Unit Treatment Value Assumption (SUTVA) implies that the observed outcome does not depend on the treatment allocation of other unemployed individuals, thus ruling out spillover effects. This assumption is plausible given the small size of the Swiss programmes relative to the labour market. Finally, exogeneity of covariates is assumed, meaning they are unaffected by the treatments. This assumption holds because all control variables included are time-invariant or measured at the start of the unemployment spell when the treatment has yet to be assigned or announced.

\subsection{Estimation}

For estimation, the MCF\footnote{The modified causal forest is provided in the \texttt{mcf} Python package. This work uses \texttt{mcf} version 0.5.0.} proposed by \cite{lechner2022modified} is applied. \cite*{Lechner:2024} derive its statistical guarantees and compare the approach to other common causal machine learning methods. The MCF builds on the Causal Forest by \cite{wager2018estimation} and augments the splitting criterion used to build the trees. In particular, it targets the minimisation of the mean squared error of the IATEs and accounts for potential selection bias in the early splits. Then, estimates of ATEs and GATEs can be obtained by aggregating the estimated IATEs from the forest. Compared to alternative CML methods, this procedure has been shown to perform robustly even if there is a strong selection into treatments \citep{Lechner:2024}. Section \ref{MCF} of the Appendix provides a more detailed description of the estimator.

In addition to effect estimation, policy trees are leveraged to yield interpretable programme assignments of the unemployed to specific programmes maximising utilitarian welfare.
The implementation for optimal policy trees provided by the MCF, is based on the algorithm suggested by \cite*{zhou2023offline}. Following the previous notation, the policy class $\Pi$ is restricted to shallow decision trees with a pre-defined number of final leaves. The algorithm conducts an exhaustive and recursive search over predefined policy variables and their values. It performs splits when the welfare within the leaves is maximised by assigning all observations in a terminal leaf node to a single treatment. The advantage of using this algorithm stems mainly from two factors. First, the policy trees based on the estimated IATEs outperform policy trees learned from doubly-robust-scores in several simulation settings \citep{hatamyar2023policy}. Second, decision trees allow optimisation within a feasible computational time and the interpretability of the result. The procedure can be  further extended by incorporating cost adjustments that are independent of the policy class. The PERM and TEMP samples are partitioned into three subsamples for training, estimation and validation to mitigate the risk of overfitting. See \cite{bodory2024enabling} and Appendix \ref{appendix_policytree} for further details on the algorithm and its implementation.

\section{Results} \label{results}

\subsection{Average Treatment Effects}

\begin{figure}[!ht]
\centering
\captionsetup{font=small}  
\caption{Evolution of ATEs for employment and earnings in the 36 months after programme start}
 \includegraphics[width=\textwidth]{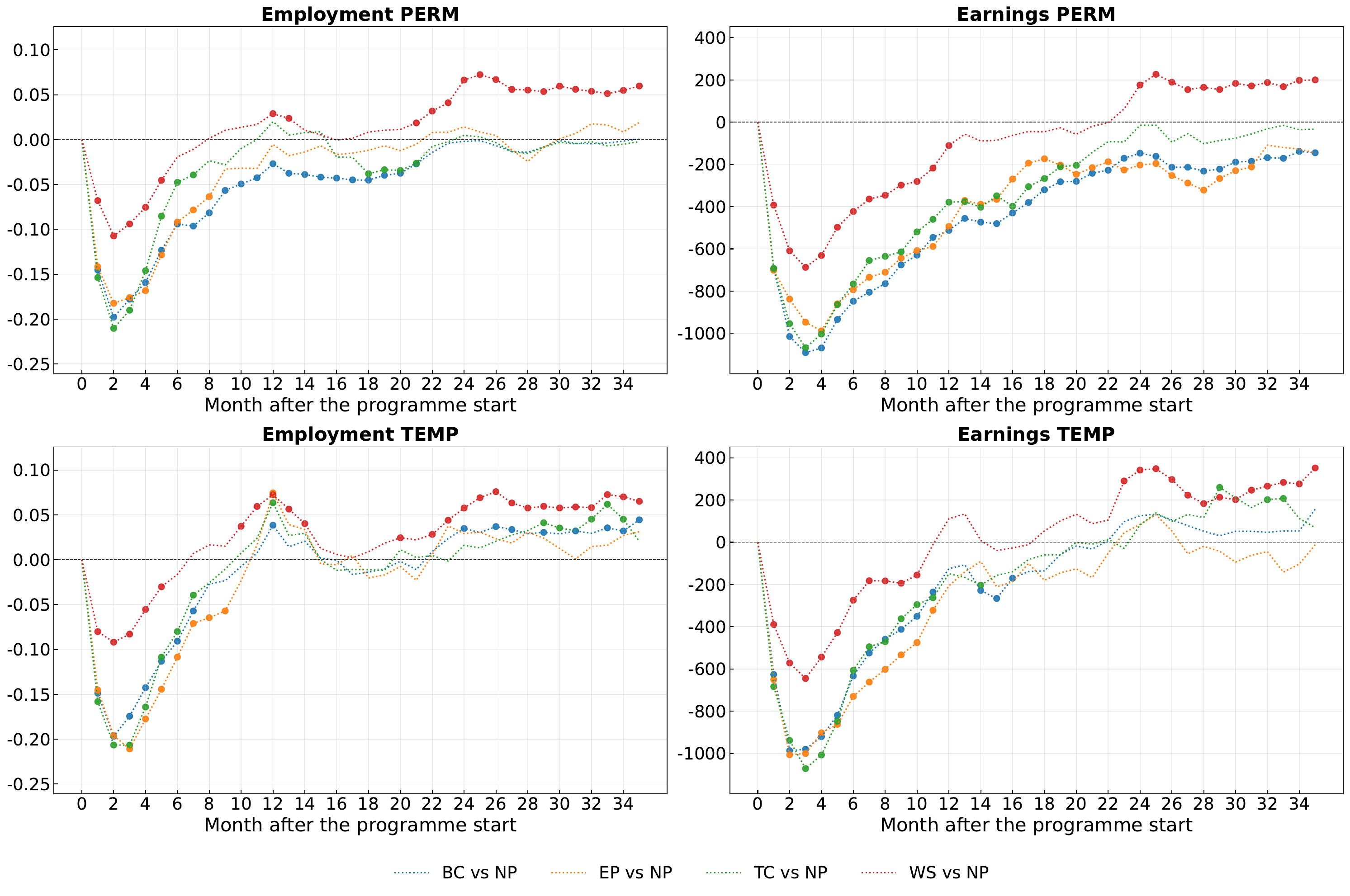}
\caption*{\footnotesize \textit{Notes:}  The vertical axis measures the ATE of participating in a programme relative to non-participation for the given outcome.
Employment is measured in months and earnings in CHF per month. The horizontal axis reports the number of months after the (pseudo) start of the programme. Dots indicate statistically significant point estimates (p-value $<$ 5\%).} 
\label{fig:ates_main_dynamics}
\end{figure}

The analysis focuses first at the aggregate level of all unemployed individuals, illustrating the programmes’ overall effectiveness and the dynamics of the effects over time. Figure \ref{fig:ates_main_dynamics} shows the evolution of the ATE over 36 months after programme start relative to individuals who do not participate in any programme for employment and earnings in both samples (PERM and TEMP). As expected, the ATE decreases in the first three months for all programmes due to the lock-in effect,\footnote{The average lock-in period might exceed the average programme duration (Table \ref{table:programme_combined}) due to individuals not finding a job directly after programme completion or participation in multiple programmes. The shares of individuals assigned to a second programme are 26\% for WS, 27\% for EP, 36\% for BC and 43\% for TC.} and starts to increase in the subsequent months. The lock-in effect arises as participants are likely to reduce their job search activities and receive fewer job offers during programme participation. It appears to be the smallest for WS, while the drop in the effect size and the subsequent recovery phase seems more pronounced for the other three programmes. This pattern holds for both samples and outcomes.

Table \ref{table:ate_both} summarises the outcomes aggregated over the entire 36-month period and specifically for the third year after the programme starts, as explained in Section \ref{out_cov}. Considering the three-year horizon in the PERM sample (columns (1) and (2)), the ATE is statistically significant and negative for all programmes except for WS, which shows a positive and statistically significant effect on the employment outcome at the 5\% significance level. Hence, none of the programmes provides sufficient earnings within three years to compensate for the lock-in effect of the programme. The fact that the effect for WS is significantly positive for employment but negative for earnings suggests this programme is effective at the extensive but not at the intensive margin over three years. The effect sizes are economically substantial. For example, participation in BC implies an average earnings loss of CHF 470 per month in the three post-treatment years. This corresponds to 10\% of median earnings in the month before the start of unemployment.

The picture changes when focusing solely on the third year (columns (3) and (4)). In this case, the ATE of WS is positive and statistically significant for both outcomes, with participation increasing employment duration by 0.74 months ($\approx$ 22 days) and earnings by CHF 2,059 over one year. BC still demonstrate a statistically significant negative effect on employment and earnings (0.13 months and CHF 2,347 less than non-participants). All other effects are close to zero and statistically insignificant. The results are confirmed in the upper two plots of Figure \ref{fig:ates_main_dynamics}. WS demonstrates statistically significant positive effects on employment and earnings starting from months 21 and 24, respectively. This suggests that engaging in the regular job market during WS benefits individuals, enhancing their prospects for reemployment, at least in the medium-term. The remaining three programmes fail to rebound from the lock-in drop, exhibiting effect sizes close to zero or negative until the end of the observed period.

\begin{table}[htpb]
    \begin{adjustbox}{width=0.9\columnwidth, center}
    \begin{threeparttable}
    \captionsetup{font=large}  
        \caption{Average Treatment Effects (ATE)}
        \centering
            \begin{tabular}{@{}lrrrrrrrr@{}}
        \toprule
        & \multicolumn{4}{c}{\textbf{Permanent residents (PERM)}} & \multicolumn{4}{c}{\textbf{Temporary residents (TEMP)}} \\
        \cmidrule(lr){2-5} \cmidrule(lr){6-9} 
        & \multicolumn{2}{c}{Months 1-36} & \multicolumn{2}{c}{Months 25-36} & \multicolumn{2}{c}{Months 1-36} & \multicolumn{2}{c}{Months 25-36} \\ \cmidrule(lr){2-3} \cmidrule(lr){4-5} \cmidrule(lr){6-7} \cmidrule(lr){8-9}
            &  \multicolumn{1}{c}{Employment} & \multicolumn{1}{c}{Earnings} & \multicolumn{1}{c}{Employment} & \multicolumn{1}{c}{Earnings} & \multicolumn{1}{c}{Employment} & \multicolumn{1}{c}{Earnings} & \multicolumn{1}{c}{Employment} & \multicolumn{1}{c}{Earnings} \\
            & \multicolumn{1}{c}{(1)} & \multicolumn{1}{c}{(2)} & \multicolumn{1}{c}{(3)} & \multicolumn{1}{c}{(4)} & \multicolumn{1}{c}{(5)} & \multicolumn{1}{c}{(6)} & \multicolumn{1}{c}{(7)} & \multicolumn{1}{c}{(8)} \\ \midrule
        WS & 
        0.27**\phantom{*(} & 
        -3,871***\phantom{(} &
        0.74***\phantom{(} &
        2,059***\phantom{(} &
        1.06***\phantom{(} &
        526\phantom{(***} &
        0.98***\phantom{(} &
        3,226***\phantom{(} \\
        &
        (0.14)\phantom{***}&
        (1,311)\phantom{***}&  
        (0.06)\phantom{***}  &
        (616)\phantom{***}  & 
        (0.24)\phantom{***}  & 
        (1,893)\phantom{***} & 
        (0.11)\phantom{***}  & 
        (823)\phantom{***} \\
        BC &  
        -2.45***\phantom{(} & 
        -16,968***\phantom{(} & 
        -0.13*\phantom{(**} & 
        -2,347***\phantom{(} & 
        -0.81***\phantom{(} &  
        -7,558***\phantom{(}  & 
        0.49***\phantom{(} & 
        904\phantom{(***}  \\
         &  
         (0.17)\phantom{***} & 
         (1,406)\phantom{***} & 
         (0.07)\phantom{***} & 
         (663)\phantom{***} &
         (0.30)\phantom{***} & 
         (2,189)\phantom{***} & 
         (0.14)\phantom{***} & 
         (968)\phantom{***}  \\
        TC & 
        -2.01***\phantom{(} &
        -13,936***\phantom{(} & 
        -0.01\phantom{(***} & 
        -1,131\phantom{(***} & 
        -0.92***\phantom{(} &  
        -6,673***\phantom{(} & 
        0.47***\phantom{(} & 
        1,737\phantom{(***}  \\
        & 
        (0.26)\phantom{***} & 
        (1,961)\phantom{***} & 
        (0.11)\phantom{***} & 
        (885)\phantom{***} & 
        (0.34)\phantom{***} &
        (2,490)\phantom{***} & 
        (0.15)\phantom{***} & 
        (1,089)\phantom{***} \\
        EP & 
        -1.45***\phantom{(} & 
        -12,849***\phantom{(} & 
        0.18\phantom{(***} & 
        -1,696\phantom{(***} & 
        -0.78*\phantom{(**} & 
        -9,539***\phantom{(} & 
        0.37*\phantom{(**} & 
        -284\phantom{(***} \\ 
        & 
        (0.28)\phantom{***} & 
        (2,246)\phantom{***} & 
        (0.12)\phantom{***}  & 
        (1,032)\phantom{***} & 
        (0.42)\phantom{***} & 
        (3,129)\phantom{***} & 
        (0.20)\phantom{***} & 
        (1,420)\phantom{***}  \\
        \bottomrule
        \end{tabular}
        \begin{tablenotes}
        \textit{Note:} ATEs of participation in different programmes compared to non-participation. WS: Wage Subsidy, BC: Basic Courses, TC: Training Courses, EP: Employment Programme. For each sample (PERM and TEMP), the two outcomes (employment indicator and earnings) are once aggregated over the full 36 months and once only over the months 25-36. Standard errors are in brackets. *, **, *** indicate the precision of the estimate by showing whether the p-value of a two-sided significance test is below 10\%, 5\%, and 1\% respectively.
        \end{tablenotes}
        \label{table:ate_both}
    \end{threeparttable}
    \end{adjustbox}
    \end{table}

In the TEMP sample, the general dynamics follow a similar pattern. The average effects over three years (columns (5) and (6) of Table \ref{table:ate_both}) are less negative compared to the PERM sample but mostly still negative and/or statistically insignificant. This implies the existence of a smaller but still persistent lock-in effect. Similarly to the PERM sample, only the employment effect of WS is statistically and economically significant and positive over the three years. Compared to non-participants, individuals in WS are employed for one month more during this period. In the third year (columns (7) and (8)), the ATEs for employment are positive and statistically significant for all programmes. Hence, individuals with temporary residence permits benefit from a broader range of programmes in the medium-term than individuals with permanent residency. Again, WS performs best among all programmes. In the third year, participants in this programme are employed for 0.98 months ($\approx$ 29 days) longer and earn 3,226 Swiss Francs more than non-participants.

Overall, the ATE results suggest that the WS programme performs best in the medium-term. However, for most programmes and outcomes, the lock-in effect is not compensated within three years. These results are qualitatively similar to previous findings for Switzerland, e.g., \citet{gerfin2002microeconometric}, \citet{lalive2008impact}, and \cite{knaus2022heterogeneous}, suggesting that programme efficiency has not improved over time. 

To understand the efficacy of the observed programme allocation, ATE is compared with ATET. The ATET represents the average treatment effects for the subpopulation participating in each programme. When allocation to programmes is randomised, the two causal estimands are expected to be the same. However, in observational studies and under the presence of heterogenous effects, this may not be the case and the difference between the two parameters can be informative about the effectiveness of the observed treatment assignment. For instance, an ATET that is larger than the ATE indicates that the assignment of the observed programme is better than random. For most of the programmes analysed, the estimates show slightly, though not significantly, larger ATETs than the ATEs. This indicates that there might be room for improvement in the caseworkers' allocations. Detailed results are available in Section \ref{ATET_app} of the Appendix.

\subsection{Effect Heterogeneity at the Group Level} \label{BGATE}

After providing an overview of the effects for the entire population, the analysis explores how these effects vary among different population subgroups while accounting for variation in other covariates. Adopting a policymaker's perspective, a set of relevant variables is identified following the existing literature. For instance, subpopulations identified by gender and education level are among the most studied \citep{crepon2013labor}. This study focuses on gender due to documented gender gaps in the effectiveness of ALMPs. Among socio-demographic characteristics, differences by the place of origin \citep[e.g.~as in][]{knaus2021machine, cockx2023priority} as well as the labour market history of each individual \citep[e.g.~as in][]{altmann2018learning, cockx2023priority} are analysed. In particular, employment duration and stability are captured by the number of months employed in the five years before becoming unemployed. Job quality is measured through mean earnings in the last three months. In addition, the sector of the last job \citep*[e.g.~as in][]{caliendo2017unobservable} and the unemployment rate of the local labour market \citep[e.g.~as in][]{10.1093/jeea/jvx028,         cockx2023priority} are considered.\footnote{The results for the unemployment rate are not shown in the paper. The reason is that the variable is discretised in the estimation procedure which yields results that lack meaningful interpretation. However, it will serve as a balancing variable.} 
Heterogeneous effects are estimated for all the mentioned variables in both samples (PERM and TEMP) for four outcome variables (earnings and months employed in the first and third year after the programme's start), using different effect measures (GATEs, BGATEs and their differences with ATE). The main findings regarding earnings in the third year for the BGATE-ATE are reported here, while detailed tables with additional results can be found in Appendix \ref{gates appendix}.   

As described above, the BGATE estimates the heterogeneity of effects within subgroups identified by a policy variable. Compared to a GATE, it balances a set of pre-determined variables, which may confound effect heterogeneity. In this application, the balancing variables are chosen as the set of remaining pre-determined policy variables plus the three additional variables age, marital status, and canton of residence.\footnote{In detail, seven policy variables are collected in set $\mathcal{Z} = \{$sector, gender, education, origin, past employment, past earnings, unemployment rate$\}$. Then for each policy variable $Z_j \in \mathcal{Z}$ the set of balancing variables is $\mathcal{W}_j = \{(\mathcal{Z} \setminus Z_j) \cup \{$age, marital status, canton$\} \}$. All variables are measured prior to the unemployment spell.} For example, if significant effect heterogeneity by gender is detected using GATEs, it raises the question of whether this variation stems from gender or correlated factors (e.g., education, marital status). However, if the BGATE also shows notable variation, indicating differing effects by gender even among individuals with identical balancing variables (i.e.~same education level, same marital status etc.), it provides more compelling evidence of gender-driven differences.

\begin{figure}[t]
\captionsetup{font=small}  
\caption{Deviations of BGATEs from ATE for the cumulative earnings in the third year comparing WS vs. Non-Participation}

\includegraphics[width=\textwidth]{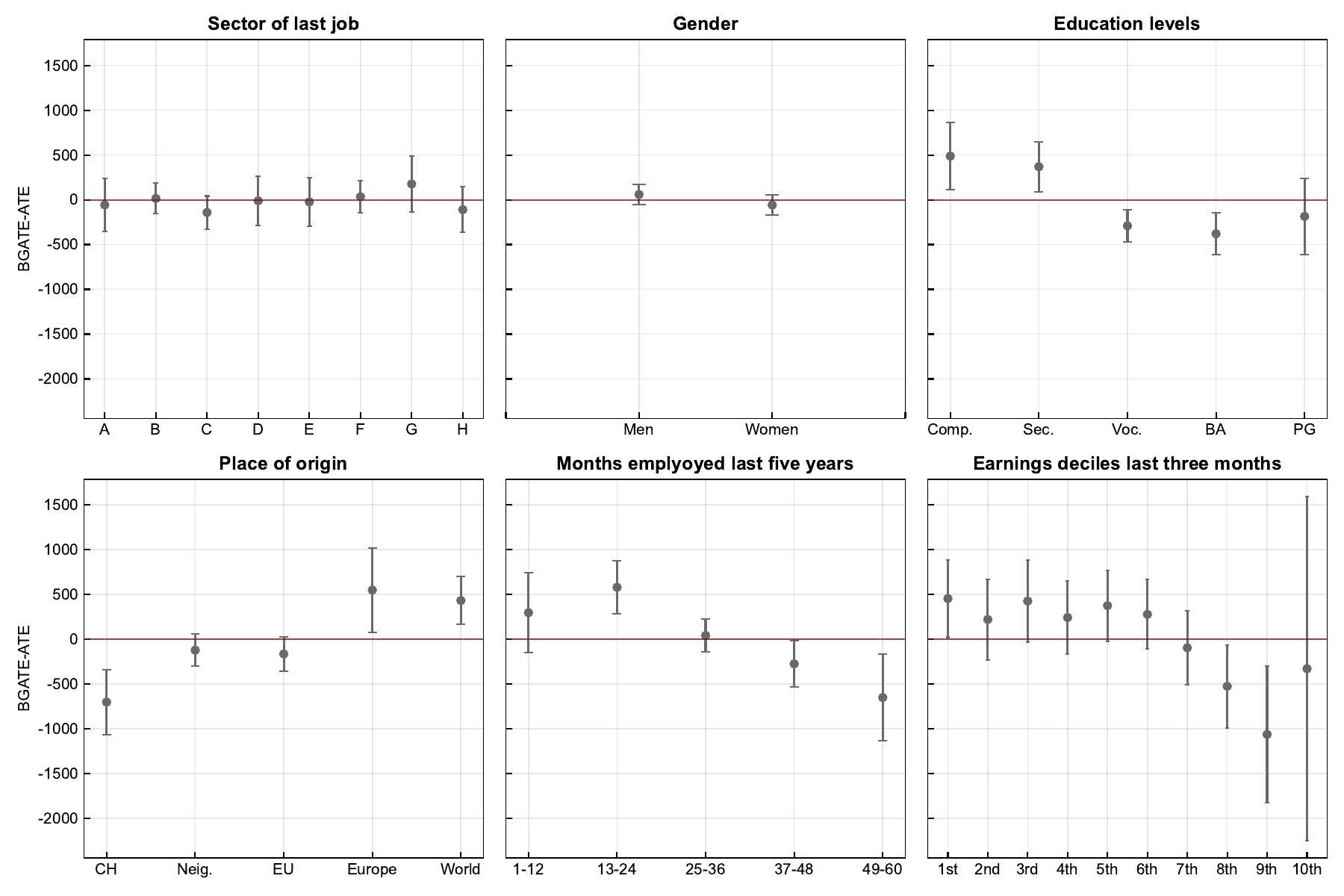}

\caption*{\textit{Note:} The vertical axes measure the difference of BGATEs to the ATE of WS programme with respect to non-participation at the 95\% confidence interval. The horizontal axes show previous job sector (categorised as A: agricultural and forestry, B: production in industry and trade, C: technical and information technology, D: construction and mining; E: transport and trade, F: hospitality and personal services, G: business, finance and law, H: education, art and science), gender, education levels (compulsory, secondary, vocational, bachelor's degree, post-graduate), place of origin (Switzerland, neighbouring countries, rest of European Union, rest of Europe and rest of the world), sum of months employed in the last five years and the average earnings within each decile of the earnings distribution in the three months before the unemployment spell.  The outcome is the cumulative earnings in the third year after the programme start and the ATE for the WS is 2,059 Swiss Francs. PERM Sample.}
\label{fig:bgates_ates}
\end{figure}

Given that WS is the largest programme, it will be the primary focus of the remainder of this section for the sake of brevity. Other programmes will be discussed only if their findings deviate.  Figure \ref{fig:bgates_ates} illustrates the differences between BGATEs and ATEs in terms of cumulative earnings in the third year following the programme's start, offering insights into which groups perform better or worse than the average in the PERM sample. This outcome is more granular than the cumulative months in employment, and the identified heterogeneity patterns manifest more distinctly for it. However, similar patterns persist consistently across both outcome variables.

 
The results indicate that subgroups of the population with compulsory or secondary education benefit more from being assigned to a programme than individuals with a higher education level.\footnote{It is worth noting that about 62\% of the Swiss and 82\% of non-Swiss citizens in the PERM sample do not hold a degree beyond secondary school levels.} However, for individuals with the highest level of education (postgraduate degree), the results are not statistically different from the ATE due to the small sample size in this group. The programmes' effectiveness also differs according to the individuals' place of origin. Individuals coming from outside of the European Union or Switzerland benefit more from being assigned to a programme. In contrast, Swiss individuals profit the least from participating in the WS programme. The fact that individuals from neighbouring countries, where a Swiss official language is spoken, profit less than other foreigners suggests that language proficiency may influence programme effectiveness. Concerning the sector of the last job and gender, there is no heterogeneity. Furthermore, the analysis shows that the WS programme is the least beneficial for individuals at the upper end of the earnings distribution and individuals who were nearly always employed during the five years before becoming unemployed. These results are comparable to the GATE-ATE results. Hence, balancing for covariates does not substantially change the patterns of the effects but only the effect size. 

Despite some exceptions, the other programmes generally exhibit the same heterogeneity patterns. In the BC programme, the heterogeneous treatment effects by earnings decile are more pronounced, while in the TC programme, there are no significant heterogeneities regarding education and earnings. Lastly, the EP programme harms men more than women, and there is almost no variation in treatment effects for education level (after balancing for other covariates). Similarly, the heterogeneous treatment effects by earnings are more pronounced in the EP programme.

The results for the same sub-groups in the TEMP sample are mostly statistically insignificant due to the smaller sample size. However, it is noteworthy that in this sample, individuals with higher levels of education benefit more from WS, while no clear patterns appear in the other programmes. The complete results are in Sections \ref{gates_app_tempw} and \ref{bgates_app_perm} in the Appendix.

The results are comparable to the results of \cite{cockx2023priority}, who also find that immigrants with lower knowledge of the local language benefit more from ALMPs. Notably, the analysis in this paper treats the population of migrants holding a temporary residence permit (TEMP) as a distinct sample. However, the PERM sample includes immigrants with settlement permits, who have a more extensive domestic job-market history and exhibit characteristics similar to Swiss citizens. In addition, unlike \citet{cockx2023priority}, this analysis balances on additional covariates in the heterogeneity assessment. The fact that heterogeneity for dimensions like education is found, even considering a more homogeneous population, may suggest that skill-based characteristics drive the heterogeneity. However, heterogeneity by origin also does not vanish by balancing other covariates, such as education. Hence, even if individuals from outside the European Union and Switzerland have the same education, they profit more from the programmes than, for example, Swiss individuals. Detailed results for the remaining programmes are provided in Section \ref{gates appendix} and \ref{bgates appendix} in the Appendix.

\subsection{Effect Heterogeneity at the Individual Level}

\begin{figure}[H]
\captionsetup{font=small}  
\caption{Sorted difference of Individualised Average Treatment Effects minus Average Treatment Effect}
	\begin{minipage}[t]{1\textwidth}
 \includegraphics[width=\textwidth]{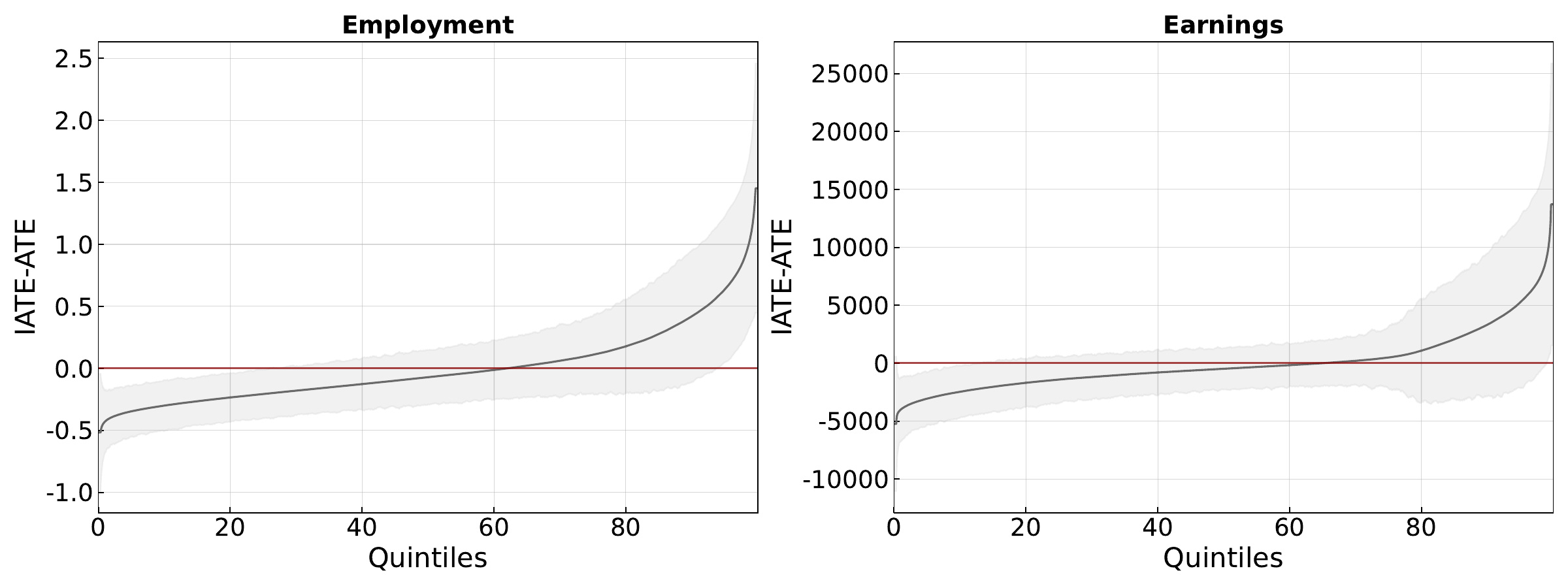}
	\end{minipage}
\caption*{\textit{Note:} The figure shows the sorted IATEs for the Wage Subsidy programme in the PERM sample. Column (1) shows the sum of months employed, and column (2) the sum of earnings over the third year after the programme starts.}
\label{fig:iates_maintxt}
\end{figure}

Finally, the most granular heterogeneity level, the IATEs, is analysed which allows to understand how each programme influences a particular individual on average given its observed characteristics. Figure \ref{fig:iates_maintxt} illustrates the differences between the IATEs and the ATE for the Temporary Wage Subsidies by the quantiles of the IATE in the PERM sample. The effects range from -0.5 to 1 month of employment and CHF -5,000 to CHF 10,000 of earnings compared to the ATE in the third year after the programme starts. However, the relatively large 95\% confidence bands indicate that it is much more challenging to get precise estimates for the granular IATEs than for the aggregated GATEs and ATEs. The patterns shown for WS hold similarly for the other programmes and the TEMP sample, as indicated in Appendix \ref{appendix:IATEs}. Figure \ref{fig:iates_densities} presents the densities of the IATEs by programme compared to non-participation in the third year after the programme started. For all the groups, there are individuals with positive and negative IATEs, indicating that some but not all individuals benefit from participation in the respective programmes.

\begin{figure}[H]
\captionsetup{font=small}  
\caption{Densities of Individualised Average Treatment Effects by programme type.}
\centering
\includegraphics[width=\textwidth]{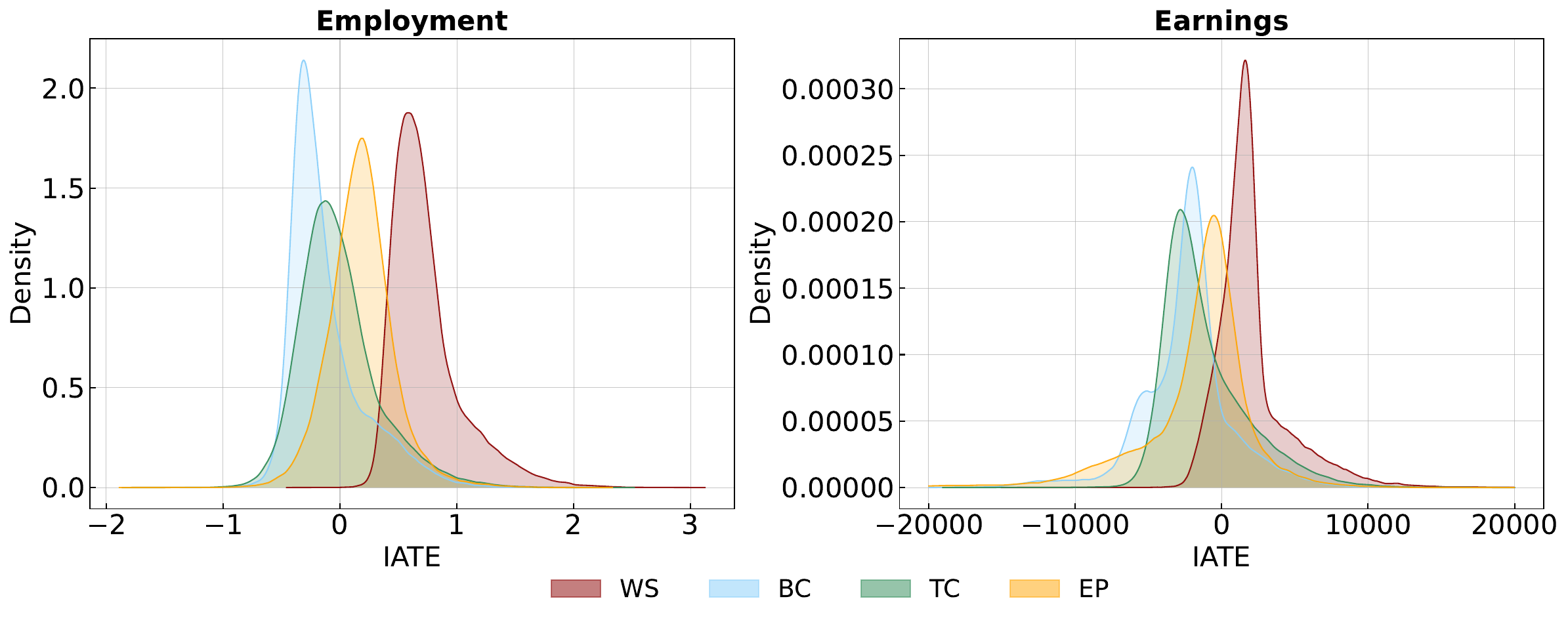}
\caption*{\textit{Notes:} The figure shows densities of IATEs of Wage subsidies (WS), Basic Courses (BC), Training Courses (TC) and Employment programmes (EP) relative to non-participation in the PERM sample. The first (second) plot shows effects on the sum of months employed (earnings) in the third year after programme start. The average standard error of the IATES for the first (second) plot is 0.15 (1,523) for WS, 0.19 (1,701) for BC, 0.27 (2,281) for TC and 0.28 (2,565) for EP.}
\label{fig:iates_densities}
\end{figure}

To enhance understanding of the characteristics of individuals benefiting from the programmes, we employ $k$-means++ clustering \citep{arthur2007k} to group individuals based on the size of their IATEs. $k$-means clustering is an unsupervised machine learning method that partitions data into $k$ distinct clusters by minimizing the variance within each cluster. The number of clusters is determined in a data-driven manner, constrained by a requirement that each cluster contains no less than 1\% of all observations. We refer to Appendix \ref{clustering_appendix} for further details on the clustering approach.

Table \ref{table:kmenas} reports covariate means of the groups obtained through clustering based on the IATEs of cumulative earnings three years after programme start. For each programme, the groups benefiting least and most are shown. In the PERM sample, the mean IATEs are negative for the least benefiting groups, while for the most benefiting clusters, they are positive for all programmes except BC. However, a similar cluster analysis on employment outcomes for the same period reveals positive IATEs also for BC, indicating that some individuals benefit from the programme, at least on the extensive margin. Table \ref{table:kmenas_appendix} in the Appendix provides further results for both samples and outcomes.

\begin{table}[!ht]
\begin{adjustbox}{width=0.95\columnwidth, center}
\begin{threeparttable}
    \caption{IATEs with respect to non-participation and descriptive statistics for groups that benefit the least and most from each programme}
    \centering
    \begin{tabular}{@{}lcccc@{}}
    \toprule
    Programme & \textbf{WS} & \textbf{BC} & \textbf{TC} & \textbf{EP} \\ 
    \midrule
    \multicolumn{1}{l}{ } & \multicolumn{4}{c}{\textbf{Permanent Residents}} \\
    Mean IATE of least and most benefiting groups & -3,252 – 9,685\phantom{-} & -9,930 – -255\phantom{,1} & -5,047 – 5,881\phantom{-} & -17,360 – 368\phantom{-17,} \\
    \midrule
    \multicolumn{1}{l}{Selected covariates} & \multicolumn{4}{c}{Values for the last and most benefiting groups by programme} \\
    \midrule
    & Least – Most &   Least – Most &  Least – Most &  Least – Most \\
    \midrule
    Age & 37 – 47 & 51 – 38 & 49 – 38 & 50 – 36 \\
    Female & 0.22 – 0.47 & 0.22 – 0.55 & 0.30 – 0.36 & 0.13 – 0.64 \\
    Education (1: compulsory – 5: post-grad.) & 3.36 – 4.31 & 3.34 – 1.99 & 2.22 – 4.19 & 3.85 – 2.23 \\
    Local language knowledge (1: basic – 7: mother tongue) & 6.64 – 5.87 & 6.60 – 5.34 & 5.90 – 6.07 & 6.57 – 5.43 \\
    Citizenship Swiss & 0.83 – 0.51 & 0.90 – 0.48 & 0.70 – 0.54 & 0.80 – 0.47 \\
    Neighbouring countries & 0.08 – 0.25 & 0.09 – 0.12 & 0.12 – 0.26 & 0.12 – 0.13 \\
    Rest of European Union & 0.04 – 0.09 & 0.02 – 0.14 & 0.09 – 0.09 & 0.03 – 0.15 \\
    Rest of Europe & 0.01 – 0.01 & 0.00 – 0.09 & 0.04 – 0.01 & 0.00 – 0.08 \\
    Rest of the world & 0.03 – 0.14 & 0.02 – 0.17 & 0.05 – 0.11 & 0.05 – 0.16 \\
    Months in employment 10 years before & 105 – 104 & 113 – 78\phantom{1} & 109 – 96\phantom{1} & 113 – 80\phantom{1} \\
    Cumulative earnings 10 years before & \phantom{1,}690,963 – 1,631,465 & 1,304,895 – 290,765\phantom{,1} & \phantom{1,}813,079 – 1,084,605 & 1,821,614 – 279,768\phantom{1,} \\
    \midrule
    \% of obs. in cluster & 2 – 2 & \phantom{1}2 – 16 & 6 – 2 & \phantom{1}1 – 18 \\
    \midrule
     \multicolumn{1}{l}{ } & \multicolumn{4}{c}{\textbf{Temporary Residents}} \\
    Mean IATE of least and most benefiting groups  & \phantom{4}1,139 – 14,529 & -5,494 – 2,501\phantom{-} & –2,463 – 7,393\phantom{-} & –7,562 – 3,260\phantom{-}\\
    \midrule
    \multicolumn{1}{l}{Selected covariates} & \multicolumn{4}{c}{Values for the last and most benefiting groups by programme} \\
    & Least – Most &   Least – Most &  Least – Most &  Least – Most \\
    \midrule
    Age &  36 – 41  & 38 – 35 & 36 – 38 & 39 – 35  \\
    Female  & 0.36 – 0.27 & 0.34 – 0.31  & 0.48 – 0.30 & 0.31  – 0.36 \\
    Education (1: compulsory – 5: post–grad.) & 1.95 – 4.13 & 3.66 – 2.55 & 2.11	– 3.36  & 3.72 – 2.98 \\
    Local language knowledge (1: basic – 7: mother tongue)  & 3.89 – 5.00 & 4.22 – 4.58 & 3.84 – 5.22 & 5.11	– 4.20\\
    Neighbouring countries  & 0.30 – 0.54 & 0.48 – 0.43  & 0.31– 0.59
 & 0.63 – 0.31 \\
    Rest of European Union  & 0.41 – 0.22  & 0.23 – 0.31 & 0.38	– 0.22 & 0.18	– 0.38 \\
    Rest of Europe  & 0.09 – 0.01  & 0.03 – 0.08 &  0.08 – 0.02
 & 0.02 – 0.07 \\
    Rest of the World  & 0.20 – 0.24  & 0.26 – 0.17 & 0.22 – 0.17 & 0.17 – 0.24 \\
    Months in employment 10 years before  & 45 – 62  & 41 – 50 & 37 – 59 & 49 – 49\\
    Cumulative earnings 10 years before & 178,411 – 917,883 & 465,075 – 270,689  & 141,518 – 570,942\phantom{1,} & 537,176 – 204,995\\
     \midrule
      \% of obs. in cluster  & 49 – 2\phantom{4} & \phantom{4}2 – 11 & 12 – 4\phantom{4} &  3 – 5\\
      \bottomrule
    \end{tabular}
\begin{tablenotes}
\textit{Note:} The outcome variable is the cumulative earnings after three years from the program start. Clusters are formed using the IATEs vs the NP group, and the cluster's mean is reported. Clusters are based on the $k$-means++ algorithm and computed by the mcf package. WS: Wage Subsidy, BC: Basic Courses, TC: Training Courses, EP: Employment Programme. The percentage of observations in each cluster varies by programme and the number of clusters formed. 
\end{tablenotes}
\label{table:kmenas}
\end{threeparttable}
\end{adjustbox}
\end{table}

The results of the cluster analysis for the four programmes highlight distinct demographic and socioeconomic profiles associated with the most benefiting groups. For the WS programme, the cluster that derives the greatest benefit predominantly comprises older, non-Swiss nationals and individuals with relatively high earnings. For BC and EP, young, female participants with low education levels, non-Swiss nationality, and lower earnings benefit most. TC appears to be most beneficial for young individuals with higher levels of education. For this programme, those benefiting the most also exhibit higher average earnings than the least-benefiting group, even though their months in employment are lower. Non-Swiss citizens generally benefit more than Swiss citizens across all programmes, confirming the patterns observed in the GATEs.

The TEMP sample reveals a slightly different picture. While women in the PERM sample benefit more than men across all programmes, this trend is reversed in the TEMP sample for the WS, BC, and TC programmes. Additionally, in the TEMP sample, individuals with a larger number of months of previous employment benefit more from these programmes, whereas in the PERM sample, those with fewer months of previous employment were more likely to benefit. Overall, in the TEMP sample, individuals with higher levels of education and coming from the neighbouring countries, gain more from WS and TC, while the reverse holds true for BC and EP.

\subsection{Simulated Allocations of Unemployed to Programmes} \label{optimal_policy} 

The overall heterogeneity analysis indicates that specific sub-groups within the population could benefit from ALMPs. Consequently, the existing caseworker assignment mechanism is evaluated to determine whether it effectively uses the available information. Potential improvements in the allocation to ALMPs are evaluated by simulating programme assignments using different methods, such as random, best-score and policy tree allocations. Policy trees provide intuitive and interpretable decision rules, which can be preferable for public policies due to their transparency. In contrast, the best-score method assigns each individual to the programme with the highest estimated potential outcome. While this procedure is expected to yield the largest improvement (at least in the training sample), it lacks interpretability, making it hard to justify its use in practice.

Table \ref{table:policy_tree} shows programme allocations of different policies for the PERM and the TEMP samples. The out-of-sample performance is measured as the average number of months employed in the third year after programme initiation under the allocation of the respective policy rule. This performance measure aligns with the primary objective of Swiss ALMPs, which is to enhance long-term stable employment.\footnote{\hyperlink{https://www.seco.admin.ch/seco/de/home/Arbeit/Arbeitslosenversicherung/oeffentliche-arbeitsvermittlung/arbeitsmarktliche-massnahmen.html}{State Secretariat for Economic Affairs, Labour Market Measures} (retrieved 29.02.2024)}

In the following, unconstrained and constrained policy trees are estimated using the same set of policy-relevant covariates employed in the analysis of effect heterogeneity (Section \ref{BGATE}) as potential splitting variables.\footnote{In the notation of Section \ref{econometrics}, the policy variables $V$ contain all policy variables from $\mathcal{Z} = \{$sector, gender, education, origin, past employment, past earnings, unemployment rate$\}$}
Trees with depth-2 (4 leaves), depth-3 (8 final leaves) and depth-4 (16 final leaves) are implemented. Additionally, sequential deeper trees of depth-4+1 (32 final leaves) and depth-4+2 (64 final leaves) and depth-4+3 (128 final leaves) are estimated. The sequential trees first build an optimal tree with a depth of 4, followed by a second tree of depth 1, 2 and 3, respectively, within the strata obtained by the leaves of the first tree. Although the combined, deeper trees are no longer optimal, their performance in this application remains comparable to that of the shallower optimal policy trees, while providing more granular decision rules at reduced computational time. For the constrained policy trees, the observed participation shares are applied as maximum capacity constraints to all programmes except the control group (NP), which remains unconstrained. To reduce the risk of overfitting, out-of-sample predictions are computed. This approach helps to reduce the ``optimistic bias'' associated with in-sample predictions.

\begin{table}[ht]
\begin{adjustbox}{width=0.95\columnwidth, center}
\begin{threeparttable}
\captionsetup{font=large}  
    \caption{Out-of-sample treatment allocations per policy}
    \centering
        \begin{tabular}{@{}lcrrrrrcrrrrr@{}}
    \toprule
    & \multicolumn{6}{c}{\textbf{Permanent Residents}} & \multicolumn{5}{c}{\textbf{Temporary Residents}} \\
    \cmidrule(r){2-7} \cmidrule(l){8-13}
    & \multicolumn{1}{c}{\textbf{\textbf{Emp.}}} & \multicolumn{5}{c}{\textbf{Shares per programme}} & \multicolumn{1}{c}{\textbf{\textbf{Emp.}}}  & \multicolumn{5}{c}{\textbf{Shares per programme}} \\
    \cmidrule(r){3-7} \cmidrule(l){9-13}
    \textbf{Policy}  & \multicolumn{1}{c}{\textbf{\textbf{Months}}} & NP & WS & BC & TC & EP & \multicolumn{1}{c}{\textbf{\textbf{Months}}} & NP & WS & BC & TC & EP  \\
    \midrule
    Observed  & 9.04 & 42.81 & 26.20 & 19.34 & 6.38 & 5.28 & 7.99 & 39.80 & 27.98 &  15.33 &  10.86 &   6.03\\ 
    Random (Observed shares) &  9.02 & 43.13 & 26.39 & 18.79 & 6.25 & 5.45 & 7.99 & 38.74 &  28.54 &  15.46 &  10.79 & 6.46 \\
    \midrule
    Best-score & 9.52 & 0.00 & 99.7 & 0.00 & 0.10 & 0.16 & 8.59 & 0.00 & 96.13 & 0.07 & 0.00 & 3.81\\
    Best-score (Constrained) & 9.15 & 52.94 & 28.01 & 2.04 & 11.01 & 6.01 & 8.08 & 39.3 & 28.01 & 15.03 & 10.99 & 6.03 \\
    \midrule
    Shallow Tree & & & & & & &&&&&\\
    ~~~~Depth-2 & 9.52 & 0.00 & 100.00 & 0.00 & 0.00 & 0.00 &  8.59  & 0.00 & 100.00 &  0.00 &  0.00 & 0.00\\
    ~~~~Depth-3 & 9.52 & 0.00 & 100.00 & 0.00 & 0.00 & 0.00 & 8.59 & 0.00 & 100.00 & 0.00 & 0.00 & 0.00\\ \midrule
    Shallow Tree (Constrained) & & & & & & &&&&&  \\
    ~~~~Depth-2  & 9.06  &  69.60 & 23.85 & 0.00 &  6.55 & 0.00  & 7.92 & 72.48 &  27.52 &   0.00 & 0.00 &  0.00\\
    ~~~~Depth-3  & 9.07  & 67.28 &  23.12 &  0.00 & 4.78 & 4.82  & 7.95 & 69.01 &  29.17 &   1.82 & 0.00 &  0.00\\
    ~~~~Depth-4  & 9.06  & 64.67 & 21.92 &  2.08 &  7.11 &  4.22 &  7.92  &   70.99 &  25.99 &  3.01 &   0.00 &  0.00\\
    \midrule
    Sequential Deeper Trees (Constrained) & & & & & & &&&&&\\
    ~~~~Depth-4+1 & 9.06  & 63.19 &  20.39 &   3.18 &  9.43 & 3.81 & 7.95 &  66.16 &  27.45 &   4.74 &   1.66 &   0.00 \\
    ~~~~Depth-4+2 & 9.08 & 62.30 &   25.41 &  1.66 &  4.22 &   6.41 & 7.97 & 61.52 &  27.78 &   9.57 &   1.13 &  0.00\\
   ~~~~Depth-4+3  & 9.09 & 63.79 &	21.26 &	2.18 & 7.54	& 5.23 & 7.96 & 62.95 &	28.05 & 9.01 & 0.00 & 0.00 \\
    \midrule
    Additional Scenarios with Depth-3 Tree & & & & & & &&&&&\\
    ~~~~Double share of WS & 9.24 &  49.07 &  50.93 & 0.00 &  0.00 & 0.00 & 8.22 & 42.09 &  57.91 & 0.00 & 0.00 &  0.00\\
    ~~~~Triple share of WS & 9.39 &  23.17 &  74.62 & 0.00 &  0.00 & 2.22 & 8.46 & 12.98 &  82.72 & 4.30 &  0.00 &  0.00\\
    \bottomrule
    \end{tabular}
    \begin{tablenotes}
    \textit{Note:} 
    The table displays the simulated and observed allocations of individuals per program under different policies. The performance measure is the average sum of months in employment in the third year after the program start. In the constrained simulations, the programme shares are occasionally higher than the observed shares.
This occurs because the figures shown are based on the validation sample, whereas all constraints are met in-sample. 
    \end{tablenotes}
    \label{table:policy_tree}
\end{threeparttable}
\end{adjustbox}
\end{table}

The results indicate that observed caseworker assignments marginally outperform random allocations\footnote{The random allocations respect the observed programme shares.} regarding the number of months in employment for the PERM sample and perform similarly to random allocations for the TEMP sample. For both samples, the unconstrained best-score algorithm and the policy trees achieve higher average months in employment by predominantly allocating individuals to the WS programme. The improvement compared to the observed allocations is considerable, with an average of approximately half a month of additional employment.
In the case of shallow constrained trees for the PERM sample, the average months in employment decrease compared to unconstrained allocations because fewer individuals can be assigned to the WS programme. There is a marginal improvement over the observed allocations. For instance, the constrained depth-3 tree results in a gain of 0.03 months (less than one day) compared to the observed allocation. Thus, given the observed treatment shares, there is minimal room for improvement. For the TEMP sample, the constrained shallow trees result in even lower average months in employment compared to the observed allocation. Finally, also the constrained sequential deeper trees lead to only marginal improvements in terms of average months in employment. However, they do result in a shift in the allocation of some individuals to programmes other than WS. Specifically, there are increased assignments to TC and EP for the PERM sample and BC for the TEMP sample. Illustrations of the policy trees and the corresponding assignment rules can be found in Appendix \ref{policy_tree_appendix}.

Based on the results indicating that the WS programme outperforms the others on average, two additional scenarios using policy trees of depth-3 are implemented for both samples. In the first scenario, the policy tree is allowed to allocate twice the observed share of individuals to the WS programme. In the second scenario, the policy tree is allowed to allocate three times the observed share. The results indicate that increasing the maximum share of individuals that can be assigned to WS increases the average months in employment. For example, doubling the share of assignments to WS from 25\% to 50\% in the PERM sample leads to an additional 0.2 months (6 days) in employment, on average. The simulations under the two scenarios indicate that increasing the allocation of individuals to WS predominantly reassigns participants from other programs rather than incorporating previously non-participating individuals. The increase in the average months employed under the scenarios is even larger for the TEMP than for the PERM sample. Using the best-score method, the WS program demonstrates clear superiority for 99.7\% of individuals in the PERM sample and 96.1\% in the TEMP sample. 

Based on these results, it is recommended to expand the assignment to the WS programme while reducing allocations to other programmes. To avoid market distortions when expanding the WS programme, a gradual approach is advisable, accompanied by continuous monitoring to ensure that location- and occupation-specific reference salaries remain adequate and that subsidised employment retains its temporary nature. Alternatively, if budget constraints or other considerations by policy-makers necessitate the maintenance of programmes other than WS, using deeper decision trees can still help optimise the allocations. Even though these deeper trees do not reach the same level of employment as in scenarios where WS shares could be increased, they still indicate an improvement with respect to the observed allocations, at least in the PERM sample.

\section{Sensitivity Analysis} \label{sensitivity_analysis}

The aforementioned results indicate that Temporary Wage Subsidies stand out as the sole effective programme, contrasting with the other programmes that exhibit no or even adverse average effects. 
Several sensitivity analyses are performed to address potential concerns surrounding these findings. Firstly, further evidence is provided that the identification strategy is credible and the results are unlikely to be plagued by selection bias. Secondly, it is demonstrated that the results are robust with respect to different definitions of the control group. More specifically, it is demonstrated that the construction of the control group successfully eliminates individuals who are likely to find a job before being treated, hence, mitigating downward biases in the results. Finally, it is shown that the results of the simulated allocations are not sensitive to the use of additional policy variables.

A placebo study is conducted to estimate the ATEs of participation in future programmes within a previous unemployment spell. It can be expected that future participation will not affect the lagged pre-treatment outcomes, further supporting the CIA.\footnote{Following \citet{imbens2009recent}, if instead, an effect of such a placebo treatment is found, it would imply that the distributions of the potential outcomes $Y^0$ are not comparable between the respective treatment groups, raising doubts about the plausibility of the CIA.} The design of the placebo mainly follows \cite{cockx2023priority}. In particular, from all the individuals in the main analysis, only those who already became unemployed between 2010 and 2012 are kept and the effects of the treatment status assigned in the main analysis are estimated. Differently from \cite{cockx2023priority} who drop all individuals treated at the time of the placebo, this approach retains them except those assigned to the same programme at the time of the analysis. This choice allows to have a slightly larger sample size (14,662 individuals) and to avoid estimating the effect of a treatment eventually assigned. The outcome of interest is the sum of months in employment one year after the (pseudo-) programme start observed for the main treatment. This analysis focuses on the first year to avoid overlap with the outcomes examined in the main analysis. Table \ref{table:placebo2} reports the ATEs and their standard errors, showing that all estimates are close to zero and statistically insignificant. In the main analysis, the ATEs over the same period are statistically significant and show a larger magnitude.

\begin{table}[ht!]
\begin{adjustbox}{width=0.7\columnwidth, center}
\begin{threeparttable}
\captionsetup{font=large}  
    \caption{Average Treatment Effects (ATE) on 12 cumulative months employed}
    \centering
        \begin{tabular}{@{}rrrrrrrr@{}}
    \toprule
    \multicolumn{4}{c}{\textbf{Placebo sample}} & \multicolumn{4}{c}{\textbf{Permanent Residents}} \\
    \cmidrule(lr){1-4} \cmidrule(lr){5-8} 
        \multicolumn{1}{c}{WS} & \multicolumn{1}{c}{BC} & \multicolumn{1}{c}{TC} & \multicolumn{1}{c}{EP} & \multicolumn{1}{c}{WS} & \multicolumn{1}{c}{BC} & \multicolumn{1}{c}{TC} & \multicolumn{1}{c}{EP} \\ \midrule
    -0.01\phantom{***(} & 
    0.15\phantom{***(} &  
    -0.17\phantom{***(} & 
    0.27\phantom{***(} & 
    -0.64***\phantom{(} & 
    -1.68***\phantom{(} & 
    -1.56***\phantom{(} & 
    -1.42***\phantom{(} \\
    (0.12)\phantom{***}  & 
    (0.18)\phantom{***}   & 
    (0.26)\phantom{***}   & 
    (0.20)\phantom{***}   & 
    (0.05)\phantom{***} & 
    (0.06)\phantom{***} & 
    (0.10)\phantom{***} & 
    (0.12)\phantom{***}\\
    \bottomrule
    \end{tabular}
    \begin{tablenotes}
    \textit{Note:} ATEs of participation in different programmes compared to non-participation for the placebo sample and the Permanent Residents sample. WS: Wage subsidy, BC: Basic Courses, TC: Training Courses, EP: Employment Programme. Standard errors are presented in parentheses. The symbols *, ** and *** indicate the level of precision for each estimate, denoting if the p-value of a two-sided significance test is below 10\%, 5\%, and 1\%, respectively.
    \end{tablenotes}
    \label{table:placebo2}
\end{threeparttable}
\end{adjustbox}
\end{table}

As an additional sensitivity check, the ATE for employment duration is estimated in the third year following the start of the unemployment spell, rather than the start of the programme. This alternative outcome measure does not require estimating pseudo-start dates, as it is observed for all individuals, including non-participants. In the main analysis, non-participants who were no longer unemployed at the time of their predicted pseudo-programme start date were eliminated. Without this correction, it is expected to observe smaller average treatment effects since the control group contains more individuals who find a job before being treated. This hypothesis is confirmed in Table \ref{table:robu_nodates}, showing ATEs for both samples (PERM and TEMP) with the alternative outcome. As expected, all the programmes demonstrate lower average treatment effects than those estimated using the pseudo dates (Table \ref{table:ate_both}). Thus, despite being somewhat more negative, these results are qualitatively in line with the main results.

\begin{table}[ht]
\begin{adjustbox}{width=0.7\columnwidth, center}
\begin{threeparttable}
\captionsetup{font=large}  
    \caption{Average Treatment Effects (ATE) on third year employment without pseudo start dates}
    \centering
        \begin{tabular}{@{}rrrrrrrr@{}}
    \toprule
    \multicolumn{4}{c}{\textbf{Permanent Residents}} & \multicolumn{4}{c}{\textbf{Temporary Residents}} \\
    \cmidrule(lr){1-4} \cmidrule(lr){5-8} 
        \multicolumn{1}{c}{WS} & \multicolumn{1}{c}{BC} & \multicolumn{1}{c}{TC} & \multicolumn{1}{c}{EP} & \multicolumn{1}{c}{WS} & \multicolumn{1}{c}{BC} & \multicolumn{1}{c}{TC} & \multicolumn{1}{c}{EP} \\ \midrule
    0.38***\phantom{(} & 
    -0.30***\phantom{(} &  
    -0.09\phantom{(***} & 
    -0.38***\phantom{(} & 
    0.55***\phantom{(} & 
    0.10\phantom{(***} & 
    0.20**\phantom{(*} & 
    0.08\phantom{(***}\\
    (0.03)\phantom{***}  & 
    (0.04)\phantom{***}   & 
    (0.06)\phantom{***} & 
    (0.06)\phantom{***}   & 
    (0.06)\phantom{***}  & 
    (0.08)\phantom{***} &  
    (0.10)\phantom{***}  & 
    (0.12)\phantom{***} \\
    \bottomrule
    \end{tabular}
    \begin{tablenotes}
    \textit{Note:} ATEs of participation in different programmes compared to non-participation. WS: Wage subsidy, BC: Basic Courses, TC: Training Courses, EP: Employment Programme. The outcomes are computed from the start of the unemployment spell. Standard errors are presented in parentheses. The symbols *, ** and *** indicate the level of precision for each estimate, denoting if the p-value of a two-sided significance test is below 10\%, 5\%, and 1\%, respectively.
    \end{tablenotes}
    \label{table:robu_nodates}
\end{threeparttable}
\end{adjustbox}
\end{table}

A final sensitivity analysis assesses whether the simulated policy allocations from the policy trees can be improved through data-driven selection of policy variables. Additional covariates are chosen based on a variable importance ranking that identifies features predictive of treatment effect heterogeneity and thus potentially valuable for targeting.\footnote{Variable importance is measured as the decrease in $R^2$ when a given covariate is randomly permuted in a random forest regression of the IATEs on the covariates.} The extended covariate set includes all variables from the main analysis, complemented by the population size of the unemployed individual's municipality, canton of residence, local language proficiency, and employment history five and ten years prior to unemployment. While these additions may enhance targeting and increase welfare, they also raise computational cost.

The results for the PERM sample, reported in Table~\ref{table:policy_tree_rob}, show that treatment assignments and associated welfare gains remain largely stable compared to those based on the more limited set of policy variables in Table~\ref{table:policy_tree}. The share of individuals assigned to WS and NP increases slightly, accompanied by a marginal improvement in overall welfare. Consistent with the findings from the main analysis, the depth-4 tree assigns a larger share of individuals to BC, TC, and EP compared to shallower trees, while achieving nearly identical welfare outcomes. This confirms that the result of prioritising Temporary Wage Subsidies in the PERM sample is not sensitive to the different sets of covariates, at least within the specifications analysed.

\begin{table}[ht]
\begin{adjustbox}{width=0.7\columnwidth, center}
\begin{threeparttable}
\captionsetup{font=large}
    \caption{Out-of-sample treatment allocations per policy}
    \centering
    \begin{tabular}{@{}lcrrrrr@{}}
    \toprule
     \textbf{Policy} & \makecell{\textbf{Emp.}\\\textbf{Months}} & \multicolumn{5}{c}{\textbf{Shares per programme}} \\
    \cmidrule(r){1-7}
     &  & NP & WS & BC & TC & EP \\
     \midrule
    Shallow Tree (Constrained) & & & & & & \\
    ~~~~Depth-2  & 9.07	& 71.76 & 24.67 & 0.00 & 0.00 & 3.57 \\
    ~~~~Depth-3  & 9.08	& 72.20 & 24.83 & 0.00 & 0.00 & 2.97 \\
    ~~~~Depth-4  & 9.07 & 67.56 & 22.63 & 4.04 & 3.22 & 2.56 \\
    \bottomrule
    \end{tabular}
    \begin{tablenotes}
    \textit{Note:} 
    The table displays the simulated allocations of individuals per program under different policies for permanent residents. The performance measure is the average sum of months in employment in the third year after the program start. In the constrained simulations, programme shares may exceed observed shares due to being based on the validation sample, while constraints are enforced in-sample.
    \end{tablenotes}
    \label{table:policy_tree_rob}
\end{threeparttable}
\end{adjustbox}
\end{table}

\section{Conclusion} \label{discussion}
This study evaluates the effectiveness of Switzerland's active labour market policies by examining their impact on future re-employment probabilities and earnings. Causal machine learning methods were used to analyse a vast Swiss administrative dataset from 2004 to 2018.

The findings reveal a strong lock-in effect for all programmes in the first year. For the sample of permanent residents, the findings indicate that, on average, the programmes have mixed effects in the medium-term regarding employment and earnings. The Temporary Wage Subsidy programme is the only one demonstrating positive and significant effects. In the sample of individuals with a temporary residence permit the average effects on employment in the medium-term are positive and significant for all the programmes. The heterogeneity analysis, conducted on the cumulative earnings in the third year, reveals that programmes are more effective for individuals with a place of origin outside the European Union or Switzerland. The results align with findings from \cite{cockx2023priority} in Belgium, but they are extended to ``long-term'' immigrants, which present more similar characteristics to the native Swiss population. By analysing balanced heterogeneous treatment effects, the results suggest that the effect heterogeneity for Temporary Wage Subsidies is most likely driven by education, origin, and employment history rather than merely being correlated to these factors. We do not detect significant heterogeneity in any dimension in the sample of recent immigrants, most probably due to the relatively small sample size.
Finally, by using policy learning methods, it can be concluded that increasing the share of individuals eligible for Temporary Wage Subsidies would improve the employment duration by up to half a month in the third year after the start of the first programme. However, if capacity constraints limit eligibility for Wage Subsidies at the observed levels, improvements in caseworker allocations can still be achieved for permanent residents following the allocations suggested by deeper sequential decision trees.

This paper faces some limitations. Due to low participation rates in some programmes, aggregate programme categories had to be constructed. Consequently, small-scale programmes tailored to specific sub-populations cannot be evaluated, as it is impossible to construct credible counterfactuals.
Finally, only a partial solution to the dynamic assignment problem by predicting the pseudo-start dates is provided. A promising future research avenue to alleviate this issue would be to extend the approach proposed by \citet{van2022long} to causal machine learning estimation.

\newpage
\bibliographystyle{apacite}
\bibliography{lib}

\begin{thebibliography}{}

\bibitem [\protect \citeauthoryear {%
Altmann%
, Falk%
, J{\"a}ger%
\BCBL {}\ \BBA {} Zimmermann%
}{%
Altmann%
\ \protect \BOthers {.}}{%
{\protect \APACyear {2018}}%
}]{%
altmann2018learning}
\APACinsertmetastar {%
altmann2018learning}%
\begin{APACrefauthors}%
Altmann, S.%
, Falk, A.%
, J{\"a}ger, S.%
\BCBL {}\ \BBA {} Zimmermann, F.%
\end{APACrefauthors}%
\unskip\
\newblock
\APACrefYearMonthDay{2018}{}{}.
\newblock
{\BBOQ}\APACrefatitle {{Learning about job search: A field experiment with job seekers in Germany}} {{Learning about job search: A field experiment with job seekers in Germany}}.{\BBCQ}
\newblock
\APACjournalVolNumPages{Journal of Public Economics}{164}{}{33--49}.
\PrintBackRefs{\CurrentBib}

\bibitem [\protect \citeauthoryear {%
Arthur%
\ \BBA {} Vassilvitskii%
}{%
Arthur%
\ \BBA {} Vassilvitskii%
}{%
{\protect \APACyear {2007}}%
}]{%
arthur2007k}
\APACinsertmetastar {%
arthur2007k}%
\begin{APACrefauthors}%
Arthur, D.%
\BCBT {}\ \BBA {} Vassilvitskii, S.%
\end{APACrefauthors}%
\unskip\
\newblock
\APACrefYearMonthDay{2007}{}{}.
\newblock
{\BBOQ}\APACrefatitle {K-means++ the advantages of careful seeding} {K-means++ the advantages of careful seeding}.{\BBCQ}
\newblock
\BIn{} \APACrefbtitle {Proceedings of the eighteenth annual ACM-SIAM symposium on Discrete algorithms} {Proceedings of the eighteenth annual acm-siam symposium on discrete algorithms}\ (\BPGS\ 1027--1035).
\PrintBackRefs{\CurrentBib}

\bibitem [\protect \citeauthoryear {%
Athey%
, Tibshirani%
\BCBL {}\ \BBA {} Wager%
}{%
Athey%
\ \protect \BOthers {.}}{%
{\protect \APACyear {2019}}%
}]{%
athey2019generalized}
\APACinsertmetastar {%
athey2019generalized}%
\begin{APACrefauthors}%
Athey, S.%
, Tibshirani, J.%
\BCBL {}\ \BBA {} Wager, S.%
\end{APACrefauthors}%
\unskip\
\newblock
\APACrefYearMonthDay{2019}{}{}.
\newblock
{\BBOQ}\APACrefatitle {Generalized random forests} {Generalized random forests}.{\BBCQ}
\newblock
\APACjournalVolNumPages{The Annals of Statistics}{47}{2}{1148--1178}.
\PrintBackRefs{\CurrentBib}

\bibitem [\protect \citeauthoryear {%
Baiardi%
\ \BBA {} Naghi%
}{%
Baiardi%
\ \BBA {} Naghi%
}{%
{\protect \APACyear {2024}}%
}]{%
Baiardi:2024}
\APACinsertmetastar {%
Baiardi:2024}%
\begin{APACrefauthors}%
Baiardi, A.%
\BCBT {}\ \BBA {} Naghi, A\BPBI A.%
\end{APACrefauthors}%
\unskip\
\newblock
\APACrefYearMonthDay{2024}{}{}.
\newblock
{\BBOQ}\APACrefatitle {The value added of machine learning to causal inference: Evidence from revisited studies} {The value added of machine learning to causal inference: Evidence from revisited studies}.{\BBCQ}
\newblock
\APACjournalVolNumPages{The Econometrics Journal}{27}{2}{213--234}.
\PrintBackRefs{\CurrentBib}

\bibitem [\protect \citeauthoryear {%
Bearth%
\ \BBA {} Lechner%
}{%
Bearth%
\ \BBA {} Lechner%
}{%
{\protect \APACyear {2025}}%
}]{%
bearth2025causal}
\APACinsertmetastar {%
bearth2025causal}%
\begin{APACrefauthors}%
Bearth, N.%
\BCBT {}\ \BBA {} Lechner, M.%
\end{APACrefauthors}%
\unskip\
\newblock
\APACrefYearMonthDay{2025}{}{}.
\newblock
{\BBOQ}\APACrefatitle {Causal machine learning for moderation effects} {Causal machine learning for moderation effects}.{\BBCQ}
\newblock
\APACjournalVolNumPages{Journal of Business \& Economic Statistics}{}{just-accepted}{1--21}.
\PrintBackRefs{\CurrentBib}

\bibitem [\protect \citeauthoryear {%
Bergemann%
\ \BBA {} Van~den Berg%
}{%
Bergemann%
\ \BBA {} Van~den Berg%
}{%
{\protect \APACyear {2008}}%
}]{%
c56cf7f9-3aeb-3d5d-9780-3a616776b70c}
\APACinsertmetastar {%
c56cf7f9-3aeb-3d5d-9780-3a616776b70c}%
\begin{APACrefauthors}%
Bergemann, A.%
\BCBT {}\ \BBA {} Van~den Berg, G\BPBI J.%
\end{APACrefauthors}%
\unskip\
\newblock
\APACrefYearMonthDay{2008}{}{}.
\newblock
{\BBOQ}\APACrefatitle {{Active labor market policy effects for women in Europe — a survey}} {{Active labor market policy effects for women in Europe — a survey}}.{\BBCQ}
\newblock
\APACjournalVolNumPages{Annales d'Economie et de Statistique}{}{}{385--408}.
\PrintBackRefs{\CurrentBib}

\bibitem [\protect \citeauthoryear {%
Bodory%
, Busshoff%
\BCBL {}\ \BBA {} Lechner%
}{%
Bodory%
\ \protect \BOthers {.}}{%
{\protect \APACyear {2022}}%
}]{%
bodory2022high}
\APACinsertmetastar {%
bodory2022high}%
\begin{APACrefauthors}%
Bodory, H.%
, Busshoff, H.%
\BCBL {}\ \BBA {} Lechner, M.%
\end{APACrefauthors}%
\unskip\
\newblock
\APACrefYearMonthDay{2022}{}{}.
\newblock
{\BBOQ}\APACrefatitle {High resolution treatment effects estimation: Uncovering effect heterogeneities with the modified causal forest} {High resolution treatment effects estimation: Uncovering effect heterogeneities with the modified causal forest}.{\BBCQ}
\newblock
\APACjournalVolNumPages{Entropy}{24}{8}{1039}.
\PrintBackRefs{\CurrentBib}

\bibitem [\protect \citeauthoryear {%
Bodory%
, Mascolo%
\BCBL {}\ \BBA {} Lechner%
}{%
Bodory%
\ \protect \BOthers {.}}{%
{\protect \APACyear {2024}}%
}]{%
bodory2024enabling}
\APACinsertmetastar {%
bodory2024enabling}%
\begin{APACrefauthors}%
Bodory, H.%
, Mascolo, F.%
\BCBL {}\ \BBA {} Lechner, M.%
\end{APACrefauthors}%
\unskip\
\newblock
\APACrefYearMonthDay{2024}{}{}.
\newblock
{\BBOQ}\APACrefatitle {Enabling Decision-Making with the Modified Causal Forest: Policy Trees for Treatment Assignment} {Enabling decision-making with the modified causal forest: Policy trees for treatment assignment}.{\BBCQ}
\newblock
\APACjournalVolNumPages{arXiv preprint arXiv:2406.02241}{}{}{}.
\PrintBackRefs{\CurrentBib}

\bibitem [\protect \citeauthoryear {%
Burlat%
}{%
Burlat%
}{%
{\protect \APACyear {2024}}%
}]{%
burlat2024everybody}
\APACinsertmetastar {%
burlat2024everybody}%
\begin{APACrefauthors}%
Burlat, H.%
\end{APACrefauthors}%
\unskip\
\newblock
\APACrefYearMonthDay{2024}{}{}.
\newblock
{\BBOQ}\APACrefatitle {Everybody’s got to learn sometime? A causal machine learning evaluation of training programmes for jobseekers in France} {Everybody’s got to learn sometime? a causal machine learning evaluation of training programmes for jobseekers in france}.{\BBCQ}
\newblock
\APACjournalVolNumPages{Labour Economics}{}{}{102573}.
\PrintBackRefs{\CurrentBib}

\bibitem [\protect \citeauthoryear {%
Caliendo%
, Mahlstedt%
\BCBL {}\ \BBA {} Mitnik%
}{%
Caliendo%
\ \protect \BOthers {.}}{%
{\protect \APACyear {2017}}%
}]{%
caliendo2017unobservable}
\APACinsertmetastar {%
caliendo2017unobservable}%
\begin{APACrefauthors}%
Caliendo, M.%
, Mahlstedt, R.%
\BCBL {}\ \BBA {} Mitnik, O\BPBI A.%
\end{APACrefauthors}%
\unskip\
\newblock
\APACrefYearMonthDay{2017}{}{}.
\newblock
{\BBOQ}\APACrefatitle {Unobservable, but unimportant? The relevance of usually unobserved variables for the evaluation of labor market policies} {Unobservable, but unimportant? the relevance of usually unobserved variables for the evaluation of labor market policies}.{\BBCQ}
\newblock
\APACjournalVolNumPages{Labour Economics}{46}{}{14--25}.
\PrintBackRefs{\CurrentBib}

\bibitem [\protect \citeauthoryear {%
Card%
, Kluve%
\BCBL {}\ \BBA {} Weber%
}{%
Card%
\ \protect \BOthers {.}}{%
{\protect \APACyear {2018}}%
}]{%
10.1093/jeea/jvx028}
\APACinsertmetastar {%
10.1093/jeea/jvx028}%
\begin{APACrefauthors}%
Card, D.%
, Kluve, J.%
\BCBL {}\ \BBA {} Weber, A.%
\end{APACrefauthors}%
\unskip\
\newblock
\APACrefYearMonthDay{2018}{10}{}.
\newblock
{\BBOQ}\APACrefatitle {{What Works? A Meta Analysis of Recent Active Labor Market Program Evaluations}} {{What Works? A Meta Analysis of Recent Active Labor Market Program Evaluations}}.{\BBCQ}
\newblock
\APACjournalVolNumPages{Journal of the European Economic Association}{16}{3}{894-931}.
\newblock
\begin{APACrefDOI} \doi{10.1093/jeea/jvx028} \end{APACrefDOI}
\PrintBackRefs{\CurrentBib}

\bibitem [\protect \citeauthoryear {%
Cockx%
, Lechner%
\BCBL {}\ \BBA {} Bollens%
}{%
Cockx%
\ \protect \BOthers {.}}{%
{\protect \APACyear {2023}}%
}]{%
cockx2023priority}
\APACinsertmetastar {%
cockx2023priority}%
\begin{APACrefauthors}%
Cockx, B.%
, Lechner, M.%
\BCBL {}\ \BBA {} Bollens, J.%
\end{APACrefauthors}%
\unskip\
\newblock
\APACrefYearMonthDay{2023}{}{}.
\newblock
{\BBOQ}\APACrefatitle {{Priority to unemployed immigrants? A causal machine learning evaluation of training in Belgium}} {{Priority to unemployed immigrants? A causal machine learning evaluation of training in Belgium}}.{\BBCQ}
\newblock
\APACjournalVolNumPages{Labour Economics}{80}{}{102306}.
\PrintBackRefs{\CurrentBib}

\bibitem [\protect \citeauthoryear {%
Cr{\'e}pon%
, Duflo%
, Gurgand%
, Rathelot%
\BCBL {}\ \BBA {} Zamora%
}{%
Cr{\'e}pon%
\ \protect \BOthers {.}}{%
{\protect \APACyear {2013}}%
}]{%
crepon2013labor}
\APACinsertmetastar {%
crepon2013labor}%
\begin{APACrefauthors}%
Cr{\'e}pon, B.%
, Duflo, E.%
, Gurgand, M.%
, Rathelot, R.%
\BCBL {}\ \BBA {} Zamora, P.%
\end{APACrefauthors}%
\unskip\
\newblock
\APACrefYearMonthDay{2013}{}{}.
\newblock
{\BBOQ}\APACrefatitle {Do labor market policies have displacement effects? Evidence from a clustered randomized experiment} {Do labor market policies have displacement effects? evidence from a clustered randomized experiment}.{\BBCQ}
\newblock
\APACjournalVolNumPages{The Quarterly Journal of Economics}{128}{2}{531--580}.
\PrintBackRefs{\CurrentBib}

\bibitem [\protect \citeauthoryear {%
D’Amour%
, Ding%
, Feller%
, Lei%
\BCBL {}\ \BBA {} Sekhon%
}{%
D’Amour%
\ \protect \BOthers {.}}{%
{\protect \APACyear {2021}}%
}]{%
d2021overlap}
\APACinsertmetastar {%
d2021overlap}%
\begin{APACrefauthors}%
D’Amour, A.%
, Ding, P.%
, Feller, A.%
, Lei, L.%
\BCBL {}\ \BBA {} Sekhon, J.%
\end{APACrefauthors}%
\unskip\
\newblock
\APACrefYearMonthDay{2021}{}{}.
\newblock
{\BBOQ}\APACrefatitle {Overlap in observational studies with high-dimensional covariates} {Overlap in observational studies with high-dimensional covariates}.{\BBCQ}
\newblock
\APACjournalVolNumPages{Journal of Econometrics}{221}{2}{644--654}.
\PrintBackRefs{\CurrentBib}

\bibitem [\protect \citeauthoryear {%
Gerfin%
\ \BBA {} Lechner%
}{%
Gerfin%
\ \BBA {} Lechner%
}{%
{\protect \APACyear {2002}}%
}]{%
gerfin2002microeconometric}
\APACinsertmetastar {%
gerfin2002microeconometric}%
\begin{APACrefauthors}%
Gerfin, M.%
\BCBT {}\ \BBA {} Lechner, M.%
\end{APACrefauthors}%
\unskip\
\newblock
\APACrefYearMonthDay{2002}{}{}.
\newblock
{\BBOQ}\APACrefatitle {A microeconometric evaluation of the active labour market policy in Switzerland} {A microeconometric evaluation of the active labour market policy in switzerland}.{\BBCQ}
\newblock
\APACjournalVolNumPages{The Economic Journal}{112}{482}{854--893}.
\PrintBackRefs{\CurrentBib}

\bibitem [\protect \citeauthoryear {%
Goller%
, Harrer%
, Lechner%
\BCBL {}\ \BBA {} Wolff%
}{%
Goller%
\ \protect \BOthers {.}}{%
{\protect \APACyear {2021}}%
}]{%
goller2021active}
\APACinsertmetastar {%
goller2021active}%
\begin{APACrefauthors}%
Goller, D.%
, Harrer, T.%
, Lechner, M.%
\BCBL {}\ \BBA {} Wolff, J.%
\end{APACrefauthors}%
\unskip\
\newblock
\APACrefYearMonthDay{2021}{}{}.
\newblock
{\BBOQ}\APACrefatitle {Active labour market policies for the long-term unemployed: New evidence from causal machine learning} {Active labour market policies for the long-term unemployed: New evidence from causal machine learning}.{\BBCQ}
\newblock
\APACjournalVolNumPages{arXiv preprint arXiv:2106.10141}{}{}{}.
\PrintBackRefs{\CurrentBib}

\bibitem [\protect \citeauthoryear {%
Hatamyar%
\ \BBA {} Kreif%
}{%
Hatamyar%
\ \BBA {} Kreif%
}{%
{\protect \APACyear {2023}}%
}]{%
hatamyar2023policy}
\APACinsertmetastar {%
hatamyar2023policy}%
\begin{APACrefauthors}%
Hatamyar, J.%
\BCBT {}\ \BBA {} Kreif, N.%
\end{APACrefauthors}%
\unskip\
\newblock
\APACrefYearMonthDay{2023}{}{}.
\newblock
{\BBOQ}\APACrefatitle {Policy Learning with Rare Outcomes} {Policy learning with rare outcomes}.{\BBCQ}
\newblock
\APACjournalVolNumPages{arXiv preprint arXiv:2302.05260}{}{}{}.
\PrintBackRefs{\CurrentBib}

\bibitem [\protect \citeauthoryear {%
Heckman%
, LaLonde%
\BCBL {}\ \BBA {} Smith%
}{%
Heckman%
\ \protect \BOthers {.}}{%
{\protect \APACyear {1999}}%
}]{%
heckman1999economics}
\APACinsertmetastar {%
heckman1999economics}%
\begin{APACrefauthors}%
Heckman, J\BPBI J.%
, LaLonde, R\BPBI J.%
\BCBL {}\ \BBA {} Smith, J\BPBI A.%
\end{APACrefauthors}%
\unskip\
\newblock
\APACrefYearMonthDay{1999}{}{}.
\newblock
{\BBOQ}\APACrefatitle {The economics and econometrics of active labor market programs} {The economics and econometrics of active labor market programs}.{\BBCQ}
\newblock
\BIn{} \APACrefbtitle {{Handbook of Labor Economics}} {{Handbook of Labor Economics}}\ (\BVOL~3, \BPGS\ 1865--2097).
\newblock
\APACaddressPublisher{}{Elsevier}.
\PrintBackRefs{\CurrentBib}

\bibitem [\protect \citeauthoryear {%
Imbens%
}{%
Imbens%
}{%
{\protect \APACyear {2000}}%
}]{%
imbens2000role}
\APACinsertmetastar {%
imbens2000role}%
\begin{APACrefauthors}%
Imbens, G\BPBI W.%
\end{APACrefauthors}%
\unskip\
\newblock
\APACrefYearMonthDay{2000}{}{}.
\newblock
{\BBOQ}\APACrefatitle {The role of the propensity score in estimating dose-response functions} {The role of the propensity score in estimating dose-response functions}.{\BBCQ}
\newblock
\APACjournalVolNumPages{Biometrika}{87}{3}{706--710}.
\PrintBackRefs{\CurrentBib}

\bibitem [\protect \citeauthoryear {%
Imbens%
\ \BBA {} Wooldridge%
}{%
Imbens%
\ \BBA {} Wooldridge%
}{%
{\protect \APACyear {2009}}%
}]{%
imbens2009recent}
\APACinsertmetastar {%
imbens2009recent}%
\begin{APACrefauthors}%
Imbens, G\BPBI W.%
\BCBT {}\ \BBA {} Wooldridge, J\BPBI M.%
\end{APACrefauthors}%
\unskip\
\newblock
\APACrefYearMonthDay{2009}{}{}.
\newblock
{\BBOQ}\APACrefatitle {Recent developments in the econometrics of program evaluation} {Recent developments in the econometrics of program evaluation}.{\BBCQ}
\newblock
\APACjournalVolNumPages{Journal of Economic Literature}{47}{1}{5--86}.
\PrintBackRefs{\CurrentBib}

\bibitem [\protect \citeauthoryear {%
Knaus%
}{%
Knaus%
}{%
{\protect \APACyear {2022}}%
}]{%
knaus2022double}
\APACinsertmetastar {%
knaus2022double}%
\begin{APACrefauthors}%
Knaus, M\BPBI C.%
\end{APACrefauthors}%
\unskip\
\newblock
\APACrefYearMonthDay{2022}{}{}.
\newblock
{\BBOQ}\APACrefatitle {Double machine learning-based programme evaluation under unconfoundedness} {Double machine learning-based programme evaluation under unconfoundedness}.{\BBCQ}
\newblock
\APACjournalVolNumPages{The Econometrics Journal}{25}{3}{602--627}.
\PrintBackRefs{\CurrentBib}

\bibitem [\protect \citeauthoryear {%
Knaus%
, Lechner%
\BCBL {}\ \BBA {} Strittmatter%
}{%
Knaus%
\ \protect \BOthers {.}}{%
{\protect \APACyear {2021}}%
}]{%
knaus2021machine}
\APACinsertmetastar {%
knaus2021machine}%
\begin{APACrefauthors}%
Knaus, M\BPBI C.%
, Lechner, M.%
\BCBL {}\ \BBA {} Strittmatter, A.%
\end{APACrefauthors}%
\unskip\
\newblock
\APACrefYearMonthDay{2021}{}{}.
\newblock
{\BBOQ}\APACrefatitle {{Machine learning estimation of heterogeneous causal effects: Empirical Monte Carlo Evidence}} {{Machine learning estimation of heterogeneous causal effects: Empirical Monte Carlo Evidence}}.{\BBCQ}
\newblock
\APACjournalVolNumPages{The Econometrics Journal}{24}{1}{134--161}.
\PrintBackRefs{\CurrentBib}

\bibitem [\protect \citeauthoryear {%
Knaus%
, Lechner%
\BCBL {}\ \BBA {} Strittmatter%
}{%
Knaus%
\ \protect \BOthers {.}}{%
{\protect \APACyear {2022}}%
}]{%
knaus2022heterogeneous}
\APACinsertmetastar {%
knaus2022heterogeneous}%
\begin{APACrefauthors}%
Knaus, M\BPBI C.%
, Lechner, M.%
\BCBL {}\ \BBA {} Strittmatter, A.%
\end{APACrefauthors}%
\unskip\
\newblock
\APACrefYearMonthDay{2022}{}{}.
\newblock
{\BBOQ}\APACrefatitle {Heterogeneous employment effects of job search programs: A machine learning approach} {Heterogeneous employment effects of job search programs: A machine learning approach}.{\BBCQ}
\newblock
\APACjournalVolNumPages{Journal of Human Resources}{57}{2}{597--636}.
\PrintBackRefs{\CurrentBib}

\bibitem [\protect \citeauthoryear {%
Lalive%
, Van~Ours%
\BCBL {}\ \BBA {} Zweim{\"u}ller%
}{%
Lalive%
\ \protect \BOthers {.}}{%
{\protect \APACyear {2008}}%
}]{%
lalive2008impact}
\APACinsertmetastar {%
lalive2008impact}%
\begin{APACrefauthors}%
Lalive, R.%
, Van~Ours, J\BPBI C.%
\BCBL {}\ \BBA {} Zweim{\"u}ller, J.%
\end{APACrefauthors}%
\unskip\
\newblock
\APACrefYearMonthDay{2008}{}{}.
\newblock
{\BBOQ}\APACrefatitle {{The impact of active labour market programmes on the duration of unemployment in Switzerland}} {{The impact of active labour market programmes on the duration of unemployment in Switzerland}}.{\BBCQ}
\newblock
\APACjournalVolNumPages{The Economic Journal}{118}{525}{235--257}.
\PrintBackRefs{\CurrentBib}

\bibitem [\protect \citeauthoryear {%
Lechner%
}{%
Lechner%
}{%
{\protect \APACyear {1999}}%
}]{%
Lechner:1999}
\APACinsertmetastar {%
Lechner:1999}%
\begin{APACrefauthors}%
Lechner, M.%
\end{APACrefauthors}%
\unskip\
\newblock
\APACrefYearMonthDay{1999}{}{}.
\newblock
{\BBOQ}\APACrefatitle {{Earnings and employment effects of continuous off-the-job training in east Germany after unification}} {{Earnings and employment effects of continuous off-the-job training in east Germany after unification}}.{\BBCQ}
\newblock
\APACjournalVolNumPages{Journal of Business \& Economic Statistics}{17}{1}{74--90}.
\PrintBackRefs{\CurrentBib}

\bibitem [\protect \citeauthoryear {%
Lechner%
}{%
Lechner%
}{%
{\protect \APACyear {2001}}%
}]{%
lechner2001identification}
\APACinsertmetastar {%
lechner2001identification}%
\begin{APACrefauthors}%
Lechner, M.%
\end{APACrefauthors}%
\unskip\
\newblock
\APACrefYear{2001}.
\newblock
\APACrefbtitle {Identification and estimation of causal effects of multiple treatments under the conditional independence assumption} {Identification and estimation of causal effects of multiple treatments under the conditional independence assumption}.
\newblock
\APACaddressPublisher{}{Springer}.
\PrintBackRefs{\CurrentBib}

\bibitem [\protect \citeauthoryear {%
Lechner%
}{%
Lechner%
}{%
{\protect \APACyear {2018}}%
}]{%
lechner2022modified}
\APACinsertmetastar {%
lechner2022modified}%
\begin{APACrefauthors}%
Lechner, M.%
\end{APACrefauthors}%
\unskip\
\newblock
\APACrefYearMonthDay{2018}{}{}.
\newblock
{\BBOQ}\APACrefatitle {Modified causal forest} {Modified causal forest}.{\BBCQ}
\newblock
\APACjournalVolNumPages{arXiv preprint arXiv:1812.09487}{}{}{}.
\PrintBackRefs{\CurrentBib}

\bibitem [\protect \citeauthoryear {%
Lechner%
\ \BBA {} Mareckova%
}{%
Lechner%
\ \BBA {} Mareckova%
}{%
{\protect \APACyear {2024}}%
}]{%
Lechner:2024}
\APACinsertmetastar {%
Lechner:2024}%
\begin{APACrefauthors}%
Lechner, M.%
\BCBT {}\ \BBA {} Mareckova, J.%
\end{APACrefauthors}%
\unskip\
\newblock
\APACrefYearMonthDay{2024}{}{}.
\newblock
{\BBOQ}\APACrefatitle {Comprehensive Causal Machine Learning} {Comprehensive causal machine learning}.{\BBCQ}
\newblock
\APACjournalVolNumPages{arXiv preprint arXiv:2405.10198}{}{}{}.
\PrintBackRefs{\CurrentBib}

\bibitem [\protect \citeauthoryear {%
Lechner%
\ \BBA {} Smith%
}{%
Lechner%
\ \BBA {} Smith%
}{%
{\protect \APACyear {2007}}%
}]{%
lechner2007value}
\APACinsertmetastar {%
lechner2007value}%
\begin{APACrefauthors}%
Lechner, M.%
\BCBT {}\ \BBA {} Smith, J.%
\end{APACrefauthors}%
\unskip\
\newblock
\APACrefYearMonthDay{2007}{}{}.
\newblock
{\BBOQ}\APACrefatitle {What is the value added by caseworkers?} {What is the value added by caseworkers?}{\BBCQ}
\newblock
\APACjournalVolNumPages{Labour Economics}{14}{2}{135--151}.
\PrintBackRefs{\CurrentBib}

\bibitem [\protect \citeauthoryear {%
Lechner%
\ \BBA {} Strittmatter%
}{%
Lechner%
\ \BBA {} Strittmatter%
}{%
{\protect \APACyear {2019}}%
}]{%
lechner2019practical}
\APACinsertmetastar {%
lechner2019practical}%
\begin{APACrefauthors}%
Lechner, M.%
\BCBT {}\ \BBA {} Strittmatter, A.%
\end{APACrefauthors}%
\unskip\
\newblock
\APACrefYearMonthDay{2019}{}{}.
\newblock
{\BBOQ}\APACrefatitle {Practical procedures to deal with common support problems in matching estimation} {Practical procedures to deal with common support problems in matching estimation}.{\BBCQ}
\newblock
\APACjournalVolNumPages{Econometric Reviews}{38}{2}{193--207}.
\PrintBackRefs{\CurrentBib}

\bibitem [\protect \citeauthoryear {%
Lechner%
\ \BBA {} Wunsch%
}{%
Lechner%
\ \BBA {} Wunsch%
}{%
{\protect \APACyear {2013}}%
}]{%
lechner2013sensitivity}
\APACinsertmetastar {%
lechner2013sensitivity}%
\begin{APACrefauthors}%
Lechner, M.%
\BCBT {}\ \BBA {} Wunsch, C.%
\end{APACrefauthors}%
\unskip\
\newblock
\APACrefYearMonthDay{2013}{}{}.
\newblock
{\BBOQ}\APACrefatitle {Sensitivity of matching-based program evaluations to the availability of control variables} {Sensitivity of matching-based program evaluations to the availability of control variables}.{\BBCQ}
\newblock
\APACjournalVolNumPages{Labour Economics}{21}{}{111--121}.
\PrintBackRefs{\CurrentBib}

\bibitem [\protect \citeauthoryear {%
Martin%
\ \BBA {} Grubb%
}{%
Martin%
\ \BBA {} Grubb%
}{%
{\protect \APACyear {2001}}%
}]{%
RePEc:hhs:ifauwp:2001_014}
\APACinsertmetastar {%
RePEc:hhs:ifauwp:2001_014}%
\begin{APACrefauthors}%
Martin, J.%
\BCBT {}\ \BBA {} Grubb, D.%
\end{APACrefauthors}%
\unskip\
\newblock
\APACrefYearMonthDay{2001}{}{}.
\newblock
\APACrefbtitle {{What works and for whom: a review of OECD countries' experiences with active labour market policies}} {{What works and for whom: a review of OECD countries' experiences with active labour market policies}}\ \APACbVolEdTR {}{Working Paper Series\ \BNUM\ 2001:14}.
\newblock
\APACaddressInstitution{}{IFAU - Institute for Evaluation of Labour Market and Education Policy}.
\PrintBackRefs{\CurrentBib}

\bibitem [\protect \citeauthoryear {%
Morlok%
, Liechti%
, Moser%
\BCBL {}\ \BBA {} Suri%
}{%
Morlok%
\ \protect \BOthers {.}}{%
{\protect \APACyear {2018}}%
}]{%
Seco:2024a}
\APACinsertmetastar {%
Seco:2024a}%
\begin{APACrefauthors}%
Morlok, M.%
, Liechti, D.%
, Moser, N.%
\BCBL {}\ \BBA {} Suri, M.%
\end{APACrefauthors}%
\unskip\
\newblock
\APACrefYearMonthDay{2018}{}{}.
\newblock
\APACrefbtitle {{Die Wirkung von arbeitsmarktlichen Massnahmen}} {{Die Wirkung von arbeitsmarktlichen Massnahmen}}\ \APACbVolEdTR{}{\BTR{}}.
\newblock
\APACaddressInstitution{}{{State Secretariat for Economic Affairs, Arbeitsmarktpolitik No 54}}.
\PrintBackRefs{\CurrentBib}

\bibitem [\protect \citeauthoryear {%
{OECD}%
}{%
{OECD}%
}{%
{\protect \APACyear {{2024}}}%
}]{%
OECD:2024}
\APACinsertmetastar {%
OECD:2024}%
\begin{APACrefauthors}%
{OECD}.%
\end{APACrefauthors}%
\unskip\
\newblock
\APACrefYearMonthDay{{2024}}{}{}.
\newblock
\APACrefbtitle {Active Labour Market Policies: Connecting People with Jobs.} {Active labour market policies: Connecting people with jobs.}
\newblock
\APAChowpublished {{https://www.oecd.org/employment/activation.htm, visited 2024-04-26}}.
\PrintBackRefs{\CurrentBib}

\bibitem [\protect \citeauthoryear {%
Pedregosa%
\ \protect \BOthers {.}}{%
Pedregosa%
\ \protect \BOthers {.}}{%
{\protect \APACyear {2011}}%
}]{%
pedregosa2011scikit}
\APACinsertmetastar {%
pedregosa2011scikit}%
\begin{APACrefauthors}%
Pedregosa, F.%
, Varoquaux, G.%
, Gramfort, A.%
, Michel, V.%
, Thirion, B.%
, Grisel, O.%
\BDBL {}others%
\end{APACrefauthors}%
\unskip\
\newblock
\APACrefYearMonthDay{2011}{}{}.
\newblock
{\BBOQ}\APACrefatitle {Scikit-learn: Machine Learning in Python} {Scikit-learn: Machine learning in python}.{\BBCQ}
\newblock
\APACjournalVolNumPages{The Journal of Machine Learning Research}{12}{}{2825--2830}.
\PrintBackRefs{\CurrentBib}

\bibitem [\protect \citeauthoryear {%
Rosenbaum%
\ \BBA {} Rubin%
}{%
Rosenbaum%
\ \BBA {} Rubin%
}{%
{\protect \APACyear {1983}}%
}]{%
10.1093/biomet/70.1.41}
\APACinsertmetastar {%
10.1093/biomet/70.1.41}%
\begin{APACrefauthors}%
Rosenbaum, P\BPBI R.%
\BCBT {}\ \BBA {} Rubin, D\BPBI B.%
\end{APACrefauthors}%
\unskip\
\newblock
\APACrefYearMonthDay{1983}{04}{}.
\newblock
{\BBOQ}\APACrefatitle {{The central role of the propensity score in observational studies for causal effects}} {{The central role of the propensity score in observational studies for causal effects}}.{\BBCQ}
\newblock
\APACjournalVolNumPages{Biometrika}{70}{1}{41-55}.
\newblock
\begin{APACrefDOI} \doi{10.1093/biomet/70.1.41} \end{APACrefDOI}
\PrintBackRefs{\CurrentBib}

\bibitem [\protect \citeauthoryear {%
Rosenbaum%
\ \BBA {} Rubin%
}{%
Rosenbaum%
\ \BBA {} Rubin%
}{%
{\protect \APACyear {1985}}%
}]{%
rosenbaum1985constructing}
\APACinsertmetastar {%
rosenbaum1985constructing}%
\begin{APACrefauthors}%
Rosenbaum, P\BPBI R.%
\BCBT {}\ \BBA {} Rubin, D\BPBI B.%
\end{APACrefauthors}%
\unskip\
\newblock
\APACrefYearMonthDay{1985}{}{}.
\newblock
{\BBOQ}\APACrefatitle {Constructing a control group using multivariate matched sampling methods that incorporate the propensity score} {Constructing a control group using multivariate matched sampling methods that incorporate the propensity score}.{\BBCQ}
\newblock
\APACjournalVolNumPages{The American Statistician}{39}{1}{33--38}.
\PrintBackRefs{\CurrentBib}

\bibitem [\protect \citeauthoryear {%
Rubin%
}{%
Rubin%
}{%
{\protect \APACyear {1974}}%
}]{%
rubin1974estimating}
\APACinsertmetastar {%
rubin1974estimating}%
\begin{APACrefauthors}%
Rubin, D\BPBI B.%
\end{APACrefauthors}%
\unskip\
\newblock
\APACrefYearMonthDay{1974}{}{}.
\newblock
{\BBOQ}\APACrefatitle {Estimating causal effects of treatments in randomized and nonrandomized studies} {Estimating causal effects of treatments in randomized and nonrandomized studies}.{\BBCQ}
\newblock
\APACjournalVolNumPages{Journal of Educational Psychology}{66}{5}{688}.
\PrintBackRefs{\CurrentBib}

\bibitem [\protect \citeauthoryear {%
{State Secretariat for Economic Affairs}%
}{%
{State Secretariat for Economic Affairs}%
}{%
{\protect \APACyear {{2024}}}%
}]{%
Seco:2024}
\APACinsertmetastar {%
Seco:2024}%
\begin{APACrefauthors}%
{State Secretariat for Economic Affairs}.%
\end{APACrefauthors}%
\unskip\
\newblock
\APACrefYearMonthDay{{2024}}{}{}.
\newblock
\APACrefbtitle {Arbeitsmarktliche Massnahmen.} {Arbeitsmarktliche massnahmen.}
\newblock
\APAChowpublished {{https://www.seco.admin.ch/seco/de/home/Arbeit/Arbeitslosenversicherung/oeffentliche-arbeitsvermittlung/arbeitsmarktliche-massnahmen.html}}.
\PrintBackRefs{\CurrentBib}

\bibitem [\protect \citeauthoryear {%
Van~den Berg%
\ \BBA {} Vikstr{\"o}m%
}{%
Van~den Berg%
\ \BBA {} Vikstr{\"o}m%
}{%
{\protect \APACyear {2022}}%
}]{%
van2022long}
\APACinsertmetastar {%
van2022long}%
\begin{APACrefauthors}%
Van~den Berg, G\BPBI J.%
\BCBT {}\ \BBA {} Vikstr{\"o}m, J.%
\end{APACrefauthors}%
\unskip\
\newblock
\APACrefYearMonthDay{2022}{}{}.
\newblock
{\BBOQ}\APACrefatitle {Long-Run Effects of Dynamically Assigned Treatments: A New Methodology and an Evaluation of Training Effects on Earnings} {Long-run effects of dynamically assigned treatments: A new methodology and an evaluation of training effects on earnings}.{\BBCQ}
\newblock
\APACjournalVolNumPages{Econometrica}{90}{3}{1337--1354}.
\PrintBackRefs{\CurrentBib}

\bibitem [\protect \citeauthoryear {%
Vooren%
, Haelermans%
, Groot%
\BCBL {}\ \BBA {} Maassen van~den Brink%
}{%
Vooren%
\ \protect \BOthers {.}}{%
{\protect \APACyear {2019}}%
}]{%
vooren2019effectiveness}
\APACinsertmetastar {%
vooren2019effectiveness}%
\begin{APACrefauthors}%
Vooren, M.%
, Haelermans, C.%
, Groot, W.%
\BCBL {}\ \BBA {} Maassen van~den Brink, H.%
\end{APACrefauthors}%
\unskip\
\newblock
\APACrefYearMonthDay{2019}{}{}.
\newblock
{\BBOQ}\APACrefatitle {The effectiveness of active labor market policies: a meta-analysis} {The effectiveness of active labor market policies: a meta-analysis}.{\BBCQ}
\newblock
\APACjournalVolNumPages{Journal of Economic Surveys}{33}{1}{125--149}.
\PrintBackRefs{\CurrentBib}

\bibitem [\protect \citeauthoryear {%
Wager%
\ \BBA {} Athey%
}{%
Wager%
\ \BBA {} Athey%
}{%
{\protect \APACyear {2018}}%
}]{%
wager2018estimation}
\APACinsertmetastar {%
wager2018estimation}%
\begin{APACrefauthors}%
Wager, S.%
\BCBT {}\ \BBA {} Athey, S.%
\end{APACrefauthors}%
\unskip\
\newblock
\APACrefYearMonthDay{2018}{}{}.
\newblock
{\BBOQ}\APACrefatitle {Estimation and inference of heterogeneous treatment effects using random forests} {Estimation and inference of heterogeneous treatment effects using random forests}.{\BBCQ}
\newblock
\APACjournalVolNumPages{Journal of the American Statistical Association}{113}{523}{1228--1242}.
\PrintBackRefs{\CurrentBib}

\bibitem [\protect \citeauthoryear {%
Zhou%
, Athey%
\BCBL {}\ \BBA {} Wager%
}{%
Zhou%
\ \protect \BOthers {.}}{%
{\protect \APACyear {2023}}%
}]{%
zhou2023offline}
\APACinsertmetastar {%
zhou2023offline}%
\begin{APACrefauthors}%
Zhou, Z.%
, Athey, S.%
\BCBL {}\ \BBA {} Wager, S.%
\end{APACrefauthors}%
\unskip\
\newblock
\APACrefYearMonthDay{2023}{}{}.
\newblock
{\BBOQ}\APACrefatitle {Offline multi-action policy learning: Generalization and optimization} {Offline multi-action policy learning: Generalization and optimization}.{\BBCQ}
\newblock
\APACjournalVolNumPages{Operations Research}{71}{1}{148--183}.
\PrintBackRefs{\CurrentBib}

\end{thebibliography}

\newpage
\begin{appendices}
\renewcommand\thetable{\thesection.\arabic{table}} 
\counterwithin{table}{section}
\renewcommand\thefigure{\thesection.\arabic{figure}} 
\counterwithin{figure}{section}

\section{Appendix: Data} \label{Appendix_data}

\subsection{Definition of Unemployed Individuals}
Unemployed individuals are defined as those eligible for unemployment benefits between January 2014 and June 2015. Within this group, exclusions apply to those who did not receive any benefit throughout this period. The start of the first unemployment spell is the first month an individual registers as unemployed after being employed in the preceding three months. Participants are categorised based on the first programme (taking longer than five business days) they enter within six months from the initiation of the unemployment spell.

\subsection{Descriptive Statistics} \label{Descriptive}

Tables \ref{table: descr_swiss+1} to \ref{table: descr_b2} present the mean and standardised difference (in parentheses) of conditioning variables for each programme. The standardised difference is calculated as 
\begin{align*}
    \Delta=\frac{\left|\bar{X}_j-\bar{X}_\text{NP}\right|}{\sqrt{1 / 2\left(\operatorname{Var}\left(\bar{X}_j\right)+\operatorname{Var}\left(\bar{X}_\text{NP}\right)\right)}} \cdot 100
\end{align*}
where $\bar{X}_j$ and $\bar{X}_\text{NP}$ indicate the sample mean of variable $X$ among individuals in the programme $j$ and the control group, respectively. Hence, each programme group is compared to the group of non-participants. As in \cite{10.1093/biomet/70.1.41}, a standardised difference higher than 20 indicates imbalances in the subgroups relative to the control group. Imbalances among groups concerning a specific variable further emphasise the importance of conditioning on that particular variable. Given the granularity of the variables, we do not report the descriptive statistics of the monthly employment, unemployment, earnings and unemployment benefits history in the 36 months before the programmes start, the monthly outcomes for earnings and employment, the REO codes, and the unemployment start dates.

\begin{table}[ht!]\centering
    \caption{Descriptive statistics by programme: Covariates PERM sample (1)}
    \label{table: descr_swiss+1}
    \begin{adjustbox}{max width=0.92\textwidth}  
    \begin{threeparttable}
    \begin{tabular}{lrrrrrrrrrr} \toprule
        \textbf{Variable} & \multicolumn{1}{c}{NP} & \multicolumn{2}{c}{WS} & \multicolumn{2}{c}{BC} & \multicolumn{2}{c}{TC} & \multicolumn{2}{c}{EP} \\
        & Mean & Mean & Std. diff. & Mean & Std. diff. & Mean & Std. diff. & Mean & Std. diff. \\ \midrule
\textbf{Sociodemographics} & & & & & & & & & \\ \midrule
Age & 38.50 & 39.72 & (13.64) & 40.03 & (17.20) & 40.62 & (24.39) & 40.02 & (16.91)\\
Female & 0.46 & 0.50 & (8.90) & 0.46 & (0.26) & 0.50 & (8.74) & 0.47 & (2.63)\\
Compulsory education & 0.15 & 0.23 & (19.86) & 0.17 & (6.30) & 0.19 & (12.09) & 0.25 & (24.76)\\
Secondary education & 0.57 & 0.58 & (3.10) & 0.59 & (5.62) & 0.55 & (4.12) & 0.57 & (0.11)\\
Tertiary education & 0.08 & 0.05 & (12.55) & 0.08 & (2.30) & 0.07 & (6.10) & 0.06 & (8.63)\\
Bachelor & 0.07 & 0.05 & (8.63) & 0.06 & (5.77) & 0.06 & (5.76) & 0.04 & (12.21)\\
Post-Grad.& 0.13 & 0.09 & (12.92) & 0.10 & (9.60) & 0.14 & (1.51) & 0.08 & (15.64)\\
Proficiency local language (1 to 7) & 6.24 & 5.90 & (20.80) & 6.16 & (4.98) & 5.67 & (31.42) & 5.90 & (20.41)\\
Best other languages (1 to 7) & 0.51 & 0.49 & (1.29) & 0.53 & (1.52) & 0.56 & (3.57) & 0.39 & (8.87)\\
English level (1 to 7) & 2.71 & 2.01 & (30.37) & 2.45 & (10.95) & 2.52 & (7.65) & 1.97 & (32.40)\\
German level (1 to 7) & 3.83 & 3.29 & (15.66) & 3.70 & (3.82) & 1.77 & (63.99) & 3.36 & (13.94)\\
French level (1 to 7) & 3.06 & 2.64 & (15.24) & 2.71 & (12.90) & 3.61 & (19.20) & 2.77 & (10.71) \\
Italian level (1 to 7) & 0.89 & 1.02 & (6.22) & 0.87 & (1.20) & 1.39 & (22.53) & 0.95 & (3.11) \\
Swiss & 0.69 & 0.65 & (9.48) & 0.66 & (7.46) & 0.64 & (11.56) & 0.64 & (11.26)\\
Neighbour countries  & 0.12 & 0.11 & (3.50) & 0.12 & (2.13) & 0.12 & (0.36) & 0.10 & (6.13)\\
Rest of European Union & 0.09 & 0.10 & (2.59) & 0.07  & (8.53)  & 0.10  & (1.66) & 0.10  & (2.05)\\
Rest of Europe & 0.04 & 0.06 & (10.34) & 0.06 & (11.37) & 0.04 & (2.49) & 0.06 & (11.54)\\
Rest of the world & 0.06 & 0.09 & (9.82) & 0.09 & (10.17) & 0.11 & (16.71) & 0.10 & (14.99) \\
Settlement permit type & 0.31 & 0.35 & (9.41) & 0.34 & (7.50) & 0.36 & (11.50) & 0.36 & (11.29)\\
Single & 0.43 & 0.37 & (12.23) & 0.37 & (12.65) & 0.34 & (18.92) & 0.38 & (11.04) \\
Married & 0.45 & 0.48 & (5.90) & 0.48 & (7.38) & 0.52 & (14.66) & 0.47 & (5.73) \\
Widow & 0.00 & 0.01 & (3.22) & 0.01 & (1.26) & 0.01 & (3.34) & 0.01 & (3.23)\\
Divorced & 0.12 & 0.15 & (8.23) & 0.14 & (7.15) & 0.13 & (4.79) & 0.14 & (6.84) \\
Age youngest child & 3.50 & 3.59 & (7.85) & 3.62 & (10.58) & 3.61 & (9.94) & 3.64 & (13.35)\\
Num. of kids & 0.17 & 0.13 & (8.52) & 0.14 & (6.87) & 0.14 & (7.17) & 0.13 & (8.16) \\
Num. allowances per kid & 0.18 & 0.11 & (13.55) & 0.22 & (5.91) & 0.21 & (5.75) & 0.23 & (7.39) \\
Pregnant & 0.02 & 0.01 & (6.64) & 0.01 & (9.47) & 0.01 & (9.94) & 0.01 & (10.74)\\
Maternity & 0.01 & 0.00 & (3.49) & 0.00 & (5.06) & 0.00 & (7.25) & 0.01 & (2.72) \\
Highest degree disability (prev. 2 years, \%) & 0.36 & 0.33 & (0.57) & 0.30 & (1.15) & 0.23 & (2.55) & 0.31 & (0.85)\\ \midrule
\textbf{Cantons} & & & & & & & & & \\ \midrule
Argau & 0.05 & 0.07 & (5.08) & 0.17 & (37.99) & 0.03 & (12.40) & 0.01 & (25.29)\\
Appenzell Innerrhoden & 0.00 & 0.00 & (1.49) & 0.00 & (1.46) & 0.00 & (2.25) & 0.00 & (4.33)\\
Appenzell Ausserrhoden & 0.01 & 0.01 & (0.05) & 0.00 & (7.77) & 0.00 & (3.43) & 0.00 & (6.43)\\
Bern & 0.13 & 0.12 & (1.29) & 0.00 & (51.26) & 0.02 & (41.40) & 0.44 & (73.57)\\
Basel Land  & 0.04 & 0.03 & (1.35) & 0.03 & (1.19) & 0.01 & (17.29) & 0.01 & (15.61)\\
Basel Stadt & 0.03 & 0.03 & (0.23) & 0.02 & (7.84) & 0.01 & (16.05) & 0.01 & (20.25)\\
Freiburg-Fribourg & 0.03 & 0.03 & (1.48) & 0.04 & (1.80) & 0.03 & (1.66) & 0.05 & (10.19)\\
Geneva & 0.07 & 0.06 & (4.28) & 0.06 & (4.12) & 0.13 & (19.98) & 0.05 & (12.08)\\
Glarus  & 0.00 & 0.00 & (1.90) & 0.00 & (1.78) & 0.00 & (1.54) & 0.00 & (0.70)\\
Graubünden & 0.02 & 0.02 & (0.44) & 0.00 & (15.00) & 0.03 & (2.61) & 0.02 & (1.03)\\
Jura & 0.01 & 0.01 & (1.98) & 0.00 & (7.26) & 0.01 & (0.90) & 0.02 & (13.62)\\
Luzern  & 0.04 & 0.04 & (1.79) & 0.03 & (6.63) & 0.05 & (8.04) & 0.03 & (4.96)\\
Neuchâtel & 0.02 & 0.02 & (1.58) & 0.03 & (3.59) & 0.04 & (7.26) & 0.01 & (10.15)\\
Nidwalden & 0.00 & 0.00 & (0.68) & 0.00 & (0.36) & 0.00 & (1.76) & 0.01 & (6.04)\\
Obwalden  & 0.00 & 0.00 & (1.89) & 0.00 & (1.36) & 0.00 & (1.67) & 0.00 & (3.94)\\
Sankt Gallen  & 0.05 & 0.06 & (3.75) & 0.07 & (11.17) & 0.04 & (1.73) & 0.04 & (3.68)\\
Schaffhausen & 0.01 & 0.01 & (2.82) & 0.02 & (8.49) & 0.01 & (4.74) & 0.01 & (0.47)\\
Solothurn & 0.02 & 0.03 & (5.78) & 0.07 & (23.85) & 0.01 & (13.36) & 0.05 & (14.42)\\
Schwyz & 0.01 & 0.01 & (0.37) & 0.02 & (9.15) & 0.00 & (10.10) & 0.01 & (1.83)\\
Turgau  & 0.03 & 0.03 & (0.65) & 0.02 & (5.85) & 0.02 & (6.97) & 0.02 & (9.89)\\
Ticino & 0.03 & 0.05 & (10.59) & 0.04 & (3.92) & 0.12 & (32.82) & 0.06 & (14.17)\\
Uri  & 0.00 & 0.00 & (1.67) & 0.00 & (0.92) & 0.00 & (4.51) & 0.00 & (0.50)\\
Vaud  & 0.11 & 0.09 & (4.04) & 0.11 & (0.04) & 0.21 & (29.78) & 0.07 & (12.37)\\
Valais & 0.06 & 0.06 & (3.11) & 0.03 & (10.98) & 0.05 & (1.91) & 0.05 & (1.86)\\
Zug & 0.01 & 0.01 & (0.10) & 0.02 & (4.19) & 0.01 & (6.77) & 0.01 & (2.69)\\
Zurich & 0.22 & 0.17 & (10.67) & 0.20 & (4.40) & 0.15 & (16.49) & 0.02 & (63.49)\\
\midrule 
Number of observations & 38,464 & \multicolumn{2}{c}{22,761} &\multicolumn{2}{c}{16,690}  &\multicolumn{2}{c}{5,508}  & \multicolumn{2}{c}{4,761}  \\
 \bottomrule 
\end{tabular}
    \begin{tablenotes}
        \small \item \textit{Note:} This table shows the mean and the standardised difference in comparison to the non-participants for some covariates by programmes for the Permanent Residents. WS: Wage Subsidy, BC: Basic Course, TC: Technical Course, EP: Employment Programme, NP: No programme. Countries with Swiss languages: Austria, France, Germany, Italy, Liechtenstein; EU: Belgium, Bulgaria, Croatia, Cyprus, Serbia, Czech Republic, Denmark, Finland, Estonia, Latvia, Lithuania,  Greece, Hungary, Ireland, Malta, Netherlands, Poland, Portugal, Romania, Slovakia, Slovenia, Spain, Sweden, Rest of Europe: Albania, Andorra, Kosovo, Belarus, Bosnia and Herzegovina, Georgia, Iceland, Moldova, Norway, Serbia, Montenegro, Ukraine, Rest of the world: countries outside Europe. Language Proficiency: 1:basic to 7:mother tongue.
    \end{tablenotes}
    \end{threeparttable}
    \end{adjustbox}
\end{table}

\begin{table}[htbp!]\centering
    \caption{Descriptive statistics by programme: Covariates PERM sample (2)}
    \label{table: descr_swiss+2}
    \begin{adjustbox}{max width=0.88\textwidth}
    \begin{threeparttable}
    \begin{tabular}{lrrrrrrrrrr} \toprule
    \textbf{Variable} & \multicolumn{1}{c}{NP} & \multicolumn{2}{c}{WS} & \multicolumn{2}{c}{BC} & \multicolumn{2}{c}{TC} & \multicolumn{2}{c}{EP} \\
        & Mean & Mean & Std. diff. & Mean & Std. diff. & Mean & Std. diff. & Mean & Std. diff. \\ \midrule
 
\textbf{Desired Job characteristics and macro} & & & & & & & & & \\ \midrule
Desired degree of employment & 92.37 & 92.68 & (1.94) & 94.60 & (14.67) & 94.44 & (13.46) & 92.81 & (2.73)\\
Skill requirements: none & 0.04 & 0.04 & (1.11) & 0.04 & (3.35) & 0.04 & (4.87) & 0.05 & (9.33) \\
Skill requirements: basic & 0.93 & 0.92 & (1.70) & 0.91 & (4.38) & 0.91 & (6.51) & 0.95 & (8.84) \\
Skill requirements: professional  & 0.04 & 0.04 & (1.26) & 0.04 & (2.70) & 0.04 & (4.13) & 0.03 & (3.14) \\
Employability (1: easy to 3: difficult) & 1.81 & 1.87 & (12.41) & 1.87 & (12.32) & 1.89 & (14.54) & 2.01 & (44.08) \\
Population municipality & 57,097 & 52,183 & (5.02) & 48,776 & (8.35) & 51,799 & (5.47) & 30,926 & (32.55) \\
Open sector positions in cant.(per 100k) & 6.68 & 7.21 & (3.12) & 6.84 & (0.94) & 7.10 & (2.72) & 4.99 & (12.55) \\
UE rate region (UE spell) & 3.41 & 3.35 & (4.83) & 3.33 & (6.34) & 3.86 & (34.51) & 3.05 & (28.33) \\
UE rate region yoy (UE spell) & -0.06 & -0.07 & (5.45) & -0.07 & (4.61) & -0.11 & (18.18) & -0.05 & (3.14) \\
Labor force (region level) & 89977 & 88932 & (1.85) & 97964 & (13.96) & 78265 & (21.56) & 75268 & (27.70) \\
Urban level (1: rural to 3: urban) & 1.41 & 1.41 & (0.26) & 1.38 & (5.23) & 1.40 & (2.11) & 1.47 & (7.69) \\
Availab. to commute: also abroad & 0.01 & 0.00 & (4.87) & 0.00 & (5.79) & 0.00 & (4.25) & 0.01 & (1.00)\\
Availab. to commute: every day & 0.95 & 0.96 & (2.21) & 0.97 & (10.43) & 0.97 & (10.19) & 0.91 & (16.73) \\
Availab. to commute: in parts of Sw.& 0.02 & 0.02 & (0.79) & 0.01 & (9.99) & 0.01 & (7.53) & 0.05 & (15.05)\\
Availab. to commute: not mobile & 0.01 & 0.01 & (1.21) & 0.01 & (0.43) & 0.01 & (1.81) & 0.01 & (1.62)\\
Availab. to commute: throughout Sw. & 0.01 & 0.01 & (1.41) & 0.00 & (4.31) & 0.00 & (6.15) & 0.02 & (9.65) \\ 
\midrule 
\textbf{Labour market history} & & & & & & & & & \\ \midrule
Sum of months in UE five years before UE spell & 5.82 & 7.05 & (13.68) & 4.31 & (19.09) & 5.00 & (9.98) & 6.44 & (7.21) \\
Sum of months in UE ten years before UE spell & 10.45 & 12.60 & (14.86) & 9.02 & (11.07) & 9.78 & (5.00) & 12.21 & (12.64) \\
Sum of months in emp. five years before UE spell  & 50.81 & 49.99 & (7.17) & 52.56 & (15.93) & 51.93 & (10.05) & 50.05 & (6.66) \\
Sum of months in emp. ten years before UE spell & 94.67 & 93.61 & (4.42) & 97.14 & (10.29) & 94.89 & (0.88) & 93.09 & (6.48) \\
Average of earnings three months before UE spell & 5,503 & 4,675 & (29.34) & 5,583 & (2.62) & 5,605 &  (3.29) & 4,735 & (26.58)\\
Sum of earnings five years before UE spell & 291,634 & 236,766 & (27.97) & 306,596 & (6.62) & 299,635 & (3.56) & 239,903 & (26.48) \\
Sum of earnings ten years before UE spell  & 509,002 & 418,462 & (25.28) & 539,520 & (7.28) & 518,918 & (2.43) & 425,509 & (23.42)\\
Sum of ue benefits five years before UE spell & 230,10 & 271,301 & (10.45) & 16,323 & (19.42) & 18,622 & (12.39) & 22,819 & (0.53)\\
Sum of ue benefits ten years before UE spell & 385,61 & 451,48 & (10.91) & 309,38 & (14.55) & 33,679 & (9.07) & 40,255 & (3.06) \\
Sum of earnings one year before UE spell & 580,258 & 361,300 & (27.77) & 634,033 & (6.16) & 637,222 & (6.51) & 377,027 & (25.56)\\
Sum of months in supplementary benefits two years before UE spell & 0.02 & 0.05 & (2.81) & 0.03 & (0.86) & 0.04 & (1.60) & 0.07 & (4.57) \\
Supplementary benefits at the time of UE spell & 0.00 & 0.00 & (2.97) & 0.00 & (0.89) & 0.00 & (1.75) & 0.00 & (4.43) \\
Sum of months in IV two years before UE spell & 0.08 & 0.07 & (0.77) & 0.07 & (1.41) & 0.05 & (3.71) & 0.10 & (1.52) \\
Sum of months in SH two years before UE spell & 0.52 & 0.60 & (2.93) & 0.43 & (3.22) & 0.48 & (1.33) & 0.90 & (11.58)\\
Sum of months in HV two years before UE spell & 0.08 & 0.12 & (2.43) & 0.09 & (0.72) & 0.11 & (2.11) & 0.10 & (1.39) \\
Sum of months in HE two years before UE spell & 0.00 & 0.00 & (0.00) & 0.00 & (1.09) & 0.00 & (0.00) & 0.00 & (0.00) \\
Sum of months in WS one year before UE spell & 0.14 & 0.43 & (26.45) & 0.06 & (12.08) & 0.09 & (7.34) & 0.17 & (3.84) \\
Sum of months in WS five year before UE spell & 1.17 & 2.86 & (34.43) & 0.97 & (5.87) & 1.08 & (2.61) & 1.55 & (9.85) \\
Sum of months in WS ten year before UE spell & 2.06 & 4.63 & (34.86) & 1.91 & (2.83) & 2.05 & (0.26) & 2.77 & (11.97) \\
Sum of months in ALMPs one year before UE spell  & 0.15 & 0.15 & (0.39) & 0.07 & (12.76) & 0.14 & (0.66) & 0.19 & (5.28) \\
Sum of months in ALMPs five year before UE spell  & 1.07 & 1.23 & (5.72) & 0.94 & (5.01) & 1.34 & (9.60) & 1.64 & (19.46) \\
Sum of months in ALMPs ten year before UE spell  & 1.68 & 1.92 & (6.83) & 1.65 & (0.75) & 2.17 & (12.99) & 2.58 & (23.20) \\
Waiting days & 4.67 & 3.58 & (24.89) & 5.15 & (9.91) & 4.83 & (3.35) & 3.93 & (16.21) \\
First year in the individual income accounts register & 1996 & 1996 & (6.97) & 1995 & (12.92) & 1996 & (7.99) & 1995 & (10.44)\\
Min 1 month missing in payment system for UE insurance funds & 0.00 & 0.00 & (0.00) & 0.01 & (16.27) & 0.02 & (22.33) & 0.05 & (33.64) \\
Missing for 1 year in the individual income accounts register & 0.29 & 0.30 & (2.79) & 0.29 & (0.64) & 0.31 & (2.83) & 0.31 & (3.69) \\
Job effort assessment & 0.33 & 0.41 & (7.66) & 0.25 & (7.48) & 0.30 & (2.39) & 0.44 & (10.50) \\
Work experience (last job) & 2.56 & 2.54 & (2.81) & 2.62 & (7.03) & 2.64 & (9.95) & 2.54 & (2.15) \\ 
Degree of employment (last job) & 89.98 & 89.12 & (4.72) & 91.67 & (9.73) & 91.14 & (6.48) & 89.06 & (4.99) \\
Avg. monthly sickness days (last UE spell) & 0.15 & 0.16 & (2.02) & 0.13 & (2.23) & 0.12 & (5.34) & 0.20 & (10.41)\\
Avg. monthly sanction days (last UE spell) & 0.63 & 0.60 & (2.06) & 0.60 & (1.61) & 0.43 & (12.58) & 0.64 & (0.19) \\
Last job function (other) & 0.01 & 0.01 & (0.17) & 0.01 & (2.73) & 0.01 & (3.01) & 0.01 & (0.18) \\
Last job function (auxiliary) & 0.20 & 0.30 & (23.40) & 0.22 & (6.11) & 0.23 & (8.03) & 0.35 & (35.28) \\
Last job function (management) & 0.09 & 0.04 & (18.38) & 0.09 & (0.09) & 0.07 & (7.71) & 0.04 & (16.99) \\
Last job function (technical) & 0.71 & 0.65 & (11.90) & 0.68 & (4.81) & 0.70 & (2.11) & 0.59 & (23.84) \\
Full-time job, not terminated  & 0.00 & 0.00 & (1.36) & 0.00 & (2.02) & 0.00 & (1.96) & 0.00 & (0.70) \\
Full-time job, terminated & 0.19 & 0.19 & (1.25) & 0.23 & (10.42) & 0.20 & (1.63) & 0.29 & (22.42) \\ 
Looking for a job (first time)  & 0.00 & 0.00 & (0.60) & 0.00 & (0.72) & 0.00 & (2.90) & 0.00 & (4.24)  \\
Partially UE, not terminated  & 0.01 & 0.06 & (27.76) & 0.00 & (5.28) & 0.01 & (3.71) & 0.01 & (2.53) \\
Partially UE, terminated, limited & 0.04 & 0.13 & (30.90) & 0.04 & (1.66) & 0.04 & (1.95) & 0.06 & (7.76) \\
Previously employed & 0.75 & 0.62 & (28.49) & 0.71 & (7.94) & 0.75 & (0.47) & 0.64 & (23.66) \\
Re-entering the labor force  & 0.00 & 0.00 & (1.06) & 0.00 & (0.65) & 0.00 & (2.92) & 0.00 & (0.36) \\
\midrule 
\textbf{Sector of the last job}  &&&&&&&&&\\
\midrule
Agricultural and forestry & 0.01 & 0.02 & (1.47) & 0.01 & (5.80) & 0.01 & (4.51) & 0.01 & (1.68) \\
Production in industry and trade & 0.10 & 0.11 & (6.36) & 0.14 & (12.85) & 0.11 & (4.31) & 0.14 & (15.05) \\
Technical and information technolog & 0.09 & 0.06 & (13.13) & 0.08 & (3.27) & 0.11 & (8.49) & 0.07 & (8.15) \\
Construction and mining & 0.08 & 0.11 & (9.09) & 0.05 & (14.11) & 0.05 & (11.96) & 0.07 & (3.72) \\
Transport and trade & 0.16 & 0.14 & (5.22) & 0.19 & (8.32)  & 0.18  & (3.70) & 0.17 & (0.81)\\
Hospitality and personal services  & 0.13 & 0.20 & (19.51) & 0.12 & (4.45) & 0.14 & (3.20) & 0.18 & (13.75) \\
Business, finance and law & 0.27 & 0.17 & (24.01) & 0.29 & (4.97) & 0.30 & (6.28) & 0.20 & (16.20) \\
Education, art and science & 0.13 & 0.15 & (6.47) & 0.09 & (14.10) & 0.07 & (18.47) & 0.11 & (6.25)\\
Non-classifiable & 0.03 & 0.04 & (5.03) & 0.04 & (7.29) & 0.03 & (1.55) & 0.05 & (10.68) \\
\midrule 
Number of observations & 38,464 & \multicolumn{2}{c}{22,761} &\multicolumn{2}{c}{16,690}  &\multicolumn{2}{c}{5,508}  & \multicolumn{2}{c}{4,761}  \\
 \bottomrule
\end{tabular}
    \begin{tablenotes}
        \small \item \textit{Note:} This table shows the mean and the standardised difference in comparison to the non-participants for some covariates by programmes for the Permanent Residents. WS: Wage Subsidy, BC: Basic Course, TC: Technical Course, EP: Employment Programme, NP: No programme. 
        UE: unemployment.
        IV: disability insurance benefits, SH: social assistance benefits, HV: survivors' insurance benefits, HE: additional benefit for major disabilities.   
        Waiting days: number of days before receiving the benefit in the first month of unemployment. Job effort assessment: -1: not available, 1:sufficient, 2:insufficient, 3: absent. Work experience (last job): 3: more than 3 years, 2: from 1 to 3 years, 1: less than 1 year, 0: no experience.
    \end{tablenotes}
    \end{threeparttable}
    \end{adjustbox}
\end{table}

\begin{table}[htbp!]\centering
    \caption{Descriptive statistics by programme: Covariates TEMP sample (1)}
    \label{table: descr_b1}
    \begin{adjustbox}{max width=0.88\textwidth}
    \begin{threeparttable}
    \begin{tabular}{lrrrrrrrrrr} \toprule
 \textbf{Variable} & \multicolumn{1}{c}{NP} & \multicolumn{2}{c}{WS} & \multicolumn{2}{c}{BC} & \multicolumn{2}{c}{TC} & \multicolumn{2}{c}{EP} \\
        & Mean & Mean & Std. diff. & Mean & Std. diff. & Mean & Std. diff. & Mean & Std. diff. \\ \midrule
\textbf{Sociodemographics} & & & & & & & & & \\ \midrule
Age & 36.36 & 36.96 & (7.58) & 36.56 & (2.51) & 36.34 & (0.35) & 36.39 & (0.34)\\
Female & 0.36 & 0.37 & (2.66) & 0.42 & (12.84) & 0.36 & (0.14) & 0.36 & (1.26)\\
Compulsory edu. & 0.36 & 0.47 & (20.98) & 0.33 & (8.03) & 0.41 & (9.73) & 0.50 & (26.69)\\
Secondary edu. & 0.37 & 0.41 & (8.75) & 0.42 & (9.89) & 0.33 & (7.52) & 0.35 & (2.99)\\
Tertiary edu. & 0.03 & 0.02 & (7.53) & 0.04 & (1.68) & 0.02 & (6.48) & 0.02 & (5.19)\\
Bachelor &  0.04 & 0.02 & (10.58) & 0.05 & (5.47) & 0.04 & (1.20) & 0.02 & (9.20)\\
Post-Grad. &  0.19 & 0.08 & (34.00) & 0.17 & (6.41) & 0.19 & (0.84) & 0.10 & (26.08)\\
Proficiency local language (1 to 7) & 4.46 & 3.91 & (23.57) & 4.62 & (6.73) & 2.92 & (66.04) & 3.94 & (22.62)\\
Best other languages (1 to 7) & 0.85 & 0.77 & (4.81) & 0.86 & (0.73) & 0.74 & (6.59) & 0.64 & (12.29)\\
English level (1 to 7) & 2.55 & 1.59 & (39.98) & 2.42 & (5.19) & 2.14 & (15.77) & 1.79 & (30.91)\\
German level (1 to 7) & 1.71 & 1.09 & (22.72) & 1.64 & (2.10) & 0.32 & (59.90) & 0.85 & (33.13)\\ 
French level (1 to 7) & 2.03 & 1.76 & (11.06) & 1.99 & (1.48) & 1.76 & (11.02) & 1.73 & (12.17)\\
Italian level (1 to 7)  & 1.09 & 1.18 & (3.90) & 0.96 & (5.94) & 1.37 & (11.74) & 1.54 & (18.88)\\ 
Neighbouring countries & 0.39 & 0.29 & (21.22) & 0.38 & (0.56) & 0.25 & (30.23) & 0.27 & (24.98)\\
Rest of European Union & 0.39 & 0.42 & (6.47) & 0.32 & (14.06) & 0.40 & (3.51) & 0.38 & (1.11)\\
Rest of Europe & 0.06 & 0.09 & (10.05) & 0.09 & (13.13) & 0.09 & (12.84) & 0.09 & (11.52)\\
Rest of the world & 0.17 & 0.21 & (10.80) & 0.20 & (9.02) & 0.25 & (21.74) & 0.26 & (22.60)\\
Single &  0.39 & 0.32 & (14.81) & 0.36 & (6.17) & 0.33 & (11.98) & 0.32 & (14.20)\\
Married  & 0.52 & 0.58 & (11.37) & 0.54 & (3.33) & 0.59 & (14.36) & 0.59 & (13.12)\\
Widow & 0.00 & 0.01 & (2.95) & 0.00 & (1.20) & 0.00 & (4.07) & 0.01 & (3.29)\\
Divorced & 0.08 & 0.10 & (4.23) & 0.10 & (4.34) & 0.07 & (4.29) & 0.08 & (0.13)\\
Age youngest child & 3.54 & 3.58 & (3.57) & 3.52 & (2.37) & 3.59 & (4.54) & 3.52 & (1.69)\\
Num. of kids & 0.11 & 0.09 & (5.47) & 0.11 & (0.54) & 0.08 & (9.11) & 0.12 & (1.97)\\
Num. allowances per kid & 0.21 & 0.15 & (10.63) & 0.24 & (5.18) & 0.28 & (10.32) & 0.31 & (15.08)\\
Pregnant  & 0.02 & 0.01 & (7.31) & 0.02 & (1.74) & 0.01 & (6.86) & 0.01 & (5.88)\\
Maternity & 0.01 & 0.01 & (0.03) & 0.00 & (1.80) & 0.00 & (1.14) & 0.01 & (2.89)\\
Highest degree disability (prev. 2 years, \%) & 0.09 & 0.07 & (0.79) & 0.09 & (0.12) & 0.08 & (0.24) & 0.16 & (1.97)\\ \midrule

\textbf{Cantons} & & & & & & & & & \\ \midrule
Argau & 0.04 & 0.06 & (7.67) & 0.16 & (38.79) & 0.01 & (22.68) & 0.00 & (27.57)\\
Appenzell Innerrhoden & 0.00 & 0.00 & (1.08) & 0.00 & (3.66) & 0.00 & (3.66) & 0.00 & (3.66)\\
Appenzell Ausserrhoden & 0.00 & 0.00 & (1.94) & 0.00 & (2.32) & 0.00 & (3.56) & 0.00 & (0.63)\\
Bern & 0.08 & 0.08 & (0.07) & 0.00 & (39.18) & 0.01 & (32.68) & 0.30 & (57.33)\\
Basel Land & 0.02 & 0.02 & (3.20) & 0.02 & (0.45) & 0.02 & (2.53) & 0.00 & (13.05)\\
Basel Stadt & 0.04 & 0.03 & (3.77) & 0.02 & (11.56) & 0.03 & (2.27) & 0.01 & (20.46)\\
Freiburg-Fribourg & 0.03 & 0.04 & (2.04) & 0.04 & (3.85) & 0.02 & (10.06) & 0.06 & (11.64)\\
Geneva & 0.08 & 0.08 & (1.98) & 0.07 & (4.27) & 0.13 & (14.55) & 0.04 & (17.06)\\
Glarus & 0.00 & 0.00 & (0.74) & 0.00 & (2.38) & 0.00 & (1.19) & 0.01 & (7.64)\\
Graubünden & 0.06 & 0.04 & (12.14) & 0.01 & (30.44) & 0.05 & (7.44) & 0.06 & (1.66)\\
Jura & 0.01 & 0.00 & (1.08) & 0.00 & (7.77) & 0.00 & (1.66) & 0.01 & (8.66)\\
Luzern & 0.02 & 0.03 & (5.23) & 0.04 & (11.03) & 0.02 & (0.51) & 0.03 & (0.48)\\
Neuchâtel & 0.02 & 0.03 & (5.02) & 0.03 & (4.31) & 0.03 & (4.38) & 0.01 & (8.67)\\
Nidwalden & 0.00 & 0.00 & (1.89) & 0.00 & (0.78) & 0.00 & (1.42) & 0.01 & (6.44)\\
Obwalden & 0.00 & 0.00 & (0.40) & 0.00 & (1.21) & 0.00 & (1.16) & 0.00 & (4.41)\\
Sankt Gallen & 0.03 & 0.05 & (7.36) & 0.06 & (12.21) & 0.06 & (13.81) & 0.02 & (7.97)\\
Schaffhausen  & 0.01 & 0.01 & (0.68) & 0.01 & (5.63) & 0.02 & (10.00) & 0.00 & (7.79)\\
Solothurn  & 0.01 & 0.02 & (5.68) & 0.03 & (14.97) & 0.00 & (8.33) & 0.07 & (29.62)\\
Schwyz & 0.01 & 0.01 & (0.32) & 0.02 & (8.12) & 0.00 & (6.18) & 0.01 & (3.10)\\
Turgau & 0.03 & 0.03 & (2.14) & 0.04 & (5.33) & 0.02 & (2.50) & 0.01 & (9.41)\\
Ticino & 0.04 & 0.05 & (8.02) & 0.05 & (7.71) & 0.06 & (12.75) & 0.12 & (30.50)\\
Uri & 0.00 & 0.01 & (0.91) & 0.00 & (1.64) & 0.00 & (9.80) & 0.01 & (6.11)\\
Vaud & 0.15 & 0.13 & (5.30) & 0.15 & (1.48) & 0.19 & (12.98) & 0.11 & (9.67)\\
Valais & 0.09 & 0.09 & (1.77) & 0.05 & (14.78) & 0.06 & (8.12) & 0.08 & (2.71)\\
Zug & 0.01 & 0.01 & (2.58) & 0.01 & (3.08) & 0.00 & (7.99) & 0.00 & (7.75)\\
Zurich & 0.22 & 0.18 & (7.67) & 0.19 & (6.82) & 0.23 & (3.79) & 0.03 & (60.26)\\
\midrule
Number of observations & 10,445 & \multicolumn{2}{c}{7,177}  &\multicolumn{2}{c}{4,037}  &\multicolumn{2}{c}{2,850}   &\multicolumn{2}{c}{1,597}\\ \bottomrule 
\end{tabular}
    \begin{tablenotes}
        \small \item \textit{Note:} This table shows the mean and the standardised difference in comparison to the non-participants for some covariates by programmes for the Temporary Residents sample.WS: Wage Subsidy, BC: Basic Course, TC: Technical Course, EP: Employment Programe, NP: No programme.  Countries with Swiss languages: Austria, France, Germany, Italy, Liechtenstein; EU: Belgium, Bulgaria, Croatia, Cyprus, Serbia, Czech Republic, Denmark, Finland, Estonia, Latvia, Lithuania,  Greece, Hungary, Ireland, Malta, Netherlands, Poland, Portugal, Romania, Slovakia, Slovenia, Spain, Sweden, Rest of Europe: Albania, Andorra, Kosovo, Belarus, Bosnia and Herzegovina, Georgia, Iceland, Moldova, Norway, Serbia, Montenegro, Ukraine, Rest of the world: countries outside Europe. Language proficiency: 1:basic to 7:mother tongue.
    \end{tablenotes}
    \end{threeparttable}
    \end{adjustbox}
\end{table}

\begin{table}[htbp!]\centering
    \caption{Descriptive statistics by programme: Covariates TEMP sample (2)}
    \label{table: descr_b2}
    \begin{adjustbox}{max width=0.88\textwidth}
    \begin{threeparttable}
    \begin{tabular}{lrrrrrrrrrr} \toprule
  \textbf{Variable} & \multicolumn{1}{c}{NP} & \multicolumn{2}{c}{WS} & \multicolumn{2}{c}{BC} & \multicolumn{2}{c}{TC} & \multicolumn{2}{c}{EP} \\
        & Mean & Mean & Std. diff. & Mean & Std. diff. & Mean & Std. diff. & Mean & Std. diff. \\ \midrule
 
\textbf{Job related and macro} & & & & & & & & & \\ \midrule
Desired degree of employment & 97.14 & 97.19 & (0.51) & 97.13 & (0.03) & 98.35 & (12.74) & 97.28 & (1.38)\\
Skill requirments: none & 0.07 & 0.08 & (2.17) & 0.08 & (2.84) & 0.10 & (8.77) & 0.04 & (14.27)\\
Skill requirments: basic & 0.88 & 0.88 & (1.35) & 0.87 & (3.01) & 0.86 & (7.00) & 0.92 & (12.97)\\
Skill requirments: professional & 0.04 & 0.04 & (0.67) & 0.05 & (1.10) & 0.04 & (0.55) & 0.04 & (2.74)\\
Employability (1 to 3) & 1.83 & 1.89 & (11.69) & 1.90 & (13.56) & 1.93 & (18.51) & 2.02 & (36.63)\\
Population municipality & 67,584 & 57,864 & (9.23) & 59,976 & (6.96) & 74,478 & (6.24) & 33,609 & (38.31)\\
Open sector positions in cant.(per 100k) & 10.00 & 9.40 & (2.94) & 8.96 & (5.04) & 8.51 & (8.60) & 7.33 & (15.92)\\
UE rate region (UE spell) & 3.66 & 3.59 & (5.07) & 3.51 & (11.19) & 3.83 & (12.52) & 3.27 & (27.71)\\
UE rate region yoy (UE spell) & -0.08 & -0.08 & (1.78) & -0.09 & (2.82) & -0.10 & (5.66) & -0.10 & (6.26)\\
Labor force (region level) & 84,539 & 86,405 & (3.14) & 95,976 & (18.91) & 86,026& (2.53) & 69,123 & (27.36) \\
Urban level (1 to 3) & 1.41 & 1.38 & (3.53) & 1.36 & (7.00) & 1.31 & (14.74) & 1.47 & (8.22)\\
Availab. to commute: also abroad & 0.02 & 0.00 & (13.46) & 0.01 & (10.79) & 0.01 & (9.37) & 0.01 & (2.96)\\
Availab. to commute: every day  & 0.94 & 0.96 & (6.59) & 0.96 & (10.09) & 0.97 & (11.74) & 0.90 & (14.11)\\
Availab. to commute: in parts of Sw. & 0.02 & 0.02 & (1.23) & 0.01 & (2.99) & 0.02 & (1.76) & 0.04 & (13.36)\\
Availab. to commute: not mobile & 0.01 & 0.01 & (1.17) & 0.01 & (0.20) & 0.01 & (4.43) & 0.01 & (0.21)\\
Availab. to commute: throughout Sw. & 0.01 & 0.01 & (3.98) & 0.01 & (7.09) & 0.00 & (8.74) & 0.03 & (12.91)\\
 \midrule 
 \textbf{Labour market history} & & & & & & & & & \\ \midrule
Sum of months in ue five years before UE spell & 4.76 & 5.64 & (11.17) & 3.02 & (25.75) & 2.76 & (29.85) & 5.27 & (6.70)\\
Sum of months in ue ten years before UE spell & 6.16 & 7.38 & (11.27) & 4.38 & (18.71) & 3.61 & (27.68) & 6.99 & (7.91)\\
Sum of months in emp. five years before UE spell & 36.95 & 35.76 & (8.27) & 37.03 & (0.59) & 34.13 & (18.99) & 36.00 & (6.67)\\
Sum of months in emp. ten years before UE spell & 48.20 & 47.08 & (4.14) & 46.79 & (5.17) & 40.85 & (28.05) & 47.32 & (3.27)\\
Average of earnings three months before UE spell & 5216.83 & 4149.55 & (47.62) & 4827.81 & (15.39) & 4791.90 & (17.00) & 4243.75 & (41.46)\\
Sum of earnings five years before UE spell & 198,733 & 145,381 & (31.17) & 184,143 & (7.07) & 172,006 & (13.46) & 147,059 & (29.75)\\
Sum of earnings ten years before UE spell & 246,858 & 186,130 & (27.52) & 224,138 & (8.92) & 201,238 & (18.18) & 188,317 & (25.92)\\
Average earnings three months before the UE spell & 5,967 & 4,131 & (19.60) & 5,310 & (5.98) & 5,350 & (6.07) & 4,322 & (15.97)\\
Sum of ue benefits five years before UE spell & 16,716 & 194,689 & (8.91) & 9,386 & (28.12) & 8,394 & (32.31) & 167,378 & (0.08)\\
Sum of ue benefits ten years before UE spell & 20,750 & 24,144 & (8.56) & 12,704 & (23.62) & 10,403 & (31.43) & 21,219 & (1.24)\\
Sum of months in supplementary benefits two years before UE spell  & 0.14 & 0.51 & (33.08) & 0.06 & (13.10) & 0.10 & (6.16) & 0.18 & (5.02)\\
Supplementary benefits at the time of UE spell & 0.00 & 0.00  & (1.87)& 0.00  & (1.68) & 0.00  & (0.13) & 0.00  & (2.48)\\
Sum of months in IV two years before UE spell & 0.01 & 0.01 & (0.80) & 0.02 & (1.38) & 0.01 & (0.10) & 0.01 & (0.47)\\
Sum of months in SH two years before UE spell & 0.55 & 0.75 & (6.36) & 0.53 & (0.73) & 0.50 & (1.96) & 0.82 & (8.38)\\
Sum of months in HV two years before UE spell & 0.02 & 0.02 & (0.09) & 0.03 & (0.38) & 0.02 & (0.20) & 0.05 & (2.91)\\
Sum of months in HE two years before UE spell (last 2 years) & 0.00 & 0.00 & (0.00) & 0.00 & (0.00) & 0.00 & (0.00) & 0.00 & (0.00)\\
Sum of months in WS one year before UE spell & 0.14 & 0.51 & (33.08) & 0.06 & (13.10) & 0.10 & (6.16) & 0.18 & (5.02)\\
Sum of months in WS five year before UE spell & 0.80 & 2.13 & (34.13) & 0.64 & (6.57) & 0.55 & (10.10) & 1.09 & (9.55)\\
Sum of months in WS ten year before UE spell & 1.05 & 2.58 & (31.53) & 0.88 & (5.08) & 0.73 & (9.76) & 1.37 & (8.41)\\
Sum of months in ALMPs one year before UE spell & 0.18 & 0.21 & (3.29) & 0.07 & (15.73) & 0.14 & (4.57) & 0.27 & (10.55)\\
Sum of months in ALMPs five year before UE spell & 0.89 & 1.06 & (7.09) & 0.67 & (10.10) & 0.74 & (6.51) & 1.42 & (19.93)\\
Sum of months in ALMPs ten year before UE spell & 1.10 & 1.32 & (7.67) & 0.91 & (7.16) & 0.90 & (7.40) & 1.76 & (20.12)\\
Waiting days  & 4.14 & 2.60 & (38.15) & 4.14 & (0.08) & 3.93 & (4.69) & 2.73 & (34.00)\\
First-year in the individual income accounts register & 2008 & 2007 & (1.96) & 2008 & (6.60) & 2009 & (25.75) & 2007 & (1.93)\\
Min 1 month missing in payment system for UE insurance funds & 0.00 & 0.00 & (0.00) & 0.02 & (19.32) & 0.03 & (24.49) & 0.06 & (36.73)\\
Missing for 1 year in individual income accounts register & 0.17 & 0.18 & (2.33) & 0.17 & (0.72) & 0.13 & (11.95) & 0.18 & (1.74)\\
Job effort assessment 
& 0.21 & 0.35 & (12.12) & 0.05 & (13.88) & 0.02 & (15.79) & 0.35 & (12.51)\\
Work experience (last job) & 2.48 & 2.39 & (10.48) & 2.46 & (2.41) & 2.38 & (12.23) & 2.32 & (18.90)\\
Degree of employment (last job) & 93.98 & 92.12 & (12.22) & 93.62 & (2.42) & 94.20 & (1.48) & 93.42 & (3.71)\\
Avg. monthly sickness days (last UE spell) & 0.09 & 0.10 & (1.32) & 0.06 & (7.22) & 0.06 & (6.84) & 0.11 & (4.95)\\
Avg. monthly sanction days (last UE spell) & 0.40 & 0.38 & (1.60) & 0.30 & (7.70) & 0.24 & (12.21) & 0.41 & (0.76)\\
Last job function (other)  & 0.01 & 0.00 & (2.31) & 0.00 & (3.65) & 0.00 & (7.95) & 0.01 & (0.80)\\
Last job function (auxiliary) & 0.35 & 0.49 & (29.06) & 0.37 & (4.63) & 0.42 & (14.73) & 0.58 & (46.42)\\
Last job function (management) & 0.05 & 0.01 & (22.20) & 0.06 & (3.06) & 0.03 & (9.46) & 0.01 & (22.74)\\
Last job function (technical) & 0.59 & 0.49 & (20.59) & 0.57 & (5.42) & 0.55 & (9.75) & 0.41 & (38.09)\\
Full-time job, not terminated & 0.00 & 0.00 & (0.94) & 0.00 & (2.36) & 0.00 & (2.98) & 0.00 & (5.48)\\
Full-time job, terminated & 0.17 & 0.17 & (1.37) & 0.22 & (11.56) & 0.19 & (5.49) & 0.28 & (26.77)\\
Looking for a job (first time) & 0.00 & 0.00 & (2.44) & 0.00 & (0.46) & 0.01 & (1.24) & 0.00 & (0.03)\\
Partially ue, not terminated & 0.01 & 0.05 & (26.03) & 0.01 & (0.60) & 0.00 & (2.05) & 0.01 & (1.58)\\
Partially ue, terminated, limited & 0.03 & 0.11 & (31.55) & 0.02 & (4.01) & 0.02 & (4.07) & 0.03 & (1.42)\\
Previously employed & 0.79 & 0.67 & (25.90) & 0.75 & (9.66) & 0.77 & (3.81) & 0.67 & (26.62)\\
Re-entering the labor force  & 0.00 & 0.00 & (1.40) & 0.00 & (1.89) & 0.00 & (0.92) & 0.00 & (3.09)\\
\midrule
\textbf{Sector of the last job}  &&&&&&&&&\\
\midrule
Agricultural and forestry & 0.03 & 0.03 & (3.16) & 0.02 & (2.57) & 0.03 & (4.67) & 0.03 & (2.44)\\
Production in industry and trade  & 0.09 & 0.11 & (8.41) & 0.13 & (14.62) & 0.12 & (9.50) & 0.12 & (11.19)\\
Technical and information technology  & 0.08 & 0.04 & (18.22) & 0.08 & (0.91) & 0.08 & (0.25) & 0.06 & (7.77)\\
Construction and mining  & 0.20 & 0.25 & (11.33) & 0.11 & (25.19) & 0.19 & (3.00) & 0.17 & (7.79)\\
Transport and trade  & 0.11 & 0.08 & (7.39) & 0.15 & (13.51) & 0.11 & (1.58) & 0.10 & (3.59)\\
Hospitality and personal services  & 0.25 & 0.33 & (17.39) & 0.25 & (0.16) & 0.27 & (3.48) & 0.35 & (21.33)\\
Business, finance and law & 0.11 & 0.05 & (23.94) & 0.13 & (5.30) & 0.10 & (3.57) & 0.05 & (23.69)\\
Education, art and science & 0.10 & 0.07 & (9.98) & 0.07 & (10.21) & 0.07 & (14.56) & 0.06 & (10.40)\\
Non-classifiable  & 0.03 & 0.03 & (0.67) & 0.05 & (9.84) & 0.03 & (3.45) & 0.05 & (10.14)\\
\midrule
Number of observations & 10,445 & \multicolumn{2}{c}{7,177}  &\multicolumn{2}{c}{4,037}  &\multicolumn{2}{c}{2,850}   &\multicolumn{2}{c}{1,597}\\ \bottomrule 
\end{tabular}
    \begin{tablenotes}
        \small \item \textit{Note:} This table shows the mean and the standardised difference in comparison to the non-participants for some covariates by programmes for the Temporary Residents sample. WS: Wage Subsidy, BC: Basic Course, TC: Technical Course, EP: Employment Programme, NP: No programme.  UE: unemployment.
        IV: disability insurance benefits, SH: social assistance benefits, HV: survivors' insurance benefits, HE: additional benefit for major disabilities.   
        Waiting days: number of days before receiving the benefit in the first month of unemployment. Job effort assessment: -1: not available, 1:sufficient, 2:insufficient, 3:absent. Work experience (last job): 3: more than 3 years, 2: from 1 to 3 years, 1: less than 1 year, 0: no experience.
    \end{tablenotes}
    \end{threeparttable}
    \end{adjustbox}
\end{table}

\clearpage

\section{Appendix: Estimation Procedure} \label{Appendix_estimator}
\subsection{Modified Causal Forest Estimator} \label{MCF}

We apply the Modified Causal Forest (MCF) estimator proposed by \cite{lechner2022modified}. The estimator builds on the Causal Forest estimator of \cite{wager2018estimation} by improving the splitting criterion proposed, reducing the selection bias in the early splits of the trees. In particular, the MCF splitting procedure targets the minimisation of the mean squared error (MSE) of the IATE directly, i.e. $$\operatorname{MSE}(\widehat{\operatorname{IATE}}(m, l ; x))=\operatorname{MSE}\left(\hat{\mu}_m(x)\right)+\operatorname{MSE}\left(\hat{\mu}_l(x)\right)-2 \operatorname{MCE}\left(\hat{\mu}_m(x), \hat{\mu}_l(x)\right)\text{,}$$
where $\hat{\mu}_d(x)$ are estimates of the conditional means of the potential outcomes $\hat{\mu}_d(x) = \hat \E [Y^d|X=x]$.
This decomposition suggests that there may be an advantage of finding splits that would tie the estimators together in a way such that the correlation of their estimation errors in the daughter leaves becomes positive to some extent.
 
Additionally, a penalty term is added to the parameter MSE to penalise possible splits if the treatment probabilities in the two daughter leaves are similar. The motivation for the penalty term is to reduce bias in the estimation of the parameter MSE caused by non-random allocation into treatment. The combination of the parameter MSE and the penalty term leads to splits that are predictive for the outcome and the conditional treatment assignment probabilities. 

To find the best split at a given node of a tree, the MSE and MCE (mean correlated error) need to be estimated in the daughter leaves. The within-leaf MSE is estimated as the mean squared difference between the observed and the mean outcomes of individuals with the same treatment in this leaf. The within-leaf MCE is computed using the closest (in terms of features) neighbour in the other treatment group. 
Once the sample splits are obtained, the treatment effect is locally calculated within each leaf as the difference in the mean outcomes between the different treatments. Moreover, the estimator provides a unified inference procedure across various aggregation levels. The inference procedure leverages that forest estimators can be viewed as a weighted average of observed outcomes. After obtaining weights for the lowest aggregation level, they can be aggregated to conduct inference at the GATE or ATE levels.
The modified causal forest is provided in the \texttt{mcf} Python package described in \cite*{bodory2022high}. This work uses \texttt{mcf} version 0.5.0. The analysis applies the default parameters, resulting in 1,000 causal trees, 87 covariates for the splits and a minimum leaf size of 12.

\subsection{Policy Tree Estimator} \label{appendix_policytree}
The policy tree implemented in the \texttt{mcf} package is a policy learner based on the algorithm proposed by \cite{zhou2023offline} but introduces three key modifications: (1) a different global approximation parameter, (2) the implementation of constraints, and (3) a different handling of categorical variables. Firstly, the approximation parameter in \cite{zhou2023offline} enhances computational efficiency, allowing the skipping of splitting points at various tree levels. The MCF-policy tree advances by utilising a finer grid at higher levels, resulting in longer computational times but more precise splits. Secondly, capacity constraints are integrated by deriving treatment-specific cost values regarding the outcome, which are then subtracted from the policy scores and, therefore, do not depend on the policy class. These costs can also be defined as constraints on the treatment shares. Lastly, the MCF-policy tree diverges from conventional approaches in the literature by not necessitating one-hot encoding for categorical variables. This deviation allows consideration of all potential splits, facilitating a more comprehensive exploration of the data space. However, the number of combinations can be high for large categorical and continuous variables, making computations intensive. In such cases, only a subset of combinations will be considered. A detailed explanation of the algorithm is provided in \cite{bodory2024enabling}.

This analysis employs optimal trees with a depth of 2 (4 final leaves), 3 (8 final leaves) and  4 (16 final leaves) and sequential deeper trees with depth-4+1 (32 final leaves), 4+2 (64 final leaves) and 4+3 (128 final leaves), and the default value for the approximation parameter in both samples. 
The original sample is partitioned into three subsamples: 40\% is allocated for training the MCF, and another 40\% for estimating the effects and policy scores. This sample is additionally used to find the optimal policy rules. The remaining 20\% of the sample is used for out-of-sample evaluation of the policy trees. This splitting of the data helps to estimate the treatment effect efficiently and to mitigate the risk of overfitting.

\subsection{Common Support} \label{Common Support}
Among the advantages of using CML estimators is their ability to handle many covariates, increasing the probability of controlling for or proxying unmeasured confounders and, therefore, sustaining the CIA. On the other hand, a large covariate set makes it harder to satisfy the common support assumption. To cope with potential common support violations, we run our estimations adopting a min-max trimming rule. This rule finds the minimum and maximum treatment probability within each treatment group. Then, the largest minimum (smallest maximum) propensity is identified, and observations with lower (larger) propensities from the opposing treatment groups are deleted because they are considered out of support. As a robustness check, we also run the analysis enforcing alternative trimming rules where instead of the maximum and minimum the 99.9th and 0.01st quantiles of the propensity score distributions are used as thresholds. The propensity scores are computed using a Random Forest with the same specification and sub-sample used to build the causal trees.

\begin{table}[h!]\centering
    \caption{ATEs relative to no programme with different CS rules (PERM sample)}
    \begin{adjustbox}{max width=0.8\textwidth}
    \label{table:cs_robustness}
    \begin{threeparttable}
    \begin{tabular}{@{}lrrrrc@{}} \toprule
        \textbf{CS Rule} & \multicolumn{1}{c}{\textbf{WS}} & \multicolumn{1}{c}{\textbf{BC}} & \multicolumn{1}{c}{\textbf{TC}}    &\multicolumn{1}{c}{\textbf{EP}}& \textbf{\% of obs. deleted} \\ \midrule
        min/max - max/min & 0.74*** & -0.13*\phantom{**} &  -0.01\phantom{***}  & 0.18  &   \phantom{3}4.71\%  \\
        min/q99.9 - max/q0.01 & 0.73*** & -0.07\phantom{***}  &  \phantom{-}0.04\phantom{***}  & 0.10  &  38.27\% \\
     \bottomrule
    \end{tabular}
    \begin{tablenotes}
        \small \item \textit{Note:} The table shows the ATE relative to the non-participants. The outcome is the sum of months in employment in the third year after the program start, WS: Wage Subsidy, BC: Basic Courses, TC: Training Courses, EP: Employment Programme. *, **, *** indicate the precision of the estimate by showing whether the p-value of a two-sided significance test is below 10\%, 5\%, and 1\%, respectively.
      
    \end{tablenotes}
    \end{threeparttable}
    \end{adjustbox}
\end{table}

Table \ref{table:cs_robustness} shows the ATE estimates for the sum of months in employment in the third year after the programme start using the alternative trimming rule. Compared to the min-max rule (first row, identical to column (3) of Table \ref{table:ate_both}), a substantially larger proportion of observations is trimmed when applying the quantile-based trimming rule (second row). This indicates a suboptimal level of overlap, an issue commonly encountered when estimating propensity scores with a large number of covariates \citep{d2021overlap}. Despite the smaller sample size under the stricter trimming rule, the estimated effect sizes and significance levels remain similar. Therefore, the results for the larger sample are presented in the main text of the paper. Table \ref{sd_main_dropped}  provides additional descriptive statistics for key covariates of the observations excluded in the two analysed samples.

\begin{table}[htbp!]
\begin{adjustbox}{width=0.8\columnwidth, center}
\begin{threeparttable}
  \captionsetup{font=large}
    \caption{Descriptive statistics of selected variables for PERM (LHS) and TEMP (RHS) samples for observations out of common support.}
    \label{sd_main_dropped}
    \centering
    \begin{tabular}{@{}lrrrrrr@{}}
    \toprule
     && \multicolumn{2}{c}{\textbf{PERM}} & \multicolumn{2}{c}{\textbf{TEMP}} \\
    \midrule
    Variable & Mean main & Mean dropped & Std. Diff. &  Mean main & Mean dropped & Std. Diff. \\
    \midrule
    Age & 39.22 & 40.92 & 19.38    &  36.49 & 36.54 & 0.65\\
    Female  & 0.47 & 0.51 & 9.44        & 0.36 & 0.44 & 17.56\\
    Education  & 2.35 & 2.52 & 13.45      & 2.18 & 2.20 & 1.44\\
    Proficiency in local language  & 6.11 & 5.59 & 28.54      & 4.20 & 3.98 & 9.10\\
    Swiss & 0.67 & 0.62 & 9.69 &&& \\
    Neig. Countries & 0.12 & 0.11 & 2.08 & 0.35 & 0.33 & 4.48 \\
    Rest of European Union & 0.09 & 0.11 & 5.89 & 0.39 & 0.36 & 5.96 \\
    Rest of Europe & 0.05 & 0.05 & 0.27  & 0.08 & 0.08 & 0.99 \\
    Rest of the World & 0.07 & 0.11 & 12.41 & 0.19 & 0.23 & 11.61 \\
    Sum of months in emp. five years  & 51.02 & 51.56 & 4.79  & 36.28 & 36.18 & 0.66\\
    Sum of months in emp. ten years & 94.94 & 95.25 & 1.28  & 46.75 & 45.63 & 4.13\\
    Sum of earn. five years & 276,881 & 379,223 & 29.70 & 177,685 & 183,040 & 2.68\\
    Sum of earn. ten yearn & 485,670 & 651,274 & 27.83  & 220,303 & 219,115 & 0.49\\
    Permit type & 0.33 & 0.38 & 9.65 &&&\\ 
    Months in ALMPs five years bef. & 1.14 & 0.92 & 8.60 & 0.92 & 0.66 & 11.93 \\
    Months in WS five years before & 1.48 & 2.96 & 25.56  & 1.11 & 0.92 & 5.54\\
    Civil status & 1.87 & 1.92 & 5.09  & 1.82 & 1.80 & 2.26\\
    First-year register & 1996 & 1996 & 1.19 & 2008 & 2008 & 11.45\\
    \midrule
    N. of obs. & 86,395 & 4,268 &&  24,222 & 3,161  & \\
    \bottomrule
    \end{tabular}
    \begin{tablenotes}
\textit{Note:} The table shows the unconditional means and standardised differences for selected variables for the observation on support and the observation out of support.
\end{tablenotes}
\end{threeparttable}
\end{adjustbox}
\end{table}

\newpage
\section{Appendix: Further Results} \label{Appendix_results}
\subsection{Average Treatment Effects on the Treated}\label{ATET_app}
\begin{table}[h!]
\begin{adjustbox}{width=0.8\columnwidth, center}
\begin{threeparttable}
  \captionsetup{font=large}
    \caption{Average Treatment effects on the treated (ATET)}
    \label{ATET}
    \centering
    \begin{tabular}{@{}lrrrrrrrr@{}}
    \toprule
     && \multicolumn{3}{c}{\textbf{PERM}} & \multicolumn{4}{c}{\textbf{TEMP}} \\
     & \multicolumn{1}{c}{ATE} & \multicolumn{3}{c}{ATET} & \multicolumn{1}{c}{ATE} & \multicolumn{3}{c}{ATET}\\ \cmidrule(lr){2-2} \cmidrule(lr){3-5} \cmidrule(lr){6-6} \cmidrule(lr){7-9}
     Programme & \multicolumn{1}{c}{Effect} & \multicolumn{1}{c}{Effect} & \multicolumn{1}{c}{S.E.} & \multicolumn{1}{c}{p-value (\%)} & \multicolumn{1}{c}{Effect} & \multicolumn{1}{c}{Effect} & \multicolumn{1}{c}{S.E.} & \multicolumn{1}{c}{p-value (\%)} \\
    \midrule
    &  \multicolumn{8}{c}{\textbf{Outcome: Employment third year}} \\ 
    \midrule
    WS & 0.74*** & 0.75*** & (0.06) & 0.00   & 0.98*** & 0.95*** & (0.10) & 0.00\\
    BC & -0.13*\phantom{**} & 0.02\phantom{***} & (0.10) & 85.50     & 0.49*** & 0.59*** & (0.15) & 0.00\\
    TC & -0.01\phantom{***}& 0.16\phantom{***} & (0.12) & 18,29       & 0.47*** & 0.58*** & (0.17) & 0.07\\
    EP & 0.18\phantom{***}  & 0.26*\phantom{**} & (0.14) & 8.63       & 0.37*\phantom{**} & 0.47**\phantom{*} & (0.21) & 2.25\\
    \midrule
    &  \multicolumn{8}{c}{\textbf{Outcome: Earnings third year}} \\ 
    \midrule
    WS & 2,059*** & 2,101*** & (570) & 0.20      & 3,226*** & 2,709*** & (720) & 0.02\\
    BC & -2,347*** & -1,414\phantom{***} & (1,182) & 23.19   &  904\phantom{***} & 909\phantom{***} & (1,259) & 47.04\\
    TC & -1,131\phantom{***} & 403\phantom{***} & (1,314) & 75.90   &  1,737\phantom{***} & 1,865\phantom{***} & (1,480 ) & 20.75\\
    EP & -1,696\phantom{***} & -365\phantom{***} & (1,652) & 82.93   & -284\phantom{***} &  171\phantom{***} & (1,563) & 91.28\\
    \bottomrule
    \end{tabular}
    \begin{tablenotes}
       \textit{Note:} The table shows the ATE and the ATET with their respective standard errors. *, **, *** indicate the precision of the estimate by showing whether the p-value of a two-sided significance test is below 10\%, 5\%, and 1\% respectively. Standard errors and p-values for ATEs can be found in the main body of the paper.
    \end{tablenotes}
\end{threeparttable}
\end{adjustbox}
\end{table}

\clearpage
\subsection{Group Average Treatment Effects} \label{gates appendix}
\subsubsection{Permanent Residents Sample}

\begin{figure}[htbp!]
\captionsetup{font=small}  
\caption{Differences of GATEs to ATE of WS programme with respect to non-participants for PERM sample.}

\includegraphics[width=0.95\textwidth]{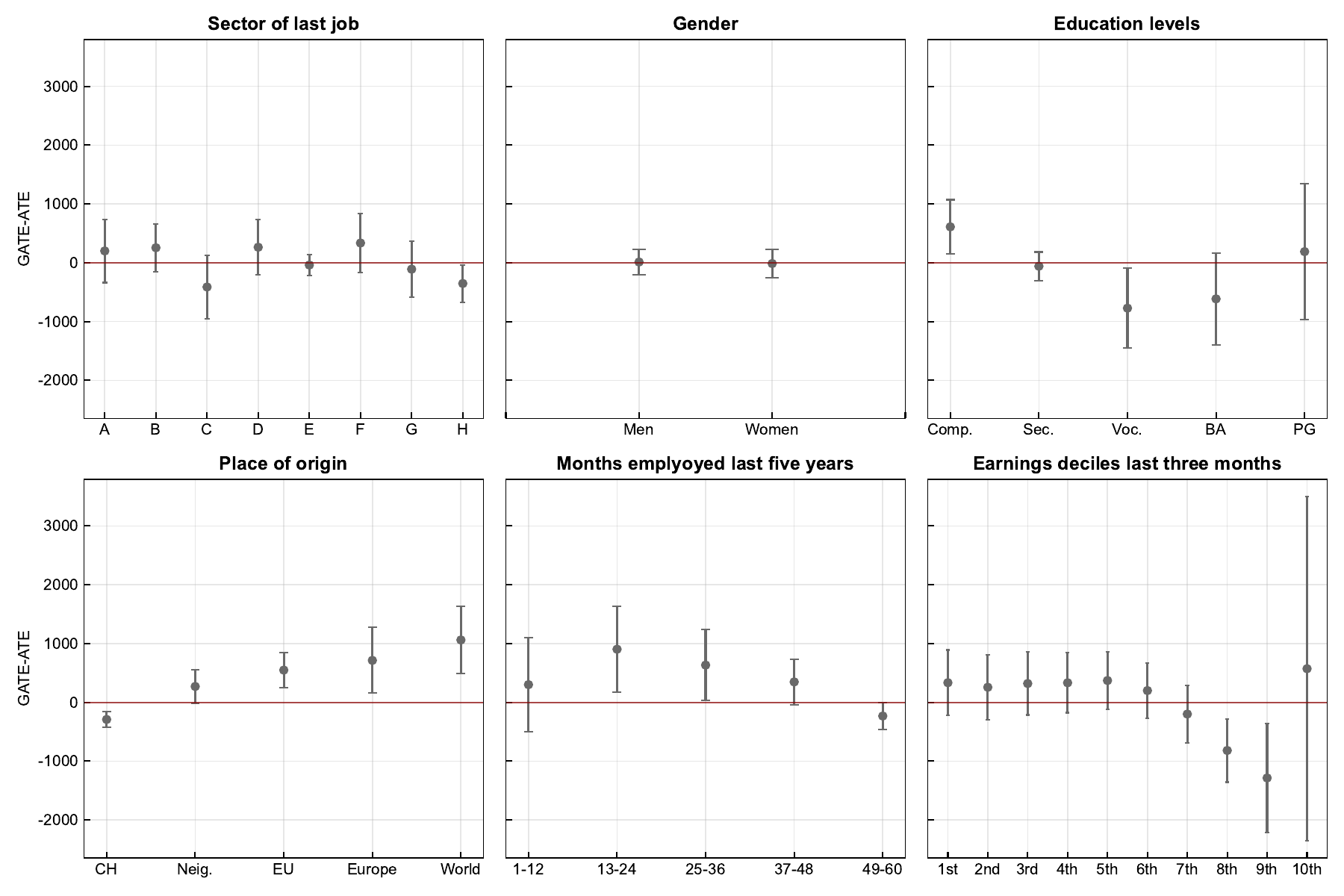}
\caption*{\textit{Note:} The vertical axes measure the difference of GATEs to the ATE of WS programme with respect to non-participation at the 95\% confidence interval. The horizontal axes show previous job sector (categorised as  A: agricultural and forestry, B: production in industry and trade, C: technical and information technology, D: construction and mining; E: transport and trade, F: hospitality and personal services, G: business, finance and Law, H: education, art and science), gender, education levels, place of origin (Switzerland, neighbouring countries, rest of European Union, rest of Europe and rest of the world), sum of months employed in the last five years and the average earnings within each decile of the earnings distribution in the three months before the unemployment spell. The outcome is the cumulative earnings in the third year after the programme start and the ATE for the WS is 2,059 Swiss Francs.}
\label{fig:gates_ates}
\end{figure}
\begin{table}[htbp!]
\begin{adjustbox}{width=0.9\columnwidth, center} 
\begin{threeparttable}
  \captionsetup{font=large}
    \caption{GATEs for last job sector, PERM sample}
    \label{table:gate_sector}
    \centering
    \begin{tabular}{@{}ll|rrrrrr|rrrrrrr@{}}
    \toprule
    \multicolumn{2}{l}{\textbf{Sector of last job}} & \multicolumn{6}{c}{\textbf{First Year}} & \multicolumn{6}{c}{\textbf{Third Year}} \\
    \midrule
     & Programme & GATE & S.E. & p-value & GATE-ATE & S.E. & p-value & GATE & S.E. & p-value & GATE-ATE & S.E. & p-value \\ \midrule
    &&  \multicolumn{12}{c}{\textbf{Outcome: Employment}} \\ \midrule
    A & WS vs NP & -0.62 & (0.06) & 0.00 & 0.02 & (0.03) & 0.44 & 0.76 & (0.07) & 0.00 & 0.02 & (0.02) & 0.37\\
    B & WS vs NP & -0.57 & (0.07) & 0.00 & 0.07 & (0.02) & 0.00 & 0.75 & (0.08) & 0.00 & 0.01 & (0.02) & 0.59\\
    C & WS vs NP & -0.73 & (0.07) & 0.00 & -0.09 & (0.03) & 0.00 & 0.66 & (0.07) & 0.00 & -0.08 & (0.03) & 0.00\\
    D & WS vs NP & -0.57 & (0.07) & 0.00 & 0.07 & (0.02) & 0.01 & 0.72 & (0.07) & 0.00 & -0.02 & (0.03) & 0.52\\
    E & WS vs NP & -0.64 & (0.07) & 0.00 & 0.01 & (0.01) & 0.62 & 0.75 & (0.07) & 0.00 & 0.01 & (0.01) & 0.35\\
    F & WS vs NP & -0.62 & (0.07) & 0.00 & 0.03 & (0.03) & 0.37 & 0.83 & (0.07) & 0.00 & 0.09 & (0.03) & 0.00\\
    G & WS vs NP & -0.67 & (0.07) & 0.00 & -0.03 & (0.02) & 0.15 & 0.70 & (0.07) & 0.00 & -0.04 & (0.02) & 0.06\\
    H & WS vs NP & -0.70 & (0.06) & 0.00 & -0.06 & (0.02) & 0.00 & 0.72 & (0.07) & 0.00 & -0.01 & (0.02) & 0.41\\
    \midrule
    A & BC vs NP & -1.70 & (0.08) & 0.00 & -0.01 & (0.03) & 0.64 & -0.11 & (0.10) & 0.27 & 0.02 & (0.03) & 0.59\\
    B & BC vs NP & -1.67 & (0.08) & 0.00 & 0.01 & (0.02) & 0.67 & -0.08 & (0.09) & 0.40 & 0.05 & (0.02) & 0.06\\
    C & BC vs NP & -1.76 & (0.08) & 0.00 & -0.08 & (0.03) & 0.01 & -0.17 & (0.09) & 0.04 & -0.05 & (0.03) & 0.14\\
    D & BC vs NP & -1.72 & (0.08) & 0.00 & -0.04 & (0.03) & 0.22 & -0.08 & (0.10) & 0.41 & 0.05 & (0.04) & 0.21\\
    E & BC vs NP & -1.67 & (0.08) & 0.00 & 0.02 & (0.01) & 0.18 & -0.15 & (0.09) & 0.10 & -0.02 & (0.01) & 0.12\\
    F & BC vs NP & -1.67 & (0.09) & 0.00 & 0.01 & (0.03) & 0.73 & -0.05 & (0.10) & 0.62 & 0.07 & (0.04) & 0.06\\
    G & BC vs NP & -1.67 & (0.08) & 0.00 & 0.01 & (0.02) & 0.72 & -0.17 & (0.09) & 0.05 & -0.05 & (0.02) & 0.04\\
    H & BC vs NP & -1.68 & (0.08) & 0.00 & -0.00 & (0.02) & 0.92 & -0.15 & (0.09) & 0.08 & -0.03 & (0.02) & 0.18\\
    \midrule
    A & TC vs NP & -1.62 & (0.14) & 0.00 & -0.06 & (0.05) & 0.24 & -0.05 & (0.16) & 0.77 & -0.04 & (0.05) & 0.52\\
    B & TC vs NP & -1.59 & (0.12) & 0.00 & -0.03 & (0.03) & 0.31 & 0.02 & (0.14) & 0.88 & 0.03 & (0.04) & 0.38\\
    C & TC vs NP & -1.52 & (0.13) & 0.00 & 0.04 & (0.05) & 0.41 & 0.03 & (0.13) & 0.81 & 0.04 & (0.05) & 0.43\\
    D & TC vs NP & -1.59 & (0.13) & 0.00 & -0.03 & (0.05) & 0.57 & -0.03 & (0.15) & 0.84 & -0.02 & (0.07) & 0.76\\
    E & TC vs NP & -1.54 & (0.12) & 0.00 & 0.02 & (0.02) & 0.30 & -0.02 & (0.14) & 0.87 & -0.01 & (0.02) & 0.55\\
    F & TC vs NP & -1.62 & (0.14) & 0.00 & -0.06 & (0.06) & 0.29 & -0.05 & (0.17) & 0.74 & -0.04 & (0.06) & 0.49\\
    G & TC vs NP & -1.50 & (0.12) & 0.00 & 0.06 & (0.03) & 0.07 & 0.01 & (0.13) & 0.92 & 0.02 & (0.04) & 0.54\\
    H & TC vs NP & -1.61 & (0.13) & 0.00 & -0.04 & (0.03) & 0.18 & -0.04 & (0.15) & 0.79 & -0.03 & (0.04) & 0.42\\
    \midrule
    A & EP vs NP & -1.32 & (0.14) & 0.00 & 0.10 & (0.06) & 0.09 & 0.24 & (0.16) & 0.12 & 0.06 & (0.05) & 0.26\\
    B & EP vs NP & -1.47 & (0.14) & 0.00 & -0.04 & (0.04) & 0.29 & 0.21 & (0.15) & 0.15 & 0.03 & (0.04) & 0.42\\
    C & EP vs NP & -1.58 & (0.16) & 0.00 & -0.16 & (0.07) & 0.02 & 0.10 & (0.16) & 0.50 & -0.08 & (0.06) & 0.20\\
    D & EP vs NP & -1.38 & (0.14) & 0.00 & 0.05 & (0.06) & 0.41 & 0.22 & (0.16) & 0.16 & 0.04 & (0.07) & 0.57\\
    E & EP vs NP & -1.41 & (0.14) & 0.00 & 0.01 & (0.02) & 0.55 & 0.17 & (0.14) & 0.23 & -0.01 & (0.02) & 0.66\\
    F & EP vs NP & -1.27 & (0.14) & 0.00 & 0.15 & (0.07) & 0.02 & 0.29 & (0.16) & 0.07 & 0.11 & (0.07) & 0.10\\
    G & EP vs NP & -1.48 & (0.15) & 0.00 & -0.05 & (0.04) & 0.21 & 0.11 & (0.15) & 0.45 & -0.07 & (0.04) & 0.07\\
    H & EP vs NP & -1.43 & (0.15) & 0.00 & -0.00 & (0.04) & 0.98 & 0.18 & (0.15) & 0.25 & -0.01 & (0.04) & 0.89\\
    \midrule
    &&  \multicolumn{12}{c}{\textbf{Outcome: Earnings}} \\ \midrule
    A & WS vs NP & -3213 & (628) & 0.00 & 303 & (380) & 0.43 & 2205 & (862) & 0.01 & -1021 & (551) & 0.06\\
    B & WS vs NP & -3128 & (686) & 0.00 & 387 & (262) & 0.14 & 2675 & (918) & 0.00 & -551 & (398) & 0.17\\
    C & WS vs NP & -4697 & (1220) & 0.00 & -1182 & (699) & 0.09 & 4495 & (1777) & 0.01 & 1269 & (1036) & 0.22\\
    D & WS vs NP & -3435 & (649) & 0.00 & 81 & (287) & 0.78 & 2681 & (881) & 0.00 & -545 & (444) & 0.22\\
    E & WS vs NP & -3511 & (838) & 0.00 & 5 & (195) & 0.98 & 3708 & (1199) & 0.00 & 482 & (310) & 0.12\\
    F & WS vs NP & -3128 & (597) & 0.00 & 387 & (374) & 0.30 & 2368 & (797) & 0.00 & -858 & (536) & 0.11\\
    G & WS vs NP & -4121 & (1409) & 0.00 & -606 & (884) & 0.49 & 5747 & (2102) & 0.01 & 2521 & (1365) & 0.06\\
    H & WS vs NP & -3908 & (913) & 0.00 & -393 & (335) & 0.24 & 3946 & (1368) & 0.00 & 720 & (605) & 0.23\\
    \midrule
    A & BC vs NP & -6371 & (746) & 0.00 & 545 & (422) & 0.20 & 709 & (1060) & 0.50 & -195 & (633) & 0.76\\
    B & BC vs NP & -6650 & (785) & 0.00 & 267 & (266) & 0.32 & 1220 & (1089) & 0.26 & 316 & (419) & 0.45\\
    C & BC vs NP & -8227 & (1316) & 0.00 & -1311 & (733) & 0.07 & 527 & (1959) & 0.79 & -377 & (1127) & 0.74\\
    D & BC vs NP & -7398 & (820) & 0.00 & -482 & (327) & 0.14 & 1122 & (1113) & 0.31 & 218 & (494) & 0.66\\
    E & BC vs NP & -6815 & (915) & 0.00 & 101 & (200) & 0.61 & 810 & (1358) & 0.55 & -93 & (338) & 0.78\\
    F & BC vs NP & -6115 & (714) & 0.00 & 802 & (417) & 0.05 & 802 & (1001) & 0.42 & -101 & (614) & 0.87\\
    G & BC vs NP & -7607 & (1420) & 0.00 & -690 & (855) & 0.42 & 488 & (2197) & 0.82 & -416 & (1375) & 0.76\\
    H & BC vs NP & -7126 & (1004) & 0.00 & -210 & (357) & 0.56 & 1087 & (1524) & 0.48 & 184 & (641) & 0.77\\
        \midrule
    A & TC vs NP & -6335 & (810) & 0.00 & 713 & (478) & 0.14 & 813 & (1131) & 0.47 & -925 & (715) & 0.20\\
    B & TC vs NP & -6836 & (857) & 0.00 & 212 & (304) & 0.48 & 1393 & (1199) & 0.25 & -344 & (490) & 0.48\\
    C & TC vs NP & -8529 & (1509) & 0.00 & -1480 & (864) & 0.09 & 2896 & (2275) & 0.20 & 1158 & (1342) & 0.39\\
    D & TC vs NP & -7399 & (901) & 0.00 & -350 & (380) & 0.36 & 1529 & (1253) & 0.22 & -209 & (569) & 0.71\\
    E & TC vs NP & -7098 & (1017) & 0.00 & -50 & (224) & 0.82 & 1988 & (1536) & 0.20 & 251 & (392) & 0.52\\
    F & TC vs NP & -6218 & (786) & 0.00 & 831 & (473) & 0.08 & 1092 & (1111) & 0.33 & -645 & (714) & 0.37\\
    G & TC vs NP & -7651 & (1612) & 0.00 & -602 & (1004) & 0.55 & 3097 & (2551) & 0.22 & 1359 & (1623) & 0.40\\
    H & TC vs NP & -7402 & (1146) & 0.00 & -354 & (426) & 0.41 & 2176 & (1764) & 0.22 & 438 & (773) & 0.57\\
        \midrule

    A & EP vs NP & -6094 & (994) & 0.00 & 1455 & (575) & 0.01 & 490 & (1349) & 0.72 & 774 & (981) & 0.43\\
    B & EP vs NP & -7164 & (997) & 0.00 & 385 & (353) & 0.28 & 287 & (1468) & 0.84 & 571 & (679) & 0.40\\
    C & EP vs NP & -9953 & (1834) & 0.00 & -2404 & (1076) & 0.03 & -1073 & (3389) & 0.75 & -789 & (2028) & 0.70\\
    D & EP vs NP & -7750 & (1069) & 0.00 & -201 & (458) & 0.66 & -439 & (1589) & 0.78 & -155 & (807) & 0.85\\
    E & EP vs NP & -7852 & (1182) & 0.00 & -303 & (266) & 0.25 & -575 & (1974) & 0.77 & -291 & (495) & 0.56\\
    F & EP vs NP & -5899 & (972) & 0.00 & 1650 & (588) & 0.01 & -130 & (1321) & 0.92 & 154 & (999) & 0.88\\
    G & EP vs NP & -9716 & (1911) & 0.00 & -2167 & (1188) & 0.07 & -1154 & (3656) & 0.75 & -870 & (2326) & 0.71\\
    H & EP vs NP & -8424 & (1342) & 0.00 & -875 & (541) & 0.11 & 574 & (2530) & 0.82 & 858 & (1183) & 0.47\\
  
      \bottomrule
    \end{tabular}
    \begin{tablenotes}
\textit{Note:} The table shows the GATEs for the different programmes in comparison to no programme (NP: Non Participation, WS: Wage Subsidy, BC: Basic courses, TC: Training Courses, EP: Employment Programme) and the different sub-groups (A: Agricultural and forestry, B: Production in industry and trade, C: Technical and information technology, D: Construction and mining; E: Transport and Trade, F: Hospitality and personal services, G: Business, Finance and Law, H: Education, Art and Science). 
\end{tablenotes}
\end{threeparttable}
\end{adjustbox}
\end{table}

\begin{table}[htbp!]
\begin{adjustbox}{width=0.95\columnwidth, center}
\begin{threeparttable}
  \captionsetup{font=large}
    \caption{GATEs for place of origin, PERM sample}
    \centering
    \begin{tabular}{@{}ll|rrrrrr|rrrrrrr@{}}
    \toprule
    \multicolumn{2}{l}{\textbf{Origin}}& \multicolumn{6}{c}{\textbf{First Year}} & \multicolumn{6}{c}{\textbf{Third Year}} \\
    \midrule
      & Programme & GATE & S.E. & p-value & GATE-ATE & S.E. & p-value & GATE & S.E. & p-value & GATE-ATE & S.E. & p-value \\ \midrule
   &&  \multicolumn{12}{c}{\textbf{Outcome: Employment}} \\ \midrule
     Swiss & WS vs NP & -0.05 & (0.01) & 0.00 & -0.05 & (0.01) & 0.00 & -0.04 & (0.01) & 0.00 & -0.04 & (0.01) & 0.00\\
    Neig. countries & WS vs NP & 0.02 & (0.01) & 0.08 & 0.02 & (0.01) & 0.08 & 0.02 & (0.02) & 0.23 & 0.02 & (0.02) & 0.23\\
    EU & WS vs NP & 0.07 & (0.02) & 0.00 & 0.07 & (0.02) & 0.00 & 0.07 & (0.02) & 0.00 & 0.07 & (0.02) & 0.00\\
    Rest of Europe & WS vs NP & 0.16 & (0.03) & 0.00 & 0.16 & (0.03) & 0.00 & 0.11 & (0.03) & 0.00 & 0.11 & (0.03) & 0.00\\
    Rest of the world & WS vs NP & 0.22 & (0.03) & 0.00 & 0.22 & (0.03) & 0.00 & 0.17 & (0.03) & 0.00 & 0.17 & (0.03) & 0.00\\
    \midrule
    Swiss & BC vs NP & -0.03 & (0.01) & 0.00 & -0.03 & (0.01) & 0.00 & -0.05 & (0.01) & 0.00 & -0.05 & (0.01) & 0.00\\
    Neig. countries & BC vs NP & 0.01 & (0.01) & 0.50 & 0.01 & (0.01) & 0.50 & 0.03 & (0.02) & 0.15 & 0.03 & (0.02) & 0.15\\
    EU & BC vs NP & 0.02 & (0.02) & 0.41 & 0.02 & (0.02) & 0.41 & 0.08 & (0.03) & 0.00 & 0.08 & (0.03) & 0.00\\
    Rest of Europe & BC vs NP & 0.09 & (0.03) & 0.00 & 0.09 & (0.03) & 0.00 & 0.15 & (0.04) & 0.00 & 0.15 & (0.04) & 0.00\\
    Rest of the world & BC vs NP & 0.18 & (0.03) & 0.00 & 0.18 & (0.03) & 0.00 & 0.17 & (0.04) & 0.00 & 0.17 & (0.04) & 0.00\\
        \midrule
    Swiss & TC vs NP & -0.02 & (0.01) & 0.09 & -0.02 & (0.01) & 0.09 & -0.03 & (0.01) & 0.08 & -0.03 & (0.01) & 0.08\\
    Neig. countries & TC vs NP & 0.04 & (0.02) & 0.10 & 0.04 & (0.02) & 0.10 & 0.05 & (0.03) & 0.11 & 0.05 & (0.03) & 0.11\\
    EU & TC vs NP & 0.03 & (0.04) & 0.38 & 0.03 & (0.04) & 0.38 & 0.04 & (0.04) & 0.36 & 0.04 & (0.04) & 0.36\\
    Rest of Europe & TC vs NP & 0.00 & (0.05) & 1.00 & 0.00 & (0.05) & 1.00 & 0.04 & (0.06) & 0.43 & 0.04 & (0.06) & 0.43\\
    Rest of the world & TC vs NP & 0.08 & (0.04) & 0.08 & 0.08 & (0.04) & 0.08 & 0.08 & (0.06) & 0.16 & 0.08 & (0.06) & 0.16\\
    \midrule
    Swiss & EP vs NP & -0.05 & (0.01) & 0.00 & -0.05 & (0.01) & 0.00 & -0.05 & (0.02) & 0.00 & -0.05 & (0.02) & 0.00\\
    Neig. countries & EP vs NP & 0.01 & (0.03) & 0.79 & 0.01 & (0.03) & 0.79 & 0.01 & (0.03) & 0.69 & 0.01 & (0.03) & 0.69\\
    EU & EP vs NP & 0.14 & (0.04) & 0.00 & 0.14 & (0.04) & 0.00 & 0.13 & (0.05) & 0.00 & 0.13 & (0.05) & 0.00\\
    Rest of Europe & EP vs NP & 0.14 & (0.06) & 0.02 & 0.14 & (0.06) & 0.02 & 0.11 & (0.06) & 0.05 & 0.11 & (0.06) & 0.05\\
    Rest of the world & EP vs NP & 0.20 & (0.05) & 0.00 & 0.20 & (0.05) & 0.00 & 0.15 & (0.06) & 0.01 & 0.15 & (0.06) & 0.01\\
    \midrule
   &&  \multicolumn{12}{c}{\textbf{Outcome: Earnings}} \\ \midrule
    Swiss & WS vs NP & -6188 & (402) & 0.00 & -386 & (36) & 0.00 & 1767 & (708) & 0.01 & -291 & (78) & 0.00\\
    Neig. countries & WS vs NP & -6188 & (458) & 0.00 & -386 & (110) & 0.00 & 2328 & (815) & 0.00 & 270 & (175) & 0.12\\
    EU & WS vs NP & -4822 & (340) & 0.00 & 980 & (116) & 0.00 & 2607 & (659) & 0.00 & 548 & (183) & 0.00\\
    Rest of Europe & WS vs NP & -4101 & (371) & 0.00 & 1702 & (204) & 0.00 & 2772 & (854) & 0.00 & 713 & (340) & 0.04\\
    Rest of the world & WS vs NP & -4172 & (403) & 0.00 & 1631 & (150) & 0.00 & 3119 & (937) & 0.00 & 1061 & (345) & 0.00\\ \midrule
    Swiss & BC vs NP & -11154 & (425) & 0.00 & -420 & (37) & 0.00 & -2709 & (766) & 0.00 & -363 & (83) & 0.00\\
    Neig. countries & BC vs NP & -11591 & (464) & 0.00 & -856 & (106) & 0.00 & -2394 & (866) & 0.01 & -47 & (189) & 0.80\\
    EU & BC vs NP & -9715 & (377) & 0.00 & 1019 & (116) & 0.00 & -1588 & (730) & 0.03 & 759 & (204) & 0.00\\
    Rest of Europe & BC vs NP & -8328 & (403) & 0.00 & 2407 & (199) & 0.00 & -881 & (904) & 0.33 & 1465 & (359) & 0.00\\
    Rest of the world & BC vs NP & -8605 & (426) & 0.00 & 2130 & (156) & 0.00 & -916 & (978) & 0.35 & 1431 & (359) & 0.00\\ \midrule
    Swiss & TC vs NP & -9911 & (640) & 0.00 & -268 & (53) & 0.00 & -1311 & (1058) & 0.22 & -179 & (111) & 0.11\\
    Neig. countries & TC vs NP & -10418 & (656) & 0.00 & -776 & (142) & 0.00 & -660 & (1165) & 0.57 & 471 & (257) & 0.07\\
    EU & TC vs NP & -8864 & (587) & 0.00 & 778 & (169) & 0.00 & -990 & (991) & 0.32 & 142 & (315) & 0.65\\
    Rest of Europe & TC vs NP & -7994 & (588) & 0.00 & 1648 & (278) & 0.00 & -1053 & (1094) & 0.34 & 79 & (499) & 0.87\\
    Rest of the world & TC vs NP & -8153 & (588) & 0.00 & 1490 & (220) & 0.00 & -470 & (1157) & 0.68 & 662 & (450) & 0.14\\ \midrule
    Swiss & EP vs NP & -7037 & (841) & 0.00 & -140 & (76) & 0.07 & -2109 & (1247) & 0.09 & -414 & (127) & 0.00\\
    Neig. countries & EP vs NP & -7333 & (868) & 0.00 & -437 & (187) & 0.02 & -1910 & (1372) & 0.16 & -215 & (312) & 0.49\\
    EU & EP vs NP & -6603 & (662) & 0.00 & 293 & (242) & 0.23 & -394 & (1085) & 0.72 & 1302 & (384) & 0.00\\
    Rest of Europe & EP vs NP & -6156 & (617) & 0.00 & 740 & (415) & 0.07 & -187 & (1127) & 0.87 & 1509 & (630) & 0.02\\
    Rest of the world & EP vs NP & -5850 & (650) & 0.00 & 1046 & (315) & 0.00 & -236 & (1215) & 0.85 & 1459 & (523) & 0.01\\ 
 
    \bottomrule
    \end{tabular}
\begin{tablenotes}
\textit{Note:} The table shows the GATEs for the different programmes in comparison to no programme (NP: Non Participation, WS: Wage Subsidy, BC: Basic courses, TC: Training Courses, EP: Employment Programme) and the different sub-groups. 
\end{tablenotes}
\label{table:only_gate_origin}
\end{threeparttable}
\end{adjustbox}
\end{table}
\begin{table}[htbp!]
\begin{adjustbox}{width=0.95\columnwidth, center}
\begin{threeparttable}
  \captionsetup{font=large}
    \caption{GATEs for education levels, PERM sample}
    \label{table:merged}
    \centering
    \begin{tabular}{@{}ll|rrrrrr|rrrrrrr@{}}
    \toprule
     \multicolumn{2}{l}{\textbf{Education}} & \multicolumn{6}{c}{\textbf{First Year}} & \multicolumn{6}{c}{\textbf{Third Year}} \\
    \midrule
     & Programme & GATE & S.E. & p-value & GATE-ATE & S.E. & p-value & GATE & S.E. & p-value & GATE-ATE & S.E. & p-value \\ \midrule
    &&  \multicolumn{12}{c}{\textbf{Outcome: Employment}} \\ \midrule
    Compulsory & WS vs NP & 0.10 & (0.02) & 0.00 & 0.10 & (0.02) & 0.00 & 0.10 & (0.02) & 0.00 & 0.10 & (0.02) & 0.00\\
    Secondary & WS vs NP & -0.01 & (0.01) & 0.15 & -0.01 & (0.01) & 0.15 & -0.02 & (0.01) & 0.01 & -0.02 & (0.01) & 0.01\\
    Vocational & WS vs NP & -0.14 & (0.03) & 0.00 & -0.14 & (0.03) & 0.00 & -0.10 & (0.03) & 0.00 & -0.10 & (0.03) & 0.00\\
    Bachelor & WS vs NP & -0.08 & (0.03) & 0.01 & -0.08 & (0.03) & 0.01 & -0.06 & (0.03) & 0.03 & -0.06 & (0.03) & 0.03\\
    Post-Grad & WS vs NP & 0.04 & (0.04) & 0.27 & 0.04 & (0.04) & 0.27 & 0.02 & (0.03) & 0.54 & 0.02 & (0.03) & 0.54\\
    \midrule
    Compulsory & BC vs NP & 0.04 & (0.03) & 0.09 & 0.04 & (0.03) & 0.09 & 0.11 & (0.03) & 0.00 & 0.11 & (0.03) & 0.00\\
    Secondary & BC vs NP & -0.02 & (0.01) & 0.00 & -0.02 & (0.01) & 0.00 & -0.01 & (0.01) & 0.05 & -0.01 & (0.01) & 0.05\\
    Vocational & BC vs NP & -0.08 & (0.03) & 0.02 & -0.08 & (0.03) & 0.02 & -0.08 & (0.03) & 0.01 & -0.08 & (0.03) & 0.01\\
    Bachelor & BC vs NP & -0.01 & (0.03) & 0.85 & -0.01 & (0.03) & 0.85 & -0.06 & (0.03) & 0.05 & -0.06 & (0.03) & 0.05\\
    Post-Grad & BC vs NP & 0.11 & (0.04) & 0.00 & 0.11 & (0.04) & 0.00 & -0.02 & (0.04) & 0.62 & -0.02 & (0.04) & 0.62\\
    \midrule
    Compulsory & TC vs NP & -0.02 & (0.04) & 0.61 & -0.02 & (0.04) & 0.61 & 0.02 & (0.05) & 0.75 & 0.02 & (0.05) & 0.75\\
    Secondary & TC vs NP & -0.04 & (0.01) & 0.00 & -0.04 & (0.01) & 0.00 & -0.02 & (0.01) & 0.09 & -0.02 & (0.01) & 0.09\\
    Vocational & TC vs NP & 0.05 & (0.06) & 0.39 & 0.05 & (0.06) & 0.39 & 0.01 & (0.05) & 0.80 & 0.01 & (0.05) & 0.80\\
    Bachelor & TC vs NP & 0.09 & (0.05) & 0.08 & 0.09 & (0.05) & 0.08 & 0.02 & (0.05) & 0.62 & 0.02 & (0.05) & 0.62\\
    Post-Grad & TC vs NP & 0.16 & (0.06) & 0.00 & 0.16 & (0.06) & 0.00 & 0.06 & (0.06) & 0.27 & 0.06 & (0.06) & 0.27\\
    \midrule
    Compulsory & EP vs NP & 0.12 & (0.05) & 0.03 & 0.12 & (0.05) & 0.03 & 0.12 & (0.05) & 0.02 & 0.12 & (0.05) & 0.02\\
    Secondary & EP vs NP & -0.03 & (0.02) & 0.11 & -0.03 & (0.02) & 0.11 & -0.01 & (0.01) & 0.57 & -0.01 & (0.01) & 0.57\\
    Vocational & EP vs NP & -0.14 & (0.08) & 0.06 & -0.14 & (0.08) & 0.06 & -0.12 & (0.06) & 0.05 & -0.12 & (0.06) & 0.05\\
    Bachelor & EP vs NP & -0.05 & (0.07) & 0.48 & -0.05 & (0.07) & 0.48 & -0.08 & (0.05) & 0.14 & -0.08 & (0.05) & 0.14\\
    Post-Grad & EP vs NP & 0.08 & (0.08) & 0.29 & 0.08 & (0.08) & 0.29 & -0.02 & (0.06) & 0.77 & -0.02 & (0.06) & 0.77\\
    \midrule
    &&  \multicolumn{12}{c}{\textbf{Outcome: Earnings}} \\ \midrule
     Compulsory & WS vs NP & -4186 & (334) & 0.00 & 1616 & (194) & 0.00 & 2670 & (695) & 0.00 & 611 & (280) & 0.03\\
    Secondary & WS vs NP & -5442 & (353) & 0.00 & 361 & (93) & 0.00 & 1999 & (652) & 0.00 & -60 & (147) & 0.69\\
    Vocational & WS vs NP & -8358 & (598) & 0.00 & -2555 & (292) & 0.00 & 1286 & (918) & 0.16 & -773 & (413) & 0.06\\
    Bachelor & WS vs NP & -8034 & (606) & 0.00 & -2232 & (313) & 0.00 & 1444 & (973) & 0.14 & -615 & (476) & 0.20\\
    Post-Grad & WS vs NP & -7454 & (747) & 0.00 & -1652 & (454) & 0.00 & 2248 & (1213) & 0.06 & 189 & (704) & 0.79\\ \midrule
    Compulsory & BC vs NP & -8506 & (375) & 0.00 & 2229 & (188) & 0.00 & -1124 & (762) & 0.14 & 1223 & (304) & 0.00\\
    Secondary & BC vs NP & -10119 & (388) & 0.00 & 616 & (84) & 0.00 & -2077 & (718) & 0.00 & 269 & (158) & 0.09\\
    Vocational & BC vs NP & -14145 & (592) & 0.00 & -3410 & (285) & 0.00 & -4007 & (973) & 0.00 & -1661 & (441) & 0.00\\
    Bachelor & BC vs NP & -13818 & (596) & 0.00 & -3083 & (298) & 0.00 & -3962 & (1036) & 0.00 & -1615 & (511) & 0.00\\
    Post-Grad & BC vs NP & -13716 & (701) & 0.00 & -2982 & (411) & 0.00 & -3740 & (1306) & 0.00 & -1393 & (757) & 0.07\\ \midrule
    Compulsory & TC vs NP & -7955 & (575) & 0.00 & 1687 & (269) & 0.00 & -1282 & (978) & 0.19 & -151 & (458) & 0.74\\
    Secondary & TC vs NP & -9219 & (597) & 0.00 & 424 & (117) & 0.00 & -1455 & (967) & 0.13 & -323 & (234) & 0.17\\
    Vocational & TC vs NP & -12038 & (841) & 0.00 & -2396 & (424) & 0.00 & -686 & (1439) & 0.63 & 445 & (707) & 0.53\\
    Bachelor & TC vs NP & -11929 & (857) & 0.00 & -2287 & (427) & 0.00 & -381 & (1579) & 0.81 & 751 & (817) & 0.36\\
    Post-Grad & TC vs NP & -11826 & (944) & 0.00 & -2184 & (549) & 0.00 & 112 & (1797) & 0.95 & 1244 & (1081) & 0.25\\ \midrule
    Compulsory & EP vs NP & -6011 & (614) & 0.00 & 885 & (400) & 0.03 & -217 & (1022) & 0.83 & 1479 & (594) & 0.01\\
    Secondary & EP vs NP & -7024 & (697) & 0.00 & -128 & (170) & 0.45 & -1359 & (1080) & 0.21 & 336 & (288) & 0.24\\
    Vocational & EP vs NP & -7797 & (1229) & 0.00 & -901 & (662) & 0.17 & -4176 & (1871) & 0.03 & -2480 & (960) & 0.01\\
    Bachelor & EP vs NP & -7231 & (1235) & 0.00 & -335 & (656) & 0.61 & -3609 & (1961) & 0.07 & -1914 & (1018) & 0.06\\
    Post-Grad & EP vs NP & -6891 & (1331) & 0.00 & 5 & (770) & 0.99 & -3139 & (2217) & 0.16 & -1443 & (1373) & 0.29\\
    \bottomrule
    \end{tabular}
\begin{tablenotes}
\textit{Note:} The table shows the GATEs for the different programmes in comparison to no programme (NP: Non Participation, WS: Wage Subsidy, BC: Basic courses, TC: Training Courses, EP: Employment Programme) and the different sub-groups. 
\end{tablenotes}
\label{table:only_gate_education}
\end{threeparttable}
\end{adjustbox}
\end{table}

\begin{table}[htbp!]
\begin{adjustbox}{width=0.95\columnwidth, center}
\begin{threeparttable}
  \captionsetup{font=large}
    \caption{GATEs for gender, PERM sample}
    \label{table:gates_gender}
    \centering
    \begin{tabular}{@{}ll|rrrrrr|rrrrrrr@{}}
    \toprule
     \multicolumn{2}{l}{\textbf{Gender}}& \multicolumn{6}{c}{\textbf{First Year}} & \multicolumn{6}{c}{\textbf{Third Year}} \\
    \midrule
     & Programme & GATE & S.E. & p-value & GATE-ATE & S.E. & p-value & GATE & S.E. & p-value & GATE-ATE & S.E. & p-value \\
    \midrule
 &&  \multicolumn{12}{c}{\textbf{Outcome: Employment}} \\ \midrule
    Man & WS vs NP & -0.64 & (0.07) & 0.00 & 0.00 & (0.01) & 0.87 & 0.71 & (0.07) & 0.00 & -0.03 & (0.01) & 0.03\\
    Woman & WS vs NP & -0.65 & (0.07) & 0.00 & -0.00 & (0.01) & 0.89 & 0.77 & (0.07) & 0.00 & 0.03 & (0.01) & 0.03\\ \midrule
    Man & BC vs NP & -1.70 & (0.08) & 0.00 & -0.02 & (0.01) & 0.25 & -0.12 & (0.09) & 0.15 & 0.00 & (0.02) & 0.90\\
    Woman & BC vs NP & -1.66 & (0.08) & 0.00 & 0.02 & (0.02) & 0.24 & -0.13 & (0.09) & 0.17 & -0.00 & (0.02) & 0.92\\ \midrule
    Man & TC vs NP & -1.53 & (0.12) & 0.00 & 0.03 & (0.02) & 0.26 & 0.01 & (0.13) & 0.95 & 0.02 & (0.03) & 0.52\\
    Woman & TC vs NP & -1.59 & (0.13) & 0.00 & -0.03 & (0.03) & 0.26 & -0.03 & (0.15) & 0.83 & -0.02 & (0.03) & 0.51\\ \midrule
    Man & EP vs NP & -1.46 & (0.14) & 0.00 & -0.03 & (0.03) & 0.24 & 0.16 & (0.14) & 0.28 & -0.02 & (0.03) & 0.40\\
    Woman & EP vs NP & -1.39 & (0.14) & 0.00 & 0.04 & (0.03) & 0.25 & 0.21 & (0.15) & 0.16 & 0.03 & (0.03) & 0.41\\ \midrule
    &&  \multicolumn{12}{c}{\textbf{Outcome: Earnings}} \\ \midrule
    Man & WS vs NP & -6396 & (453) & 0.00 & -594 & (93) & 0.00 & 2071 & (801) & 0.01 & 12 & (131) & 0.93\\
    Woman & WS vs NP & -5120 & (338) & 0.00 & 682 & (107) & 0.00 & 2045 & (671) & 0.00 & -14 & (149) & 0.93\\ \midrule
    Man & BC vs NP & -11814 & (463) & 0.00 & -1079 & (95) & 0.00 & -2752 & (850) & 0.00 & -405 & (144) & 0.01\\
    Woman & BC vs NP & -9495 & (380) & 0.00 & 1240 & (110) & 0.00 & -1884 & (742) & 0.01 & 462 & (165) & 0.01\\ \midrule
    Man & TC vs NP & -10431 & (669) & 0.00 & -789 & (141) & 0.00 & -1030 & (1148) & 0.37 & 102 & (217) & 0.64\\
    Woman & TC vs NP & -8736 & (601) & 0.00 & 906 & (162) & 0.00 & -1247 & (993) & 0.21 & -116 & (247) & 0.64\\ \midrule
    Man & EP vs NP & -7385 & (884) & 0.00 & -489 & (184) & 0.01 & -2364 & (1369) & 0.08 & -668 & (281) & 0.02\\
    Woman & EP vs NP & -6334 & (698) & 0.00 & 562 & (211) & 0.01 & -934 & (1098) & 0.40 & 762 & (320) & 0.02\\
     \bottomrule
    \end{tabular}
    \begin{tablenotes}
\textit{Note:} The table shows the GATEs for the different programmes in comparison to no programme (NP: Non Participation, WS: Wage Subsidy, BC: Basic courses, TC: Training Courses, EP: Employment Programme) and the different sub-groups. 
\end{tablenotes}
\end{threeparttable}
\end{adjustbox}
\end{table}

\begin{table}[htbp!]
\begin{adjustbox}{width=0.95\columnwidth, center}
\begin{threeparttable}
  \captionsetup{font=large}
    \caption{GATEs for months in employment, PERM sample}
    \label{table:gates_employment_months}
    \centering
    \begin{tabular}{@{}ll|rrrrrr|rrrrrrr@{}}
    \toprule
     \multicolumn{2}{l}{\textbf{Months in employment}}& \multicolumn{6}{c}{\textbf{First Year}} & \multicolumn{6}{c}{\textbf{Third Year}} \\
    \midrule
     & Programme & GATE & S.E. & p-value & GATE-ATE & S.E. & p-value & GATE & S.E. & p-value & GATE-ATE & S.E. & p-value \\
    \midrule
 &&  \multicolumn{12}{c}{\textbf{Outcome: Employment}} \\ \midrule
    12 & WS vs NP & -0.44 & (0.09) & 0.00 & 0.20 & (0.07) & 0.00 & 0.91 & (0.12) & 0.00 & 0.17 & (0.09) & 0.05\\
    24 & WS vs NP & -0.36 & (0.09) & 0.00 & 0.29 & (0.06) & 0.00 & 0.98 & (0.11) & 0.00 & 0.24 & (0.07) & 0.00\\
    36 & WS vs NP & -0.41 & (0.08) & 0.00 & 0.23 & (0.04) & 0.00 & 0.90 & (0.10) & 0.00 & 0.16 & (0.05) & 0.00\\
    48 & WS vs NP & -0.49 & (0.07) & 0.00 & 0.16 & (0.03) & 0.00 & 0.82 & (0.08) & 0.00 & 0.08 & (0.03) & 0.01\\
    60 & WS vs NP & -0.73 & (0.07) & 0.00 & -0.09 & (0.02) & 0.00 & 0.68 & (0.07) & 0.00 & -0.06 & (0.02) & 0.00\\ \midrule
    12 & BC vs NP & -1.26 & (0.13) & 0.00 & 0.42 & (0.10) & 0.00 & 0.11 & (0.17) & 0.50 & 0.24 & (0.12) & 0.06\\
    24 & BC vs NP & -1.30 & (0.11) & 0.00 & 0.38 & (0.07) & 0.00 & 0.13 & (0.14) & 0.36 & 0.26 & (0.10) & 0.01\\
    36 & BC vs NP & -1.41 & (0.10) & 0.00 & 0.27 & (0.06) & 0.00 & 0.04 & (0.13) & 0.72 & 0.17 & (0.07) & 0.02\\
    48 & BC vs NP & -1.53 & (0.08) & 0.00 & 0.15 & (0.03) & 0.00 & -0.05 & (0.10) & 0.62 & 0.07 & (0.04) & 0.06\\
    60 & BC vs NP & -1.78 & (0.08) & 0.00 & -0.10 & (0.02) & 0.00 & -0.18 & (0.08) & 0.03 & -0.06 & (0.03) & 0.02\\ \midrule
    12 & TC vs NP & -1.59 & (0.20) & 0.00 & -0.03 & (0.15) & 0.83 & 0.09 & (0.27) & 0.74 & 0.10 & (0.20) & 0.62\\
    24 & TC vs NP & -1.49 & (0.17) & 0.00 & 0.07 & (0.11) & 0.53 & 0.12 & (0.23) & 0.59 & 0.13 & (0.15) & 0.38\\
    36 & TC vs NP & -1.52 & (0.15) & 0.00 & 0.04 & (0.09) & 0.60 & 0.09 & (0.20) & 0.64 & 0.10 & (0.12) & 0.38\\
    48 & TC vs NP & -1.51 & (0.13) & 0.00 & 0.05 & (0.06) & 0.41 & -0.00 & (0.16) & 1.00 & 0.01 & (0.06) & 0.89\\
    60 & TC vs NP & -1.58 & (0.12) & 0.00 & -0.02 & (0.03) & 0.48 & -0.03 & (0.13) & 0.80 & -0.02 & (0.04) & 0.55\\ \midrule
    12 & EP vs NP & -1.07 & (0.21) & 0.00 & 0.36 & (0.17) & 0.03 & 0.15 & (0.24) & 0.54 & -0.03 & (0.19) & 0.86\\
    24 & EP vs NP & -1.07 & (0.18) & 0.00 & 0.35 & (0.13) & 0.01 & 0.27 & (0.21) & 0.20 & 0.08 & (0.14) & 0.56\\
    36 & EP vs NP & -1.07 & (0.16) & 0.00 & 0.35 & (0.10) & 0.00 & 0.24 & (0.19) & 0.19 & 0.06 & (0.11) & 0.59\\
    48 & EP vs NP & -1.19 & (0.14) & 0.00 & 0.23 & (0.07) & 0.00 & 0.23 & (0.16) & 0.15 & 0.05 & (0.06) & 0.46\\
    60 & EP vs NP & -1.56 & (0.15) & 0.00 & -0.14 & (0.04) & 0.00 & 0.15 & (0.15) & 0.29 & -0.03 & (0.04) & 0.51\\ \midrule
    &&  \multicolumn{12}{c}{\textbf{Outcome: Earnings}} \\ \midrule
    12 & WS vs NP & -3391 & (366) & 0.00 & 2412 & (293) & 0.00 & 2359 & (801) & 0.00 & 301 & (485) & 0.54\\
    24 & WS vs NP & -3354 & (398) & 0.00 & 2448 & (265) & 0.00 & 2962 & (792) & 0.00 & 903 & (446) & 0.04\\
    36 & WS vs NP & -3740 & (359) & 0.00 & 2063 & (226) & 0.00 & 2692 & (767) & 0.00 & 633 & (366) & 0.08\\
    48 & WS vs NP & -4442 & (345) & 0.00 & 1361 & (157) & 0.00 & 2406 & (705) & 0.00 & 347 & (236) & 0.14\\
    60 & WS vs NP & -6630 & (443) & 0.00 & -828 & (91) & 0.00 & 1826 & (757) & 0.02 & -233 & (139) & 0.09\\ \midrule
    12 & BC vs NP & -6635 & (470) & 0.00 & 4100 & (356) & 0.00 & -606 & (976) & 0.53 & 1741 & (637) & 0.01\\
    24 & BC vs NP & -6898 & (458) & 0.00 & 3837 & (292) & 0.00 & -404 & (899) & 0.65 & 1942 & (523) & 0.00\\
    36 & BC vs NP & -7475 & (415) & 0.00 & 3260 & (242) & 0.00 & -918 & (863) & 0.29 & 1429 & (426) & 0.00\\
    48 & BC vs NP & -8701 & (391) & 0.00 & 2034 & (159) & 0.00 & -1574 & (785) & 0.04 & 772 & (265) & 0.00\\
    60 & BC vs NP & -12009 & (454) & 0.00 & -1274 & (95) & 0.00 & -2872 & (808) & 0.00 & -526 & (159) & 0.00\\ \midrule
    12 & TC vs NP & -6707 & (749) & 0.00 & 2935 & (563) & 0.00 & -543 & (1336) & 0.68 & 589 & (954) & 0.54\\
    24 & TC vs NP & -6883 & (657) & 0.00 & 2759 & (441) & 0.00 & -210 & (1173) & 0.86 & 921 & (765) & 0.23\\
    36 & TC vs NP & -7295 & (639) & 0.00 & 2347 & (376) & 0.00 & -750 & (1116) & 0.50 & 382 & (630) & 0.54\\
    48 & TC vs NP & -8180 & (609) & 0.00 & 1462 & (243) & 0.00 & -1236 & (1042) & 0.24 & -104 & (393) & 0.79\\
    60 & TC vs NP & -10558 & (658) & 0.00 & -916 & (147) & 0.00 & -1198 & (1113) & 0.28 & -67 & (237) & 0.78\\ \midrule
    12 & EP vs NP & -4454 & (733) & 0.00 & 2443 & (649) & 0.00 & -527 & (1315) & 0.69 & 1169 & (1035) & 0.26\\
    24 & EP vs NP & -4888 & (675) & 0.00 & 2008 & (562) & 0.00 & -20 & (1193) & 0.99 & 1675 & (868) & 0.05\\
    36 & EP vs NP & -5304 & (658) & 0.00 & 1593 & (486) & 0.00 & -350 & (1143) & 0.76 & 1346 & (754) & 0.07\\
    48 & EP vs NP & -6082 & (665) & 0.00 & 815 & (329) & 0.01 & -676 & (1107) & 0.54 & 1020 & (476) & 0.03\\
    60 & EP vs NP & -7475 & (887) & 0.00 & -578 & (192) & 0.00 & -2267 & (1336) & 0.09 & -572 & (286) & 0.05\\
     \bottomrule
    \end{tabular}
    \begin{tablenotes}
\textit{Note:} The table shows the GATEs for the different programmes in comparison to no programme (NP: Non Participation, WS: Wage Subsidy, BC: Basic courses, TC: Training Courses, EP: Employment Programme) and the different sub-groups. The ranges for months in employment are: 12: 1-12, 24: 13-24, 36: 25-36, 48: 37-48, 60: 49-60.
\end{tablenotes}
\end{threeparttable}
\end{adjustbox}
\end{table}

\begin{table}[htbp!]
\begin{adjustbox}{width=0.78\columnwidth, center} 
\begin{threeparttable}
  \captionsetup{font=large}
    \caption{GATEs for earnings decile, PERM sample}
    \label{table:earnings_decile}
    \centering
    \begin{tabular}{@{}ll|rrrrrr|rrrrrrr@{}}
    \toprule
     \multicolumn{2}{l}{\textbf{Earnings decile}}& \multicolumn{6}{c}{\textbf{First Year}} & \multicolumn{6}{c}{\textbf{Third Year}} \\
    \midrule
     & Programme & GATE & S.E. & p-value & GATE-ATE & S.E. & p-value & GATE & S.E. & p-value & GATE-ATE & S.E. & p-value \\
    \midrule
    &&  \multicolumn{12}{c}{\textbf{Outcome: Employment}} \\ \midrule
    1192 & WS vs NP & -0.52 & (0.08) & 0.00 & 0.13 & (0.04) & 0.00 & 0.91 & (0.09) & 0.00 & 0.17 & (0.05) & 0.00\\
    2445 & WS vs NP & -0.57 & (0.07) & 0.00 & 0.08 & (0.04) & 0.04 & 0.85 & (0.08) & 0.00 & 0.11 & (0.04) & 0.01\\
    3307 & WS vs NP & -0.57 & (0.07) & 0.00 & 0.07 & (0.03) & 0.02 & 0.81 & (0.08) & 0.00 & 0.08 & (0.03) & 0.02\\
    3969 & WS vs NP & -0.58 & (0.07) & 0.00 & 0.06 & (0.03) & 0.02 & 0.78 & (0.08) & 0.00 & 0.04 & (0.02) & 0.08\\
    4538 & WS vs NP & -0.58 & (0.07) & 0.00 & 0.06 & (0.02) & 0.01 & 0.75 & (0.07) & 0.00 & 0.01 & (0.02) & 0.50\\
    5090 & WS vs NP & -0.61 & (0.07) & 0.00 & 0.04 & (0.02) & 0.10 & 0.71 & (0.07) & 0.00 & -0.03 & (0.02) & 0.20\\
    5698 & WS vs NP & -0.66 & (0.07) & 0.00 & -0.02 & (0.02) & 0.48 & 0.66 & (0.07) & 0.00 & -0.07 & (0.02) & 0.00\\
    6500 & WS vs NP & -0.74 & (0.07) & 0.00 & -0.10 & (0.03) & 0.00 & 0.62 & (0.07) & 0.00 & -0.11 & (0.03) & 0.00\\
    7919 & WS vs NP & -0.86 & (0.08) & 0.00 & -0.22 & (0.05) & 0.00 & 0.60 & (0.07) & 0.00 & -0.14 & (0.04) & 0.00\\
    12487 & WS vs NP & -0.74 & (0.12) & 0.00 & -0.10 & (0.09) & 0.29 & 0.68 & (0.10) & 0.00 & -0.06 & (0.07) & 0.41\\ \midrule
    1192 & BC vs NP & -1.50 & (0.10) & 0.00 & 0.18 & (0.06) & 0.00 & -0.05 & (0.13) & 0.70 & 0.08 & (0.08) & 0.33\\
    2445 & BC vs NP & -1.58 & (0.09) & 0.00 & 0.10 & (0.05) & 0.03 & -0.10 & (0.12) & 0.38 & 0.02 & (0.06) & 0.71\\
    3307 & BC vs NP & -1.61 & (0.09) & 0.00 & 0.08 & (0.04) & 0.04 & -0.10 & (0.10) & 0.35 & 0.03 & (0.04) & 0.53\\
    3969 & BC vs NP & -1.64 & (0.08) & 0.00 & 0.05 & (0.03) & 0.13 & -0.07 & (0.10) & 0.47 & 0.05 & (0.03) & 0.09\\
    4538 & BC vs NP & -1.66 & (0.08) & 0.00 & 0.02 & (0.03) & 0.44 & -0.07 & (0.09) & 0.48 & 0.06 & (0.03) & 0.04\\
    5090 & BC vs NP & -1.71 & (0.08) & 0.00 & -0.03 & (0.03) & 0.27 & -0.08 & (0.09) & 0.35 & 0.04 & (0.03) & 0.15\\
    5698 & BC vs NP & -1.77 & (0.08) & 0.00 & -0.09 & (0.03) & 0.00 & -0.13 & (0.09) & 0.13 & -0.01 & (0.03) & 0.88\\
    6500 & BC vs NP & -1.82 & (0.08) & 0.00 & -0.14 & (0.04) & 0.00 & -0.19 & (0.09) & 0.03 & -0.06 & (0.04) & 0.12\\
    7919 & BC vs NP & -1.85 & (0.09) & 0.00 & -0.17 & (0.06) & 0.01 & -0.24 & (0.09) & 0.01 & -0.12 & (0.06) & 0.03\\
    12487 & BC vs NP & -1.67 & (0.11) & 0.00 & 0.01 & (0.09) & 0.90 & -0.22 & (0.11) & 0.04 & -0.10 & (0.08) & 0.23\\ \midrule
    1192 & TC vs NP & -1.59 & (0.16) & 0.00 & -0.03 & (0.09) & 0.73 & -0.03 & (0.21) & 0.90 & -0.01 & (0.13) & 0.91\\
    2445 & TC vs NP & -1.60 & (0.15) & 0.00 & -0.04 & (0.07) & 0.57 & -0.03 & (0.19) & 0.88 & -0.02 & (0.11) & 0.86\\
    3307 & TC vs NP & -1.59 & (0.14) & 0.00 & -0.03 & (0.06) & 0.57 & -0.03 & (0.17) & 0.88 & -0.02 & (0.07) & 0.82\\
    3969 & TC vs NP & -1.59 & (0.13) & 0.00 & -0.03 & (0.05) & 0.50 & -0.06 & (0.16) & 0.70 & -0.05 & (0.05) & 0.33\\
    4538 & TC vs NP & -1.60 & (0.13) & 0.00 & -0.04 & (0.04) & 0.37 & -0.05 & (0.15) & 0.73 & -0.04 & (0.05) & 0.39\\
    5090 & TC vs NP & -1.61 & (0.13) & 0.00 & -0.05 & (0.04) & 0.26 & -0.03 & (0.14) & 0.85 & -0.02 & (0.05) & 0.73\\
    5698 & TC vs NP & -1.62 & (0.13) & 0.00 & -0.06 & (0.05) & 0.20 & -0.01 & (0.14) & 0.94 & -0.00 & (0.05) & 0.99\\
    6500 & TC vs NP & -1.62 & (0.13) & 0.00 & -0.06 & (0.07) & 0.33 & 0.02 & (0.14) & 0.87 & 0.03 & (0.07) & 0.63\\
    7919 & TC vs NP & -1.50 & (0.15) & 0.00 & 0.06 & (0.10) & 0.56 & 0.03 & (0.14) & 0.84 & 0.04 & (0.09) & 0.67\\
    12487 & TC vs NP & -1.24 & (0.17) & 0.00 & 0.32 & (0.14) & 0.02 & 0.08 & (0.16) & 0.60 & 0.10 & (0.12) & 0.45\\ \midrule
    1192 & EP vs NP & -1.24 & (0.16) & 0.00 & 0.18 & (0.10) & 0.07 & 0.23 & (0.19) & 0.24 & 0.05 & (0.12) & 0.70\\
    2445 & EP vs NP & -1.27 & (0.15) & 0.00 & 0.15 & (0.09) & 0.07 & 0.24 & (0.18) & 0.19 & 0.06 & (0.10) & 0.58\\
    3307 & EP vs NP & -1.29 & (0.14) & 0.00 & 0.13 & (0.07) & 0.06 & 0.23 & (0.16) & 0.16 & 0.05 & (0.07) & 0.50\\
    3969 & EP vs NP & -1.30 & (0.14) & 0.00 & 0.12 & (0.06) & 0.06 & 0.25 & (0.16) & 0.11 & 0.07 & (0.06) & 0.22\\
    4538 & EP vs NP & -1.33 & (0.14) & 0.00 & 0.09 & (0.06) & 0.10 & 0.26 & (0.15) & 0.09 & 0.08 & (0.05) & 0.12\\
    5090 & EP vs NP & -1.41 & (0.14) & 0.00 & 0.01 & (0.05) & 0.80 & 0.24 & (0.15) & 0.10 & 0.06 & (0.05) & 0.21\\
    5698 & EP vs NP & -1.52 & (0.15) & 0.00 & -0.09 & (0.06) & 0.10 & 0.20 & (0.15) & 0.18 & 0.02 & (0.05) & 0.66\\
    6500 & EP vs NP & -1.65 & (0.17) & 0.00 & -0.23 & (0.08) & 0.00 & 0.14 & (0.16) & 0.38 & -0.04 & (0.07) & 0.53\\
    7919 & EP vs NP & -1.72 & (0.21) & 0.00 & -0.30 & (0.14) & 0.03 & 0.03 & (0.18) & 0.87 & -0.15 & (0.10) & 0.14\\
    12487 & EP vs NP & -1.49 & (0.25) & 0.00 & -0.07 & (0.20) & 0.75 & -0.02 & (0.21) & 0.92 & -0.20 & (0.15) & 0.19\\ \midrule
    &&  \multicolumn{12}{c}{\textbf{Outcome: Earnings}} \\ \midrule
     1192 & WS vs NP & -3569 & (330) & 0.00 & 2234 & (245) & 0.00 & 2393 & (776) & 0.00 & 334 & (338) & 0.32\\
    2445 & WS vs NP & -3674 & (317) & 0.00 & 2128 & (252) & 0.00 & 2315 & (671) & 0.00 & 256 & (337) & 0.45\\
    3307 & WS vs NP & -3872 & (320) & 0.00 & 1931 & (244) & 0.00 & 2379 & (659) & 0.00 & 320 & (325) & 0.32\\
    3969 & WS vs NP & -4153 & (326) & 0.00 & 1650 & (232) & 0.00 & 2391 & (656) & 0.00 & 333 & (309) & 0.28\\
    4538 & WS vs NP & -4512 & (331) & 0.00 & 1291 & (221) & 0.00 & 2430 & (653) & 0.00 & 371 & (299) & 0.21\\
    5090 & WS vs NP & -4969 & (352) & 0.00 & 834 & (212) & 0.00 & 2259 & (643) & 0.00 & 200 & (286) & 0.48\\
    5698 & WS vs NP & -5664 & (393) & 0.00 & 139 & (219) & 0.53 & 1859 & (671) & 0.01 & -200 & (297) & 0.50\\
    6500 & WS vs NP & -6951 & (496) & 0.00 & -1148 & (237) & 0.00 & 1240 & (753) & 0.10 & -819 & (328) & 0.01\\
    7919 & WS vs NP & -9854 & (697) & 0.00 & -4051 & (419) & 0.00 & 772 & (996) & 0.44 & -1287 & (562) & 0.02\\
    12487 & WS vs NP & -10988 & (1468) & 0.00 & -5185 & (1237) & 0.00 & 2631 & (2195) & 0.23 & 572 & (1780) & 0.75\\  \midrule
    1192 & BC vs NP & -7177 & (397) & 0.00 & 3557 & (273) & 0.00 & -987 & (860) & 0.25 & 1360 & (396) & 0.00\\
    2445 & BC vs NP & -7290 & (389) & 0.00 & 3445 & (272) & 0.00 & -1055 & (766) & 0.17 & 1291 & (385) & 0.00\\
    3307 & BC vs NP & -7638 & (378) & 0.00 & 3096 & (252) & 0.00 & -1120 & (747) & 0.13 & 1227 & (362) & 0.00\\
    3969 & BC vs NP & -8122 & (379) & 0.00 & 2612 & (234) & 0.00 & -1210 & (742) & 0.10 & 1137 & (341) & 0.00\\
    4538 & BC vs NP & -8680 & (383) & 0.00 & 2054 & (221) & 0.00 & -1288 & (734) & 0.08 & 1058 & (325) & 0.00\\
    5090 & BC vs NP & -9594 & (402) & 0.00 & 1140 & (218) & 0.00 & -1552 & (726) & 0.03 & 795 & (318) & 0.01\\
    5698 & BC vs NP & -10825 & (452) & 0.00 & -90 & (235) & 0.70 & -2178 & (757) & 0.00 & 168 & (335) & 0.62\\
    6500 & BC vs NP & -12488 & (551) & 0.00 & -1753 & (277) & 0.00 & -3313 & (848) & 0.00 & -966 & (397) & 0.01\\
    7919 & BC vs NP & -15916 & (717) & 0.00 & -5181 & (451) & 0.00 & -4532 & (1092) & 0.00 & -2185 & (654) & 0.00\\
    12487 & BC vs NP & -19998 & (1286) & 0.00 & -9263 & (1079) & 0.00 & -6465 & (2170) & 0.00 & -4119 & (1757) & 0.02\\ \midrule
    1192 & TC vs NP & -6959 & (593) & 0.00 & 2683 & (406) & 0.00 & -1001 & (1107) & 0.37 & 130 & (626) & 0.84\\
    2445 & TC vs NP & -7002 & (591) & 0.00 & 2641 & (404) & 0.00 & -1292 & (1021) & 0.21 & -160 & (602) & 0.79\\
    3307 & TC vs NP & -7329 & (603) & 0.00 & 2313 & (372) & 0.00 & -1392 & (994) & 0.16 & -260 & (557) & 0.64\\
    3969 & TC vs NP & -7787 & (611) & 0.00 & 1855 & (341) & 0.00 & -1554 & (989) & 0.12 & -422 & (531) & 0.43\\
    4538 & TC vs NP & -8382 & (609) & 0.00 & 1261 & (325) & 0.00 & -1664 & (1009) & 0.10 & -533 & (508) & 0.29\\
    5090 & TC vs NP & -9137 & (638) & 0.00 & 505 & (333) & 0.13 & -1694 & (1023) & 0.10 & -562 & (490) & 0.25\\
    5698 & TC vs NP & -10073 & (729) & 0.00 & -431 & (380) & 0.26 & -1722 & (1111) & 0.12 & -590 & (547) & 0.28\\
    6500 & TC vs NP & -11605 & (820) & 0.00 & -1963 & (435) & 0.00 & -1542 & (1247) & 0.22 & -411 & (636) & 0.52\\
    7919 & TC vs NP & -13371 & (1056) & 0.00 & -3728 & (704) & 0.00 & -623 & (1647) & 0.71 & 509 & (1068) & 0.63\\
    12487 & TC vs NP & -14915 & (1729) & 0.00 & -5272 & (1484) & 0.00 & 1375 & (2988) & 0.65 & 2506 & (2511) & 0.32\\ \midrule
    1192 & EP vs NP & -4504 & (624) & 0.00 & 2392 & (546) & 0.00 & -92 & (1133) & 0.93 & 1603 & (764) & 0.04\\
    2445 & EP vs NP & -4748 & (622) & 0.00 & 2149 & (536) & 0.00 & -95 & (1062) & 0.93 & 1601 & (758) & 0.03\\
    3307 & EP vs NP & -5332 & (623) & 0.00 & 1564 & (511) & 0.00 & -203 & (1034) & 0.84 & 1493 & (720) & 0.04\\
    3969 & EP vs NP & -5933 & (620) & 0.00 & 963 & (477) & 0.04 & -260 & (1038) & 0.80 & 1436 & (688) & 0.04\\
    4538 & EP vs NP & -6554 & (632) & 0.00 & 342 & (456) & 0.45 & -197 & (1068) & 0.85 & 1498 & (652) & 0.02\\
    5090 & EP vs NP & -7323 & (686) & 0.00 & -427 & (439) & 0.33 & -355 & (1135) & 0.75 & 1340 & (624) & 0.03\\
    5698 & EP vs NP & -8239 & (801) & 0.00 & -1343 & (463) & 0.00 & -825 & (1271) & 0.52 & 870 & (643) & 0.18\\
    6500 & EP vs NP & -9475 & (1003) & 0.00 & -2578 & (595) & 0.00 & -2239 & (1565) & 0.15 & -543 & (813) & 0.50\\
    7919 & EP vs NP & -10275 & (1699) & 0.00 & -3379 & (1145) & 0.00 & -4837 & (2140) & 0.02 & -3141 & (1376) & 0.02\\
    12487 & EP vs NP & -6265 & (2582) & 0.02 & 631 & (2123) & 0.77 & -8284 & (4112) & 0.04 & -6588 & (3491) & 0.06\\
     \bottomrule
    \end{tabular}
    \begin{tablenotes}
\textit{Note:} The table shows the GATEs for the different programmes in comparison to no programme (NP: Non Participation, WS: Wage Subsidy, BC: Basic courses, TC: Training Courses, EP: Employment Programme) and the different sub-groups. The Earnings decile is the average earnings within each decile of the earnings distribution in the three months before the unemployment spell.
\end{tablenotes}
\end{threeparttable}
\end{adjustbox}
\end{table}

\begin{table}[htbp!]
\begin{adjustbox}{width=0.95\columnwidth, center}
\begin{threeparttable}
  \captionsetup{font=large}
    \caption{GATEs for ue rate, PERM sample}
    \centering
    \begin{tabular}{@{}ll|rrrrrr|rrrrrrr@{}}
    \toprule
     \multicolumn{2}{l}{\textbf{Ue rate (\%)}} & \multicolumn{6}{c}{\textbf{First Year}} & \multicolumn{6}{c}{\textbf{Third Year}} \\
    \midrule
  & Programme & GATE & S.E. & p-value & GATE-ATE & S.E. & p-value & GATE & S.E. & p-value & GATE-ATE & S.E. & p-value \\
    \midrule
    &&  \multicolumn{12}{c}{\textbf{Outcome: Employment}} \\ \midrule
    2 & WS vs NP & -1 & (0.07) & 0.00 & -0 & (0.02) & 0.00 & 1 & (0.07) & 0.00 & -0 & (0.02) & 0.00\\
    3 & WS vs NP & -1 & (0.07) & 0.00 & -0 & (0.01) & 0.00 & 1 & (0.07) & 0.00 & -0 & (0.01) & 0.00\\
    4 & WS vs NP & -1 & (0.07) & 0.00 & 0 & (0.01) & 0.00 & 1 & (0.07) & 0.00 & 0 & (0.01) & 0.24\\
    7 & WS vs NP & -0 & (0.07) & 0.00 & 0 & (0.03) & 0.00 & 1 & (0.08) & 0.00 & 0 & (0.03) & 0.00\\ \midrule
    2 & BC vs NP & -2 & (0.08) & 0.00 & -0 & (0.03) & 0.00 & -0 & (0.09) & 0.09 & -0 & (0.02) & 0.29\\
    3 & BC vs NP & -2 & (0.08) & 0.00 & -0 & (0.01) & 0.02 & -0 & (0.09) & 0.12 & -0 & (0.01) & 0.50\\
    4 & BC vs NP & -2 & (0.08) & 0.00 & 0 & (0.02) & 0.00 & -0 & (0.09) & 0.15 & -0 & (0.02) & 0.85\\
    7 & BC vs NP & -2 & (0.08) & 0.00 & 0 & (0.03) & 0.00 & -0 & (0.10) & 0.38 & 0 & (0.03) & 0.22\\ \midrule
    2 & TC vs NP & -2 & (0.13) & 0.00 & -0 & (0.05) & 0.06 & -0 & (0.14) & 0.77 & -0 & (0.04) & 0.39\\
    3 & TC vs NP & -2 & (0.12) & 0.00 & -0 & (0.02) & 0.07 & -0 & (0.14) & 0.85 & -0 & (0.02) & 0.44\\
    4 & TC vs NP & -2 & (0.12) & 0.00 & 0 & (0.03) & 0.55 & -0 & (0.14) & 0.87 & -0 & (0.03) & 0.64\\
    7 & TC vs NP & -1 & (0.12) & 0.00 & 0 & (0.05) & 0.01 & 0 & (0.14) & 0.71 & 0 & (0.05) & 0.18\\ \midrule
    2 & EP vs NP & -2 & (0.14) & 0.00 & -0 & (0.05) & 0.02 & 0 & (0.15) & 0.41 & -0 & (0.04) & 0.12\\
    3 & EP vs NP & -2 & (0.14) & 0.00 & -0 & (0.03) & 0.00 & 0 & (0.14) & 0.36 & -0 & (0.03) & 0.09\\
    4 & EP vs NP & -1 & (0.14) & 0.00 & 0 & (0.03) & 0.25 & 0 & (0.15) & 0.19 & 0 & (0.02) & 0.73\\
    7 & EP vs NP & -1 & (0.14) & 0.00 & 0 & (0.06) & 0.00 & 0 & (0.16) & 0.06 & 0 & (0.06) & 0.06\\
    \midrule
    &&  \multicolumn{12}{c}{\textbf{Outcome: Earnings}} \\ \midrule
    2 & WS vs NP & -6236 & (393) & 0.00 & -434 & (83) & 0.00 & 1528 & (695) & 0.03 & -530 & (173) & 0.00\\
    3 & WS vs NP & -6303 & (399) & 0.00 & -500 & (69) & 0.00 & 1691 & (709) & 0.02 & -368 & (114) & 0.00\\
    4 & WS vs NP & -5901 & (427) & 0.00 & -98 & (74) & 0.19 & 2264 & (791) & 0.00 & 206 & (113) & 0.07\\
    7 & WS vs NP & -4706 & (403) & 0.00 & 1097 & (139) & 0.00 & 2840 & (761) & 0.00 & 781 & (251) & 0.00\\   \midrule
    2 & BC vs NP & -11122 & (423) & 0.00 & -388 & (89) & 0.00 & -2521 & (763) & 0.00 & -174 & (188) & 0.35\\
    3 & BC vs NP & -11015 & (425) & 0.00 & -280 & (70) & 0.00 & -2487 & (771) & 0.00 & -140 & (122) & 0.25\\  
    4 & BC vs NP & -10795 & (443) & 0.00 & -60 & (77) & 0.43 & -2310 & (841) & 0.01 & 37 & (123) & 0.76\\
    7 & BC vs NP & -9956 & (423) & 0.00 & 778 & (144) & 0.00 & -2040 & (816) & 0.01 & 307 & (270) & 0.26\\ \midrule
    2 & TC vs NP & -9744 & (641) & 0.00 & -102 & (145) & 0.48 & -1638 & (1047) & 0.12 & -506 & (259) & 0.05\\
    3 & TC vs NP & -9873 & (632) & 0.00 & -231 & (89) & 0.01 & -1415 & (1065) & 0.18 & -284 & (164) & 0.08\\
    4 & TC vs NP & -9867 & (636) & 0.00 & -224 & (94) & 0.02 & -1070 & (1109) & 0.33 & 61 & (176) & 0.73\\
    7 & TC vs NP & -9074 & (606) & 0.00 & 568 & (199) & 0.00 & -328 & (1075) & 0.76 & 803 & (364) & 0.03\\ \midrule
    2 & EP vs NP & -7005 & (845) & 0.00 & -108 & (178) & 0.54 & -2385 & (1192) & 0.05 & -689 & (295) & 0.02\\
    3 & EP vs NP & -7020 & (853) & 0.00 & -123 & (112) & 0.27 & -2248 & (1235) & 0.07 & -553 & (201) & 0.01\\
    4 & EP vs NP & -6996 & (859) & 0.00 & -99 & (106) & 0.35 & -1629 & (1297) & 0.21 & 66 & (184) & 0.72\\
    7 & EP vs NP & -6553 & (749) & 0.00 & 344 & (216) & 0.11 & -409 & (1245) & 0.74 & 1287 & (413) & 0.00\\
    \bottomrule
    \bottomrule
    \end{tabular}
    \begin{tablenotes}
\textit{Note:} The table shows the GATEs for the different programmes in comparison to no programme (NP: Non Participation, WS: Wage Subsidy, BC: Basic courses, TC: Training Courses, EP: Employment Programme) and the different sub-groups. The unemployment rate values approximate the closest integer.
\end{tablenotes}
\end{threeparttable}
\end{adjustbox}
\end{table}
\clearpage

\subsubsection{Temporary Residents Sample} \label{gates_app_tempw}
\begin{table}[htbp!]
\begin{adjustbox}{width=0.95\columnwidth, center}
\begin{threeparttable}
  \captionsetup{font=large}
    \caption{GATEs for place of origin, TEMP sample}
    \label{table:gate_b_origin}
    \centering
    \begin{tabular}{@{}ll|rrrrrr|rrrrrrr@{}}
    \toprule
     \multicolumn{2}{l}{\textbf{Origin}}& \multicolumn{6}{c}{\textbf{First Year}} & \multicolumn{6}{c}{\textbf{Third Year}} \\
    \midrule
     & Programme & GATE & S.E. & p-value & GATE-ATE & S.E. & p-value & GATE & S.E. & p-value & GATE-ATE & S.E. & p-value \\
    \midrule
     &&  \multicolumn{12}{c}{\textbf{Outcome: Employment}} \\ \midrule
    Neig. countries & WS vs NP & -0.42 & (0.11) & 0.00 & -0.01 & (0.02) & 0.80 & 1.02 & (0.14) & 0.00 & 0.04 & (0.04) & 0.29\\
    EU & WS vs NP & -0.45 & (0.10) & 0.00 & -0.04 & (0.02) & 0.02 & 0.92 & (0.12) & 0.00 & -0.06 & (0.02) & 0.01\\
    Rest of Europe & WS vs NP & -0.37 & (0.11) & 0.00 & 0.04 & (0.03) & 0.14 & 0.94 & (0.13) & 0.00 & -0.04 & (0.03) & 0.21\\
    Rest of the world & WS vs NP & -0.34 & (0.12) & 0.00 & 0.08 & (0.03) & 0.00 & 1.05 & (0.14) & 0.00 & 0.07 & (0.03) & 0.04\\
    \midrule
    Neig. countries & BC vs NP & -1.30 & (0.13) & 0.00 & -0.00 & (0.03) & 0.97 & 0.47 & (0.17) & 0.01 & -0.02 & (0.05) & 0.64\\
    EU & BC vs NP & -1.33 & (0.13) & 0.00 & -0.03 & (0.02) & 0.09 & 0.49 & (0.16) & 0.00 & -0.01 & (0.03) & 0.86\\
    Rest of Europe & BC vs NP & -1.30 & (0.14) & 0.00 & -0.00 & (0.03) & 0.92 & 0.53 & (0.17) & 0.00 & 0.03 & (0.04) & 0.42\\
    Rest of the world & BC vs NP & -1.23 & (0.13) & 0.00 & 0.07 & (0.03) & 0.02 & 0.53 & (0.17) & 0.00 & 0.04 & (0.04) & 0.38\\
    \midrule
    Neig. countries & TC vs NP & -1.41 & (0.15) & 0.00 & -0.04 & (0.03) & 0.21 & 0.51 & (0.20) & 0.01 & 0.03 & (0.06) & 0.58\\
    EU & TC vs NP & -1.39 & (0.15) & 0.00 & -0.02 & (0.02) & 0.31 & 0.43 & (0.18) & 0.02 & -0.04 & (0.04) & 0.21\\
    Rest of Europe & TC vs NP & -1.34 & (0.15) & 0.00 & 0.03 & (0.03) & 0.38 & 0.46 & (0.18) & 0.01 & -0.02 & (0.05) & 0.73\\
    Rest of the world & TC vs NP & -1.26 & (0.15) & 0.00 & 0.11 & (0.03) & 0.00 & 0.51 & (0.19) & 0.01 & 0.04 & (0.05) & 0.46\\
    \midrule
    Neig. countries & EP vs NP & -1.35 & (0.19) & 0.00 & -0.07 & (0.04) & 0.07 & 0.35 & (0.26) & 0.19 & -0.03 & (0.07) & 0.71\\
    EU & EP vs NP & -1.26 & (0.19) & 0.00 & 0.02 & (0.03) & 0.55 & 0.36 & (0.24) & 0.13 & -0.01 & (0.04) & 0.77\\
    Rest of Europe & EP vs NP & -1.26 & (0.19) & 0.00 & 0.01 & (0.04) & 0.74 & 0.39 & (0.24) & 0.10 & 0.01 & (0.06) & 0.81\\
    Rest of the world & EP vs NP & -1.18 & (0.19) & 0.00 & 0.10 & (0.04) & 0.03 & 0.45 & (0.25) & 0.07 & 0.07 & (0.07) & 0.28\\
    \midrule
     &&  \multicolumn{12}{c}{\textbf{Outcome: Earnings}} \\ \midrule
    Neig. countries & WS vs NP & -3978 & (830) & 0.00 & -463 & (240) & 0.05 & 3705 & (1182) & 0.00 & 479 & (350) & 0.17\\
    EU & WS vs NP & -3451 & (629) & 0.00 & 64 & (153) & 0.67 & 2853 & (863) & 0.00 & -373 & (230) & 0.10\\
    Rest of Europe & WS vs NP & -2988 & (693) & 0.00 & 528 & (261) & 0.04 & 2566 & (926) & 0.01 & -660 & (392) & 0.09\\
    Rest of the world & WS vs NP & -2999 & (771) & 0.00 & 516 & (174) & 0.00 & 3371 & (1104) & 0.00 & 145 & (240) & 0.55\\
    \midrule
    Neig. countries & BC vs NP & -7354 & (932) & 0.00 & -438 & (255) & 0.09 & 882 & (1362) & 0.52 & -22 & (389) & 0.96\\
    EU & BC vs NP & -6884 & (750) & 0.00 & 32 & (157) & 0.84 & 903 & (1054) & 0.39 & -0 & (247) & 1.00\\
    Rest of Europe & BC vs NP & -6575 & (790) & 0.00 & 342 & (264) & 0.20 & 1102 & (1090) & 0.31 & 198 & (412) & 0.63\\
    Rest of the world & BC vs NP & -6305 & (845) & 0.00 & 612 & (202) & 0.00 & 864 & (1243) & 0.49 & -39 & (286) & 0.89\\
    \midrule
    Neig. countries & TC vs NP & -7703 & (1051) & 0.00 & -654 & (298) & 0.03 & 2120 & (1573) & 0.18 & 383 & (469) & 0.41\\
    EU & TC vs NP & -6945 & (823) & 0.00 & 104 & (185) & 0.58 & 1443 & (1168) & 0.22 & -294 & (301) & 0.33\\
    Rest of Europe & TC vs NP & -6569 & (855) & 0.00 & 479 & (299) & 0.11 & 1412 & (1184) & 0.23 & -326 & (478) & 0.50\\
    Rest of the world & TC vs NP & -6237 & (922) & 0.00 & 812 & (230) & 0.00 & 1762 & (1361) & 0.20 & 25 & (327) & 0.94\\
    \midrule
    Neig. countries & EP vs NP & -8543 & (1243) & 0.00 & -994 & (363) & 0.01 & -669 & (2188) & 0.76 & -385 & (685) & 0.57\\
    EU & EP vs NP & -7173 & (985) & 0.00 & 376 & (224) & 0.09 & -172 & (1500) & 0.91 & 112 & (422) & 0.79\\
    Rest of Europe & EP vs NP & -6930 & (996) & 0.00 & 619 & (362) & 0.09 & 10 & (1432) & 0.99 & 294 & (679) & 0.66\\
    Rest of the world & EP vs NP & -6724 & (1076) & 0.00 & 825 & (261) & 0.00 & 84 & (1725) & 0.96 & 368 & (416) & 0.38\\

    \bottomrule
    \end{tabular}
        \begin{tablenotes}
\textit{Note:} The table shows the GATEs for the different programmes in comparison to no programme (NP: Non Participation, WS: Wage Subsidy, BC: Basic courses, TC: Training Courses, EP: Employment Programme) and the different sub-groups. 
\end{tablenotes}
\end{threeparttable}
\end{adjustbox}
\end{table}

\begin{table}[htbp!]
\begin{adjustbox}{width=0.95\columnwidth, center}
\begin{threeparttable}
  \captionsetup{font=large}
    \caption{GATEs for gender, TEMP sample}
    \label{table:gate_b_gender}
    \centering
    \begin{tabular}{@{}ll|rrrrrr|rrrrrrr@{}}
    \toprule
     \multicolumn{2}{l}{\textbf{Gender}} & \multicolumn{6}{c}{\textbf{First Year}} & \multicolumn{6}{c}{\textbf{Third Year}} \\
    \midrule
  & Programme & GATE & S.E. & p-value & GATE-ATE & S.E. & p-value & GATE & S.E. & p-value & GATE-ATE & S.E. & p-value \\
    \midrule
    &&  \multicolumn{12}{c}{\textbf{Outcome: Employment}} \\ \midrule
    Man & WS vs NP & -0.42 & (0.11) & 0.00 & -0.01 & (0.01) & 0.67 & 0.94 & (0.13) & 0.00 & -0.04 & (0.01) & 0.00\\ 
    Woman & WS vs NP & -0.40 & (0.11) & 0.00 & 0.01 & (0.02) & 0.63 & 1.06 & (0.13) & 0.00 & 0.08 & (0.03) & 0.00\\ \midrule
    Man & BC vs NP & -1.33 & (0.13) & 0.00 & -0.03 & (0.02) & 0.03 & 0.49 & (0.16) & 0.00 & -0.00 & (0.02) & 0.84\\ 
    Woman & BC vs NP & -1.24 & (0.13) & 0.00 & 0.06 & (0.03) & 0.03 & 0.50 & (0.17) & 0.00 & 0.01 & (0.04) & 0.84\\ \midrule
    Man & TC vs NP & -1.39 & (0.15) & 0.00 & -0.02 & (0.02) & 0.31 & 0.48 & (0.18) & 0.01 & 0.00 & (0.02) & 0.93\\ 
    Woman & TC vs NP & -1.34 & (0.15) & 0.00 & 0.03 & (0.03) & 0.30 & 0.47 & (0.19) & 0.01 & -0.00 & (0.04) & 0.94\\ \midrule
    Man & EP vs NP & -1.32 & (0.19) & 0.00 & -0.04 & (0.02) & 0.05 & 0.34 & (0.25) & 0.16 & -0.03 & (0.03) & 0.26\\
    Woman & EP vs NP & -1.19 & (0.19) & 0.00 & 0.08 & (0.04) & 0.05 & 0.43 & (0.25) & 0.08 & 0.06 & (0.05) & 0.27\\
    \midrule
    &&  \multicolumn{12}{c}{\textbf{Outcome: Earnings}} \\ \midrule
   Man & WS vs NP & -3653 & (718) & 0.00 & -138 & (74) & 0.06 & 3285 & (1001) & 0.00 & 59 & (104) & 0.57\\
    Woman & WS vs NP & -3266 & (699) & 0.00 & 249 & (134) & 0.06 & 3120 & (990) & 0.00 & -106 & (187) & 0.57\\ \midrule
    Man & BC vs NP & -7295 & (832) & 0.00 & -378 & (97) & 0.00 & 937 & (1187) & 0.43 & 33 & (133) & 0.80\\
    Woman & BC vs NP & -6233 & (787) & 0.00 & 684 & (175) & 0.00 & 843 & (1144) & 0.46 & -60 & (240) & 0.80\\ \midrule
    Man & TC vs NP & -7372 & (934) & 0.00 & -324 & (110) & 0.00 & 1846 & (1340) & 0.17 & 108 & (150) & 0.47\\
    Woman & TC vs NP & -6464 & (865) & 0.00 & 585 & (199) & 0.00 & 1541 & (1281) & 0.23 & -196 & (271) & 0.47\\ \midrule
    Man & EP vs NP & -7990 & (1089) & 0.00 & -441 & (138) & 0.00 & -448 & (1759) & 0.80 & -164 & (205) & 0.42\\
    Woman & EP vs NP & -6753 & (1042) & 0.00 & 796 & (249) & 0.00 & 12 & (1684) & 0.99 & 296 & (371) & 0.42\\
    \bottomrule
    \end{tabular}
    \begin{tablenotes}
\textit{Note:} The table shows the GATEs for the different programmes in comparison to no programme (NP: Non Participation, WS: Wage Subsidy, BC: Basic courses, TC: Training Courses, EP: Employment Programme) and the different sub-groups. 
\end{tablenotes}
\end{threeparttable}
\end{adjustbox}
\end{table}

\begin{table}[htbp!]
\begin{adjustbox}{width=0.9\columnwidth, center} 
\begin{threeparttable}
  \captionsetup{font=large}
    \caption{GATEs for last job sector, TEMP sample}
    \label{table:gate_b_sector}
    \centering
    \begin{tabular}{@{}ll|rrrrrr|rrrrrrr@{}}
    \toprule
     \multicolumn{2}{l}{\textbf{Sector of last job}}& \multicolumn{6}{c}{\textbf{First Year}} & \multicolumn{6}{c}{\textbf{Third Year}} \\
    \midrule
     & Programme & GATE & S.E. & p-value & GATE-ATE & S.E. & p-value & GATE & S.E. & p-value & GATE-ATE & S.E. & p-value \\
    \midrule
    &&  \multicolumn{12}{c}{\textbf{Outcome: Employment}} \\ \midrule
    A & WS vs NP & -0.47 & (0.11) & 0.00 & -0.06 & (0.04) & 0.13 & 0.87 & (0.13) & 0.00 & -0.12 & (0.05) & 0.02\\
    B & WS vs NP & -0.40 & (0.11) & 0.00 & 0.01 & (0.03) & 0.57 & 0.96 & (0.13) & 0.00 & -0.02 & (0.03) & 0.44\\
    C & WS vs NP & -0.36 & (0.13) & 0.00 & 0.05 & (0.05) & 0.35 & 1.06 & (0.15) & 0.00 & 0.08 & (0.07) & 0.24\\
    D & WS vs NP & -0.43 & (0.11) & 0.00 & -0.02 & (0.03) & 0.63 & 0.82 & (0.13) & 0.00 & -0.16 & (0.04) & 0.00\\
    E & WS vs NP & -0.34 & (0.12) & 0.00 & 0.07 & (0.02) & 0.00 & 1.07 & (0.14) & 0.00 & 0.08 & (0.03) & 0.00\\
    F & WS vs NP & -0.52 & (0.11) & 0.00 & -0.11 & (0.04) & 0.01 & 0.95 & (0.13) & 0.00 & -0.03 & (0.05) & 0.51\\
    G & WS vs NP & -0.25 & (0.14) & 0.08 & 0.16 & (0.07) & 0.02 & 1.17 & (0.16) & 0.00 & 0.19 & (0.08) & 0.01\\
    H & WS vs NP & -0.35 & (0.12) & 0.00 & 0.06 & (0.04) & 0.10 & 1.16 & (0.14) & 0.00 & 0.18 & (0.05) & 0.00\\ \midrule
    A & BC vs NP & -1.33 & (0.14) & 0.00 & -0.03 & (0.05) & 0.61 & 0.46 & (0.18) & 0.01 & -0.03 & (0.07) & 0.67\\
    B & BC vs NP & -1.31 & (0.14) & 0.00 & -0.01 & (0.03) & 0.74 & 0.53 & (0.17) & 0.00 & 0.04 & (0.03) & 0.23\\
    C & BC vs NP & -1.29 & (0.14) & 0.00 & 0.01 & (0.06) & 0.86 & 0.51 & (0.18) & 0.01 & 0.02 & (0.08) & 0.83\\
    D & BC vs NP & -1.37 & (0.14) & 0.00 & -0.07 & (0.05) & 0.13 & 0.48 & (0.17) & 0.00 & -0.01 & (0.06) & 0.80\\
    E & BC vs NP & -1.24 & (0.13) & 0.00 & 0.06 & (0.03) & 0.02 & 0.51 & (0.17) & 0.00 & 0.02 & (0.04) & 0.65\\
    F & BC vs NP & -1.34 & (0.14) & 0.00 & -0.04 & (0.05) & 0.44 & 0.43 & (0.17) & 0.01 & -0.06 & (0.06) & 0.34\\
    G & BC vs NP & -1.16 & (0.15) & 0.00 & 0.14 & (0.07) & 0.05 & 0.53 & (0.19) & 0.01 & 0.04 & (0.09) & 0.68\\
    H & BC vs NP & -1.24 & (0.14) & 0.00 & 0.06 & (0.04) & 0.11 & 0.57 & (0.17) & 0.00 & 0.08 & (0.05) & 0.13\\ \midrule
    A & TC vs NP & -1.39 & (0.16) & 0.00 & -0.02 & (0.06) & 0.68 & 0.38 & (0.19) & 0.04 & -0.09 & (0.07) & 0.19\\
    B & TC vs NP & -1.39 & (0.15) & 0.00 & -0.02 & (0.03) & 0.56 & 0.47 & (0.19) & 0.01 & -0.00 & (0.04) & 0.94\\
    C & TC vs NP & -1.36 & (0.16) & 0.00 & 0.01 & (0.07) & 0.89 & 0.61 & (0.21) & 0.00 & 0.14 & (0.09) & 0.13\\
    D & TC vs NP & -1.42 & (0.16) & 0.00 & -0.04 & (0.05) & 0.41 & 0.40 & (0.19) & 0.03 & -0.07 & (0.06) & 0.25\\
    E & TC vs NP & -1.33 & (0.15) & 0.00 & 0.04 & (0.03) & 0.14 & 0.53 & (0.20) & 0.01 & 0.06 & (0.04) & 0.15\\
    F & TC vs NP & -1.41 & (0.16) & 0.00 & -0.04 & (0.06) & 0.48 & 0.38 & (0.20) & 0.05 & -0.09 & (0.07) & 0.19\\
    G & TC vs NP & -1.22 & (0.17) & 0.00 & 0.15 & (0.08) & 0.06 & 0.66 & (0.22) & 0.00 & 0.18 & (0.10) & 0.07\\
    H & TC vs NP & -1.33 & (0.16) & 0.00 & 0.04 & (0.05) & 0.34 & 0.58 & (0.20) & 0.00 & 0.11 & (0.07) & 0.11\\ \midrule
    A & EP vs NP & -1.21 & (0.20) & 0.00 & 0.06 & (0.07) & 0.36 & 0.41 & (0.24) & 0.10 & 0.03 & (0.10) & 0.73\\
    B & EP vs NP & -1.28 & (0.19) & 0.00 & -0.01 & (0.04) & 0.80 & 0.42 & (0.24) & 0.08 & 0.05 & (0.05) & 0.35\\
    C & EP vs NP & -1.36 & (0.20) & 0.00 & -0.09 & (0.09) & 0.32 & 0.41 & (0.29) & 0.16 & 0.04 & (0.13) & 0.78\\
    D & EP vs NP & -1.32 & (0.19) & 0.00 & -0.05 & (0.07) & 0.46 & 0.29 & (0.25) & 0.24 & -0.08 & (0.09) & 0.35\\
    E & EP vs NP & -1.26 & (0.19) & 0.00 & 0.01 & (0.04) & 0.81 & 0.41 & (0.26) & 0.11 & 0.03 & (0.05) & 0.52\\
    F & EP vs NP & -1.19 & (0.20) & 0.00 & 0.08 & (0.08) & 0.29 & 0.32 & (0.25) & 0.21 & -0.06 & (0.10) & 0.56\\
    G & EP vs NP & -1.31 & (0.21) & 0.00 & -0.04 & (0.10) & 0.70 & 0.43 & (0.30) & 0.15 & 0.06 & (0.14) & 0.68\\
    H & EP vs NP & -1.30 & (0.20) & 0.00 & -0.02 & (0.06) & 0.70 & 0.54 & (0.28) & 0.05 & 0.17 & (0.10) & 0.08\\
    \midrule
&&  \multicolumn{12}{c}{\textbf{Outcome: Earnings}} \\ \midrule
A & WS vs NP & -3213 & (628) & 0.00 & 303 & (380) & 0.43 & 2205 & (862) & 0.01 & -1021 & (551) & 0.06\\
    B & WS vs NP & -3128 & (686) & 0.00 & 387 & (262) & 0.14 & 2675 & (918) & 0.00 & -551 & (398) & 0.17\\
    C & WS vs NP & -4697 & (1220) & 0.00 & -1182 & (699) & 0.09 & 4495 & (1777) & 0.01 & 1269 & (1036) & 0.22\\
    D & WS vs NP & -3435 & (649) & 0.00 & 81 & (287) & 0.78 & 2681 & (881) & 0.00 & -545 & (444) & 0.22\\
    E & WS vs NP & -3511 & (838) & 0.00 & 5 & (195) & 0.98 & 3708 & (1199) & 0.00 & 482 & (310) & 0.12\\
    F & WS vs NP & -3128 & (597) & 0.00 & 387 & (374) & 0.30 & 2368 & (797) & 0.00 & -858 & (536) & 0.11\\
    G & WS vs NP & -4121 & (1409) & 0.00 & -606 & (884) & 0.49 & 5747 & (2102) & 0.01 & 2521 & (1365) & 0.06\\
    H & WS vs NP & -3908 & (913) & 0.00 & -393 & (335) & 0.24 & 3946 & (1368) & 0.00 & 720 & (605) & 0.23\\ \midrule
    A & BC vs NP & -6371 & (746) & 0.00 & 545 & (422) & 0.20 & 709 & (1060) & 0.50 & -195 & (633) & 0.76\\
    B & BC vs NP & -6650 & (785) & 0.00 & 267 & (266) & 0.32 & 1220 & (1089) & 0.26 & 316 & (419) & 0.45\\
    C & BC vs NP & -8227 & (1316) & 0.00 & -1311 & (733) & 0.07 & 527 & (1959) & 0.79 & -377 & (1127) & 0.74\\
    D & BC vs NP & -7398 & (820) & 0.00 & -482 & (327) & 0.14 & 1122 & (1113) & 0.31 & 218 & (494) & 0.66\\
    E & BC vs NP & -6815 & (915) & 0.00 & 101 & (200) & 0.61 & 810 & (1358) & 0.55 & -93 & (338) & 0.78\\
    F & BC vs NP & -6115 & (714) & 0.00 & 802 & (417) & 0.05 & 802 & (1001) & 0.42 & -101 & (614) & 0.87\\
    G & BC vs NP & -7607 & (1420) & 0.00 & -690 & (855) & 0.42 & 488 & (2197) & 0.82 & -416 & (1375) & 0.76\\
    H & BC vs NP & -7126 & (1004) & 0.00 & -210 & (357) & 0.56 & 1087 & (1524) & 0.48 & 184 & (641) & 0.77\\ \midrule
    A & TC vs NP & -6335 & (810) & 0.00 & 713 & (478) & 0.14 & 813 & (1131) & 0.47 & -925 & (715) & 0.20\\
    B & TC vs NP & -6836 & (857) & 0.00 & 212 & (304) & 0.48 & 1393 & (1199) & 0.25 & -344 & (490) & 0.48\\
    C & TC vs NP & -8529 & (1509) & 0.00 & -1480 & (864) & 0.09 & 2896 & (2275) & 0.20 & 1158 & (1342) & 0.39\\
    D & TC vs NP & -7399 & (901) & 0.00 & -350 & (380) & 0.36 & 1529 & (1253) & 0.22 & -209 & (569) & 0.71\\
    E & TC vs NP & -7098 & (1017) & 0.00 & -50 & (224) & 0.82 & 1988 & (1536) & 0.20 & 251 & (392) & 0.52\\
    F & TC vs NP & -6218 & (786) & 0.00 & 831 & (473) & 0.08 & 1092 & (1111) & 0.33 & -645 & (714) & 0.37\\
    G & TC vs NP & -7651 & (1612) & 0.00 & -602 & (1004) & 0.55 & 3097 & (2551) & 0.22 & 1359 & (1623) & 0.40\\
    H & TC vs NP & -7402 & (1146) & 0.00 & -354 & (426) & 0.41 & 2176 & (1764) & 0.22 & 438 & (773) & 0.57\\ \midrule
    A & EP vs NP & -6094 & (994) & 0.00 & 1455 & (575) & 0.01 & 490 & (1349) & 0.72 & 774 & (981) & 0.43\\
    B & EP vs NP & -7164 & (997) & 0.00 & 385 & (353) & 0.28 & 287 & (1468) & 0.84 & 571 & (679) & 0.40\\
    C & EP vs NP & -9953 & (1834) & 0.00 & -2404 & (1076) & 0.03 & -1073 & (3389) & 0.75 & -789 & (2028) & 0.70\\
    D & EP vs NP & -7750 & (1069) & 0.00 & -201 & (458) & 0.66 & -439 & (1589) & 0.78 & -155 & (807) & 0.85\\
    E & EP vs NP & -7852 & (1182) & 0.00 & -303 & (266) & 0.25 & -575 & (1974) & 0.77 & -291 & (495) & 0.56\\
    F & EP vs NP & -5899 & (972) & 0.00 & 1650 & (588) & 0.01 & -130 & (1321) & 0.92 & 154 & (999) & 0.88\\
    G & EP vs NP & -9716 & (1911) & 0.00 & -2167 & (1188) & 0.07 & -1154 & (3656) & 0.75 & -870 & (2326) & 0.71\\
    H & EP vs NP & -8424 & (1342) & 0.00 & -875 & (541) & 0.11 & 574 & (2530) & 0.82 & 858 & (1183) & 0.47\\
    \bottomrule
    \end{tabular}
    \begin{tablenotes}
\textit{Note:} The table shows the GATEs for the different programmes in comparison to no programme (NP: Non Participation, WS: Wage Subsidy, BC: Basic courses, TC: Training Courses, EP: Employment Programme) and the different sub-groups (A: Agricultural and forestry, B: Production in industry and trade, C: Technical and information technology, D: Construction and mining; E: Transport and Trade, F: Hospitality and personal services, G: Business, Finance and Law, H: Education, Art and Science.)

\end{tablenotes}
\end{threeparttable}
\end{adjustbox}
\end{table}

\begin{table}[htbp!]
\begin{adjustbox}{width=0.95\columnwidth, center}
\begin{threeparttable}
  \captionsetup{font=large}
    \caption{GATEs for education levels, TEMP sample}
    \label{table:gate_b_education}
    \centering
    \begin{tabular}{@{}ll|rrrrrr|rrrrrrr@{}}
    \toprule
     \multicolumn{2}{l}{\textbf{Education}}& \multicolumn{6}{c}{\textbf{First Year}} & \multicolumn{6}{c}{\textbf{Third Year}} \\
    \midrule
    
     & Programme & GATE & S.E. & p-value & GATE-ATE & S.E. & p-value & GATE & S.E. & p-value & GATE-ATE & S.E. & p-value \\
    \midrule
    &&  \multicolumn{12}{c}{\textbf{Outcome: Employment}} \\ \midrule
    Compulsory & WS vs NP & -0.46 & (0.11) & 0.00 & -0.05 & (0.02) & 0.06 & 0.91 & (0.13) & 0.00 & -0.07 & (0.03) & 0.03\\
    Secondary & WS vs NP & -0.45 & (0.11) & 0.00 & -0.03 & (0.01) & 0.01 & 0.94 & (0.13) & 0.00 & -0.04 & (0.02) & 0.02\\
    Vocational & WS vs NP & -0.43 & (0.11) & 0.00 & -0.02 & (0.03) & 0.55 & 1.04 & (0.14) & 0.00 & 0.06 & (0.04) & 0.13\\
    Bachelor & WS vs NP & -0.31 & (0.13) & 0.02 & 0.11 & (0.05) & 0.03 & 1.14 & (0.15) & 0.00 & 0.16 & (0.06) & 0.01\\
    Post-Grad & WS vs NP & -0.24 & (0.14) & 0.09 & 0.17 & (0.07) & 0.01 & 1.21 & (0.17) & 0.00 & 0.23 & (0.08) & 0.01\\
        \midrule
    Compulsory & BC vs NP & -1.34 & (0.13) & 0.00 & -0.04 & (0.03) & 0.16 & 0.48 & (0.17) & 0.00 & -0.01 & (0.04) & 0.72\\
    Secondary & BC vs NP & -1.33 & (0.13) & 0.00 & -0.03 & (0.01) & 0.05 & 0.47 & (0.16) & 0.00 & -0.03 & (0.02) & 0.19\\
    Vocational & BC vs NP & -1.33 & (0.13) & 0.00 & -0.03 & (0.04) & 0.42 & 0.49 & (0.17) & 0.00 & -0.00 & (0.05) & 0.97\\
    Bachelor & BC vs NP & -1.24 & (0.14) & 0.00 & 0.07 & (0.05) & 0.19 & 0.53 & (0.18) & 0.00 & 0.04 & (0.07) & 0.56\\
    Post-Grad & BC vs NP & -1.15 & (0.15) & 0.00 & 0.16 & (0.07) & 0.02 & 0.58 & (0.19) & 0.00 & 0.09 & (0.09) & 0.32\\
        \midrule
    Compulsory & TC vs NP & -1.39 & (0.15) & 0.00 & -0.02 & (0.03) & 0.59 & 0.40 & (0.18) & 0.03 & -0.08 & (0.05) & 0.10\\
    Secondary & TC vs NP & -1.43 & (0.15) & 0.00 & -0.05 & (0.02) & 0.00 & 0.45 & (0.19) & 0.02 & -0.02 & (0.03) & 0.36\\
    Vocational & TC vs NP & -1.38 & (0.16) & 0.00 & -0.01 & (0.04) & 0.76 & 0.55 & (0.20) & 0.01 & 0.07 & (0.06) & 0.24\\
    Bachelor & TC vs NP & -1.28 & (0.16) & 0.00 & 0.09 & (0.06) & 0.13 & 0.64 & (0.21) & 0.00 & 0.17 & (0.08) & 0.04\\
    Post-Grad & TC vs NP & -1.22 & (0.17) & 0.00 & 0.15 & (0.08) & 0.05 & 0.67 & (0.22) & 0.00 & 0.20 & (0.11) & 0.06\\
        \midrule
    Compulsory & EP vs NP & -1.23 & (0.19) & 0.00 & 0.05 & (0.04) & 0.26 & 0.35 & (0.24) & 0.14 & -0.03 & (0.06) & 0.67\\
    Secondary & EP vs NP & -1.30 & (0.19) & 0.00 & -0.03 & (0.02) & 0.15 & 0.34 & (0.24) & 0.16 & -0.03 & (0.03) & 0.28\\
    Vocational & EP vs NP & -1.33 & (0.20) & 0.00 & -0.06 & (0.06) & 0.31 & 0.37 & (0.27) & 0.17 & -0.00 & (0.09) & 0.98\\
    Bachelor & EP vs NP & -1.31 & (0.20) & 0.00 & -0.04 & (0.08) & 0.61 & 0.42 & (0.28) & 0.14 & 0.04 & (0.11) & 0.71\\
    Post-Grad & EP vs NP & -1.30 & (0.21) & 0.00 & -0.03 & (0.10) & 0.79 & 0.51 & (0.30) & 0.09 & 0.14 & (0.15) & 0.36\\
\midrule
&&  \multicolumn{12}{c}{\textbf{Outcome: Earnings}} \\ \midrule
   Compulsory & WS vs NP & -3157 & (610) & 0.00 & 359 & (291) & 0.22 & 2500 & (837) & 0.00 & -727 & (451) & 0.11\\
    Secondary & WS vs NP & -3534 & (655) & 0.00 & -19 & (143) & 0.89 & 2806 & (890) & 0.00 & -420 & (265) & 0.11\\
    Vocational & WS vs NP & -4319 & (923) & 0.00 & -804 & (381) & 0.04 & 3882 & (1277) & 0.00 & 656 & (521) & 0.21\\
    Bachelor & WS vs NP & -3966 & (1137) & 0.00 & -451 & (575) & 0.43 & 4815 & (1684) & 0.00 & 1589 & (887) & 0.07\\
    Post-Grad & WS vs NP & -4109 & (1372) & 0.00 & -594 & (844) & 0.48 & 5568 & (2160) & 0.01 & 2342 & (1451) & 0.11\\
    \midrule
    Compulsory & BC vs NP & -6583 & (724) & 0.00 & 333 & (295) & 0.26 & 930 & (1028) & 0.37 & 26 & (480) & 0.96\\
    Secondary & BC vs NP & -6916 & (774) & 0.00 & 0 & (136) & 1.00 & 948 & (1078) & 0.38 & 44 & (269) & 0.87\\
    Vocational & BC vs NP & -7614 & (1039) & 0.00 & -698 & (420) & 0.10 & 705 & (1476) & 0.63 & -198 & (592) & 0.74\\
    Bachelor & BC vs NP & -7526 & (1192) & 0.00 & -610 & (576) & 0.29 & 540 & (1831) & 0.77 & -364 & (939) & 0.70\\
    Post-Grad & BC vs NP & -7474 & (1397) & 0.00 & -558 & (823) & 0.50 & 855 & (2264) & 0.71 & -49 & (1476) & 0.97\\
     \midrule
    Compulsory & TC vs NP & -6576 & (792) & 0.00 & 472 & (347) & 0.17 & 1179 & (1124) & 0.29 & -558 & (570) & 0.33\\
    Secondary & TC vs NP & -7175 & (859) & 0.00 & -126 & (157) & 0.42 & 1686 & (1223) & 0.17 & -51 & (311) & 0.87\\
    Vocational & TC vs NP & -7854 & (1178) & 0.00 & -806 & (495) & 0.10 & 2761 & (1721) & 0.11 & 1024 & (724) & 0.16\\
    Bachelor & TC vs NP & -7583 & (1354) & 0.00 & -535 & (680) & 0.43 & 2941 & (2115) & 0.16 & 1204 & (1102) & 0.27\\
    Post-Grad & TC vs NP & -7641 & (1592) & 0.00 & -593 & (967) & 0.54 & 2778 & (2631) & 0.29 & 1040 & (1736) & 0.55\\
      \midrule
    Compulsory & EP vs NP & -6516 & (948) & 0.00 & 1033 & (423) & 0.01 & -123 & (1350) & 0.93 & 160 & (838) & 0.85\\
    Secondary & EP vs NP & -7437 & (1012) & 0.00 & 112 & (190) & 0.55 & -289 & (1521) & 0.85 & -6 & (442) & 0.99\\
    Vocational & EP vs NP & -8872 & (1400) & 0.00 & -1323 & (598) & 0.03 & -670 & (2456) & 0.79 & -386 & (1083) & 0.72\\
    Bachelor & EP vs NP & -9299 & (1604) & 0.00 & -1750 & (830) & 0.04 & -1221 & (2987) & 0.68 & -937 & (1623) & 0.56\\
    Post-Grad & EP vs NP & -9739 & (1885) & 0.00 & -2190 & (1184) & 0.06 & -373 & (3762) & 0.92 & -89 & (2527) & 0.97\\
  \bottomrule
    \end{tabular}
    \begin{tablenotes}
\textit{Note:} The table shows the GATEs for the different programmes in comparison to no programme (NP: Non Participation, WS: Wage Subsidy, BC: Basic courses, TC: Training Courses, EP: Employment Programme) and the different sub-groups. 
\end{tablenotes}
\end{threeparttable}
\end{adjustbox}
\end{table}

\begin{table}[htbp!]
\begin{adjustbox}{width=0.95\columnwidth, center}
\begin{threeparttable}
  \captionsetup{font=large}
    \caption{GATEs for months into employment, TEMP sample}
    \label{table:gate_b_employment_months}
    \centering
    \begin{tabular}{@{}ll|rrrrrr|rrrrrrr@{}}
    \toprule
     \multicolumn{2}{l}{\textbf{Months into employment}} & \multicolumn{6}{c}{\textbf{First Year}} & \multicolumn{6}{c}{\textbf{Third Year}} \\
    \midrule
  & Programme & GATE & S.E. & p-value & GATE-ATE & S.E. & p-value & GATE & S.E. & p-value & GATE-ATE & S.E. & p-value \\
    \midrule
    &&  \multicolumn{12}{c}{\textbf{Outcome: Employment}} \\ \midrule
    12 & WS vs NP & -0.50 & (0.14) & 0.00 & -0.09 & (0.08) & 0.31 & 1.04 & (0.17) & 0.00 & 0.06 & (0.10) & 0.56\\
    24 & WS vs NP & -0.48 & (0.12) & 0.00 & -0.07 & (0.05) & 0.11 & 0.99 & (0.14) & 0.00 & 0.01 & (0.06) & 0.85\\
    36 & WS vs NP & -0.41 & (0.11) & 0.00 & 0.00 & (0.01) & 1.00 & 0.94 & (0.13) & 0.00 & -0.05 & (0.02) & 0.01\\
    48 & WS vs NP & -0.37 & (0.11) & 0.00 & 0.05 & (0.02) & 0.02 & 0.96 & (0.13) & 0.00 & -0.02 & (0.02) & 0.32\\
    60 & WS vs NP & -0.38 & (0.12) & 0.00 & 0.03 & (0.04) & 0.47 & 1.04 & (0.14) & 0.00 & 0.06 & (0.05) & 0.30\\ \midrule
    12 & BC vs NP & -1.27 & (0.17) & 0.00 & 0.03 & (0.11) & 0.80 & 0.45 & (0.22) & 0.04 & -0.04 & (0.14) & 0.75\\
    24 & BC vs NP & -1.31 & (0.15) & 0.00 & -0.01 & (0.06) & 0.93 & 0.53 & (0.18) & 0.00 & 0.03 & (0.08) & 0.66\\
    36 & BC vs NP & -1.28 & (0.13) & 0.00 & 0.02 & (0.02) & 0.32 & 0.48 & (0.17) & 0.00 & -0.01 & (0.02) & 0.56\\
    48 & BC vs NP & -1.28 & (0.13) & 0.00 & 0.03 & (0.03) & 0.32 & 0.47 & (0.17) & 0.01 & -0.03 & (0.03) & 0.44\\
    60 & BC vs NP & -1.34 & (0.14) & 0.00 & -0.04 & (0.05) & 0.42 & 0.51 & (0.17) & 0.00 & 0.01 & (0.07) & 0.82\\ \midrule
    12 & TC vs NP & -1.50 & (0.18) & 0.00 & -0.13 & (0.12) & 0.27 & 0.44 & (0.24) & 0.07 & -0.04 & (0.15) & 0.81\\
    24 & TC vs NP & -1.47 & (0.16) & 0.00 & -0.10 & (0.06) & 0.13 & 0.46 & (0.20) & 0.02 & -0.01 & (0.08) & 0.90\\
    36 & TC vs NP & -1.35 & (0.15) & 0.00 & 0.02 & (0.02) & 0.39 & 0.45 & (0.18) & 0.01 & -0.02 & (0.03) & 0.39\\
    48 & TC vs NP & -1.29 & (0.15) & 0.00 & 0.08 & (0.03) & 0.01 & 0.48 & (0.19) & 0.01 & 0.01 & (0.04) & 0.87\\
    60 & TC vs NP & -1.36 & (0.16) & 0.00 & 0.01 & (0.06) & 0.86 & 0.51 & (0.20) & 0.01 & 0.03 & (0.08) & 0.67\\ \midrule
    12 & EP vs NP & -1.34 & (0.24) & 0.00 & -0.07 & (0.14) & 0.62 & 0.47 & (0.31) & 0.13 & 0.09 & (0.20) & 0.64\\
    24 & EP vs NP & -1.40 & (0.21) & 0.00 & -0.13 & (0.08) & 0.10 & 0.44 & (0.27) & 0.10 & 0.06 & (0.11) & 0.56\\
    36 & EP vs NP & -1.27 & (0.19) & 0.00 & 0.00 & (0.02) & 0.93 & 0.33 & (0.24) & 0.17 & -0.05 & (0.03) & 0.17\\
    48 & EP vs NP & -1.20 & (0.19) & 0.00 & 0.08 & (0.03) & 0.02 & 0.36 & (0.24) & 0.13 & -0.01 & (0.05) & 0.78\\
    60 & EP vs NP & -1.23 & (0.20) & 0.00 & 0.05 & (0.07) & 0.52 & 0.36 & (0.27) & 0.18 & -0.01 & (0.10) & 0.92\\

    \midrule
    &&  \multicolumn{12}{c}{\textbf{Outcome: Earnings}} \\ \midrule
   12 & WS vs NP & -3391 & (366) & 0.00 & 2412 & (293) & 0.00 & 2359 & (801) & 0.00 & 301 & (485) & 0.54\\
    24 & WS vs NP & -3354 & (398) & 0.00 & 2448 & (265) & 0.00 & 2962 & (792) & 0.00 & 903 & (446) & 0.04\\
    36 & WS vs NP & -3740 & (359) & 0.00 & 2063 & (226) & 0.00 & 2692 & (767) & 0.00 & 633 & (366) & 0.08\\
    48 & WS vs NP & -4442 & (345) & 0.00 & 1361 & (157) & 0.00 & 2406 & (705) & 0.00 & 347 & (236) & 0.14\\
    60 & WS vs NP & -6630 & (443) & 0.00 & -828 & (91) & 0.00 & 1826 & (757) & 0.02 & -233 & (139) & 0.09\\ \midrule
    12 & BC vs NP & -6635 & (470) & 0.00 & 4100 & (356) & 0.00 & -606 & (976) & 0.53 & 1741 & (637) & 0.01\\
    24 & BC vs NP & -6898 & (458) & 0.00 & 3837 & (292) & 0.00 & -404 & (899) & 0.65 & 1942 & (523) & 0.00\\
    36 & BC vs NP & -7475 & (415) & 0.00 & 3260 & (242) & 0.00 & -918 & (863) & 0.29 & 1429 & (426) & 0.00\\
    48 & BC vs NP & -8701 & (391) & 0.00 & 2034 & (159) & 0.00 & -1574 & (785) & 0.04 & 772 & (265) & 0.00\\
    60 & BC vs NP & -12009 & (454) & 0.00 & -1274 & (95) & 0.00 & -2872 & (808) & 0.00 & -526 & (159) & 0.00\\ \midrule
    12 & TC vs NP & -6707 & (749) & 0.00 & 2935 & (563) & 0.00 & -543 & (1336) & 0.68 & 589 & (954) & 0.54\\
    24 & TC vs NP & -6883 & (657) & 0.00 & 2759 & (441) & 0.00 & -210 & (1173) & 0.86 & 921 & (765) & 0.23\\
    36 & TC vs NP & -7295 & (639) & 0.00 & 2347 & (376) & 0.00 & -750 & (1116) & 0.50 & 382 & (630) & 0.54\\
    48 & TC vs NP & -8180 & (609) & 0.00 & 1462 & (243) & 0.00 & -1236 & (1042) & 0.24 & -104 & (393) & 0.79\\
    60 & TC vs NP & -10558 & (658) & 0.00 & -916 & (147) & 0.00 & -1198 & (1113) & 0.28 & -67 & (237) & 0.78\\ \midrule
    12 & EP vs NP & -4454 & (733) & 0.00 & 2443 & (649) & 0.00 & -527 & (1315) & 0.69 & 1169 & (1035) & 0.26\\
    24 & EP vs NP & -4888 & (675) & 0.00 & 2008 & (562) & 0.00 & -20 & (1193) & 0.99 & 1675 & (868) & 0.05\\
    36 & EP vs NP & -5304 & (658) & 0.00 & 1593 & (486) & 0.00 & -350 & (1143) & 0.76 & 1346 & (754) & 0.07\\
    48 & EP vs NP & -6082 & (665) & 0.00 & 815 & (329) & 0.01 & -676 & (1107) & 0.54 & 1020 & (476) & 0.03\\
    60 & EP vs NP & -7475 & (887) & 0.00 & -578 & (192) & 0.00 & -2267 & (1336) & 0.09 & -572 & (286) & 0.05\\
    \bottomrule
    \end{tabular}
    \begin{tablenotes}
\textit{Note:} The table shows the GATEs for the different programmes in comparison to no programme (NP: Non Participation, WS: Wage Subsidy, BC: Basic courses, TC: Training Courses, EP: Employment Programme) and the different sub-groups. The ranges for months in employment are: 12: 1-12, 24: 13-24, 36: 25-36, 48: 37-48, 60: 49-60.
\end{tablenotes}
\end{threeparttable}
\end{adjustbox}
\end{table}

\begin{table}[htbp!]
\begin{adjustbox}{width=0.8\columnwidth, center} 
\begin{threeparttable}
  \captionsetup{font=large}
    \caption{GATEs for earnings decile, TEMP sample}
    \label{table:gate_b_earnings_decile}
    \centering
    \begin{tabular}{@{}ll|rrrrrr|rrrrrrr@{}}
    \toprule
     \multicolumn{2}{l}{\textbf{Earnings decile}} & \multicolumn{6}{c}{\textbf{First Year}} & \multicolumn{6}{c}{\textbf{Third Year}} \\
    \midrule
  & Programme & GATE & S.E. & p-value & GATE-ATE & S.E. & p-value & GATE & S.E. & p-value & GATE-ATE & S.E. & p-value \\
    \midrule
    &&  \multicolumn{12}{c}{\textbf{Outcome: Employment}} \\ \midrule
1317 & WS vs NP & -0.32 & (0.13) & 0.01 & 0.09 & (0.07) & 0.19 & 1.15 & (0.16) & 0.00 & 0.16 & (0.09) & 0.07\\
    2542 & WS vs NP & -0.45 & (0.12) & 0.00 & -0.04 & (0.05) & 0.50 & 1.02 & (0.14) & 0.00 & 0.04 & (0.06) & 0.54\\
    3293 & WS vs NP & -0.48 & (0.11) & 0.00 & -0.06 & (0.04) & 0.13 & 0.95 & (0.13) & 0.00 & -0.04 & (0.05) & 0.47\\
    3767 & WS vs NP & -0.50 & (0.11) & 0.00 & -0.09 & (0.03) & 0.01 & 0.90 & (0.13) & 0.00 & -0.08 & (0.04) & 0.04\\
    4170 & WS vs NP & -0.47 & (0.11) & 0.00 & -0.05 & (0.03) & 0.05 & 0.89 & (0.13) & 0.00 & -0.09 & (0.03) & 0.01\\
    4596 & WS vs NP & -0.45 & (0.11) & 0.00 & -0.03 & (0.03) & 0.23 & 0.88 & (0.13) & 0.00 & -0.10 & (0.03) & 0.00\\
    5041 & WS vs NP & -0.44 & (0.11) & 0.00 & -0.03 & (0.03) & 0.36 & 0.87 & (0.13) & 0.00 & -0.11 & (0.03) & 0.00\\
    5624 & WS vs NP & -0.41 & (0.11) & 0.00 & 0.00 & (0.04) & 0.96 & 0.90 & (0.13) & 0.00 & -0.09 & (0.04) & 0.05\\
    6610 & WS vs NP & -0.39 & (0.12) & 0.00 & 0.03 & (0.05) & 0.63 & 1.00 & (0.15) & 0.00 & 0.02 & (0.07) & 0.77\\
    10583 & WS vs NP & -0.22 & (0.18) & 0.21 & 0.19 & (0.13) & 0.13 & 1.29 & (0.21) & 0.00 & 0.30 & (0.15) & 0.04\\ \midrule
    1317 & BC vs NP & -1.11 & (0.16) & 0.00 & 0.19 & (0.08) & 0.03 & 0.48 & (0.21) & 0.02 & -0.01 & (0.12) & 0.91\\
    2542 & BC vs NP & -1.23 & (0.15) & 0.00 & 0.07 & (0.07) & 0.32 & 0.48 & (0.19) & 0.01 & -0.01 & (0.08) & 0.88\\
    3293 & BC vs NP & -1.28 & (0.14) & 0.00 & 0.02 & (0.05) & 0.75 & 0.47 & (0.18) & 0.01 & -0.03 & (0.06) & 0.69\\
    3767 & BC vs NP & -1.36 & (0.14) & 0.00 & -0.06 & (0.04) & 0.18 & 0.47 & (0.17) & 0.01 & -0.03 & (0.05) & 0.62\\
    4170 & BC vs NP & -1.38 & (0.13) & 0.00 & -0.07 & (0.03) & 0.02 & 0.47 & (0.17) & 0.00 & -0.02 & (0.04) & 0.68\\
    4596 & BC vs NP & -1.39 & (0.13) & 0.00 & -0.09 & (0.03) & 0.01 & 0.49 & (0.17) & 0.00 & 0.00 & (0.04) & 0.94\\
    5041 & BC vs NP & -1.40 & (0.14) & 0.00 & -0.10 & (0.04) & 0.01 & 0.49 & (0.17) & 0.00 & -0.00 & (0.05) & 0.97\\
    5624 & BC vs NP & -1.38 & (0.14) & 0.00 & -0.07 & (0.05) & 0.14 & 0.47 & (0.17) & 0.01 & -0.02 & (0.06) & 0.76\\
    6610 & BC vs NP & -1.33 & (0.15) & 0.00 & -0.03 & (0.07) & 0.68 & 0.52 & (0.18) & 0.00 & 0.03 & (0.09) & 0.71\\
    10583 & BC vs NP & -1.14 & (0.18) & 0.00 & 0.17 & (0.12) & 0.17 & 0.57 & (0.23) & 0.01 & 0.08 & (0.16) & 0.62\\ \midrule
    1317 & TC vs NP & -1.24 & (0.17) & 0.00 & 0.14 & (0.10) & 0.16 & 0.43 & (0.23) & 0.06 & -0.04 & (0.14) & 0.75\\
    2542& TC vs NP & -1.33 & (0.17) & 0.00 & 0.04 & (0.08) & 0.63 & 0.39 & (0.21) & 0.06 & -0.08 & (0.10) & 0.38\\
    3293 & TC vs NP & -1.39 & (0.16) & 0.00 & -0.02 & (0.06) & 0.78 & 0.39 & (0.20) & 0.05 & -0.08 & (0.07) & 0.25\\
    3767 & TC vs NP & -1.45 & (0.15) & 0.00 & -0.08 & (0.05) & 0.08 & 0.40 & (0.19) & 0.04 & -0.07 & (0.06) & 0.21\\
    4170 & TC vs NP & -1.44 & (0.15) & 0.00 & -0.07 & (0.04) & 0.06 & 0.42 & (0.19) & 0.02 & -0.05 & (0.05) & 0.30\\
    4596 & TC vs NP & -1.43 & (0.15) & 0.00 & -0.06 & (0.04) & 0.09 & 0.43 & (0.19) & 0.02 & -0.04 & (0.05) & 0.39\\
    5041 & TC vs NP & -1.45 & (0.16) & 0.00 & -0.07 & (0.04) & 0.09 & 0.43 & (0.19) & 0.02 & -0.04 & (0.05) & 0.41\\
    5624 & TC vs NP & -1.40 & (0.16) & 0.00 & -0.03 & (0.06) & 0.57 & 0.46 & (0.20) & 0.02 & -0.01 & (0.07) & 0.85\\
    6610 & TC vs NP & -1.38 & (0.17) & 0.00 & -0.01 & (0.08) & 0.95 & 0.62 & (0.21) & 0.00 & 0.14 & (0.10) & 0.15\\
    10583 & TC vs NP & -1.19 & (0.20) & 0.00 & 0.18 & (0.14) & 0.18 & 0.76 & (0.27) & 0.01 & 0.28 & (0.19) & 0.13\\ \midrule
    1317 & EP vs NP & -0.97 & (0.22) & 0.00 & 0.31 & (0.12) & 0.01 & 0.56 & (0.28) & 0.05 & 0.18 & (0.17) & 0.29\\
    2542 & EP vs NP & -1.09 & (0.21) & 0.00 & 0.18 & (0.10) & 0.06 & 0.41 & (0.26) & 0.11 & 0.04 & (0.12) & 0.75\\
    3293 & EP vs NP & -1.16 & (0.21) & 0.00 & 0.11 & (0.08) & 0.15 & 0.35 & (0.25) & 0.16 & -0.02 & (0.10) & 0.84\\
    3767 & EP vs NP & -1.28 & (0.20) & 0.00 & -0.01 & (0.06) & 0.90 & 0.31 & (0.25) & 0.21 & -0.07 & (0.09) & 0.46\\
    4170 & EP vs NP & -1.35 & (0.19) & 0.00 & -0.08 & (0.05) & 0.08 & 0.32 & (0.24) & 0.19 & -0.05 & (0.07) & 0.43\\
    4596 & EP vs NP & -1.37 & (0.19) & 0.00 & -0.10 & (0.05) & 0.04 & 0.36 & (0.24) & 0.14 & -0.01 & (0.06) & 0.84\\
    5041 & EP vs NP & -1.39 & (0.20) & 0.00 & -0.12 & (0.05) & 0.02 & 0.35 & (0.25) & 0.16 & -0.02 & (0.07) & 0.75\\
    5624 & EP vs NP & -1.37 & (0.20) & 0.00 & -0.10 & (0.07) & 0.18 & 0.34 & (0.26) & 0.19 & -0.03 & (0.09) & 0.76\\
    6610 & EP vs NP & -1.36 & (0.21) & 0.00 & -0.09 & (0.10) & 0.37 & 0.37 & (0.30) & 0.22 & -0.01 & (0.15) & 0.97\\
    10583 & EP vs NP & -1.35 & (0.26) & 0.00 & -0.08 & (0.18) & 0.67 & 0.38 & (0.40) & 0.34 & 0.00 & (0.27) & 0.99\\

    \midrule
    &&  \multicolumn{12}{c}{\textbf{Outcome: Earnings}} \\ \midrule
    1317 & WS vs NP & -2253 & (731) & 0.00 & 1262 & (470) & 0.01 & 2434 & (992) & 0.01 & -792 & (629) & 0.21\\
    2542 & WS vs NP & -2816 & (656) & 0.00 & 699 & (427) & 0.10 & 2282 & (867) & 0.01 & -944 & (584) & 0.11\\
    3293 & WS vs NP & -2960 & (641) & 0.00 & 556 & (403) & 0.17 & 2333 & (854) & 0.01 & -893 & (555) & 0.11\\
    3767 & WS vs NP & -3096 & (631) & 0.00 & 419 & (386) & 0.28 & 2391 & (827) & 0.00 & -835 & (538) & 0.12\\
    4170 & WS vs NP & -3203 & (657) & 0.00 & 312 & (357) & 0.38 & 2559 & (862) & 0.00 & -667 & (501) & 0.18\\
    4596 & WS vs NP & -3271 & (658) & 0.00 & 244 & (347) & 0.48 & 2813 & (868) & 0.00 & -413 & (484) & 0.39\\
    5041 & WS vs NP & -3511 & (679) & 0.00 & 4 & (332) & 0.99 & 2838 & (886) & 0.00 & -388 & (471) & 0.41\\
    5624 & WS vs NP & -3694 & (744) & 0.00 & -178 & (353) & 0.61 & 2831 & (960) & 0.00 & -395 & (473) & 0.40\\
    6610 & WS vs NP & -4676 & (1066) & 0.00 & -1161 & (613) & 0.06 & 3180 & (1451) & 0.03 & -46 & (794) & 0.95\\
    10583 & WS vs NP & -5541 & (2416) & 0.02 & -2025 & (2019) & 0.32 & 8589 & (3705) & 0.02 & 5363 & (3104) & 0.08\\ \midrule
    1317 & BC vs NP & -4846 & (824) & 0.00 & 2070 & (545) & 0.00 & 614 & (1156) & 0.60 & -290 & (746) & 0.70\\
    2542 & BC vs NP & -5400 & (762) & 0.00 & 1516 & (496) & 0.00 & 723 & (1058) & 0.49 & -181 & (693) & 0.79\\
    3293 & BC vs NP & -5728 & (750) & 0.00 & 1189 & (461) & 0.01 & 866 & (1043) & 0.41 & -37 & (647) & 0.95\\
    3767 & BC vs NP & -6294 & (736) & 0.00 & 623 & (432) & 0.15 & 896 & (1021) & 0.38 & -8 & (617) & 0.99\\
    4170 & BC vs NP & -6865 & (760) & 0.00 & 52 & (389) & 0.89 & 870 & (1048) & 0.41 & -34 & (553) & 0.95\\
    4596 & BC vs NP & -7271 & (786) & 0.00 & -355 & (381) & 0.35 & 1108 & (1066) & 0.30 & 204 & (532) & 0.70\\
    5041 & BC vs NP & -7596 & (843) & 0.00 & -679 & (376) & 0.07 & 1080 & (1122) & 0.34 & 177 & (544) & 0.75\\
    5624 & BC vs NP & -7701 & (957) & 0.00 & -785 & (454) & 0.08 & 1087 & (1261) & 0.39 & 184 & (609) & 0.76\\
    6610 & BC vs NP & -8224 & (1285) & 0.00 & -1308 & (768) & 0.09 & 1543 & (1785) & 0.39 & 640 & (1049) & 0.54\\
    10583 & BC vs NP & -8978 & (2319) & 0.00 & -2062 & (1894) & 0.28 & 179 & (3785) & 0.96 & -725 & (3111) & 0.82\\ \midrule
    1317 & TC vs NP & -5102 & (891) & 0.00 & 1947 & (633) & 0.00 & 781 & (1247) & 0.53 & -957 & (860) & 0.27\\
    2542 & TC vs NP & -5639 & (831) & 0.00 & 1410 & (574) & 0.01 & 714 & (1166) & 0.54 & -1023 & (796) & 0.20\\
    3293 & TC vs NP & -6000 & (813) & 0.00 & 1049 & (527) & 0.05 & 910 & (1143) & 0.43 & -828 & (755) & 0.27\\
    3767 & TC vs NP & -6464 & (805) & 0.00 & 584 & (495) & 0.24 & 1025 & (1123) & 0.36 & -712 & (719) & 0.32\\
    4170 & TC vs NP & -6798 & (835) & 0.00 & 251 & (448) & 0.57 & 1189 & (1159) & 0.30 & -548 & (650) & 0.40\\
    4596 & TC vs NP & -7066 & (869) & 0.00 & -18 & (444) & 0.97 & 1462 & (1193) & 0.22 & -276 & (626) & 0.66\\
    5041 & TC vs NP & -7577 & (932) & 0.00 & -528 & (452) & 0.24 & 1608 & (1267) & 0.20 & -129 & (629) & 0.84\\
    5624 & TC vs NP & -7951 & (1072) & 0.00 & -903 & (519) & 0.08 & 1958 & (1456) & 0.18 & 221 & (698) & 0.75\\
    6610 & TC vs NP & -8759 & (1475) & 0.00 & -1711 & (897) & 0.06 & 3646 & (2121) & 0.09 & 1908 & (1245) & 0.13\\
    10583 & TC vs NP & -8873 & (2796) & 0.00 & -1824 & (2280) & 0.42 & 3927 & (4452) & 0.38 & 2189 & (3688) & 0.55\\ \midrule
    1317 & EP vs NP & -4450 & (1081) & 0.00 & 3099 & (754) & 0.00 & 600 & (1443) & 0.68 & 884 & (1141) & 0.44\\
    2542 & EP vs NP & -5034 & (1034) & 0.00 & 2516 & (702) & 0.00 & 350 & (1372) & 0.80 & 634 & (1088) & 0.56\\
    3293 & EP vs NP & -5533 & (1003) & 0.00 & 2016 & (639) & 0.00 & 203 & (1371) & 0.88 & 487 & (1045) & 0.64\\
    3767 & EP vs NP & -6284 & (982) & 0.00 & 1265 & (605) & 0.04 & 90 & (1370) & 0.95 & 374 & (1003) & 0.71\\
    4170 & EP vs NP & -7181 & (983) & 0.00 & 368 & (544) & 0.50 & 285 & (1454) & 0.84 & 569 & (915) & 0.53\\
    4596 & EP vs NP & -7748 & (1026) & 0.00 & -199 & (530) & 0.71 & 622 & (1536) & 0.69 & 906 & (882) & 0.30\\
    5041 & EP vs NP & -8394 & (1085) & 0.00 & -844 & (524) & 0.11 & 253 & (1593) & 0.87 & 537 & (803) & 0.50\\
    5624 & EP vs NP & -8662 & (1267) & 0.00 & -1113 & (646) & 0.08 & -487 & (1919) & 0.80 & -203 & (965) & 0.83\\
    6610 & EP vs NP & -9816 & (1754) & 0.00 & -2267 & (1113) & 0.04 & -1621 & (2911) & 0.58 & -1337 & (1677) & 0.43\\
    10583 & EP vs NP & -12002 & (3494) & 0.00 & -4453 & (2879) & 0.12 & -3059 & (6636) & 0.64 & -2776 & (5417) & 0.61\\
    \bottomrule
    \end{tabular}
    \begin{tablenotes}
\textit{Note:} The table shows the GATEs for the different programmes in comparison to no programme (NP: Non Participation, WS: Wage Subsidy, BC: Basic courses, TC: Training Courses, EP: Employment Programme) and the different sub-groups. The Earnings decile is the average earnings within each decile of the earnings distribution in the three months before the
unemployment spell.
\end{tablenotes}
\end{threeparttable}
\end{adjustbox}
\end{table}

\begin{table}[htbp!]
\begin{adjustbox}{width=0.95\columnwidth, center}
\begin{threeparttable}
  \captionsetup{font=large}
    \caption{GATEs for ue rate, TEMP sample}
    \centering
    \begin{tabular}{@{}ll|rrrrrr|rrrrrrr@{}}
    \toprule
     \multicolumn{2}{l}{\textbf{Ue rate (\%)}} & \multicolumn{6}{c}{\textbf{First Year}} & \multicolumn{6}{c}{\textbf{Third Year}} \\
    \midrule
  & Programme & GATE & S.E. & p-value & GATE-ATE & S.E. & p-value & GATE & S.E. & p-value & GATE-ATE & S.E. & p-value \\
    \midrule
    &&  \multicolumn{12}{c}{\textbf{Outcome: Employment}} \\ \midrule
    3 & WS vs NP & -0 & (0.11) & 0.00 & -0 & (0.02) & 0.02 & 1 & (0.13) & 0.00 & -0 & (0.03) & 0.10\\
    4 & WS vs NP & -0 & (0.11) & 0.00 & 0 & (0.03) & 0.97 & 1 & (0.13) & 0.00 & 0 & (0.03) & 0.78\\
    7 & WS vs NP & -0 & (0.12) & 0.01 & 0 & (0.04) & 0.02 & 1 & (0.14) & 0.00 & 0 & (0.05) & 0.09\\ \midrule
    3 & BC vs NP & -1 & (0.13) & 0.00 & -0 & (0.03) & 0.15 & 0 & (0.17) & 0.01 & -0 & (0.03) & 0.37\\
    4 & BC vs NP & -1 & (0.13) & 0.00 & 0 & (0.03) & 0.71 & 0 & (0.17) & 0.01 & -0 & (0.04) & 0.38\\
    7 & BC vs NP & -1 & (0.14) & 0.00 & 0 & (0.05) & 0.20 & 1 & (0.18) & 0.00 & 0 & (0.06) & 0.12\\ \midrule
    3 & TC vs NP & -1 & (0.15) & 0.00 & -0 & (0.03) & 0.09 & 0 & (0.19) & 0.02 & -0 & (0.04) & 0.44\\
    4 & TC vs NP & -1 & (0.15) & 0.00 & -0 & (0.03) & 0.39 & 0 & (0.19) & 0.01 & -0 & (0.04) & 0.89\\
    7 & TC vs NP & -1 & (0.16) & 0.00 & 0 & (0.06) & 0.03 & 1 & (0.19) & 0.01 & 0 & (0.07) & 0.35\\ \midrule
    3 & EP vs NP & -1 & (0.19) & 0.00 & -0 & (0.03) & 0.05 & 0 & (0.24) & 0.18 & -0 & (0.05) & 0.38\\
    4 & EP vs NP & -1 & (0.19) & 0.00 & 0 & (0.03) & 0.55 & 0 & (0.25) & 0.15 & -0 & (0.05) & 0.81\\
    7 & EP vs NP & -1 & (0.20) & 0.00 & 0 & (0.07) & 0.08 & 0 & (0.26) & 0.07 & 0 & (0.09) & 0.27\\
    \midrule
    &&  \multicolumn{12}{c}{\textbf{Outcome: Earnings}} \\ \midrule
    3 & WS vs NP & -3716 & (691) & 0.00 & -201 & (109) & 0.07 & 2967 & (959) & 0.00 & -259 & (173) & 0.13\\
    4 & WS vs NP & -3737 & (763) & 0.00 & -222 & (165) & 0.18 & 3326 & (1083) & 0.00 & 100 & (206) & 0.63\\
    7 & WS vs NP & -2897 & (750) & 0.00 & 618 & (240) & 0.01 & 3655 & (1072) & 0.00 & 429 & (277) & 0.12\\   \midrule
    3 & BC vs NP & -6970 & (810) & 0.00 & -53 & (128) & 0.68 & 951 & (1145) & 0.41 & 48 & (203) & 0.81\\
    4 & BC vs NP & -7024 & (857) & 0.00 & -107 & (188) & 0.57 & 704 & (1236) & 0.57 & -200 & (239) & 0.40\\
    7 & BC vs NP & -6706 & (849) & 0.00 & 211 & (282) & 0.45 & 998 & (1231) & 0.42 & 95 & (333) & 0.78\\   \midrule
    3 & TC vs NP & -7228 & (910) & 0.00 & -179 & (140) & 0.20 & 1702 & (1303) & 0.19 & -36 & (230) & 0.88\\
    4 & TC vs NP & -7220 & (956) & 0.00 & -171 & (213) & 0.42 & 1673 & (1401) & 0.23 & -65 & (271) & 0.81\\
    7 & TC vs NP & -6522 & (941) & 0.00 & 527 & (312) & 0.09 & 1872 & (1364) & 0.17 & 134 & (397) & 0.73\\   \midrule
    3 & EP vs NP & -7763 & (1044) & 0.00 & -214 & (161) & 0.18 & -485 & (1650) & 0.77 & -201 & (289) & 0.49\\
    4 & EP vs NP & -7583 & (1115) & 0.00 & -34 & (222) & 0.88 & -355 & (1806) & 0.84 & -71 & (321) & 0.82\\
    7 & EP vs NP & -7083 & (1107) & 0.00 & 466 & (349) & 0.18 & 191 & (1783) & 0.91 & 475 & (498) & 0.34\\
    \bottomrule
    \end{tabular}
    \begin{tablenotes}
\textit{Note:} The table shows the GATEs for the different programmes in comparison to no programme (NP: Non Participation, WS: Wage Subsidy, BC: Basic courses, TC: Training Courses, EP: Employment Programme) and the different sub-groups. 
\end{tablenotes}
\end{threeparttable}
\end{adjustbox}
\end{table}

\clearpage

\subsection{Balanced Group Average Treatment Effects} \label{bgates appendix}
\subsubsection{Permanent Residents Sample} \label{bgates_app_perm}
\begin{table}[htbp]
\begin{adjustbox}{width=0.9\columnwidth, center}
\begin{threeparttable}
  \captionsetup{font=large}
    \caption{BGATEs for place of origin, PERM sample}
    \label{table:bgate_origin}
    \centering
    \begin{tabular}{@{}ll|rrrrrr|rrrrrrr@{}}
    \toprule
     \multicolumn{2}{l}{\textbf{Origin}}& \multicolumn{6}{c}{\textbf{First Year}} & \multicolumn{6}{c}{\textbf{Third Year}} \\
    \midrule
     & Programme & BGATE & S.E. & p-value & BGATE-ATE & S.E. & p-value & BGATE & S.E. & p-value & BGATE-ATE & S.E. & p-value \\
    \midrule
     &&  \multicolumn{12}{c}{\textbf{Outcome: Employment}} \\ \midrule
    Swiss & WS vs NP & -0.69 & (0.06) & 0.00 & -0.11 & (0.01) & 0.00 & 0.70 & (0.07) & 0.00 & -0.09 & (0.02) & 0.00\\
    Neig. countries & WS vs NP & -0.61 & (0.07) & 0.00 & -0.02 & (0.01) & 0.06 & 0.77 & (0.07) & 0.00 & -0.02 & (0.01) & 0.02\\
    EU & WS vs NP & -0.66 & (0.06) & 0.00 & -0.07 & (0.01) & 0.00 & 0.77 & (0.07) & 0.00 & -0.02 & (0.01) & 0.17\\
    Rest of Europe & WS vs NP & -0.48 & (0.08) & 0.00 & 0.11 & (0.02) & 0.00 & 0.85 & (0.09) & 0.00 & 0.06 & (0.02) & 0.00\\
    Rest of the world & WS vs NP & -0.51 & (0.07) & 0.00 & 0.08 & (0.01) & 0.00 & 0.86 & (0.08) & 0.00 & 0.07 & (0.02) & 0.00\\ \midrule
    Swiss & BC vs NP & -1.71 & (0.08) & 0.00 & -0.06 & (0.02) & 0.00 & -0.17 & (0.09) & 0.05 & -0.10 & (0.02) & 0.00\\
    Neig. countries & BC vs NP & -1.66 & (0.08) & 0.00 & -0.01 & (0.01) & 0.46 & -0.09 & (0.09) & 0.33 & -0.02 & (0.01) & 0.09\\
    EU & BC vs NP & -1.71 & (0.08) & 0.00 & -0.06 & (0.01) & 0.00 & -0.08 & (0.09) & 0.37 & -0.01 & (0.01) & 0.32\\
    Rest of Europe & BC vs NP & -1.59 & (0.08) & 0.00 & 0.06 & (0.02) & 0.00 & 0.01 & (0.10) & 0.90 & 0.08 & (0.02) & 0.00\\
    Rest of the world & BC vs NP & -1.58 & (0.08) & 0.00 & 0.07 & (0.01) & 0.00 & -0.01 & (0.10) & 0.89 & 0.05 & (0.02) & 0.00\\ \midrule
    Swiss & TC vs NP & -1.60 & (0.12) & 0.00 & -0.06 & (0.02) & 0.02 & -0.04 & (0.14) & 0.78 & -0.06 & (0.03) & 0.07\\
    Neig. countries & TC vs NP & -1.54 & (0.12) & 0.00 & -0.01 & (0.01) & 0.71 & 0.02 & (0.14) & 0.87 & 0.00 & (0.02) & 0.92\\
    EU & TC vs NP & -1.56 & (0.12) & 0.00 & -0.02 & (0.02) & 0.34 & -0.00 & (0.14) & 0.98 & -0.03 & (0.02) & 0.23\\
    Rest of Europe & TC vs NP & -1.50 & (0.12) & 0.00 & 0.04 & (0.02) & 0.12 & 0.07 & (0.14) & 0.63 & 0.05 & (0.03) & 0.06\\
    Rest of the world & TC vs NP & -1.49 & (0.12) & 0.00 & 0.05 & (0.02) & 0.02 & 0.05 & (0.14) & 0.71 & 0.03 & (0.02) & 0.16\\ \midrule
    Swiss & EP vs NP & -1.47 & (0.14) & 0.00 & -0.09 & (0.03) & 0.00 & 0.15 & (0.15) & 0.31 & -0.09 & (0.03) & 0.01\\
    Neig. countries & EP vs NP & -1.40 & (0.14) & 0.00 & -0.02 & (0.01) & 0.19 & 0.21 & (0.15) & 0.15 & -0.02 & (0.02) & 0.29\\
    EU & EP vs NP & -1.38 & (0.14) & 0.00 & -0.00 & (0.02) & 0.96 & 0.25 & (0.15) & 0.10 & 0.01 & (0.02) & 0.63\\
    Rest of Europe & EP vs NP & -1.33 & (0.13) & 0.00 & 0.05 & (0.03) & 0.10 & 0.28 & (0.15) & 0.06 & 0.04 & (0.03) & 0.17\\
    Rest of the world & EP vs NP & -1.31 & (0.13) & 0.00 & 0.06 & (0.02) & 0.00 & 0.30 & (0.15) & 0.05 & 0.06 & (0.02) & 0.01\\
    \midrule
     &&  \multicolumn{12}{c}{\textbf{Outcome: Earnings}} \\ \midrule
    Swiss & WS vs NP & -6125 & (378) & 0.00 & -583 & (95) & 0.00 & 1735 & (651) & 0.01 & -701 & (219) & 0.00\\
    Neig. countries & WS vs NP & -5675 & (412) & 0.00 & -133 & (44) & 0.00 & 2315 & (758) & 0.00 & -121 & (110) & 0.27\\
    EU & WS vs NP & -5779 & (384) & 0.00 & -236 & (58) & 0.00 & 2273 & (724) & 0.00 & -163 & (116) & 0.16\\
    Rest of Europe & WS vs NP & -5029 & (433) & 0.00 & 513 & (82) & 0.00 & 2985 & (1040) & 0.00 & 550 & (286) & 0.05\\
    Rest of the world & WS vs NP & -5104 & (445) & 0.00 & 438 & (75) & 0.00 & 2869 & (903) & 0.00 & 434 & (162) & 0.01\\ \midrule
    Swiss & BC vs NP & -11002 & (407) & 0.00 & -499 & (97) & 0.00 & -2572 & (722) & 0.00 & -652 & (225) & 0.00\\
    Neig. countries & BC vs NP & -10707 & (428) & 0.00 & -204 & (46) & 0.00 & -2087 & (815) & 0.01 & -166 & (115) & 0.15\\
    EU & BC vs NP & -10769 & (407) & 0.00 & -266 & (60) & 0.00 & -2091 & (790) & 0.01 & -171 & (120) & 0.16\\
    Rest of Europe & BC vs NP & -9980 & (446) & 0.00 & 524 & (83) & 0.00 & -1239 & (1070) & 0.25 & 682 & (291) & 0.02\\
    Rest of the world & BC vs NP & -10059 & (453) & 0.00 & 444 & (77) & 0.00 & -1613 & (945) & 0.09 & 307 & (166) & 0.06\\ \midrule
    Swiss & TC vs NP & -9847 & (626) & 0.00 & -307 & (117) & 0.01 & -1441 & (1020) & 0.16 & -579 & (268) & 0.03\\
    Neig. countries & TC vs NP & -9792 & (618) & 0.00 & -252 & (59) & 0.00 & -864 & (1071) & 0.42 & -2 & (146) & 0.99\\
    EU & TC vs NP & -9611 & (614) & 0.00 & -70 & (74) & 0.34 & -1033 & (1056) & 0.33 & -171 & (147) & 0.24\\
    Rest of Europe & TC vs NP & -9178 & (611) & 0.00 & 362 & (103) & 0.00 & -461 & (1238) & 0.71 & 401 & (324) & 0.22\\
    Rest of the world & TC vs NP & -9273 & (613) & 0.00 & 267 & (99) & 0.01 & -511 & (1153) & 0.66 & 351 & (197) & 0.08\\ \midrule
    Swiss & EP vs NP & -7005 & (849) & 0.00 & -76 & (139) & 0.59 & -1943 & (1188) & 0.10 & -723 & (279) & 0.01\\
    Neig. countries & EP vs NP & -7142 & (783) & 0.00 & -213 & (69) & 0.00 & -1332 & (1252) & 0.29 & -112 & (163) & 0.49\\
    EU & EP vs NP & -6906 & (780) & 0.00 & 23 & (79) & 0.77 & -1354 & (1221) & 0.27 & -133 & (163) & 0.41\\
    Rest of Europe & EP vs NP & -7061 & (695) & 0.00 & -132 & (140) & 0.35 & -587 & (1314) & 0.66 & 633 & (349) & 0.07\\
    Rest of the world & EP vs NP & -6532 & (768) & 0.00 & 397 & (110) & 0.00 & -886 & (1301) & 0.50 & 334 & (207) & 0.11\\
    \bottomrule
    \end{tabular}
        \begin{tablenotes}
\textit{Note:} The table shows the BGATEs for the different programmes in comparison to no programme (NP: Non Participation, WS: Wage Subsidy, BC: Basic courses, TC: Training Courses, EP: Employment Programme) and the different sub-groups. 
\end{tablenotes}
\end{threeparttable}
\end{adjustbox}
\end{table}

\begin{table}[htbp]
\begin{adjustbox}{width=0.9\columnwidth, center}
\begin{threeparttable}
  \captionsetup{font=large}
    \caption{BGATEs for gender, PERM sample}
    \label{table:bgate_gender}
    \centering
    \begin{tabular}{@{}ll|rrrrrr|rrrrrrr@{}}
    \toprule
     \multicolumn{2}{l}{\textbf{Gender}}& \multicolumn{6}{c}{\textbf{First Year}} & \multicolumn{6}{c}{\textbf{Third Year}} \\
    \midrule
     & Programme & BGATE & S.E. & p-value & BGATE-ATE & S.E. & p-value & BGATE & S.E. & p-value & BGATE-ATE & S.E. & p-value \\
    \midrule
     &&  \multicolumn{12}{c}{\textbf{Outcome: Employment}} \\ \midrule
    Man & WS vs NP & -0.64 & (0.07) & 0.00 & 0.01 & (0.01) & 0.15 & 0.72 & (0.07) & 0.00 & -0.01 & (0.01) & 0.13\\
    Woman & WS vs NP & -0.66 & (0.06) & 0.00 & -0.01 & (0.01) & 0.15 & 0.75 & (0.07) & 0.00 & 0.01 & (0.01) & 0.13\\ \midrule
    Man & BC vs NP & -1.68 & (0.08) & 0.00 & -0.01 & (0.01) & 0.51 & -0.12 & (0.09) & 0.16 & 0.00 & (0.01) & 0.68\\
    Woman & BC vs NP & -1.67 & (0.08) & 0.00 & 0.01 & (0.01) & 0.51 & -0.13 & (0.09) & 0.15 & -0.00 & (0.01) & 0.68\\ \midrule
    Man & TC vs NP & -1.54 & (0.12) & 0.00 & 0.01 & (0.01) & 0.31 & 0.01 & (0.14) & 0.95 & 0.01 & (0.02) & 0.44\\
    Woman & TC vs NP & -1.57 & (0.12) & 0.00 & -0.01 & (0.01) & 0.31 & -0.01 & (0.14) & 0.91 & -0.01 & (0.02) & 0.44\\ \midrule
    Man & EP vs NP & -1.45 & (0.14) & 0.00 & -0.02 & (0.02) & 0.19 & 0.16 & (0.14) & 0.27 & -0.02 & (0.02) & 0.27\\
    Woman & EP vs NP & -1.40 & (0.14) & 0.00 & 0.02 & (0.02) & 0.19 & 0.20 & (0.14) & 0.18 & 0.02 & (0.02) & 0.27\\
    \midrule
     &&  \multicolumn{12}{c}{\textbf{Outcome: Earnings}} \\ \midrule
    Man & WS vs NP & -5942 & (410) & 0.00 & -183 & (38) & 0.00 & 2097 & (765) & 0.01 & 59 & (67) & 0.38\\
    Woman & WS vs NP & -5575 & (369) & 0.00 & 183 & (38) & 0.00 & 1980 & (706) & 0.01 & -59 & (67) & 0.38\\ \midrule
    Man & BC vs NP & -11084 & (430) & 0.00 & -417 & (45) & 0.00 & -2489 & (819) & 0.00 & -125 & (78) & 0.11\\
    Woman & BC vs NP & -10250 & (399) & 0.00 & 417 & (45) & 0.00 & -2240 & (768) & 0.00 & 125 & (78) & 0.11\\ \midrule
    Man & TC vs NP & -9986 & (623) & 0.00 & -348 & (74) & 0.00 & -1018 & (1103) & 0.36 & 69 & (119) & 0.57\\
    Woman & TC vs NP & -9290 & (612) & 0.00 & 348 & (74) & 0.00 & -1155 & (1035) & 0.26 & -69 & (119) & 0.57\\ \midrule
    Man & EP vs NP & -7198 & (816) & 0.00 & -234 & (89) & 0.01 & -2073 & (1297) & 0.11 & -322 & (140) & 0.02\\
    Woman & EP vs NP & -6729 & (747) & 0.00 & 234 & (89) & 0.01 & -1429 & (1178) & 0.23 & 322 & (140) & 0.02\\
    \bottomrule
    \end{tabular}
        \begin{tablenotes}
\textit{Note:} The table shows the BGATEs for the different programmes in comparison to no programme (NP: Non Participation, WS: Wage Subsidy, BC: Basic courses, TC: Training Courses, EP: Employment Programme) and the different sub-groups. 
\end{tablenotes}
\end{threeparttable}
\end{adjustbox}
\end{table}

\begin{table}[htbp]
\begin{adjustbox}{width=0.9\columnwidth, center} 
\begin{threeparttable}
  \captionsetup{font=large}
    \caption{BGATEs for sector of last job, PERM sample}
    \label{table:bgate_sector}
    \centering
    \begin{tabular}{@{}ll|rrrrrr|rrrrrrr@{}}
    \toprule
     \multicolumn{2}{l}{\textbf{Sector}}& \multicolumn{6}{c}{\textbf{First Year}} & \multicolumn{6}{c}{\textbf{Third Year}} \\
    \midrule
     & Programme & BGATE & S.E. & p-value & BGATE-ATE & S.E. & p-value & BGATE & S.E. & p-value & BGATE-ATE & S.E. & p-value \\
    \midrule
     &&  \multicolumn{12}{c}{\textbf{Outcome: Employment}} \\ \midrule
    A & WS vs NP & -0.69 & (0.07) & 0.00 & -0.01 & (0.02) & 0.44 & 0.70 & (0.07) & 0.00 & -0.01 & (0.01) & 0.59\\
    B & WS vs NP & -0.66 & (0.07) & 0.00 & 0.01 & (0.01) & 0.24 & 0.71 & (0.07) & 0.00 & -0.00 & (0.01) & 0.94\\
    C & WS vs NP & -0.72 & (0.06) & 0.00 & -0.05 & (0.01) & 0.00 & 0.68 & (0.07) & 0.00 & -0.03 & (0.01) & 0.01\\
    D & WS vs NP & -0.68 & (0.06) & 0.00 & -0.01 & (0.02) & 0.68 & 0.67 & (0.07) & 0.00 & -0.04 & (0.02) & 0.08\\
    E & WS vs NP & -0.62 & (0.07) & 0.00 & 0.06 & (0.02) & 0.00 & 0.74 & (0.07) & 0.00 & 0.03 & (0.02) & 0.09\\
    F & WS vs NP & -0.69 & (0.06) & 0.00 & -0.02 & (0.01) & 0.26 & 0.73 & (0.07) & 0.00 & 0.03 & (0.02) & 0.09\\
    G & WS vs NP & -0.65 & (0.07) & 0.00 & 0.03 & (0.01) & 0.05 & 0.72 & (0.07) & 0.00 & 0.01 & (0.02) & 0.61\\
    H & WS vs NP & -0.73 & (0.06) & 0.00 & -0.05 & (0.02) & 0.00 & 0.70 & (0.06) & 0.00 & -0.01 & (0.02) & 0.45\\ \midrule
    A & BC vs NP & -1.77 & (0.08) & 0.00 & -0.04 & (0.02) & 0.04 & -0.18 & (0.09) & 0.04 & -0.03 & (0.02) & 0.06\\
    B & BC vs NP & -1.73 & (0.08) & 0.00 & -0.00 & (0.01) & 0.88 & -0.14 & (0.09) & 0.12 & 0.01 & (0.02) & 0.66\\
    C & BC vs NP & -1.76 & (0.08) & 0.00 & -0.03 & (0.01) & 0.02 & -0.16 & (0.08) & 0.06 & -0.01 & (0.01) & 0.28\\
    D & BC vs NP & -1.80 & (0.08) & 0.00 & -0.08 & (0.02) & 0.00 & -0.14 & (0.09) & 0.10 & -0.00 & (0.02) & 0.97\\
    E & BC vs NP & -1.67 & (0.08) & 0.00 & 0.06 & (0.02) & 0.00 & -0.14 & (0.09) & 0.10 & 0.00 & (0.02) & 0.96\\
    F & BC vs NP & -1.75 & (0.08) & 0.00 & -0.03 & (0.02) & 0.14 & -0.11 & (0.09) & 0.22 & 0.04 & (0.02) & 0.05\\
    G & BC vs NP & -1.66 & (0.08) & 0.00 & 0.07 & (0.02) & 0.00 & -0.16 & (0.09) & 0.07 & -0.01 & (0.02) & 0.41\\
    H & BC vs NP & -1.73 & (0.08) & 0.00 & -0.00 & (0.02) & 0.95 & -0.17 & (0.09) & 0.05 & -0.02 & (0.02) & 0.22\\ \midrule
    A & TC vs NP & -1.63 & (0.13) & 0.00 & -0.05 & (0.03) & 0.09 & -0.04 & (0.14) & 0.75 & -0.02 & (0.03) & 0.40\\
    B & TC vs NP & -1.60 & (0.12) & 0.00 & -0.02 & (0.02) & 0.44 & -0.01 & (0.14) & 0.96 & 0.01 & (0.02) & 0.53\\
    C & TC vs NP & -1.59 & (0.12) & 0.00 & -0.00 & (0.02) & 0.87 & -0.02 & (0.14) & 0.89 & 0.00 & (0.02) & 0.83\\
    D & TC vs NP & -1.65 & (0.13) & 0.00 & -0.06 & (0.03) & 0.05 & -0.06 & (0.14) & 0.68 & -0.04 & (0.04) & 0.37\\
    E & TC vs NP & -1.53 & (0.12) & 0.00 & 0.06 & (0.02) & 0.02 & -0.01 & (0.13) & 0.95 & 0.01 & (0.03) & 0.61\\
    F & TC vs NP & -1.59 & (0.12) & 0.00 & -0.01 & (0.03) & 0.76 & -0.04 & (0.14) & 0.78 & -0.02 & (0.03) & 0.59\\
    G & TC vs NP & -1.51 & (0.12) & 0.00 & 0.07 & (0.03) & 0.00 & 0.01 & (0.13) & 0.91 & 0.04 & (0.03) & 0.20\\
    H & TC vs NP & -1.60 & (0.13) & 0.00 & -0.02 & (0.03) & 0.41 & -0.04 & (0.14) & 0.78 & -0.02 & (0.03) & 0.50\\ \midrule
    A & EP vs NP & -1.49 & (0.14) & 0.00 & -0.02 & (0.03) & 0.61 & 0.15 & (0.15) & 0.30 & -0.02 & (0.03) & 0.43\\
    B & EP vs NP & -1.50 & (0.14) & 0.00 & -0.03 & (0.02) & 0.21 & 0.18 & (0.15) & 0.24 & 0.00 & (0.03) & 0.94\\
    C & EP vs NP & -1.53 & (0.15) & 0.00 & -0.06 & (0.02) & 0.01 & 0.16 & (0.15) & 0.28 & -0.01 & (0.02) & 0.53\\
    D & EP vs NP & -1.51 & (0.15) & 0.00 & -0.04 & (0.03) & 0.25 & 0.15 & (0.15) & 0.32 & -0.02 & (0.04) & 0.62\\
    E & EP vs NP & -1.44 & (0.14) & 0.00 & 0.03 & (0.03) & 0.28 & 0.19 & (0.14) & 0.19 & 0.01 & (0.03) & 0.64\\
    F & EP vs NP & -1.41 & (0.14) & 0.00 & 0.07 & (0.03) & 0.03 & 0.22 & (0.15) & 0.14 & 0.04 & (0.03) & 0.16\\
    G & EP vs NP & -1.45 & (0.14) & 0.00 & 0.02 & (0.03) & 0.52 & 0.14 & (0.15) & 0.34 & -0.04 & (0.03) & 0.23\\
    H & EP vs NP & -1.46 & (0.15) & 0.00 & 0.01 & (0.03) & 0.65 & 0.16 & (0.15) & 0.29 & -0.01 & (0.03) & 0.67\\

    \midrule
     &&  \multicolumn{12}{c}{\textbf{Outcome: Earnings}} \\ \midrule
    A & WS vs NP & -5571 & (332) & 0.00 & 235 & (90) & 0.01 & 1871 & (623) & 0.00 & -58 & (180) & 0.75\\
    B & WS vs NP & -5356 & (374) & 0.00 & 450 & (75) & 0.00 & 1944 & (651) & 0.00 & 15 & (104) & 0.89\\
    C & WS vs NP & -6249 & (391) & 0.00 & -443 & (79) & 0.00 & 1787 & (690) & 0.01 & -142 & (112) & 0.20\\
    D & WS vs NP & -5855 & (340) & 0.00 & -49 & (100) & 0.62 & 1918 & (635) & 0.00 & -11 & (166) & 0.95\\
    E & WS vs NP & -5730 & (396) & 0.00 & 76 & (75) & 0.31 & 1905 & (791) & 0.02 & -24 & (165) & 0.88\\
    F & WS vs NP & -5524 & (337) & 0.00 & 281 & (72) & 0.00 & 1963 & (668) & 0.00 & 34 & (111) & 0.76\\
    G & WS vs NP & -6060 & (433) & 0.00 & -254 & (100) & 0.01 & 2105 & (823) & 0.01 & 176 & (189) & 0.35\\
    H & WS vs NP & -6313 & (362) & 0.00 & -507 & (81) & 0.00 & 1819 & (649) & 0.01 & -110 & (154) & 0.47\\ 
    \midrule
    A & BC vs NP & -10182 & (376) & 0.00 & 460 & (91) & 0.00 & -2246 & (695) & 0.00 & 190 & (187) & 0.31\\
    B & BC vs NP & -10106 & (403) & 0.00 & 536 & (78) & 0.00 & -2264 & (717) & 0.00 & 171 & (115) & 0.14\\
    C & BC vs NP & -11382 & (418) & 0.00 & -740 & (81) & 0.00 & -2610 & (758) & 0.00 & -175 & (121) & 0.15\\
    D & BC vs NP & -10959 & (388) & 0.00 & -317 & (106) & 0.00 & -2310 & (712) & 0.00 & 126 & (174) & 0.47\\
    E & BC vs NP & -10642 & (416) & 0.00 & 0 & (74) & 1.00 & -2663 & (831) & 0.00 & -228 & (162) & 0.16\\
    F & BC vs NP & -10109 & (380) & 0.00 & 533 & (75) & 0.00 & -2211 & (736) & 0.00 & 224 & (117) & 0.06\\
    G & BC vs NP & -11084 & (448) & 0.00 & -442 & (99) & 0.00 & -2728 & (872) & 0.00 & -292 & (197) & 0.14\\
    H & BC vs NP & -10988 & (401) & 0.00 & -346 & (85) & 0.00 & -2682 & (728) & 0.00 & -247 & (163) & 0.13\\
    \midrule
    A & TC vs NP & -9032 & (596) & 0.00 & 566 & (128) & 0.00 & -1653 & (963) & 0.09 & -403 & (255) & 0.11\\
    B & TC vs NP & -9268 & (594) & 0.00 & 330 & (106) & 0.00 & -1386 & (994) & 0.16 & -136 & (165) & 0.41\\
    C & TC vs NP & -10310 & (634) & 0.00 & -712 & (119) & 0.00 & -1084 & (1068) & 0.31 & 167 & (172) & 0.33\\
    D & TC vs NP & -9864 & (632) & 0.00 & -266 & (157) & 0.09 & -1578 & (1016) & 0.12 & -328 & (256) & 0.20\\
    E & TC vs NP & -9441 & (608) & 0.00 & 157 & (103) & 0.13 & -1225 & (1089) & 0.26 & 25 & (207) & 0.90\\
    F & TC vs NP & -9117 & (603) & 0.00 & 481 & (109) & 0.00 & -1273 & (1010) & 0.21 & -22 & (163) & 0.89\\
    G & TC vs NP & -9933 & (639) & 0.00 & -335 & (131) & 0.01 & -658 & (1163) & 0.57 & 592 & (275) & 0.03\\
    H & TC vs NP & -9935 & (647) & 0.00 & -337 & (122) & 0.01 & -1296 & (1052) & 0.22 & -46 & (220) & 0.83\\
    \midrule
    A & EP vs NP & -6918 & (687) & 0.00 & 144 & (168) & 0.39 & -1464 & (1088) & 0.18 & 288 & (316) & 0.36\\
    B & EP vs NP & -7042 & (703) & 0.00 & 19 & (151) & 0.90 & -1536 & (1151) & 0.18 & 216 & (188) & 0.25\\
    C & EP vs NP & -7423 & (895) & 0.00 & -362 & (159) & 0.02 & -2070 & (1303) & 0.11 & -318 & (200) & 0.11\\
    D & EP vs NP & -7362 & (771) & 0.00 & -300 & (172) & 0.08 & -1565 & (1151) & 0.17 & 187 & (289) & 0.52\\
    E & EP vs NP & -6645 & (823) & 0.00 & 417 & (120) & 0.00 & -1973 & (1314) & 0.13 & -222 & (236) & 0.35\\
    F & EP vs NP & -6663 & (731) & 0.00 & 399 & (140) & 0.00 & -1523 & (1165) & 0.19 & 229 & (182) & 0.21\\
    G & EP vs NP & -7377 & (845) & 0.00 & -316 & (167) & 0.06 & -2286 & (1391) & 0.10 & -534 & (323) & 0.10\\
    H & EP vs NP & -6803 & (900) & 0.00 & 258 & (172) & 0.13 & -1899 & (1278) & 0.14 & -147 & (271) & 0.59\\
    \bottomrule
    \end{tabular}
        \begin{tablenotes}
\textit{Note:} The table shows the BGATEs for the different programmes in comparison to no programme (NP: Non Participation, WS: Wage Subsidy, BC: Basic courses, TC: Training Courses, EP: Employment Programme) and the different sub-groups. The job sectors are categorised as follow: A: agricultural and forestry, B: production in industry and trade, C: technical and information technology, D: construction and mining; E: transport and trade, F: hospitality and personal services, G: business, finance and law, H: education, art and science.
\end{tablenotes}
\end{threeparttable}
\end{adjustbox}
\end{table}

\begin{table}[htbp]
\begin{adjustbox}{width=0.9\columnwidth, center}
\begin{threeparttable}
  \captionsetup{font=large}
    \caption{BGATEs for education, PERM sample}
    \label{table:bgate_educ}
    \centering
    \begin{tabular}{@{}ll|rrrrrr|rrrrrrr@{}}
    \toprule
     \multicolumn{2}{l}{\textbf{Education}}& \multicolumn{6}{c}{\textbf{First Year}} & \multicolumn{6}{c}{\textbf{Third Year}} \\
    \midrule
     & Programme & BGATE & S.E. & p-value & BGATE-ATE & S.E. & p-value & BGATE & S.E. & p-value & BGATE-ATE & S.E. & p-value \\
    \midrule
     &&  \multicolumn{12}{c}{\textbf{Outcome: Employment}} \\ \midrule
    Compulsory & WS vs NP & -0.64 & (0.07) & 0.00 & 0.05 & (0.02) & 0.01 & 0.77 & (0.07) & 0.00 & 0.05 & (0.02) & 0.01\\
    Secondary & WS vs NP & -0.65 & (0.07) & 0.00 & 0.04 & (0.01) & 0.00 & 0.74 & (0.07) & 0.00 & 0.02 & (0.01) & 0.07\\
    Vocational & WS vs NP & -0.74 & (0.06) & 0.00 & -0.06 & (0.01) & 0.00 & 0.67 & (0.06) & 0.00 & -0.04 & (0.01) & 0.00\\
    Bachelor & WS vs NP & -0.72 & (0.07) & 0.00 & -0.04 & (0.01) & 0.00 & 0.69 & (0.07) & 0.00 & -0.03 & (0.01) & 0.00\\
    Post-Grad & WS vs NP & -0.67 & (0.07) & 0.00 & 0.01 & (0.01) & 0.37 & 0.72 & (0.07) & 0.00 & 0.01 & (0.01) & 0.60\\
    \midrule
    Compulsory & BC vs NP & -1.71 & (0.08) & 0.00 & -0.02 & (0.02) & 0.41 & -0.09 & (0.09) & 0.34 & 0.06 & (0.02) & 0.01\\
    Secondary & BC vs NP & -1.69 & (0.08) & 0.00 & 0.01 & (0.01) & 0.54 & -0.14 & (0.09) & 0.12 & 0.01 & (0.01) & 0.49\\
    Vocational & BC vs NP & -1.73 & (0.08) & 0.00 & -0.04 & (0.01) & 0.00 & -0.18 & (0.08) & 0.03 & -0.03 & (0.01) & 0.00\\
    Bachelor & BC vs NP & -1.70 & (0.08) & 0.00 & -0.00 & (0.01) & 1.00 & -0.17 & (0.09) & 0.05 & -0.02 & (0.01) & 0.04\\
    Post-Grad & BC vs NP & -1.65 & (0.08) & 0.00 & 0.04 & (0.02) & 0.01 & -0.16 & (0.09) & 0.08 & -0.01 & (0.02) & 0.49\\
   \midrule
    Compulsory & TC vs NP & -1.61 & (0.12) & 0.00 & -0.04 & (0.03) & 0.22 & -0.03 & (0.14) & 0.86 & -0.01 & (0.03) & 0.88\\
    Secondary & TC vs NP & -1.58 & (0.12) & 0.00 & -0.01 & (0.02) & 0.62 & -0.02 & (0.14) & 0.88 & -0.00 & (0.02) & 1.00\\
    Vocational & TC vs NP & -1.58 & (0.13) & 0.00 & -0.01 & (0.02) & 0.45 & -0.02 & (0.14) & 0.88 & -0.00 & (0.02) & 0.95\\
    Bachelor & TC vs NP & -1.55 & (0.12) & 0.00 & 0.02 & (0.02) & 0.22 & -0.03 & (0.14) & 0.83 & -0.01 & (0.02) & 0.62\\
    Post-Grad & TC vs NP & -1.53 & (0.13) & 0.00 & 0.04 & (0.02) & 0.14 & -0.01 & (0.14) & 0.97 & 0.01 & (0.02) & 0.54\\
    \midrule
    Compulsory & EP vs NP & -1.43 & (0.14) & 0.00 & 0.02 & (0.04) & 0.67 & 0.23 & (0.15) & 0.11 & 0.07 & (0.04) & 0.07\\
    Secondary & EP vs NP & -1.44 & (0.14) & 0.00 & 0.01 & (0.02) & 0.71 & 0.17 & (0.14) & 0.23 & 0.00 & (0.02) & 0.83\\
    Vocational & EP vs NP & -1.49 & (0.15) & 0.00 & -0.04 & (0.02) & 0.04 & 0.12 & (0.15) & 0.41 & -0.04 & (0.02) & 0.02\\
    Bachelor & EP vs NP & -1.46 & (0.15) & 0.00 & -0.01 & (0.02) & 0.54 & 0.15 & (0.15) & 0.32 & -0.02 & (0.02) & 0.23\\
    Post-Grad & EP vs NP & -1.41 & (0.15) & 0.00 & 0.04 & (0.03) & 0.25 & 0.17 & (0.15) & 0.27 & -0.00 & (0.03) & 0.97\\
    \midrule
     &&  \multicolumn{12}{c}{\textbf{Outcome: Earnings}} \\ \midrule
Compulsory & WS vs NP & -5284 & (357) & 0.00 & 922 & (128) & 0.00 & 2204 & (657) & 0.00 & 487 & (230) & 0.03\\
    Secondary & WS vs NP & -5638 & (379) & 0.00 & 567 & (77) & 0.00 & 2086 & (684) & 0.00 & 369 & (170) & 0.03\\
    Vocational & WS vs NP & -6733 & (413) & 0.00 & -528 & (62) & 0.00 & 1426 & (700) & 0.04 & -291 & (110) & 0.01\\
    Bachelor & WS vs NP & -6807 & (438) & 0.00 & -602 & (73) & 0.00 & 1337 & (769) & 0.08 & -380 & (141) & 0.01\\
    Post-Grad & WS vs NP & -6564 & (487) & 0.00 & -358 & (120) & 0.00 & 1531 & (883) & 0.08 & -186 & (258) & 0.47\\
    \midrule
    Compulsory & BC vs NP & -10056 & (389) & 0.00 & 1065 & (135) & 0.00 & -1924 & (721) & 0.01 & 776 & (273) & 0.00\\
    Secondary & BC vs NP & -10493 & (403) & 0.00 & 627 & (92) & 0.00 & -2252 & (744) & 0.00 & 449 & (208) & 0.03\\
    Vocational & BC vs NP & -11738 & (441) & 0.00 & -618 & (67) & 0.00 & -2991 & (780) & 0.00 & -291 & (124) & 0.02\\
    Bachelor & BC vs NP & -11793 & (467) & 0.00 & -672 & (85) & 0.00 & -3154 & (865) & 0.00 & -454 & (177) & 0.01\\
    Post-Grad & BC vs NP & -11522 & (510) & 0.00 & -401 & (135) & 0.00 & -3180 & (992) & 0.00 & -480 & (313) & 0.13\\
    \midrule
    Compulsory & TC vs NP & -9115 & (594) & 0.00 & 1018 & (199) & 0.00 & -1438 & (966) & 0.14 & -313 & (472) & 0.51\\
    Secondary & TC vs NP & -9463 & (607) & 0.00 & 670 & (137) & 0.00 & -1242 & (997) & 0.21 & -117 & (371) & 0.75\\
    Vocational & TC vs NP & -10651 & (668) & 0.00 & -519 & (98) & 0.00 & -1069 & (1147) & 0.35 & 56 & (191) & 0.77\\
    Bachelor & TC vs NP & -10797 & (687) & 0.00 & -664 & (131) & 0.00 & -917 & (1291) & 0.48 & 209 & (311) & 0.50\\
    Post-Grad & TC vs NP & -10638 & (716) & 0.00 & -506 & (187) & 0.01 & -961 & (1464) & 0.51 & 164 & (559) & 0.77\\
   \midrule
    Compulsory & EP vs NP & -6840 & (700) & 0.00 & 234 & (297) & 0.43 & -1225 & (1060) & 0.25 & 686 & (552) & 0.21\\
    Secondary & EP vs NP & -7172 & (744) & 0.00 & -98 & (218) & 0.65 & -1624 & (1137) & 0.15 & 287 & (428) & 0.50\\
    Vocational & EP vs NP & -7398 & (920) & 0.00 & -324 & (145) & 0.03 & -2504 & (1403) & 0.07 & -593 & (213) & 0.01\\
    Bachelor & EP vs NP & -7106 & (1010) & 0.00 & -32 & (204) & 0.88 & -2154 & (1543) & 0.16 & -243 & (356) & 0.49\\
    Post-Grad & EP vs NP & -6856 & (1073) & 0.00 & 219 & (298) & 0.46 & -2048 & (1722) & 0.23 & -137 & (657) & 0.83\\
    \bottomrule
    \end{tabular}
        \begin{tablenotes}
\textit{Note:} The table shows the BGATEs for the different programmes in comparison to no programme (NP: Non Participation, WS: Wage Subsidy, BC: Basic courses, TC: Training Courses, EP: Employment Programme) and the different sub-groups. 
\end{tablenotes}
\end{threeparttable}
\end{adjustbox}
\end{table}

\begin{table}[htbp]
\begin{adjustbox}{width=0.9\columnwidth, center}
\begin{threeparttable}
  \captionsetup{font=large}
    \caption{BGATEs for months in employment, PERM sample}
    \label{table:bgate_employment_months}
    \centering
    \begin{tabular}{@{}ll|rrrrrr|rrrrrrr@{}}
    \toprule
     \multicolumn{2}{l}{\textbf{Months in Employment}}& \multicolumn{6}{c}{\textbf{First Year}} & \multicolumn{6}{c}{\textbf{Third Year}} \\
    \midrule
     & Programme & BGATE & S.E. & p-value & BGATE-ATE & S.E. & p-value & BGATE & S.E. & p-value & BGATE-ATE & S.E. & p-value \\
    \midrule
     &&  \multicolumn{12}{c}{\textbf{Outcome: Employment}} \\ \midrule
    12 & WS vs NP & -0.35 & (0.09) & 0.00 & 0.12 & (0.04) & 0.00 & 0.93 & (0.12) & 0.00 & 0.09 & (0.05) & 0.09\\
    24 & WS vs NP & -0.33 & (0.08) & 0.00 & 0.14 & (0.02) & 0.00 & 0.94 & (0.11) & 0.00 & 0.10 & (0.03) & 0.00\\
    36 & WS vs NP & -0.46 & (0.07) & 0.00 & 0.02 & (0.02) & 0.32 & 0.85 & (0.09) & 0.00 & 0.01 & (0.02) & 0.65\\
    48 & WS vs NP & -0.53 & (0.07) & 0.00 & -0.06 & (0.02) & 0.01 & 0.79 & (0.08) & 0.00 & -0.06 & (0.03) & 0.04\\
    60 & WS vs NP & -0.69 & (0.07) & 0.00 & -0.22 & (0.04) & 0.00 & 0.70 & (0.07) & 0.00 & -0.14 & (0.05) & 0.00\\ \midrule
    12 & BC vs NP & -1.25 & (0.12) & 0.00 & 0.21 & (0.05) & 0.00 & 0.12 & (0.16) & 0.46 & 0.12 & (0.07) & 0.08\\
    24 & BC vs NP & -1.32 & (0.10) & 0.00 & 0.14 & (0.02) & 0.00 & 0.11 & (0.13) & 0.40 & 0.11 & (0.03) & 0.00\\
    36 & BC vs NP & -1.45 & (0.09) & 0.00 & 0.01 & (0.02) & 0.49 & 0.01 & (0.12) & 0.96 & 0.01 & (0.03) & 0.69\\
    48 & BC vs NP & -1.57 & (0.08) & 0.00 & -0.10 & (0.03) & 0.00 & -0.09 & (0.10) & 0.39 & -0.08 & (0.04) & 0.03\\
    60 & BC vs NP & -1.73 & (0.08) & 0.00 & -0.27 & (0.05) & 0.00 & -0.17 & (0.09) & 0.05 & -0.16 & (0.06) & 0.01\\ \midrule
    12 & TC vs NP & -1.51 & (0.18) & 0.00 & 0.01 & (0.08) & 0.92 & 0.10 & (0.25) & 0.69 & 0.05 & (0.11) & 0.66\\
    24 & TC vs NP & -1.46 & (0.15) & 0.00 & 0.06 & (0.03) & 0.07 & 0.17 & (0.20) & 0.39 & 0.12 & (0.05) & 0.01\\
    36 & TC vs NP & -1.53 & (0.14) & 0.00 & -0.01 & (0.03) & 0.65 & 0.08 & (0.18) & 0.67 & 0.02 & (0.04) & 0.52\\
    48 & TC vs NP & -1.51 & (0.13) & 0.00 & 0.00 & (0.04) & 0.94 & -0.03 & (0.15) & 0.86 & -0.08 & (0.06) & 0.14\\
    60 & TC vs NP & -1.58 & (0.12) & 0.00 & -0.06 & (0.07) & 0.42 & -0.05 & (0.14) & 0.72 & -0.10 & (0.10) & 0.30\\ \midrule
    12 & EP vs NP & -1.08 & (0.19) & 0.00 & 0.13 & (0.08) & 0.11 & 0.17 & (0.23) & 0.45 & -0.03 & (0.10) & 0.81\\
    24 & EP vs NP & -1.08 & (0.16) & 0.00 & 0.13 & (0.04) & 0.00 & 0.24 & (0.19) & 0.20 & 0.05 & (0.05) & 0.29\\
    36 & EP vs NP & -1.15 & (0.15) & 0.00 & 0.06 & (0.03) & 0.04 & 0.20 & (0.18) & 0.26 & 0.00 & (0.04) & 0.91\\
    48 & EP vs NP & -1.26 & (0.14) & 0.00 & -0.05 & (0.04) & 0.22 & 0.19 & (0.15) & 0.21 & -0.00 & (0.05) & 0.94\\
    60 & EP vs NP & -1.48 & (0.14) & 0.00 & -0.27 & (0.08) & 0.00 & 0.17 & (0.14) & 0.23 & -0.02 & (0.10) & 0.81\\
    \midrule
     &&  \multicolumn{12}{c}{\textbf{Outcome: Earnings}} \\ \midrule
    12 & WS vs NP & -3530 & (425) & 0.00 & 1085 & (155) & 0.00 & 2794 & (862) & 0.00 & 298 & (270) & 0.27\\
    24 & WS vs NP & -3882 & (518) & 0.00 & 733 & (123) & 0.00 & 3078 & (841) & 0.00 & 581 & (179) & 0.00\\
    36 & WS vs NP & -4455 & (465) & 0.00 & 160 & (59) & 0.01 & 2539 & (802) & 0.00 & 43 & (111) & 0.70\\
    48 & WS vs NP & -5153 & (366) & 0.00 & -538 & (90) & 0.00 & 2223 & (822) & 0.01 & -273 & (157) & 0.08\\
    60 & WS vs NP & -6055 & (394) & 0.00 & -1440 & (172) & 0.00 & 1847 & (703) & 0.01 & -649 & (293) & 0.03\\ \midrule
    12 & BC vs NP & -7163 & (492) & 0.00 & 1729 & (179) & 0.00 & -680 & (999) & 0.50 & 816 & (341) & 0.02\\
    24 & BC vs NP & -7876 & (553) & 0.00 & 1015 & (128) & 0.00 & -688 & (925) & 0.46 & 808 & (196) & 0.00\\
    36 & BC vs NP & -8552 & (502) & 0.00 & 340 & (61) & 0.00 & -1562 & (884) & 0.08 & -67 & (129) & 0.60\\
    48 & BC vs NP & -9735 & (404) & 0.00 & -843 & (100) & 0.00 & -1997 & (880) & 0.02 & -501 & (188) & 0.01\\
    60 & BC vs NP & -11132 & (414) & 0.00 & -2240 & (193) & 0.00 & -2552 & (763) & 0.00 & -1056 & (347) & 0.00\\ \midrule
    12 & TC vs NP & -6962 & (727) & 0.00 & 1420 & (257) & 0.00 & -165 & (1275) & 0.90 & 544 & (445) & 0.22\\
    24 & TC vs NP & -7757 & (724) & 0.00 & 625 & (158) & 0.00 & -45 & (1189) & 0.97 & 664 & (260) & 0.01\\
    36 & TC vs NP & -8191 & (688) & 0.00 & 190 & (80) & 0.02 & -796 & (1149) & 0.49 & -87 & (180) & 0.63\\
    48 & TC vs NP & -8991 & (623) & 0.00 & -609 & (133) & 0.00 & -1166 & (1127) & 0.30 & -457 & (252) & 0.07\\
    60 & TC vs NP & -10008 & (620) & 0.00 & -1626 & (298) & 0.00 & -1373 & (1053) & 0.19 & -664 & (491) & 0.18\\ \midrule
    12 & EP vs NP & -5141 & (739) & 0.00 & 1049 & (299) & 0.00 & -563 & (1292) & 0.66 & 442 & (475) & 0.35\\
    24 & EP vs NP & -5971 & (764) & 0.00 & 219 & (191) & 0.25 & -423 & (1209) & 0.73 & 582 & (271) & 0.03\\
    36 & EP vs NP & -6086 & (742) & 0.00 & 103 & (97) & 0.29 & -1003 & (1192) & 0.40 & 2 & (190) & 0.99\\
    48 & EP vs NP & -6563 & (722) & 0.00 & -374 & (148) & 0.01 & -1223 & (1236) & 0.32 & -218 & (262) & 0.41\\
    60 & EP vs NP & -7186 & (843) & 0.00 & -997 & (357) & 0.01 & -1813 & (1214) & 0.14 & -808 & (558) & 0.15\\
    \bottomrule
    \end{tabular}
        \begin{tablenotes}
\textit{Note:} The table shows the BGATEs for the different programmes in comparison to no programme (NP: Non Participation, WS: Wage Subsidy, BC: Basic courses, TC: Training Courses, EP: Employment Programme) and the different sub-groups. The ranges for months in employment are: 12: 1-12, 24: 13-24, 36: 25-36, 48: 37-48, 60:49-60 
\end{tablenotes}
\end{threeparttable}
\end{adjustbox}
\end{table}

\begin{table}[htbp]
\begin{adjustbox}{width=0.8\columnwidth, center} 
\begin{threeparttable}
  \captionsetup{font=large}
    \caption{BGATEs for earnings decile, PERM sample}
    \label{table:bgate_earnings_decile}
    \centering
    \begin{tabular}{@{}ll|rrrrrr|rrrrrrr@{}}
    \toprule
     \multicolumn{2}{l}{\textbf{Earnings decile}}& \multicolumn{6}{c}{\textbf{First Year}} & \multicolumn{6}{c}{\textbf{Third Year}} \\
    \midrule
     & Programme & BGATE & S.E. & p-value & BGATE-ATE & S.E. & p-value & BGATE & S.E. & p-value & BGATE-ATE & S.E. & p-value \\
    \midrule
     &&  \multicolumn{12}{c}{\textbf{Outcome: Employment}} \\ \midrule
    1192 & WS vs NP & -0.51 & (0.08) & 0.00 & 0.15 & (0.04) & 0.00 & 0.86 & (0.09) & 0.00 & 0.14 & (0.05) & 0.00\\
    2445 & WS vs NP & -0.61 & (0.07) & 0.00 & 0.06 & (0.03) & 0.07 & 0.80 & (0.08) & 0.00 & 0.08 & (0.04) & 0.03\\
    3307 & WS vs NP & -0.61 & (0.07) & 0.00 & 0.05 & (0.03) & 0.05 & 0.79 & (0.08) & 0.00 & 0.07 & (0.03) & 0.02\\
    3969 & WS vs NP & -0.63 & (0.07) & 0.00 & 0.03 & (0.02) & 0.16 & 0.74 & (0.07) & 0.00 & 0.02 & (0.02) & 0.31\\
    4538 & WS vs NP & -0.62 & (0.07) & 0.00 & 0.04 & (0.02) & 0.04 & 0.72 & (0.07) & 0.00 & 0.00 & (0.02) & 0.87\\
    5090& WS vs NP & -0.64 & (0.07) & 0.00 & 0.02 & (0.02) & 0.30 & 0.70 & (0.07) & 0.00 & -0.02 & (0.02) & 0.31\\
    5698 & WS vs NP & -0.69 & (0.07) & 0.00 & -0.03 & (0.02) & 0.15 & 0.66 & (0.07) & 0.00 & -0.06 & (0.02) & 0.01\\
    6500 & WS vs NP & -0.70 & (0.07) & 0.00 & -0.04 & (0.03) & 0.15 & 0.65 & (0.07) & 0.00 & -0.07 & (0.03) & 0.01\\
    7919 & WS vs NP & -0.83 & (0.08) & 0.00 & -0.16 & (0.04) & 0.00 & 0.62 & (0.07) & 0.00 & -0.10 & (0.03) & 0.00\\
    12487 & WS vs NP & -0.78 & (0.09) & 0.00 & -0.12 & (0.06) & 0.06 & 0.64 & (0.08) & 0.00 & -0.07 & (0.05) & 0.14\\ \midrule
    1192 & BC vs NP & -1.53 & (0.09) & 0.00 & 0.18 & (0.04) & 0.00 & -0.07 & (0.12) & 0.55 & 0.07 & (0.06) & 0.30\\
    2445 & BC vs NP & -1.63 & (0.09) & 0.00 & 0.08 & (0.04) & 0.05 & -0.13 & (0.11) & 0.23 & 0.00 & (0.05) & 0.94\\
    3307 & BC vs NP & -1.65 & (0.08) & 0.00 & 0.06 & (0.03) & 0.07 & -0.11 & (0.10) & 0.27 & 0.02 & (0.04) & 0.54\\
    3969 & BC vs NP & -1.68 & (0.08) & 0.00 & 0.03 & (0.03) & 0.27 & -0.12 & (0.09) & 0.20 & 0.02 & (0.03) & 0.55\\
    4538 & BC vs NP & -1.69 & (0.08) & 0.00 & 0.02 & (0.02) & 0.45 & -0.10 & (0.09) & 0.28 & 0.04 & (0.02) & 0.15\\
    5090 & BC vs NP & -1.73 & (0.08) & 0.00 & -0.02 & (0.02) & 0.44 & -0.09 & (0.09) & 0.28 & 0.04 & (0.03) & 0.11\\
    5698 & BC vs NP & -1.78 & (0.08) & 0.00 & -0.07 & (0.03) & 0.00 & -0.13 & (0.09) & 0.13 & 0.00 & (0.03) & 0.89\\
    6500 & BC vs NP & -1.78 & (0.08) & 0.00 & -0.07 & (0.03) & 0.04 & -0.15 & (0.09) & 0.08 & -0.02 & (0.03) & 0.55\\
    7919 & BC vs NP & -1.85 & (0.09) & 0.00 & -0.14 & (0.05) & 0.00 & -0.22 & (0.09) & 0.01 & -0.09 & (0.05) & 0.06\\
    12487 & BC vs NP & -1.76 & (0.10) & 0.00 & -0.05 & (0.06) & 0.40 & -0.22 & (0.09) & 0.02 & -0.08 & (0.06) & 0.15\\ \midrule
    1192 & TC vs NP & -1.54 & (0.14) & 0.00 & 0.04 & (0.07) & 0.57 & 0.01 & (0.18) & 0.94 & 0.03 & (0.11) & 0.80\\
    2445 & TC vs NP & -1.62 & (0.14) & 0.00 & -0.04 & (0.07) & 0.51 & -0.03 & (0.18) & 0.87 & -0.02 & (0.09) & 0.86\\
    3307 & TC vs NP & -1.61 & (0.13) & 0.00 & -0.03 & (0.05) & 0.57 & -0.04 & (0.16) & 0.81 & -0.03 & (0.07) & 0.69\\
    3969 & TC vs NP & -1.61 & (0.13) & 0.00 & -0.03 & (0.04) & 0.49 & -0.07 & (0.15) & 0.66 & -0.05 & (0.04) & 0.24\\
    4538 & TC vs NP & -1.61 & (0.13) & 0.00 & -0.03 & (0.04) & 0.43 & -0.07 & (0.15) & 0.62 & -0.06 & (0.04) & 0.16\\
    5090 & TC vs NP & -1.62 & (0.13) & 0.00 & -0.04 & (0.04) & 0.33 & -0.02 & (0.14) & 0.89 & -0.01 & (0.05) & 0.88\\
    5698 & TC vs NP & -1.65 & (0.13) & 0.00 & -0.07 & (0.04) & 0.10 & -0.01 & (0.14) & 0.93 & 0.00 & (0.05) & 1.00\\
    6500 & TC vs NP & -1.60 & (0.13) & 0.00 & -0.02 & (0.05) & 0.65 & 0.04 & (0.14) & 0.78 & 0.05 & (0.05) & 0.35\\
    7919 & TC vs NP & -1.54 & (0.14) & 0.00 & 0.04 & (0.08) & 0.65 & 0.01 & (0.14) & 0.92 & 0.03 & (0.07) & 0.71\\
    12487 & TC vs NP & -1.39 & (0.15) & 0.00 & 0.19 & (0.10) & 0.06 & 0.04 & (0.14) & 0.76 & 0.06 & (0.09) & 0.52\\ \midrule
    1192 & EP vs NP & -1.30 & (0.14) & 0.00 & 0.16 & (0.08) & 0.04 & 0.20 & (0.18) & 0.25 & 0.03 & (0.10) & 0.75\\
    2445 & EP vs NP & -1.33 & (0.15) & 0.00 & 0.13 & (0.08) & 0.08 & 0.22 & (0.18) & 0.22 & 0.05 & (0.09) & 0.62\\
    3307 & EP vs NP & -1.37 & (0.14) & 0.00 & 0.09 & (0.06) & 0.16 & 0.22 & (0.16) & 0.16 & 0.05 & (0.07) & 0.43\\
    3969 & EP vs NP & -1.40 & (0.14) & 0.00 & 0.06 & (0.05) & 0.21 & 0.21 & (0.15) & 0.18 & 0.03 & (0.05) & 0.47\\
    4538& EP vs NP & -1.39 & (0.14) & 0.00 & 0.07 & (0.05) & 0.13 & 0.23 & (0.15) & 0.12 & 0.06 & (0.05) & 0.17\\
    5090 & EP vs NP & -1.45 & (0.15) & 0.00 & 0.01 & (0.05) & 0.88 & 0.23 & (0.15) & 0.13 & 0.06 & (0.05) & 0.20\\
    5698 & EP vs NP & -1.52 & (0.15) & 0.00 & -0.06 & (0.05) & 0.23 & 0.20 & (0.15) & 0.19 & 0.03 & (0.05) & 0.56\\
    6500 & EP vs NP & -1.61 & (0.16) & 0.00 & -0.15 & (0.06) & 0.02 & 0.14 & (0.15) & 0.36 & -0.03 & (0.06) & 0.58\\
    7919 & EP vs NP & -1.67 & (0.19) & 0.00 & -0.21 & (0.11) & 0.05 & 0.06 & (0.17) & 0.72 & -0.11 & (0.08) & 0.18\\
    12487 & EP vs NP & -1.56 & (0.21) & 0.00 & -0.10 & (0.14) & 0.47 & 0.01 & (0.18) & 0.98 & -0.17 & (0.11) & 0.13\\
    \midrule
     &&  \multicolumn{12}{c}{\textbf{Outcome: Earnings}} \\ \midrule
     1192 & WS vs NP & -3949 & (369) & 0.00 & 1780 & (176) & 0.00 & 2323 & (814) & 0.00 & 456 & (263) & 0.08\\
    2445 & WS vs NP & -4023 & (315) & 0.00 & 1706 & (200) & 0.00 & 2088 & (692) & 0.00 & 221 & (273) & 0.42\\
    3307 & WS vs NP & -4111 & (307) & 0.00 & 1618 & (215) & 0.00 & 2295 & (642) & 0.00 & 428 & (279) & 0.13\\
    3969 & WS vs NP & -4303 & (345) & 0.00 & 1426 & (184) & 0.00 & 2111 & (642) & 0.00 & 244 & (247) & 0.32\\
    4538 & WS vs NP & -4660 & (327) & 0.00 & 1068 & (189) & 0.00 & 2244 & (648) & 0.00 & 377 & (241) & 0.12\\
    5090 & WS vs NP & -5134 & (368) & 0.00 & 594 & (166) & 0.00 & 2147 & (632) & 0.00 & 279 & (236) & 0.24\\
    5698 & WS vs NP & -5762 & (403) & 0.00 & -34 & (161) & 0.83 & 1773 & (657) & 0.01 & -94 & (249) & 0.71\\
    6500 & WS vs NP & -6562 & (500) & 0.00 & -834 & (204) & 0.00 & 1343 & (775) & 0.08 & -524 & (281) & 0.06\\
    7919 & WS vs NP & -8534 & (607) & 0.00 & -2806 & (337) & 0.00 & 807 & (908) & 0.37 & -1060 & (463) & 0.02\\ 
    12487 & WS vs NP & -10247 & (996) & 0.00 & -4519 & (797) & 0.00 & 1540 & (1548) & 0.32 & -327 & (1168) & 0.78\\ \midrule
    1192 & BC vs NP & -7950 & (414) & 0.00 & 2601 & (207) & 0.00 & -1404 & (875) & 0.11 & 1015 & (307) & 0.00\\
    2445 & BC vs NP & -7891 & (379) & 0.00 & 2660 & (223) & 0.00 & -1401 & (776) & 0.07 & 1018 & (324) & 0.00\\
    3307 & BC vs NP & -7989 & (364) & 0.00 & 2562 & (224) & 0.00 & -1273 & (729) & 0.08 & 1146 & (317) & 0.00\\
    3969 & BC vs NP & -8522 & (386) & 0.00 & 2029 & (190) & 0.00 & -1650 & (726) & 0.02 & 769 & (279) & 0.01\\
    4538 & BC vs NP & -8804 & (382) & 0.00 & 1747 & (196) & 0.00 & -1537 & (732) & 0.04 & 882 & (272) & 0.00\\
    5090 & BC vs NP & -9734 & (414) & 0.00 & 817 & (176) & 0.00 & -1755 & (719) & 0.01 & 664 & (270) & 0.01\\
    5698 & BC vs NP & -10840 & (455) & 0.00 & -289 & (181) & 0.11 & -2191 & (746) & 0.00 & 228 & (282) & 0.42\\
    6500 & BC vs NP & -11905 & (548) & 0.00 & -1354 & (237) & 0.00 & -3110 & (858) & 0.00 & -690 & (333) & 0.04\\
    7919 & BC vs NP & -14263 & (633) & 0.00 & -3712 & (362) & 0.00 & -4278 & (1003) & 0.00 & -1858 & (544) & 0.00\\
    12487 & BC vs NP & -17613 & (916) & 0.00 & -7062 & (728) & 0.00 & -5593 & (1577) & 0.00 & -3174 & (1189) & 0.01\\ \midrule
    1192 & TC vs NP & -7561 & (584) & 0.00 & 2067 & (314) & 0.00 & -805 & (1116) & 0.47 & 419 & (495) & 0.40\\
    2445 & TC vs NP & -7582 & (580) & 0.00 & 2046 & (339) & 0.00 & -1464 & (1032) & 0.16 & -240 & (522) & 0.65\\
    3307 & TC vs NP & -7688 & (608) & 0.00 & 1940 & (331) & 0.00 & -1387 & (989) & 0.16 & -164 & (491) & 0.74\\
    3969 & TC vs NP & -8092 & (581) & 0.00 & 1536 & (278) & 0.00 & -1696 & (992) & 0.09 & -472 & (433) & 0.28\\
    4538 & TC vs NP & -8564 & (612) & 0.00 & 1064 & (298) & 0.00 & -1741 & (1020) & 0.09 & -517 & (433) & 0.23\\
    5090 & TC vs NP & -9263 & (635) & 0.00 & 365 & (272) & 0.18 & -1766 & (1050) & 0.09 & -543 & (426) & 0.20\\
    5698 & TC vs NP & -10026 & (716) & 0.00 & -398 & (297) & 0.18 & -1752 & (1118) & 0.12 & -529 & (457) & 0.25\\
    6500 & TC vs NP & -11234 & (809) & 0.00 & -1606 & (368) & 0.00 & -1450 & (1231) & 0.24 & -227 & (522) & 0.66\\
    7919 & TC vs NP & -12365 & (935) & 0.00 & -2737 & (549) & 0.00 & -878 & (1531) & 0.57 & 345 & (904) & 0.70\\
    12487 & TC vs NP & -13905 & (1264) & 0.00 & -4277 & (1002) & 0.00 & 703 & (2246) & 0.75 & 1926 & (1719) & 0.26\\ \midrule
    1192 & EP vs NP & -5066 & (642) & 0.00 & 2051 & (400) & 0.00 & -759 & (1172) & 0.52 & 1002 & (554) & 0.07\\
    2445 & EP vs NP & -5229 & (625) & 0.00 & 1888 & (461) & 0.00 & -417 & (1088) & 0.70 & 1345 & (649) & 0.04\\
    3307 & EP vs NP & -5654 & (622) & 0.00 & 1463 & (451) & 0.00 & -379 & (1043) & 0.72 & 1383 & (623) & 0.03\\
    3969 & EP vs NP & -6207 & (615) & 0.00 & 910 & (397) & 0.02 & -676 & (1073) & 0.53 & 1086 & (566) & 0.05\\
    4538 & EP vs NP & -6680 & (644) & 0.00 & 437 & (410) & 0.29 & -499 & (1094) & 0.65 & 1262 & (568) & 0.03\\
    5090 & EP vs NP & -7391 & (695) & 0.00 & -274 & (363) & 0.45 & -517 & (1174) & 0.66 & 1245 & (555) & 0.02\\
    5698 & EP vs NP & -8210 & (816) & 0.00 & -1093 & (379) & 0.00 & -934 & (1271) & 0.46 & 828 & (572) & 0.15\\
    6500 & EP vs NP & -9168 & (959) & 0.00 & -2051 & (467) & 0.00 & -2197 & (1473) & 0.14 & -436 & (693) & 0.53\\
    7919 & EP vs NP & -10020 & (1274) & 0.00 & -2903 & (727) & 0.00 & -4390 & (1924) & 0.02 & -2628 & (1154) & 0.02\\
    12487 & EP vs NP & -7543 & (1935) & 0.00 & -427 & (1516) & 0.78 & -6848 & (3073) & 0.03 & -5086 & (2406) & 0.03\\
    \bottomrule
    \end{tabular}
        \begin{tablenotes}
\textit{Note:} The table shows the BGATEs for the different programmes in comparison to no programme (NP: Non Participation, WS: Wage Subsidy, BC: Basic courses, TC: Training Courses, EP: Employment Programme) and the different sub-groups. The Earning decile is the average earnings within each decile of the earnings distribution in the three months before the
unemployment spell.
\end{tablenotes}
\end{threeparttable}
\end{adjustbox}
\end{table}

\begin{table}[htbp]
\begin{adjustbox}{width=0.9\columnwidth, center}
\begin{threeparttable}
  \captionsetup{font=large}
    \caption{BGATEs for ue rate, PERM sample}
    \centering
    \begin{tabular}{@{}ll|rrrrrr|rrrrrrr@{}}
    \toprule
     \multicolumn{2}{l}{\textbf{UE rate (\%)}}& \multicolumn{6}{c}{\textbf{First Year}} & \multicolumn{6}{c}{\textbf{Third Year}} \\
    \midrule
     & Programme & BGATE & S.E. & p-value & BGATE-ATE & S.E. & p-value & BGATE & S.E. & p-value & BGATE-ATE & S.E. & p-value \\
    \midrule
     &&  \multicolumn{12}{c}{\textbf{Outcome: Employment}} \\ \midrule
    2 & WS vs NP & -0.77 & (0.07) & 0.00 & -0.14 & (0.02) & 0.00 & 0.67 & (0.07) & 0.00 & -0.08 & (0.02) & 0.00\\
    3 & WS vs NP & -0.69 & (0.07) & 0.00 & -0.05 & (0.01) & 0.00 & 0.71 & (0.07) & 0.00 & -0.04 & (0.01) & 0.00\\
    4 & WS vs NP & -0.58 & (0.07) & 0.00 & 0.06 & (0.01) & 0.00 & 0.77 & (0.07) & 0.00 & 0.02 & (0.01) & 0.13\\
    7 & WS vs NP & -0.50 & (0.07) & 0.00 & 0.13 & (0.03) & 0.00 & 0.84 & (0.08) & 0.00 & 0.10 & (0.03) & 0.00\\ \midrule
    2 & BC vs NP & -1.83 & (0.08) & 0.00 & -0.15 & (0.03) & 0.00 & -0.15 & (0.09) & 0.09 & -0.02 & (0.02) & 0.30\\
    3 & BC vs NP & -1.71 & (0.08) & 0.00 & -0.03 & (0.02) & 0.03 & -0.14 & (0.09) & 0.12 & -0.01 & (0.02) & 0.53\\
    4 & BC vs NP & -1.61 & (0.08) & 0.00 & 0.06 & (0.02) & 0.00 & -0.12 & (0.09) & 0.17 & 0.00 & (0.01) & 0.89\\
    7 & BC vs NP & -1.54 & (0.08) & 0.00 & 0.13 & (0.03) & 0.00 & -0.10 & (0.10) & 0.32 & 0.03 & (0.03) & 0.32\\ \midrule
    2 & TC vs NP & -1.64 & (0.13) & 0.00 & -0.09 & (0.05) & 0.07 & -0.04 & (0.14) & 0.75 & -0.04 & (0.03) & 0.26\\
    3 & TC vs NP & -1.59 & (0.12) & 0.00 & -0.04 & (0.02) & 0.08 & -0.04 & (0.14) & 0.80 & -0.03 & (0.02) & 0.18\\
    4 & TC vs NP & -1.54 & (0.12) & 0.00 & 0.01 & (0.03) & 0.57 & -0.02 & (0.14) & 0.88 & -0.01 & (0.02) & 0.50\\
    7 & TC vs NP & -1.44 & (0.12) & 0.00 & 0.11 & (0.05) & 0.01 & 0.08 & (0.14) & 0.59 & 0.08 & (0.04) & 0.05\\ \midrule
    2 & EP vs NP & -1.52 & (0.14) & 0.00 & -0.12 & (0.05) & 0.02 & 0.12 & (0.14) & 0.42 & -0.07 & (0.04) & 0.06\\
    3 & EP vs NP & -1.50 & (0.14) & 0.00 & -0.09 & (0.03) & 0.00 & 0.14 & (0.15) & 0.33 & -0.04 & (0.03) & 0.11\\
    4 & EP vs NP & -1.38 & (0.14) & 0.00 & 0.03 & (0.03) & 0.33 & 0.19 & (0.15) & 0.19 & 0.01 & (0.02) & 0.80\\
    7 & EP vs NP & -1.23 & (0.15) & 0.00 & 0.18 & (0.05) & 0.00 & 0.29 & (0.16) & 0.06 & 0.11 & (0.05) & 0.04\\
    \midrule
     &&  \multicolumn{12}{c}{\textbf{Outcome: Earnings}} \\ \midrule
    2 & WS vs NP & -6398 & (399) & 0.00 & -621 & (92) & 0.00 & 1413 & (692) & 0.04 & -672 & (176) & 0.00\\
    3 & WS vs NP & -6274 & (396) & 0.00 & -497 & (78) & 0.00 & 1744 & (707) & 0.01 & -341 & (123) & 0.01\\
    4 & WS vs NP & -5664 & (412) & 0.00 & 113 & (64) & 0.08 & 2313 & (758) & 0.00 & 228 & (106) & 0.03\\
    7 & WS vs NP & -4772 & (421) & 0.00 & 1005 & (148) & 0.00 & 2869 & (773) & 0.00 & 784 & (243) & 0.00\\ \midrule
    2 & BC vs NP & -11315 & (432) & 0.00 & -610 & (99) & 0.00 & -2534 & (763) & 0.00 & -240 & (193) & 0.22\\
    3 & BC vs NP & -10935 & (424) & 0.00 & -230 & (79) & 0.00 & -2396 & (771) & 0.00 & -102 & (133) & 0.44\\
    4 & BC vs NP & -10495 & (433) & 0.00 & 210 & (66) & 0.00 & -2170 & (809) & 0.01 & 124 & (116) & 0.28\\
    7 & BC vs NP & -10075 & (436) & 0.00 & 630 & (150) & 0.00 & -2076 & (821) & 0.01 & 218 & (261) & 0.40\\ \midrule
    2 & TC vs NP & -9919 & (648) & 0.00 & -322 & (158) & 0.04 & -1639 & (1057) & 0.12 & -542 & (275) & 0.05\\
    3 & TC vs NP & -9770 & (634) & 0.00 & -173 & (99) & 0.08 & -1414 & (1055) & 0.18 & -317 & (172) & 0.07\\
    4 & TC vs NP & -9611 & (624) & 0.00 & -14 & (87) & 0.87 & -1111 & (1052) & 0.29 & -13 & (155) & 0.93\\
    7 & TC vs NP & -9088 & (614) & 0.00 & 509 & (195) & 0.01 & -226 & (1072) & 0.83 & 872 & (338) & 0.01\\ \midrule
    2 & EP vs NP & -7045 & (865) & 0.00 & -139 & (189) & 0.46 & -2353 & (1180) & 0.05 & -795 & (295) & 0.01\\
    3 & EP vs NP & -7045 & (872) & 0.00 & -140 & (118) & 0.24 & -2048 & (1219) & 0.09 & -490 & (204) & 0.02\\
    4 & EP vs NP & -6942 & (792) & 0.00 & -36 & (95) & 0.70 & -1301 & (1189) & 0.27 & 257 & (153) & 0.09\\
    7 & EP vs NP & -6589 & (766) & 0.00 & 316 & (209) & 0.13 & -531 & (1240) & 0.67 & 1027 & (375) & 0.01\\
    \bottomrule
    \end{tabular}
        \begin{tablenotes}
\textit{Note:} The table shows the BGATEs for the different programmes in comparison to no programme (NP: Non Participation, WS: Wage Subsidy, BC: Basic courses, TC: Training Courses, EP: Employment Programme) and the different sub-groups. The unemployment rate values approximate the closest integer.
\end{tablenotes}
\end{threeparttable}
\end{adjustbox}
\end{table}
\clearpage

\subsubsection{Temporary Residents Sample}
\textcolor{red}{}
\begin{table}[htbp!]
\begin{adjustbox}{width=0.93\columnwidth, center}
\begin{threeparttable}
  \captionsetup{font=large}
    \caption{BGATEs for place of origin, TEMP sample}
    \label{table:bgate_b_origin}
    \centering
    \begin{tabular}{@{}ll|rrrrrr|rrrrrrr@{}}
    \toprule
     \multicolumn{2}{l}{\textbf{Origin}}& \multicolumn{6}{c}{\textbf{First Year}} & \multicolumn{6}{c}{\textbf{Third Year}} \\
    \midrule
     & Programme & BGATE & S.E. & p-value & BGATE-ATE & S.E. & p-value & BGATE & S.E. & p-value & BGATE-ATE & S.E. & p-value \\
    \midrule
     &&  \multicolumn{12}{c}{\textbf{Outcome: Employment}} \\ \midrule
    Neig. countries & WS vs NP & -0.45 & (0.11) & 0.00 & -0.06 & (0.02) & 0.01 & 0.99 & (0.13) & 0.00 & -0.01 & (0.03) & 0.85\\
    Rest of European Union & WS vs NP & -0.44 & (0.11) & 0.00 & -0.05 & (0.02) & 0.00 & 0.96 & (0.13) & 0.00 & -0.04 & (0.02) & 0.05\\
    Rest of Europe & WS vs NP & -0.34 & (0.11) & 0.00 & 0.05 & (0.02) & 0.00 & 1.00 & (0.13) & 0.00 & -0.00 & (0.02) & 0.96\\
    Rest of the world & WS vs NP & -0.34 & (0.11) & 0.00 & 0.05 & (0.02) & 0.01 & 1.04 & (0.13) & 0.00 & 0.04 & (0.02) & 0.06\\ \midrule
    Neig. countries & BC vs NP & -1.32 & (0.13) & 0.00 & -0.03 & (0.02) & 0.22 & 0.45 & (0.17) & 0.01 & -0.06 & (0.04) & 0.15\\
    Rest of European Union & BC vs NP & -1.33 & (0.13) & 0.00 & -0.04 & (0.02) & 0.05 & 0.51 & (0.16) & 0.00 & 0.01 & (0.02) & 0.71\\
    Rest of Europe & BC vs NP & -1.27 & (0.13) & 0.00 & 0.02 & (0.02) & 0.29 & 0.53 & (0.17) & 0.00 & 0.03 & (0.03) & 0.23\\
    Rest of the world & BC vs NP & -1.24 & (0.13) & 0.00 & 0.05 & (0.02) & 0.02 & 0.52 & (0.17) & 0.00 & 0.02 & (0.03) & 0.51\\ \midrule
    Neig. countries & TC vs NP & -1.41 & (0.15) & 0.00 & -0.07 & (0.03) & 0.00 & 0.46 & (0.19) & 0.02 & -0.02 & (0.05) & 0.64\\
    Rest of European Union & TC vs NP & -1.38 & (0.15) & 0.00 & -0.04 & (0.02) & 0.04 & 0.46 & (0.18) & 0.01 & -0.02 & (0.03) & 0.45\\
    Rest of Europe & TC vs NP & -1.28 & (0.15) & 0.00 & 0.05 & (0.02) & 0.02 & 0.49 & (0.19) & 0.01 & 0.02 & (0.03) & 0.58\\
    Rest of the world & TC vs NP & -1.27 & (0.15) & 0.00 & 0.06 & (0.02) & 0.00 & 0.51 & (0.19) & 0.01 & 0.03 & (0.03) & 0.41\\ \midrule
    Neig. countries & EP vs NP & -1.30 & (0.19) & 0.00 & -0.05 & (0.03) & 0.11 & 0.32 & (0.26) & 0.21 & -0.07 & (0.06) & 0.29\\
    Rest of European Union & EP vs NP & -1.28 & (0.19) & 0.00 & -0.03 & (0.03) & 0.34 & 0.35 & (0.24) & 0.15 & -0.04 & (0.03) & 0.30\\
    Rest of Europe & EP vs NP & -1.22 & (0.19) & 0.00 & 0.04 & (0.03) & 0.18 & 0.44 & (0.24) & 0.07 & 0.05 & (0.03) & 0.15\\
    Rest of the world & EP vs NP & -1.22 & (0.19) & 0.00 & 0.04 & (0.03) & 0.13 & 0.44 & (0.24) & 0.07 & 0.05 & (0.04) & 0.22\\  \midrule
     &&  \multicolumn{12}{c}{\textbf{Outcome: Earnings}} \\ \midrule
    Neig. countries & WS vs NP & -3785 & (707) & 0.00 & -363 & (165) & 0.03 & 3303 & (984) & 0.00 & 18 & (245) & 0.94\\
    Rest of European Union & WS vs NP & -3574 & (688) & 0.00 & -152 & (91) & 0.10 & 3262 & (960) & 0.00 & -23 & (124) & 0.85\\
    Rest of Europe & WS vs NP & -3070 & (767) & 0.00 & 352 & (121) & 0.00 & 3244 & (1075) & 0.00 & -42 & (182) & 0.82\\
    Rest of the world & WS vs NP & -3259 & (774) & 0.00 & 163 & (120) & 0.17 & 3332 & (1090) & 0.00 & 46 & (162) & 0.77\\
    \midrule
    Neig. countries & BC vs NP & -7135 & (817) & 0.00 & -273 & (175) & 0.12 & 835 & (1168) & 0.47 & -58 & (267) & 0.83\\
    Rest of European Union & BC vs NP & -7039 & (798) & 0.00 & -176 & (101) & 0.08 & 1006 & (1139) & 0.38 & 113 & (143) & 0.43\\
    Rest of Europe & BC vs NP & -6708 & (855) & 0.00 & 154 & (128) & 0.23 & 939 & (1226) & 0.44 & 46 & (199) & 0.82\\
    Rest of the world & BC vs NP & -6568 & (861) & 0.00 & 294 & (129) & 0.02 & 791 & (1241) & 0.52 & -101 & (177) & 0.57\\
    \midrule
    Neig. countries & TC vs NP & -7393 & (925) & 0.00 & -470 & (190) & 0.01 & 1785 & (1358) & 0.19 & 86 & (309) & 0.78\\
    Rest of European Union & TC vs NP & -7139 & (884) & 0.00 & -217 & (109) & 0.05 & 1643 & (1276) & 0.20 & -56 & (154) & 0.71\\
    Rest of Europe & TC vs NP & -6582 & (937) & 0.00 & 340 & (137) & 0.01 & 1630 & (1347) & 0.23 & -69 & (209) & 0.74\\
    Rest of the world & TC vs NP & -6576 & (940) & 0.00 & 347 & (142) & 0.01 & 1739 & (1371) & 0.20 & 40 & (202) & 0.84\\
    \midrule
    Neig. countries & EP vs NP & -7858 & (1078) & 0.00 & -390 & (213) & 0.07 & -490 & (1837) & 0.79 & -274 & (352) & 0.44\\
    Rest of European Union & EP vs NP & -7626 & (1050) & 0.00 & -158 & (125) & 0.21 & -331 & (1702) & 0.85 & -114 & (197) & 0.56\\
    Rest of Europe & EP vs NP & -7225 & (1069) & 0.00 & 243 & (157) & 0.12 & -26 & (1678) & 0.99 & 190 & (250) & 0.45\\
    Rest of the world & EP vs NP & -7162 & (1083) & 0.00 & 306 & (149) & 0.04 & -18 & (1733) & 0.99 & 199 & (224) & 0.38\\

    \bottomrule
    \end{tabular}
        \begin{tablenotes}
\textit{Note:} The table shows the BGATEs for the different programmes in comparison to no programme (NP: Non Participation, WS: Wage subsidy , BC: Basic courses, TC: Training Courses, EP: Employment Programme) and the different sub-groups. 
\end{tablenotes}
\end{threeparttable}
\end{adjustbox}
\end{table}

\begin{table}[htbp!]
\begin{adjustbox}{width=0.93\columnwidth, center}
\begin{threeparttable}
  \captionsetup{font=large}
    \caption{BGATEs for gender, TEMP sample}
    \label{table:bgate_b_gender}
    \centering
    \begin{tabular}{@{}ll|rrrrrr|rrrrrrr@{}}
    \toprule
     \multicolumn{2}{l}{\textbf{Gender}}& \multicolumn{6}{c}{\textbf{First Year}} & \multicolumn{6}{c}{\textbf{Third Year}} \\
    \midrule
     & Programme & BGATE & S.E. & p-value & BGATE-ATE & S.E. & p-value & BGATE & S.E. & p-value & BGATE-ATE & S.E. & p-value \\
    \midrule
     &&  \multicolumn{12}{c}{\textbf{Outcome: Employment}} \\ \midrule
    Man & WS vs NP & -0.43 & (0.11) & 0.00 & -0.00 & (0.01) & 0.92 & 0.95 & (0.13) & 0.00 & -0.04 & (0.01) & 0.00\\
    Woman & WS vs NP & -0.43 & (0.11) & 0.00 & 0.00 & (0.01) & 0.92 & 1.03 & (0.13) & 0.00 & 0.04 & (0.01) & 0.00\\ \midrule
    Man & BC vs NP & -1.32 & (0.13) & 0.00 & -0.02 & (0.01) & 0.18 & 0.49 & (0.16) & 0.00 & -0.01 & (0.02) & 0.77\\
    Woman & BC vs NP & -1.29 & (0.13) & 0.00 & 0.02 & (0.01) & 0.18 & 0.50 & (0.17) & 0.00 & 0.01 & (0.02) & 0.77\\  \midrule
    Man & TC vs NP & -1.39 & (0.15) & 0.00 & -0.01 & (0.02) & 0.65 & 0.46 & (0.18) & 0.01 & -0.01 & (0.02) & 0.68\\
    Woman & TC vs NP & -1.37 & (0.15) & 0.00 & 0.01 & (0.02) & 0.65 & 0.48 & (0.19) & 0.01 & 0.01 & (0.02) & 0.68\\  \midrule
    Man & EP vs NP & -1.30 & (0.19) & 0.00 & -0.02 & (0.02) & 0.34 & 0.35 & (0.24) & 0.15 & -0.03 & (0.03) & 0.30\\
    Woman & EP vs NP & -1.26 & (0.19) & 0.00 & 0.02 & (0.02) & 0.34 & 0.40 & (0.24) & 0.10 & 0.03 & (0.03) & 0.30\\
    \midrule
     &&  \multicolumn{12}{c}{\textbf{Outcome: Earnings}} \\ \midrule
    Man & WS vs NP & -3671 & (695) & 0.00 & -76 & (55) & 0.17 & 3185 & (975) & 0.00 & -14 & (80) & 0.86\\
    Woman & WS vs NP & -3518 & (713) & 0.00 & 76 & (55) & 0.17 & 3213 & (995) & 0.00 & 14 & (80) & 0.86\\ \midrule
    Man & BC vs NP & -7183 & (809) & 0.00 & -237 & (73) & 0.00 & 914 & (1154) & 0.43 & 20 & (102) & 0.84\\
    Woman & BC vs NP & -6710 & (808) & 0.00 & 237 & (73) & 0.00 & 875 & (1159) & 0.45 & -20 & (102) & 0.84\\ \midrule
    Man & TC vs NP & -7235 & (910) & 0.00 & -151 & (81) & 0.06 & 1731 & (1308) & 0.19 & -22 & (113) & 0.84\\
    Woman & TC vs NP & -6933 & (906) & 0.00 & 151 & (81) & 0.06 & 1775 & (1318) & 0.18 & 22 & (113) & 0.84\\ \midrule
    Man & EP vs NP & -7854 & (1064) & 0.00 & -220 & (99) & 0.03 & -355 & (1722) & 0.84 & -68 & (153) & 0.66\\
    Woman & EP vs NP & -7414 & (1059) & 0.00 & 220 & (99) & 0.03 & -219 & (1711) & 0.90 & 68 & (153) & 0.66\\

    \bottomrule
    \end{tabular}
        \begin{tablenotes}
\textit{Note:} The table shows the BGATEs for the different programmes in comparison to no programme (NP: Non Participation, WS: Wage subsidy , BC: Basic courses, TC: Training Courses, EP: Employment Programme) and the different sub-groups. 
\end{tablenotes}
\end{threeparttable}
\end{adjustbox}
\end{table}

\begin{table}[htbp!]
\begin{adjustbox}{width=0.95\columnwidth, center} 
\begin{threeparttable}
  \captionsetup{font=large}
    \caption{BGATEs for sector, TEMP sample}
    \label{table:bgate_b_sector}
    \centering
    \begin{tabular}{@{}ll|rrrrrr|rrrrrrr@{}}
    \toprule
     \multicolumn{2}{l}{\textbf{Sector}}& \multicolumn{6}{c}{\textbf{First Year}} & \multicolumn{6}{c}{\textbf{Third Year}} \\
    \midrule
     & Programme & BGATE & S.E. & p-value & BGATE-ATE & S.E. & p-value & GATE & S.E. & p-value & bGATE-ATE & S.E. & p-value \\
    \midrule
     &&  \multicolumn{12}{c}{\textbf{Outcome: Employment}} \\ \midrule
    A & WS vs NP & -0.39 & (0.12) & 0.00 & 0.02 & (0.04) & 0.61 & 0.92 & (0.14) & 0.00 & -0.06 & (0.04) & 0.16\\
    B & WS vs NP & -0.34 & (0.12) & 0.00 & 0.07 & (0.02) & 0.00 & 1.00 & (0.13) & 0.00 & 0.03 & (0.02) & 0.31\\
    C & WS vs NP & -0.43 & (0.11) & 0.00 & -0.01 & (0.02) & 0.45 & 0.95 & (0.13) & 0.00 & -0.03 & (0.02) & 0.26\\
    D & WS vs NP & -0.44 & (0.11) & 0.00 & -0.03 & (0.03) & 0.35 & 0.86 & (0.13) & 0.00 & -0.11 & (0.04) & 0.00\\
    E & WS vs NP & -0.42 & (0.11) & 0.00 & -0.01 & (0.02) & 0.57 & 1.01 & (0.13) & 0.00 & 0.03 & (0.02) & 0.16\\
    F & WS vs NP & -0.51 & (0.11) & 0.00 & -0.10 & (0.04) & 0.00 & 0.94 & (0.13) & 0.00 & -0.04 & (0.03) & 0.32\\
    G & WS vs NP & -0.39 & (0.11) & 0.00 & 0.03 & (0.02) & 0.28 & 1.03 & (0.13) & 0.00 & 0.05 & (0.03) & 0.07\\
    H & WS vs NP & -0.44 & (0.11) & 0.00 & -0.03 & (0.02) & 0.30 & 1.08 & (0.13) & 0.00 & 0.11 & (0.03) & 0.00\\ \midrule
    A & BC vs NP & -1.28 & (0.14) & 0.00 & 0.04 & (0.04) & 0.32 & 0.49 & (0.17) & 0.01 & -0.00 & (0.05) & 0.94\\
    B & BC vs NP & -1.27 & (0.14) & 0.00 & 0.05 & (0.03) & 0.08 & 0.55 & (0.17) & 0.00 & 0.06 & (0.03) & 0.04\\
    C & BC vs NP & -1.35 & (0.13) & 0.00 & -0.03 & (0.02) & 0.20 & 0.45 & (0.17) & 0.01 & -0.04 & (0.03) & 0.16\\
    D & BC vs NP & -1.37 & (0.13) & 0.00 & -0.05 & (0.04) & 0.15 & 0.48 & (0.17) & 0.00 & -0.01 & (0.05) & 0.84\\
    E & BC vs NP & -1.33 & (0.13) & 0.00 & -0.01 & (0.02) & 0.65 & 0.47 & (0.17) & 0.01 & -0.03 & (0.03) & 0.34\\
    F & BC vs NP & -1.39 & (0.14) & 0.00 & -0.07 & (0.04) & 0.10 & 0.42 & (0.17) & 0.01 & -0.07 & (0.05) & 0.10\\
    G & BC vs NP & -1.30 & (0.13) & 0.00 & 0.02 & (0.03) & 0.47 & 0.48 & (0.17) & 0.00 & -0.01 & (0.03) & 0.82\\
    H & BC vs NP & -1.31 & (0.13) & 0.00 & 0.01 & (0.03) & 0.84 & 0.51 & (0.17) & 0.00 & 0.02 & (0.04) & 0.59\\ \midrule
    A & TC vs NP & -1.36 & (0.16) & 0.00 & 0.03 & (0.04) & 0.48 & 0.40 & (0.19) & 0.04 & -0.08 & (0.05) & 0.15\\
    B & TC vs NP & -1.34 & (0.15) & 0.00 & 0.04 & (0.03) & 0.12 & 0.50 & (0.19) & 0.01 & 0.02 & (0.03) & 0.46\\
    C & TC vs NP & -1.38 & (0.15) & 0.00 & 0.01 & (0.03) & 0.74 & 0.52 & (0.19) & 0.01 & 0.05 & (0.03) & 0.16\\
    D & TC vs NP & -1.42 & (0.15) & 0.00 & -0.03 & (0.04) & 0.44 & 0.43 & (0.19) & 0.02 & -0.04 & (0.05) & 0.43\\
    E & TC vs NP & -1.39 & (0.15) & 0.00 & -0.00 & (0.02) & 0.94 & 0.48 & (0.19) & 0.01 & 0.01 & (0.03) & 0.76\\
    F & TC vs NP & -1.44 & (0.15) & 0.00 & -0.05 & (0.05) & 0.30 & 0.43 & (0.19) & 0.02 & -0.05 & (0.05) & 0.37\\
    G & TC vs NP & -1.39 & (0.15) & 0.00 & 0.00 & (0.03) & 0.98 & 0.49 & (0.20) & 0.01 & 0.01 & (0.04) & 0.72\\
    H & TC vs NP & -1.40 & (0.16) & 0.00 & -0.01 & (0.03) & 0.65 & 0.53 & (0.20) & 0.01 & 0.05 & (0.04) & 0.23\\ \midrule
    A & EP vs NP & -1.21 & (0.19) & 0.00 & 0.08 & (0.05) & 0.11 & 0.40 & (0.24) & 0.10 & 0.02 & (0.07) & 0.80\\
    B & EP vs NP & -1.24 & (0.19) & 0.00 & 0.06 & (0.04) & 0.09 & 0.45 & (0.24) & 0.06 & 0.07 & (0.04) & 0.07\\
    C & EP vs NP & -1.30 & (0.19) & 0.00 & -0.01 & (0.03) & 0.78 & 0.39 & (0.25) & 0.12 & 0.01 & (0.04) & 0.82\\
    D & EP vs NP & -1.33 & (0.19) & 0.00 & -0.03 & (0.06) & 0.59 & 0.31 & (0.24) & 0.21 & -0.07 & (0.08) & 0.36\\
    E & EP vs NP & -1.32 & (0.19) & 0.00 & -0.02 & (0.03) & 0.52 & 0.35 & (0.25) & 0.16 & -0.03 & (0.04) & 0.51\\
    F & EP vs NP & -1.28 & (0.20) & 0.00 & 0.02 & (0.06) & 0.80 & 0.25 & (0.25) & 0.32 & -0.13 & (0.07) & 0.06\\
    G & EP vs NP & -1.32 & (0.20) & 0.00 & -0.03 & (0.04) & 0.51 & 0.42 & (0.25) & 0.09 & 0.05 & (0.05) & 0.34\\
    H & EP vs NP & -1.34 & (0.20) & 0.00 & -0.04 & (0.04) & 0.34 & 0.46 & (0.26) & 0.07 & 0.08 & (0.06) & 0.21\\ 

    \midrule
     &&  \multicolumn{12}{c}{\textbf{Outcome: Earnings}} \\ \midrule
    A & WS vs NP & -3209 & (775) & 0.00 & 310 & (277) & 0.26 & 2330 & (1048) & 0.03 & -662 & (386) & 0.09\\
    B & WS vs NP & -3067 & (739) & 0.00 & 452 & (160) & 0.00 & 2922 & (997) & 0.00 & -70 & (225) & 0.76\\
    C & WS vs NP & -3860 & (748) & 0.00 & -341 & (188) & 0.07 & 3012 & (1042) & 0.00 & 19 & (254) & 0.94\\
    D & WS vs NP & -3424 & (630) & 0.00 & 95 & (224) & 0.67 & 2724 & (854) & 0.00 & -268 & (340) & 0.43\\
    E & WS vs NP & -3644 & (695) & 0.00 & -125 & (147) & 0.39 & 3351 & (967) & 0.00 & 359 & (210) & 0.09\\
    F & WS vs NP & -3462 & (604) & 0.00 & 57 & (194) & 0.77 & 2772 & (812) & 0.00 & -220 & (276) & 0.42\\
    G & WS vs NP & -3639 & (770) & 0.00 & -121 & (190) & 0.53 & 3422 & (1058) & 0.00 & 430 & (271) & 0.11\\
    H & WS vs NP & -3904 & (732) & 0.00 & -386 & (164) & 0.02 & 3364 & (1016) & 0.00 & 372 & (256) & 0.15\\ \midrule
    A & BC vs NP & -6420 & (859) & 0.00 & 535 & (295) & 0.07 & 793 & (1190) & 0.51 & -120 & (417) & 0.77\\
    B & BC vs NP & -6602 & (827) & 0.00 & 353 & (169) & 0.04 & 1199 & (1151) & 0.30 & 287 & (239) & 0.23\\
    C & BC vs NP & -7424 & (869) & 0.00 & -469 & (219) & 0.03 & 487 & (1227) & 0.69 & -426 & (294) & 0.15\\
    D & BC vs NP & -7240 & (780) & 0.00 & -284 & (251) & 0.26 & 1137 & (1070) & 0.29 & 225 & (372) & 0.55\\
    E & BC vs NP & -7048 & (801) & 0.00 & -93 & (159) & 0.56 & 800 & (1143) & 0.48 & -112 & (235) & 0.63\\
    F & BC vs NP & -6848 & (723) & 0.00 & 107 & (217) & 0.62 & 849 & (1018) & 0.40 & -63 & (318) & 0.84\\
    G & BC vs NP & -7107 & (871) & 0.00 & -151 & (210) & 0.47 & 611 & (1226) & 0.62 & -302 & (297) & 0.31\\
    H & BC vs NP & -7132 & (834) & 0.00 & -177 & (182) & 0.33 & 882 & (1190) & 0.46 & -31 & (294) & 0.92\\ \midrule
    A & TC vs NP & -6540 & (921) & 0.00 & 578 & (320) & 0.07 & 1170 & (1292) & 0.37 & -513 & (471) & 0.28\\
    B & TC vs NP & -6766 & (906) & 0.00 & 353 & (185) & 0.06 & 1573 & (1270) & 0.22 & -109 & (263) & 0.68\\
    C & TC vs NP & -7497 & (986) & 0.00 & -378 & (265) & 0.15 & 1799 & (1409) & 0.20 & 116 & (349) & 0.74\\
    D & TC vs NP & -7232 & (865) & 0.00 & -113 & (290) & 0.70 & 1710 & (1202) & 0.15 & 28 & (429) & 0.95\\
    E & TC vs NP & -7247 & (897) & 0.00 & -129 & (177) & 0.47 & 1699 & (1335) & 0.20 & 16 & (289) & 0.96\\
    F & TC vs NP & -6918 & (809) & 0.00 & 201 & (243) & 0.41 & 1650 & (1150) & 0.15 & -32 & (364) & 0.93\\
    G & TC vs NP & -7369 & (975) & 0.00 & -250 & (236) & 0.29 & 1696 & (1409) & 0.23 & 13 & (346) & 0.97\\
    H & TC vs NP & -7403 & (966) & 0.00 & -284 & (206) & 0.17 & 1827 & (1396) & 0.19 & 145 & (325) & 0.66\\ \midrule
    A & EP vs NP & -6550 & (1058) & 0.00 & 1026 & (363) & 0.00 & 111 & (1542) & 0.94 & 247 & (547) & 0.65\\
    B & EP vs NP & -7194 & (1041) & 0.00 & 382 & (214) & 0.07 & 231 & (1618) & 0.89 & 367 & (311) & 0.24\\
    C & EP vs NP & -7980 & (1160) & 0.00 & -404 & (308) & 0.19 & -92 & (1894) & 0.96 & 44 & (514) & 0.93\\
    D & EP vs NP & -7635 & (1038) & 0.00 & -59 & (367) & 0.87 & -256 & (1521) & 0.87 & -120 & (622) & 0.85\\
    E & EP vs NP & -7820 & (1068) & 0.00 & -244 & (196) & 0.21 & -568 & (1728) & 0.74 & -433 & (348) & 0.21\\
    F & EP vs NP & -7134 & (984) & 0.00 & 442 & (311) & 0.16 & -562 & (1463) & 0.70 & -426 & (483) & 0.38\\
    G & EP vs NP & -8123 & (1118) & 0.00 & -547 & (259) & 0.03 & -89 & (1864) & 0.96 & 47 & (433) & 0.91\\
    H & EP vs NP & -8010 & (1156) & 0.00 & -434 & (309) & 0.16 & 263 & (1952) & 0.89 & 399 & (557) & 0.47\\
    \bottomrule
    \end{tabular}
        \begin{tablenotes}
\textit{Note:} The table shows the BGATEs for the different programmes in comparison to no programme (NP: Non Participation, WS: Wage subsidy , BC: Basic courses, TC: Training Courses, EP: Employment Programme) and the different sub-groups. The job sectors are categorised as A: agricultural and forestry, B: production in industry and trade, C: technical and information technology, D: construction and mining; E: transport and trade, F: hospitality and personal services, G: business, finance and law, H: education, art and science.
\end{tablenotes}
\end{threeparttable}
\end{adjustbox}
\end{table}

\begin{table}[htbp!]
\begin{adjustbox}{width=0.95\columnwidth, center}
\begin{threeparttable}
  \captionsetup{font=large}
    \caption{BGATEs for education, TEMP sample}
    \label{table:bgate_b_educ}
    \centering
    \begin{tabular}{@{}ll|rrrrrr|rrrrrrr@{}}
    \toprule
     \multicolumn{2}{l}{\textbf{Education}}& \multicolumn{6}{c}{\textbf{First Year}} & \multicolumn{6}{c}{\textbf{Third Year}} \\
    \midrule
     & Programme & BGATE & S.E. & p-value & BGATE-ATE & S.E. & p-value & BGATE & S.E. & p-value & BGATE-ATE & S.E. & p-value \\
    \midrule
     &&  \multicolumn{12}{c}{\textbf{Outcome: Employment}} \\ \midrule
    Compulsory & WS vs NP & -0.42 & (0.11) & 0.00 & -0.03 & (0.02) & 0.21 & 0.94 & (0.13) & 0.00 & -0.08 & (0.03) & 0.01\\
    Secondary & WS vs NP & -0.44 & (0.11) & 0.00 & -0.05 & (0.02) & 0.00 & 0.96 & (0.13) & 0.00 & -0.06 & (0.02) & 0.00\\
    Vocational & WS vs NP & -0.44 & (0.11) & 0.00 & -0.05 & (0.02) & 0.01 & 1.01 & (0.13) & 0.00 & -0.01 & (0.02) & 0.55\\
    Bachelor & WS vs NP & -0.35 & (0.12) & 0.00 & 0.04 & (0.02) & 0.02 & 1.07 & (0.14) & 0.00 & 0.06 & (0.02) & 0.01\\
    Post-Grad & WS vs NP & -0.31 & (0.12) & 0.01 & 0.08 & (0.03) & 0.00 & 1.11 & (0.14) & 0.00 & 0.09 & (0.03) & 0.00\\ \midrule
    Compulsory & BC vs NP & -1.31 & (0.13) & 0.00 & -0.03 & (0.03) & 0.25 & 0.49 & (0.17) & 0.00 & 0.00 & (0.04) & 1.00\\
    Secondary & BC vs NP & -1.32 & (0.13) & 0.00 & -0.04 & (0.02) & 0.03 & 0.46 & (0.16) & 0.01 & -0.04 & (0.02) & 0.13\\
    Vocational & BC vs NP & -1.32 & (0.13) & 0.00 & -0.04 & (0.02) & 0.02 & 0.49 & (0.17) & 0.00 & -0.01 & (0.02) & 0.73\\
    Bachelor & BC vs NP & -1.25 & (0.14) & 0.00 & 0.03 & (0.02) & 0.24 & 0.51 & (0.17) & 0.00 & 0.02 & (0.03) & 0.53\\
    Post-Grad & BC vs NP & -1.19 & (0.14) & 0.00 & 0.09 & (0.03) & 0.00 & 0.52 & (0.17) & 0.00 & 0.03 & (0.04) & 0.48\\ \midrule
    Compulsory & TC vs NP & -1.36 & (0.15) & 0.00 & -0.01 & (0.03) & 0.88 & 0.42 & (0.18) & 0.02 & -0.08 & (0.05) & 0.10\\
    Secondary & TC vs NP & -1.40 & (0.15) & 0.00 & -0.04 & (0.02) & 0.04 & 0.46 & (0.19) & 0.01 & -0.04 & (0.03) & 0.08\\
    Vocational & TC vs NP & -1.39 & (0.15) & 0.00 & -0.04 & (0.02) & 0.07 & 0.49 & (0.19) & 0.01 & -0.01 & (0.03) & 0.66\\
    Bachelor & TC vs NP & -1.33 & (0.15) & 0.00 & 0.02 & (0.02) & 0.38 & 0.58 & (0.20) & 0.00 & 0.07 & (0.03) & 0.02\\
    Post-Grad & TC vs NP & -1.29 & (0.16) & 0.00 & 0.07 & (0.03) & 0.05 & 0.57 & (0.20) & 0.00 & 0.07 & (0.04) & 0.13\\ \midrule
    Compulsory & EP vs NP & -1.21 & (0.19) & 0.00 & 0.06 & (0.04) & 0.15 & 0.36 & (0.24) & 0.13 & -0.04 & (0.06) & 0.56\\
    Secondary & EP vs NP & -1.28 & (0.19) & 0.00 & -0.01 & (0.03) & 0.65 & 0.32 & (0.24) & 0.19 & -0.08 & (0.04) & 0.04\\
    Vocational & EP vs NP & -1.29 & (0.20) & 0.00 & -0.02 & (0.03) & 0.51 & 0.40 & (0.25) & 0.11 & 0.01 & (0.03) & 0.82\\
    Bachelor & EP vs NP & -1.30 & (0.19) & 0.00 & -0.02 & (0.03) & 0.39 & 0.39 & (0.26) & 0.13 & -0.00 & (0.04) & 0.91\\
    Post-Grad & EP vs NP & -1.28 & (0.20) & 0.00 & -0.01 & (0.05) & 0.90 & 0.51 & (0.27) & 0.06 & 0.11 & (0.06) & 0.06\\
    \midrule
     &&  \multicolumn{12}{c}{\textbf{Outcome: Earnings}} \\ \midrule
    Compulsory & WS vs NP & -3316 & (656) & 0.00 & 245 & (216) & 0.26 & 2762 & (897) & 0.00 & -521 & (411) & 0.21\\
    Secondary & WS vs NP & -3696 & (672) & 0.00 & -136 & (140) & 0.33 & 2952 & (915) & 0.00 & -330 & (306) & 0.28\\
    Vocational & WS vs NP & -3784 & (706) & 0.00 & -223 & (122) & 0.07 & 3267 & (985) & 0.00 & -16 & (195) & 0.94\\
    Bachelor & WS vs NP & -3446 & (824) & 0.00 & 115 & (133) & 0.39 & 3640 & (1197) & 0.00 & 357 & (251) & 0.15\\
    Post-Grad & WS vs NP & -3561 & (901) & 0.00 & -1 & (255) & 1.00 & 3791 & (1506) & 0.01 & 509 & (637) & 0.42\\ \midrule
    Compulsory & BC vs NP & -6782 & (768) & 0.00 & 122 & (230) & 0.60 & 1003 & (1082) & 0.35 & 241 & (441) & 0.58\\
    Secondary & BC vs NP & -7030 & (788) & 0.00 & -127 & (154) & 0.41 & 810 & (1104) & 0.46 & 48 & (332) & 0.88\\
    Vocational & BC vs NP & -7018 & (829) & 0.00 & -114 & (135) & 0.40 & 814 & (1180) & 0.49 & 52 & (218) & 0.81\\
    Bachelor & BC vs NP & -6928 & (918) & 0.00 & -24 & (144) & 0.87 & 445 & (1355) & 0.74 & -317 & (276) & 0.25\\
    Post-Grad & BC vs NP & -6760 & (1000) & 0.00 & 144 & (270) & 0.59 & 738 & (1636) & 0.65 & -24 & (661) & 0.97\\ \midrule
    Compulsory & TC vs NP & -6807 & (845) & 0.00 & 294 & (271) & 0.28 & 1544 & (1196) & 0.20 & -415 & (523) & 0.43\\
    Secondary & TC vs NP & -7141 & (885) & 0.00 & -41 & (170) & 0.81 & 1880 & (1253) & 0.13 & -78 & (376) & 0.83\\
    Vocational & TC vs NP & -7291 & (943) & 0.00 & -191 & (149) & 0.20 & 2060 & (1367) & 0.13 & 102 & (238) & 0.67\\
    Bachelor & TC vs NP & -7107 & (1019) & 0.00 & -7 & (166) & 0.97 & 2335 & (1551) & 0.13 & 377 & (318) & 0.24\\
    Post-Grad & TC vs NP & -7155 & (1128) & 0.00 & -55 & (301) & 0.85 & 1973 & (1853) & 0.29 & 14 & (746) & 0.98\\ \midrule
    Compulsory & EP vs NP & -6923 & (1000) & 0.00 & 798 & (323) & 0.01 & -414 & (1482) & 0.78 & -192 & (785) & 0.81\\
    Secondary & EP vs NP & -7524 & (1027) & 0.00 & 198 & (209) & 0.34 & -566 & (1598) & 0.72 & -344 & (555) & 0.54\\
    Vocational & EP vs NP & -7817 & (1121) & 0.00 & -95 & (177) & 0.59 & -243 & (1824) & 0.89 & -21 & (382) & 0.96\\
    Bachelor & EP vs NP & -8087 & (1172) & 0.00 & -366 & (190) & 0.05 & -528 & (2153) & 0.81 & -305 & (433) & 0.48\\
    Post-Grad & EP vs NP & -8257 & (1294) & 0.00 & -535 & (391) & 0.17 & 639 & (2709) & 0.81 & 862 & (1250) & 0.49\\

    \bottomrule
    \end{tabular}
        \begin{tablenotes}
\textit{Note:} The table shows the BGATEs for the different programmes in comparison to no programme (NP: Non Participation, WS: Wage subsidy , BC: Basic courses, TC: Training Courses, EP: Employment Programme) and the different sub-groups. 
\end{tablenotes}
\end{threeparttable}
\end{adjustbox}
\end{table}

\begin{table}[htbp!]
\begin{adjustbox}{width=0.95\columnwidth, center}
\begin{threeparttable}
  \captionsetup{font=large}
    \caption{BGATEs for months in employment, TEMP sample}
    \label{table:bgates_b_months_employment}
    \centering
    \begin{tabular}{@{}ll|rrrrrr|rrrrrrr@{}}
    \toprule
     \multicolumn{2}{l}{\textbf{Months in employment}}& \multicolumn{6}{c}{\textbf{First Year}} & \multicolumn{6}{c}{\textbf{Third Year}} \\
    \midrule
     & Programme & BGATE & S.E. & p-value & BGATE-ATE & S.E. & p-value & BGATE & S.E. & p-value & BGATE-ATE & S.E. & p-value \\
    \midrule
     &&  \multicolumn{12}{c}{\textbf{Outcome: Employment}} \\ \midrule
    12 & WS vs NP & -0.52 & (0.14) & 0.00 & -0.08 & (0.07) & 0.20 & 1.03 & (0.16) & 0.00 & 0.05 & (0.08) & 0.53\\
    24 & WS vs NP & -0.45 & (0.12) & 0.00 & -0.01 & (0.03) & 0.61 & 0.98 & (0.14) & 0.00 & 0.00 & (0.04) & 0.93\\
    36 & WS vs NP & -0.40 & (0.11) & 0.00 & 0.04 & (0.02) & 0.04 & 0.95 & (0.13) & 0.00 & -0.02 & (0.02) & 0.31\\
    48 & WS vs NP & -0.37 & (0.11) & 0.00 & 0.07 & (0.03) & 0.02 & 0.95 & (0.13) & 0.00 & -0.03 & (0.04) & 0.50\\
    60 & WS vs NP & -0.45 & (0.11) & 0.00 & -0.01 & (0.05) & 0.88 & 0.97 & (0.13) & 0.00 & -0.01 & (0.06) & 0.93\\ \midrule
    12 & BC vs NP & -1.30 & (0.17) & 0.00 & -0.00 & (0.08) & 0.99 & 0.48 & (0.22) & 0.03 & -0.01 & (0.11) & 0.94\\
    24 & BC vs NP & -1.29 & (0.14) & 0.00 & 0.01 & (0.04) & 0.76 & 0.53 & (0.18) & 0.00 & 0.04 & (0.05) & 0.34\\
    36 & BC vs NP & -1.26 & (0.13) & 0.00 & 0.04 & (0.02) & 0.08 & 0.49 & (0.17) & 0.00 & 0.01 & (0.03) & 0.82\\
    48 & BC vs NP & -1.27 & (0.13) & 0.00 & 0.03 & (0.04) & 0.50 & 0.45 & (0.17) & 0.01 & -0.03 & (0.05) & 0.56\\
    60 & BC vs NP & -1.38 & (0.14) & 0.00 & -0.08 & (0.06) & 0.20 & 0.47 & (0.17) & 0.01 & -0.01 & (0.07) & 0.86\\ \midrule
    12 & TC vs NP & -1.53 & (0.18) & 0.00 & -0.14 & (0.09) & 0.14 & 0.47 & (0.24) & 0.05 & 0.01 & (0.12) & 0.90\\
    24 & TC vs NP & -1.44 & (0.16) & 0.00 & -0.04 & (0.04) & 0.29 & 0.45 & (0.20) & 0.02 & -0.01 & (0.05) & 0.80\\
    36 & TC vs NP & -1.31 & (0.15) & 0.00 & 0.08 & (0.02) & 0.00 & 0.48 & (0.19) & 0.01 & 0.02 & (0.03) & 0.54\\
    48 & TC vs NP & -1.31 & (0.15) & 0.00 & 0.09 & (0.04) & 0.03 & 0.46 & (0.19) & 0.01 & 0.00 & (0.05) & 0.99\\
    60 & TC vs NP & -1.38 & (0.16) & 0.00 & 0.01 & (0.07) & 0.84 & 0.43 & (0.20) & 0.03 & -0.02 & (0.09) & 0.78\\ \midrule
    12 & EP vs NP & -1.41 & (0.23) & 0.00 & -0.12 & (0.11) & 0.28 & 0.47 & (0.31) & 0.12 & 0.08 & (0.16) & 0.59\\
    24 & EP vs NP & -1.36 & (0.20) & 0.00 & -0.07 & (0.05) & 0.13 & 0.38 & (0.26) & 0.14 & -0.00 & (0.07) & 0.98\\
    36 & EP vs NP & -1.25 & (0.19) & 0.00 & 0.04 & (0.03) & 0.16 & 0.35 & (0.24) & 0.14 & -0.04 & (0.04) & 0.40\\
    48 & EP vs NP & -1.21 & (0.19) & 0.00 & 0.08 & (0.05) & 0.10 & 0.37 & (0.24) & 0.13 & -0.02 & (0.07) & 0.79\\
    60 & EP vs NP & -1.21 & (0.20) & 0.00 & 0.07 & (0.08) & 0.36 & 0.36 & (0.26) & 0.16 & -0.03 & (0.11) & 0.79\\
    \midrule
     &&  \multicolumn{12}{c}{\textbf{Outcome: Earnings}} \\ \midrule
    12 & WS vs NP & -3742 & (751) & 0.00 & -127 & (346) & 0.71 & 2423 & (1014) & 0.02 & -570 & (465) & 0.22\\
    24 & WS vs NP & -3707 & (733) & 0.00 & -92 & (166) & 0.58 & 2652 & (1010) & 0.01 & -341 & (221) & 0.12\\
    36 & WS vs NP & -3469 & (700) & 0.00 & 146 & (103) & 0.16 & 3101 & (981) & 0.00 & 108 & (137) & 0.43\\
    48 & WS vs NP & -3406 & (699) & 0.00 & 209 & (159) & 0.19 & 3214 & (978) & 0.00 & 221 & (214) & 0.30\\
    60 & WS vs NP & -3752 & (710) & 0.00 & -137 & (269) & 0.61 & 3574 & (987) & 0.00 & 581 & (353) & 0.10\\ \midrule
    12 & BC vs NP & -6543 & (900) & 0.00 & 317 & (416) & 0.45 & 622 & (1300) & 0.63 & -216 & (607) & 0.72\\
    24 & BC vs NP & -6718 & (851) & 0.00 & 141 & (204) & 0.49 & 729 & (1190) & 0.54 & -109 & (271) & 0.69\\
    36 & BC vs NP & -6709 & (803) & 0.00 & 151 & (121) & 0.21 & 940 & (1152) & 0.41 & 102 & (170) & 0.55\\
    48 & BC vs NP & -6909 & (807) & 0.00 & -50 & (191) & 0.79 & 911 & (1154) & 0.43 & 72 & (268) & 0.79\\
    60 & BC vs NP & -7418 & (840) & 0.00 & -559 & (319) & 0.08 & 989 & (1191) & 0.41 & 151 & (436) & 0.73\\ \midrule
    12 & TC vs NP & -7341 & (959) & 0.00 & -238 & (457) & 0.60 & 1067 & (1436) & 0.46 & -493 & (672) & 0.46\\
    24 & TC vs NP & -7202 & (914) & 0.00 & -98 & (219) & 0.65 & 1010 & (1343) & 0.45 & -550 & (303) & 0.07\\
    36 & TC vs NP & -6725 & (890) & 0.00 & 378 & (129) & 0.00 & 1748 & (1289) & 0.18 & 188 & (198) & 0.34\\
    48 & TC vs NP & -6891 & (902) & 0.00 & 212 & (213) & 0.32 & 1895 & (1304) & 0.15 & 335 & (307) & 0.27\\
    60 & TC vs NP & -7357 & (967) & 0.00 & -254 & (367) & 0.49 & 2081 & (1354) & 0.12 & 521 & (489) & 0.29\\ \midrule
    12 & EP vs NP & -8026 & (1184) & 0.00 & -407 & (537) & 0.45 & 192 & (1773) & 0.91 & 379 & (788) & 0.63\\
    24 & EP vs NP & -8143 & (1083) & 0.00 & -524 & (248) & 0.03 & -290 & (1686) & 0.86 & -103 & (373) & 0.78\\
    36 & EP vs NP & -7395 & (1060) & 0.00 & 223 & (153) & 0.14 & -355 & (1707) & 0.84 & -168 & (214) & 0.43\\
    48 & EP vs NP & -7252 & (1049) & 0.00 & 366 & (237) & 0.12 & -246 & (1710) & 0.89 & -59 & (371) & 0.87\\
    60 & EP vs NP & -7276 & (1119) & 0.00 & 343 & (404) & 0.40 & -236 & (1831) & 0.90 & -49 & (598) & 0.93\\
    \bottomrule
    \end{tabular}
        \begin{tablenotes}
\textit{Note:} The table shows the BGATEs for the different programmes in comparison to no programme (NP: Non Participation, WS: Wage Subsidy, BC: Basic courses, TC: Training Courses, EP: Employment Programme) and the different sub-groups. The ranges for months in employment are: 12: 1-12, 24: 13-24, 36: 25-36, 48: 37-48, 60:49-60.
\end{tablenotes}
\end{threeparttable}
\end{adjustbox}
\end{table}

\begin{table}[htbp!]
\begin{adjustbox}{width=0.8\columnwidth, center} 
\begin{threeparttable}
  \captionsetup{font=large}
    \caption{BGATEs for earnings decile, TEMP sample}
    \label{table:bgates_b_earnings_decile}
    \centering
    \begin{tabular}{@{}ll|rrrrrr|rrrrrrr@{}}
    \toprule
     \multicolumn{2}{l}{\textbf{Earnings decile}}& \multicolumn{6}{c}{\textbf{First Year}} & \multicolumn{6}{c}{\textbf{Third Year}} \\
    \midrule
     & Programme & BGATE & S.E. & p-value & BGATE-ATE & S.E. & p-value & BGATE & S.E. & p-value & BGATE-ATE & S.E. & p-value \\
    \midrule
     &&  \multicolumn{12}{c}{\textbf{Outcome: Employment}} \\ \midrule
    1317 & WS vs NP & -0.36 & (0.12) & 0.00 & 0.05 & (0.06) & 0.34 & 1.07 & (0.15) & 0.00 & 0.09 & (0.08) & 0.26\\
    2542 & WS vs NP & -0.41 & (0.12) & 0.00 & 0.01 & (0.05) & 0.91 & 1.03 & (0.14) & 0.00 & 0.05 & (0.06) & 0.39\\
    3293 & WS vs NP & -0.47 & (0.11) & 0.00 & -0.06 & (0.04) & 0.15 & 0.95 & (0.13) & 0.00 & -0.03 & (0.05) & 0.50\\
    3767 & WS vs NP & -0.49 & (0.11) & 0.00 & -0.07 & (0.03) & 0.02 & 0.91 & (0.13) & 0.00 & -0.07 & (0.04) & 0.07\\
    4170 & WS vs NP & -0.47 & (0.11) & 0.00 & -0.05 & (0.03) & 0.11 & 0.91 & (0.13) & 0.00 & -0.07 & (0.03) & 0.05\\
    4596 & WS vs NP & -0.45 & (0.11) & 0.00 & -0.04 & (0.03) & 0.16 & 0.90 & (0.13) & 0.00 & -0.07 & (0.03) & 0.01\\
    5041 & WS vs NP & -0.44 & (0.11) & 0.00 & -0.02 & (0.03) & 0.44 & 0.93 & (0.13) & 0.00 & -0.05 & (0.03) & 0.11\\
    5624 & WS vs NP & -0.39 & (0.11) & 0.00 & 0.03 & (0.03) & 0.37 & 0.92 & (0.13) & 0.00 & -0.06 & (0.04) & 0.15\\
    6610 & WS vs NP & -0.35 & (0.12) & 0.00 & 0.07 & (0.05) & 0.18 & 1.01 & (0.14) & 0.00 & 0.03 & (0.06) & 0.62\\
    10583 & WS vs NP & -0.34 & (0.15) & 0.03 & 0.08 & (0.10) & 0.41 & 1.16 & (0.18) & 0.00 & 0.18 & (0.12) & 0.13\\ \midrule
    1317 & BC vs NP & -1.17 & (0.15) & 0.00 & 0.13 & (0.07) & 0.07 & 0.45 & (0.20) & 0.02 & -0.04 & (0.10) & 0.73\\
    2542 & BC vs NP & -1.21 & (0.15) & 0.00 & 0.09 & (0.06) & 0.14 & 0.51 & (0.18) & 0.01 & 0.03 & (0.07) & 0.67\\
    3293 & BC vs NP & -1.28 & (0.14) & 0.00 & 0.01 & (0.05) & 0.76 & 0.45 & (0.18) & 0.01 & -0.04 & (0.06) & 0.57\\
    3767 & BC vs NP & -1.35 & (0.13) & 0.00 & -0.05 & (0.04) & 0.14 & 0.48 & (0.17) & 0.00 & 0.00 & (0.05) & 1.00\\
    4170 & BC vs NP & -1.37 & (0.14) & 0.00 & -0.07 & (0.03) & 0.03 & 0.45 & (0.17) & 0.01 & -0.04 & (0.04) & 0.41\\
    4596 & BC vs NP & -1.37 & (0.13) & 0.00 & -0.07 & (0.03) & 0.01 & 0.47 & (0.17) & 0.01 & -0.02 & (0.04) & 0.68\\
    5041 & BC vs NP & -1.38 & (0.13) & 0.00 & -0.09 & (0.03) & 0.01 & 0.48 & (0.17) & 0.00 & -0.01 & (0.04) & 0.88\\
    5624 & BC vs NP & -1.32 & (0.14) & 0.00 & -0.03 & (0.04) & 0.48 & 0.49 & (0.17) & 0.00 & 0.01 & (0.06) & 0.89\\
    6610 & BC vs NP & -1.28 & (0.14) & 0.00 & 0.02 & (0.06) & 0.79 & 0.54 & (0.18) & 0.00 & 0.06 & (0.08) & 0.46\\
    10583 & BC vs NP & -1.22 & (0.16) & 0.00 & 0.07 & (0.10) & 0.46 & 0.52 & (0.21) & 0.01 & 0.03 & (0.13) & 0.80\\ \midrule
    1317 & TC vs NP & -1.27 & (0.17) & 0.00 & 0.12 & (0.08) & 0.15 & 0.42 & (0.22) & 0.06 & -0.04 & (0.12) & 0.70\\
    2542 & TC vs NP & -1.33 & (0.16) & 0.00 & 0.05 & (0.07) & 0.42 & 0.43 & (0.20) & 0.03 & -0.03 & (0.08) & 0.68\\
    3293 & TC vs NP & -1.39 & (0.16) & 0.00 & -0.00 & (0.06) & 0.94 & 0.41 & (0.20) & 0.04 & -0.05 & (0.07) & 0.44\\
    3767 & TC vs NP & -1.46 & (0.15) & 0.00 & -0.08 & (0.04) & 0.05 & 0.41 & (0.19) & 0.03 & -0.06 & (0.05) & 0.28\\
    4170 & TC vs NP & -1.44 & (0.15) & 0.00 & -0.06 & (0.04) & 0.16 & 0.40 & (0.19) & 0.03 & -0.07 & (0.05) & 0.19\\
    4596 & TC vs NP & -1.45 & (0.15) & 0.00 & -0.06 & (0.03) & 0.06 & 0.40 & (0.19) & 0.03 & -0.06 & (0.04) & 0.14\\
    5041 & TC vs NP & -1.45 & (0.15) & 0.00 & -0.06 & (0.04) & 0.09 & 0.45 & (0.19) & 0.02 & -0.01 & (0.04) & 0.74\\
    5624 & TC vs NP & -1.37 & (0.16) & 0.00 & 0.02 & (0.05) & 0.74 & 0.45 & (0.19) & 0.02 & -0.01 & (0.06) & 0.83\\
    6610 & TC vs NP & -1.37 & (0.17) & 0.00 & 0.02 & (0.07) & 0.78 & 0.60 & (0.21) & 0.00 & 0.13 & (0.09) & 0.14\\
    10583 & TC vs NP & -1.33 & (0.19) & 0.00 & 0.06 & (0.11) & 0.61 & 0.69 & (0.25) & 0.01 & 0.22 & (0.16) & 0.16\\ \midrule
    1317 & EP vs NP & -1.02 & (0.21) & 0.00 & 0.27 & (0.11) & 0.01 & 0.50 & (0.27) & 0.06 & 0.12 & (0.15) & 0.43\\
    2542 & EP vs NP & -1.09 & (0.20) & 0.00 & 0.19 & (0.08) & 0.02 & 0.43 & (0.25) & 0.08 & 0.05 & (0.11) & 0.64\\
    3293 & EP vs NP & -1.17 & (0.21) & 0.00 & 0.12 & (0.07) & 0.11 & 0.39 & (0.25) & 0.12 & 0.00 & (0.09) & 0.97\\
    3767 & EP vs NP & -1.31 & (0.20) & 0.00 & -0.02 & (0.05) & 0.67 & 0.33 & (0.25) & 0.18 & -0.05 & (0.07) & 0.46\\
    4170 & EP vs NP & -1.38 & (0.19) & 0.00 & -0.10 & (0.05) & 0.05 & 0.31 & (0.24) & 0.21 & -0.07 & (0.07) & 0.28\\
    4596 & EP vs NP & -1.35 & (0.19) & 0.00 & -0.07 & (0.04) & 0.11 & 0.39 & (0.24) & 0.11 & 0.00 & (0.06) & 0.96\\
    5041 & EP vs NP & -1.41 & (0.20) & 0.00 & -0.12 & (0.05) & 0.01 & 0.42 & (0.25) & 0.09 & 0.04 & (0.06) & 0.54\\
    5624 & EP vs NP & -1.36 & (0.20) & 0.00 & -0.08 & (0.06) & 0.21 & 0.38 & (0.26) & 0.14 & -0.00 & (0.08) & 0.98\\
    6610 & EP vs NP & -1.36 & (0.21) & 0.00 & -0.07 & (0.09) & 0.42 & 0.38 & (0.29) & 0.19 & -0.00 & (0.13) & 0.98\\
    10583 & EP vs NP & -1.41 & (0.23) & 0.00 & -0.12 & (0.15) & 0.41 & 0.31 & (0.35) & 0.38 & -0.08 & (0.23) & 0.73\\
    \midrule
     &&  \multicolumn{12}{c}{\textbf{Outcome: Earnings}} \\ \midrule
    1317 & WS vs NP & -2535 & (664) & 0.00 & 992 & (378) & 0.01 & 2322 & (904) & 0.01 & -731 & (502) & 0.15\\
    2542 & WS vs NP & -2777 & (705) & 0.00 & 751 & (379) & 0.05 & 2411 & (953) & 0.01 & -642 & (500) & 0.20\\
    3293 & WS vs NP & -2958 & (646) & 0.00 & 569 & (367) & 0.12 & 2359 & (892) & 0.01 & -694 & (496) & 0.16\\
    3767 & WS vs NP & -3209 & (611) & 0.00 & 318 & (330) & 0.33 & 2545 & (805) & 0.00 & -508 & (460) & 0.27\\
    4170 & WS vs NP & -3264 & (681) & 0.00 & 263 & (340) & 0.44 & 2646 & (886) & 0.00 & -407 & (452) & 0.37\\
    4596 & WS vs NP & -3308 & (662) & 0.00 & 219 & (315) & 0.49 & 2827 & (873) & 0.00 & -226 & (425) & 0.59\\
    5041 & WS vs NP & -3538 & (683) & 0.00 & -11 & (270) & 0.97 & 3001 & (905) & 0.00 & -52 & (371) & 0.89\\
    5624 & WS vs NP & -3555 & (773) & 0.00 & -28 & (301) & 0.93 & 2777 & (1016) & 0.01 & -276 & (386) & 0.48\\
    6610 & WS vs NP & -4233 & (1044) & 0.00 & -705 & (539) & 0.19 & 3087 & (1396) & 0.03 & 34 & (693) & 0.96\\
    10583 & WS vs NP & -5896 & (2028) & 0.00 & -2369 & (1630) & 0.15 & 6554 & (2997) & 0.03 & 3501 & (2434) & 0.15\\ \midrule
    1317 & BC vs NP & -5313 & (763) & 0.00 & 1549 & (456) & 0.00 & 535 & (1087) & 0.62 & -263 & (612) & 0.67\\
    2542 & BC vs NP & -5486 & (796) & 0.00 & 1376 & (445) & 0.00 & 753 & (1116) & 0.50 & -45 & (591) & 0.94\\
    3293 & BC vs NP & -5775 & (755) & 0.00 & 1086 & (435) & 0.01 & 646 & (1081) & 0.55 & -152 & (592) & 0.80\\
    3767 & BC vs NP & -6429 & (725) & 0.00 & 433 & (370) & 0.24 & 1008 & (1005) & 0.32 & 210 & (526) & 0.69\\
    4170 & BC vs NP & -6926 & (775) & 0.00 & -64 & (379) & 0.87 & 785 & (1063) & 0.46 & -13 & (516) & 0.98\\
    4596 & BC vs NP & -7139 & (786) & 0.00 & -278 & (353) & 0.43 & 906 & (1071) & 0.40 & 108 & (486) & 0.82\\
    5041 & BC vs NP & -7489 & (835) & 0.00 & -627 & (316) & 0.05 & 984 & (1123) & 0.38 & 186 & (435) & 0.67\\
    5624 & BC vs NP & -7387 & (963) & 0.00 & -526 & (394) & 0.18 & 1060 & (1284) & 0.41 & 262 & (515) & 0.61\\
    6610 & BC vs NP & -7681 & (1243) & 0.00 & -819 & (674) & 0.22 & 1485 & (1703) & 0.38 & 688 & (906) & 0.45\\
    10583 & BC vs NP & -8991 & (2052) & 0.00 & -2130 & (1607) & 0.18 & -184 & (3192) & 0.95 & -982 & (2560) & 0.70\\ \midrule
    1317 & TC vs NP & -5475 & (833) & 0.00 & 1629 & (516) & 0.00 & 767 & (1187) & 0.52 & -929 & (703) & 0.19\\
    2542 & TC vs NP & -5829 & (853) & 0.00 & 1275 & (496) & 0.01 & 841 & (1211) & 0.49 & -856 & (683) & 0.21\\
    3293 & TC vs NP & -6046 & (820) & 0.00 & 1057 & (490) & 0.03 & 858 & (1167) & 0.46 & -838 & (691) & 0.23\\
    3767 & TC vs NP & -6709 & (794) & 0.00 & 395 & (420) & 0.35 & 1163 & (1113) & 0.30 & -533 & (614) & 0.38\\
    4170 & TC vs NP & -6859 & (840) & 0.00 & 244 & (435) & 0.57 & 1085 & (1168) & 0.35 & -612 & (614) & 0.32\\
    4596 & TC vs NP & -7084 & (862) & 0.00 & 20 & (409) & 0.96 & 1313 & (1197) & 0.27 & -384 & (571) & 0.50\\
    5041 & TC vs NP & -7559 & (916) & 0.00 & -456 & (364) & 0.21 & 1822 & (1278) & 0.15 & 125 & (495) & 0.80\\
    5624 & TC vs NP & -7677 & (1055) & 0.00 & -574 & (445) & 0.20 & 1978 & (1476) & 0.18 & 282 & (601) & 0.64\\
    6610 & TC vs NP & -8381 & (1410) & 0.00 & -1277 & (787) & 0.10 & 3488 & (2012) & 0.08 & 1792 & (1092) & 0.10\\
    10583 & TC vs NP & -9416 & (2435) & 0.00 & -2313 & (1916) & 0.23 & 3650 & (3707) & 0.32 & 1954 & (2983) & 0.51\\ \midrule
    1317 & EP vs NP & -4860 & (1036) & 0.00 & 2747 & (638) & 0.00 & 516 & (1455) & 0.72 & 686 & (965) & 0.48\\
    2542 & EP vs NP & -5213 & (1028) & 0.00 & 2394 & (597) & 0.00 & 511 & (1435) & 0.72 & 681 & (922) & 0.46\\
    3293 & EP vs NP & -5622 & (1005) & 0.00 & 1985 & (595) & 0.00 & 515 & (1422) & 0.72 & 685 & (949) & 0.47\\
    3767 & EP vs NP & -6640 & (968) & 0.00 & 967 & (502) & 0.05 & 238 & (1391) & 0.86 & 408 & (841) & 0.63\\
    4170 & EP vs NP & -7372 & (985) & 0.00 & 235 & (518) & 0.65 & 267 & (1445) & 0.85 & 437 & (849) & 0.61\\
    4596 & EP vs NP & -7661 & (1032) & 0.00 & -54 & (490) & 0.91 & 677 & (1513) & 0.65 & 846 & (802) & 0.29\\
    5041 & EP vs NP & -8415 & (1085) & 0.00 & -808 & (440) & 0.07 & 753 & (1649) & 0.65 & 922 & (663) & 0.16\\
    5624 & EP vs NP & -8563 & (1240) & 0.00 & -956 & (551) & 0.08 & -427 & (1891) & 0.82 & -258 & (809) & 0.75\\
    6610 & EP vs NP & -9600 & (1664) & 0.00 & -1993 & (950) & 0.04 & -1662 & (2700) & 0.54 & -1492 & (1430) & 0.30\\
    10583 & EP vs NP & -12125 & (2881) & 0.00 & -4517 & (2326) & 0.05 & -3084 & (5577) & 0.58 & -2915 & (4494) & 0.52\\
    \bottomrule
    \end{tabular}
        \begin{tablenotes}
\textit{Note:} The table shows the BGATEs for the different programmes in comparison to no programme (NP: Non Participation, WS: Wage Subsidy, BC: Basic courses, TC: Training Courses, EP: Employment Programme) and the different sub-groups. The Earning decile is the average earnings within each decile of the earnings distribution in the three months before the
unemployment spell.
\end{tablenotes}
\end{threeparttable}
\end{adjustbox}
\end{table}

\begin{table}[htbp]
\begin{adjustbox}{width=0.9\columnwidth, center}
\begin{threeparttable}
  \captionsetup{font=large}
    \caption{BGATEs for ue rate, TEMP sample}
    \label{table:bgate_uerate}
    \centering
    \begin{tabular}{@{}ll|rrrrrr|rrrrrrr@{}}
    \toprule
     \multicolumn{2}{l}{\textbf{ue rate (\%)}}& \multicolumn{6}{c}{\textbf{First Year}} & \multicolumn{6}{c}{\textbf{Third Year}} \\
    \midrule
     & Programme & BGATE & S.E. & p-value & BGATE-ATE & S.E. & p-value & BGATE & S.E. & p-value & BGATE-ATE & S.E. & p-value \\
    \midrule
     &&  \multicolumn{12}{c}{\textbf{Outcome: Employment}} \\ \midrule
    3 & WS vs NP & -0.44 & (0.11) & 0.00 & -0.05 & (0.03) & 0.08 & 0.94 & (0.13) & 0.00 & -0.05 & (0.04) & 0.17\\
    4 & WS vs NP & -0.41 & (0.11) & 0.00 & -0.01 & (0.02) & 0.44 & 0.97 & (0.13) & 0.00 & -0.02 & (0.02) & 0.31\\
    7 & WS vs NP & -0.33 & (0.12) & 0.01 & 0.07 & (0.04) & 0.07 & 1.06 & (0.14) & 0.00 & 0.07 & (0.04) & 0.08\\ \midrule
    3 & BC vs NP & -1.32 & (0.13) & 0.00 & -0.03 & (0.04) & 0.36 & 0.46 & (0.17) & 0.01 & -0.04 & (0.04) & 0.36\\
    4 & BC vs NP & -1.30 & (0.13) & 0.00 & -0.01 & (0.02) & 0.50 & 0.45 & (0.17) & 0.01 & -0.05 & (0.03) & 0.06\\
    7 & BC vs NP & -1.24 & (0.14) & 0.00 & 0.05 & (0.04) & 0.27 & 0.59 & (0.18) & 0.00 & 0.09 & (0.05) & 0.06\\ \midrule
    3 & TC vs NP & -1.40 & (0.15) & 0.00 & -0.06 & (0.04) & 0.16 & 0.45 & (0.19) & 0.02 & -0.03 & (0.05) & 0.52\\
    4 & TC vs NP & -1.39 & (0.15) & 0.00 & -0.04 & (0.03) & 0.10 & 0.46 & (0.19) & 0.01 & -0.02 & (0.03) & 0.55\\
    7 & TC vs NP & -1.25 & (0.16) & 0.00 & 0.10 & (0.05) & 0.04 & 0.53 & (0.19) & 0.01 & 0.05 & (0.06) & 0.35\\ \midrule
    3 & EP vs NP & -1.31 & (0.19) & 0.00 & -0.07 & (0.05) & 0.15 & 0.33 & (0.24) & 0.18 & -0.06 & (0.06) & 0.38\\
    4 & EP vs NP & -1.23 & (0.19) & 0.00 & 0.01 & (0.03) & 0.79 & 0.36 & (0.24) & 0.14 & -0.03 & (0.04) & 0.43\\
    7 & EP vs NP & -1.18 & (0.20) & 0.00 & 0.06 & (0.06) & 0.28 & 0.47 & (0.26) & 0.07 & 0.09 & (0.07) & 0.25\\
    \midrule
     &&  \multicolumn{12}{c}{\textbf{Outcome: Earnings}} \\ \midrule
   3 & WS vs NP & -3744 & (710.88) & 0.00 & -288 & (143.40) & 0.04 & 3055 & (994.68) & 0.00 & -224 & (198.62) & 0.26\\
    4 & WS vs NP & -3637 & (738.83) & 0.00 & -182 & (119.99) & 0.13 & 3214 & (1021.32) & 0.00 & -65 & (137.71) & 0.64\\
    7 & WS vs NP & -2986 & (738.65) & 0.00 & 470 & (202.27) & 0.02 & 3567 & (1054.36) & 0.00 & 289 & (228.68) & 0.21\\ \midrule
    3 & BC vs NP & -6990 & (831.31) & 0.00 & -91 & (173.58) & 0.60 & 1004 & (1182.96) & 0.40 & 69 & (244.84) & 0.78\\
    4 & BC vs NP & -6997 & (835.43) & 0.00 & -98 & (137.97) & 0.48 & 756 & (1182.27) & 0.52 & -178 & (160.18) & 0.27\\
    7 & BC vs NP & -6710 & (841.75) & 0.00 & 189 & (236.71) & 0.43 & 1043 & (1215.92) & 0.39 & 109 & (270.99) & 0.69\\  \midrule
    3 & TC vs NP & -7246 & (929.77) & 0.00 & -269 & (197.14) & 0.17 & 1809 & (1342.60) & 0.18 & 4 & (285.60) & 0.99\\
    4 & TC vs NP & -7108 & (933.59) & 0.00 & -131 & (155.07) & 0.40 & 1679 & (1343.80) & 0.21 & -125 & (186.73) & 0.50\\
    7 & TC vs NP & -6577 & (930.64) & 0.00 & 400 & (264.16) & 0.13 & 1925 & (1351.55) & 0.15 & 121 & (325.50) & 0.71\\  \midrule
    3 & EP vs NP & -7790 & (1071.64) & 0.00 & -343 & (231.31) & 0.14 & -529 & (1760.26) & 0.76 & -286 & (350.97) & 0.42\\
    4 & EP vs NP & -7419 & (1086.37) & 0.00 & 28 & (167.12) & 0.86 & -333 & (1725.34) & 0.85 & -90 & (202.82) & 0.66\\
    7 & EP vs NP & -7133 & (1117.08) & 0.00 & 314 & (308.95) & 0.31 & 133 & (1788.37) & 0.94 & 376 & (417.02) & 0.37\\  
    \bottomrule
    \end{tabular}
        \begin{tablenotes}
\textit{Note:} The table shows the BGATEs for the different programmes in comparison to no programme (NP: Non Participation, WS: Wage Subsidy, BC: Basic courses, TC: Training Courses, EP: Employment Programme) and the different sub-groups. 
The unemployment rate is computed at regional level at the start of the individual unemployment spell. The unemployment rate values approximate the closest integer.
\end{tablenotes}
\end{threeparttable}
\end{adjustbox}
\end{table}

\clearpage
\subsection{Individualised Average Treatment Effects}
\label{appendix:IATEs}

\subsubsection{Permanent Residents Sample} \label{bgates_app_temp}
This section includes the IATEs-ATEs of the remaining programmes compared to the non-participation for the Permanent Residents sample.
\begin{figure}[H]
\caption{Sorted individualised average treatment effects}
 \includegraphics[width=0.90\textwidth]{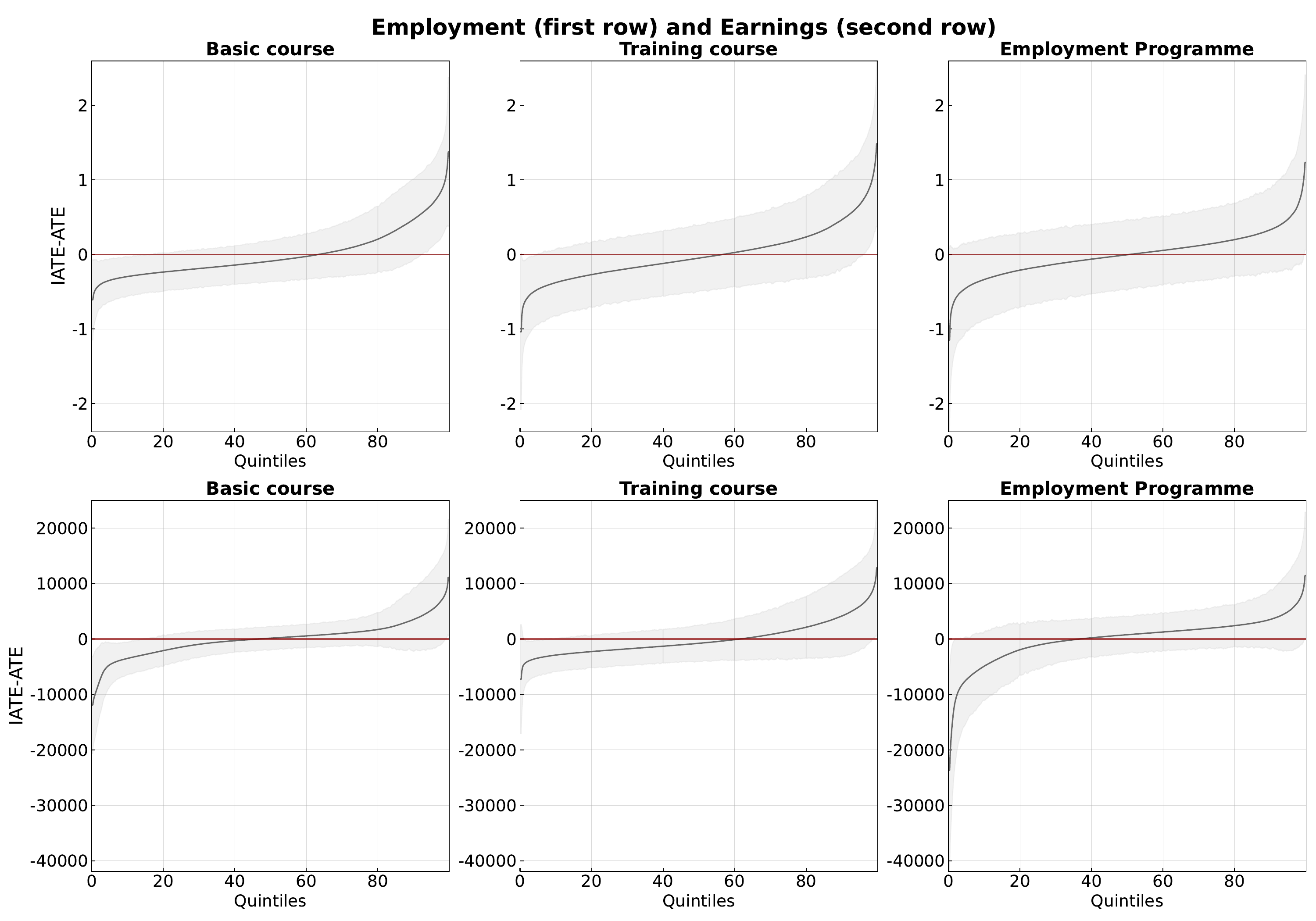}
\caption*{\textit{Note:} This figure shows the sorted IATEs for Basic Courses (BC), Training Courses (TC) and the Employment Programme (EP) in comparison to non-participants. Outcomes: Sum of months employed and sum of earnings over the third year after the program start. The x-axes indicate the average gain of participating in each programme. Sample: PERM sample.}
\label{fig:iates_main_dens}
\end{figure}

\subsubsection{Temporary Residents Sample}

This section includes the IATEs-ATEs and the density plots of the estimated IATEs of each programme compared to the non-participation programme for the TEMP sample. The outcomes are the cumulative earnings and months employed in the third year after the programmes start.

\begin{figure}[H]
    \captionsetup{font=small}  
    \caption{Sorted Individualised Average Treatment Effects for the TEMP sample}
 \begin{minipage}[t]{1\textwidth}
 \centering
 \includegraphics[width=0.68\textwidth]{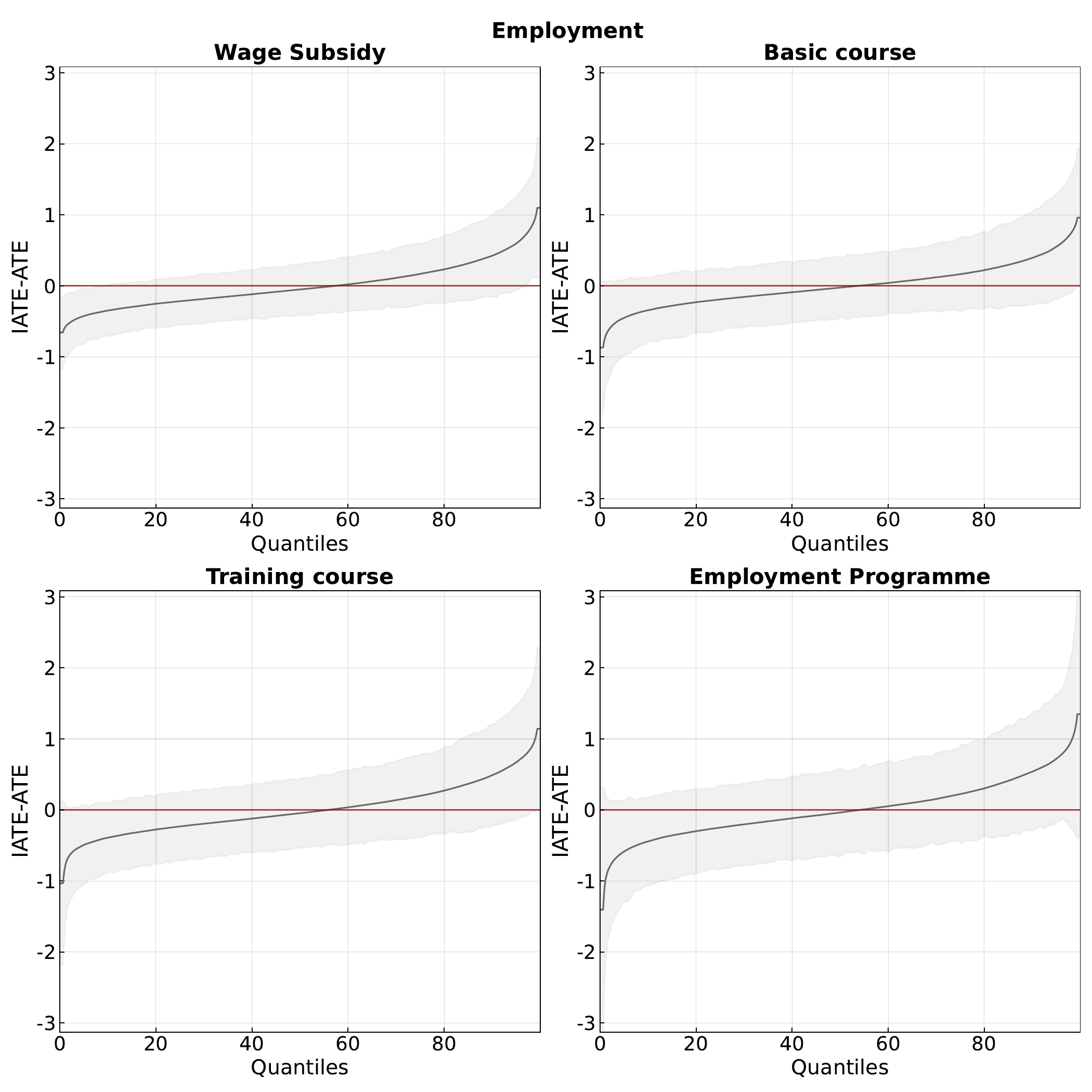}
	\end{minipage} 
	\begin{minipage}[t]{1\textwidth}
  \centering
 \includegraphics[width=0.75\textwidth]{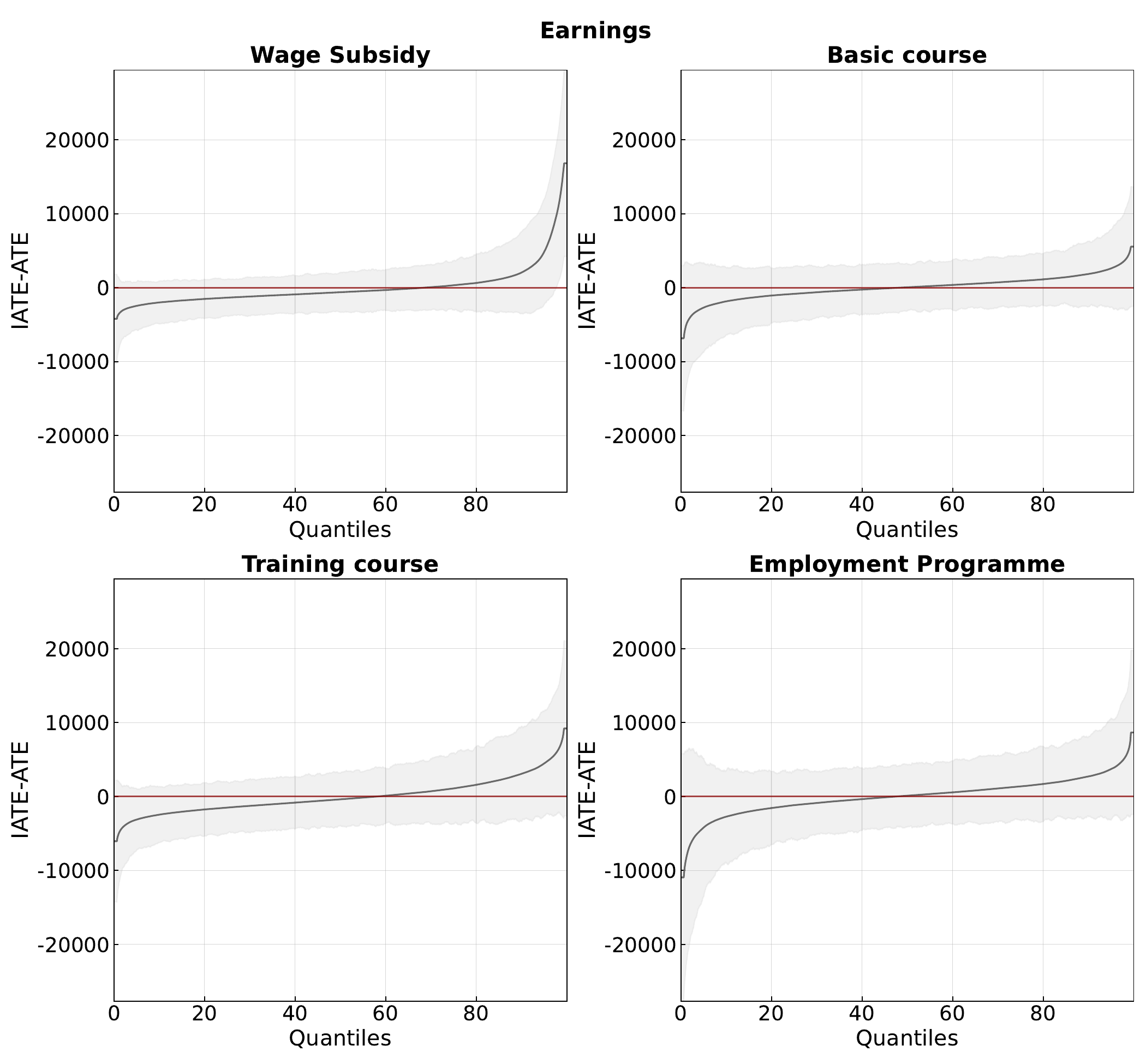}
	\end{minipage}  
\caption*{\textit{Note:} This figure shows the sorted IATEs for the Wage subsidy (WS), Basic Courses (BC), Training Courses (TC) and Employment Programme (EP) in comparison to the non-participants. Outcome: Sum of earnings over the third year after the program starts. On the x-axis indicates the average gain of participating in each programme. Sample: TEMP.}
\label{fig:iates_main}
\end{figure}

\begin{figure}[H]
\centering
\captionsetup{font=small}  
\caption{Sorted distribution of Individualised Average Treatment in the Temporary Residents sample}\includegraphics[width=1\textwidth]{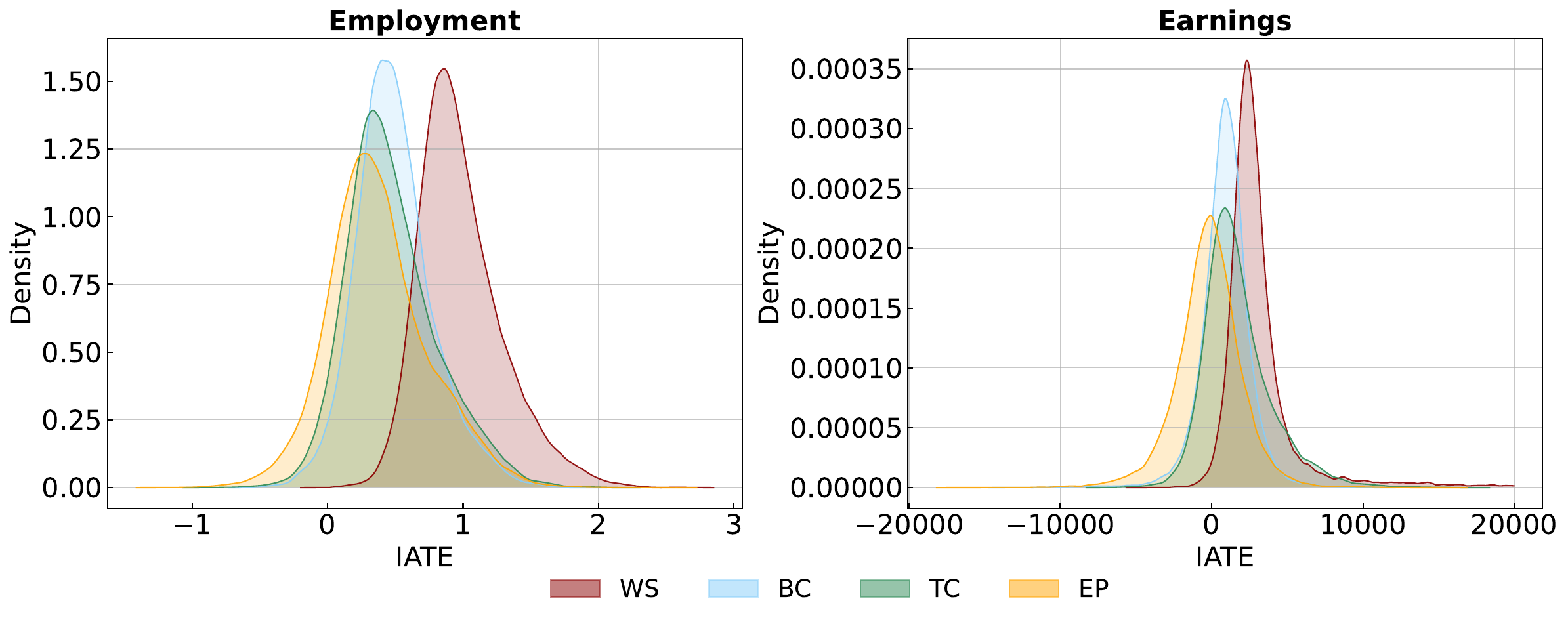}
\caption*{\textit{Note:} The figure shows densities of IATEs of Wage subsidies (WS), Basic Courses (BC), Training Courses
(TC) and Employment programmes (EP) relative to non-participation in the TEMP sample. The first
(second) plot shows effects on the sum of months employed (earnings) in the third year after programme start.  The average standard error of the IATES for the first (second) plot is 0.24 (1,885) for WS, 0.28 (12,149) for BC, 0.31 (2,392) for TC and 0.38 (2,887) for EP.}
\label{fig:iates_densb}
\end{figure}

\subsection{K-means Clustering}  \label{clustering_appendix}
To gain closer insights on which types of individuals profit from the programmes, we cluster individuals into groups according to the size of their IATEs using $k$-means++ clustering \citep{arthur2007k}, as provided in the scikit-learn Python package \citep{pedregosa2011scikit} and integrated in the \texttt{mcf} package, version 0.7.0. For a given IATE $\hat\tau_{d, d'}(X)$, the procedure partitions the observations into $k$ clusters such that the within-cluster variance is minimised. Each cluster is characterised by the mean of its observations, also called the centroid.

The algorithm begins by randomly selecting $k$ observations from the data to serve as initial centroids. Next, each observation is assigned to the nearest centroid. The centroids are then updated by calculating the mean of all observations assigned to the same centroid. This process is repeated, with observations being reassigned and centroids recalculated, until the change in centroid values falls below a specified threshold. A problem with the procedure is that it may converge to a local rather than a global minimum depending on the starting values. To mitigate this issue, the $k$-means++ algorithm initializes the centroids multiple times, ensuring that the initial centroids are spaced far apart. The number of clusters is determined by running the algorithm with various cluster counts and selecting the number that yields the largest distance between clusters, as quantified by the so-called silhouette coefficient.

Table \ref{table:kmenas_appendix} shows the results of the $k$-means clustering for the cumulative months into employment three years after the programme start for the PERM and TEMP sample.

\begin{table}[h]
\begin{adjustbox}{width=0.95\columnwidth, center}
\begin{threeparttable}
    \caption{IATEs with respect to non-participation and descriptive statistics for groups that benefit the least and most from each programme}
    \centering
    \begin{tabular}{@{}lcccc@{}}
    \toprule
    Programme & \textbf{WS} & \textbf{BC} & \textbf{TC} & \textbf{EP} \\ 
    \midrule
    \multicolumn{1}{l}{ } & \multicolumn{4}{c}{\textbf{Permanent Residents}} \\
    IATEs of the least and most beneficial groups & 0.26 - 1.11 & -0.63	- 0.28 & -0.59 - 0.15
 & -0.64 - 0.46 \\
    \midrule
    \multicolumn{1}{l}{Selected covariates} & \multicolumn{4}{c}{Values for the last and most beneficial groups by programme} \\
    \midrule
    & Least - Most &   Least - Most &  Least - Most &  Least - Most \\
    \midrule
    Age & 42.7 - 40.7  & 44.47 - 40.54 & 42.57 - 38.97 & 45.91 - 37.84 \\
    Female  & 0.32 - 0.55 & 0.38 - 0.41 & 0.58 - 0.39 & 0.28 - 0.56  \\
    Education (1: compulsory – 5: post-grad.) & 2.38 - 2.10 & 2.52 - 1.98 & 1.95 - 2.42 & 2.5 - 2.14 \\
    Local language knowledge (1: basic – 7: mother tongue)  & 6.50 - 5.03 & 6.66 - 4.74 & 5.80 - 5.92 & 6.48 - 4.80 \\
    Citizenship Swiss  & 0.83 - 0.44 & 0.89 - 0.32 &  0.69 - 0.60 &  0.82 - 0.29 \\
    Neighbouring countries  & 0.06 - 0.13 & 0.05 - 0.15 & 0.12 - 0.13 & 0.08 - 0.16 \\
    Rest of European Union  & 0.04 - 0.15 & 0.02 - 0.19 & 0.09 - 0.11 & 0.04 - 0.20\\
    Rest of Europe & 0.03 - 0.08 & 0.01	- 0.10 & 0.04 - 0.06 & 0.02	- 0.09\\
    Rest of the world & 0.03 - 0.20 & 0.02- 0.23 & 0.06	- 0.10 & 0.03 - 0.25 \\
    Months into employment 10 years before  & 110 - 75 & 110 - 76 & 98 - 91 &  107 - 81\\
    Cumulative earnings 10 years before  & 723,878 - 321,540 & 810,934 - 336,599 & 411,083 - 5,112,301 & 798,781 - 337,871 \\
    \midrule
    \% of obs. in cluster & 8 - 5 & 5 - 6 & 3 - 4 & 5 - 4 \\
    \midrule
     \multicolumn{1}{l}{ } & \multicolumn{4}{c}{\textbf{Temporary Residents}} \\
    IATEs of the least and most beneficial groups & 0.51 - 1.37 & -0.04 - 0.76 & -0.38 - 	1.12 & -0.60 - 1.12 \\
    \midrule
    \multicolumn{1}{l}{Selected covariates} & \multicolumn{4}{c}{Values for the last and most beneficial groups by programme} \\
    & Least - Most &   Least - Most &  Least - Most &  Least - Most \\
    \midrule
    Age & 38 - 36 & 38 - 35 & 37 - 35 & 37 - 35\\
    Female & 0.18 - 0.53 & 0.38	- 0.40 & 0.61 - 0.28 & 0.29	- 0.47 \\
    Education (1: compulsory – 5: post-grad.) & 1.66 - 3.05 & 2.02 - 2.91 & 1.77 - 3.60 & 2.03 - 3.15 \\
    Local language knowledge (1: basic – 7: mother tongue) & 3.64 - 4.79 & 4.28	- 4.64 & 3.77 - 5.42 & 4.92 - 4.18 \\
    Neighbouring countries & 0.28 - 0.41 & 0.40	- 0.40 & 0.33 - 0.56 & 0.56	- 0.27\\
    Rest of European Union & 0.48 -	0.29 & 0.36 - 0.33 & 0.37 - 0.23 & 0.29 - 0.36\\
    Rest of Europe & 0.14 - 0.25 & 0.18 - 0.20 & 0.20 - 0.18 & 0.12 - 0.33\\
    Rest of the World  & 0.14 - 0.25 & 0.18	- 0.20 & 0.20 - 0.18 & 0.12	- 0.33\\
    Months into employment 10 years before  & 49 - 45 & 46 - 48 & 48 - 53 & 49 - 43\\
    Cumulative earnings 10 years before & 208,652 - 282,512 & 205,898 - 272,831 & 137,308 - 435,215 & 242,519 - 203,515\\
     \midrule
      \% of obs. in cluster & 16 - 14 & 4 - 3 & 3 - 2 & 3 - 2\\
      \bottomrule
    \end{tabular}
\begin{tablenotes}
\textit{Note:} The outcome variable is the months into employment after three years from the program start. Clusters are formed using the IATEs vs the NP group, and the cluster's mean is reported. Clusters are based on the $k$-means++ algorithm and computed by the \texttt{mcf} package. WS: Wage Subsidy, BC: Basic Courses, TC: Training Courses, EP: Employment Programme. The percentage of observations in each cluster varies by programme and the number of clusters formed. Clusters that reach at least 1\% of the total observations are reported.
\end{tablenotes}
\label{table:kmenas_appendix}
\end{threeparttable}
\end{adjustbox}
\end{table}

\subsection{Optimal Policy} \label{policy_tree_appendix}
\begin{figure}[H]
    \captionsetup{font=small}  
    \caption{Depth-3 constrained Policy Tree, PERM sample, months employed in third year}
	\begin{minipage}[t]{\textwidth}
 \includegraphics[width=0.95\textwidth]{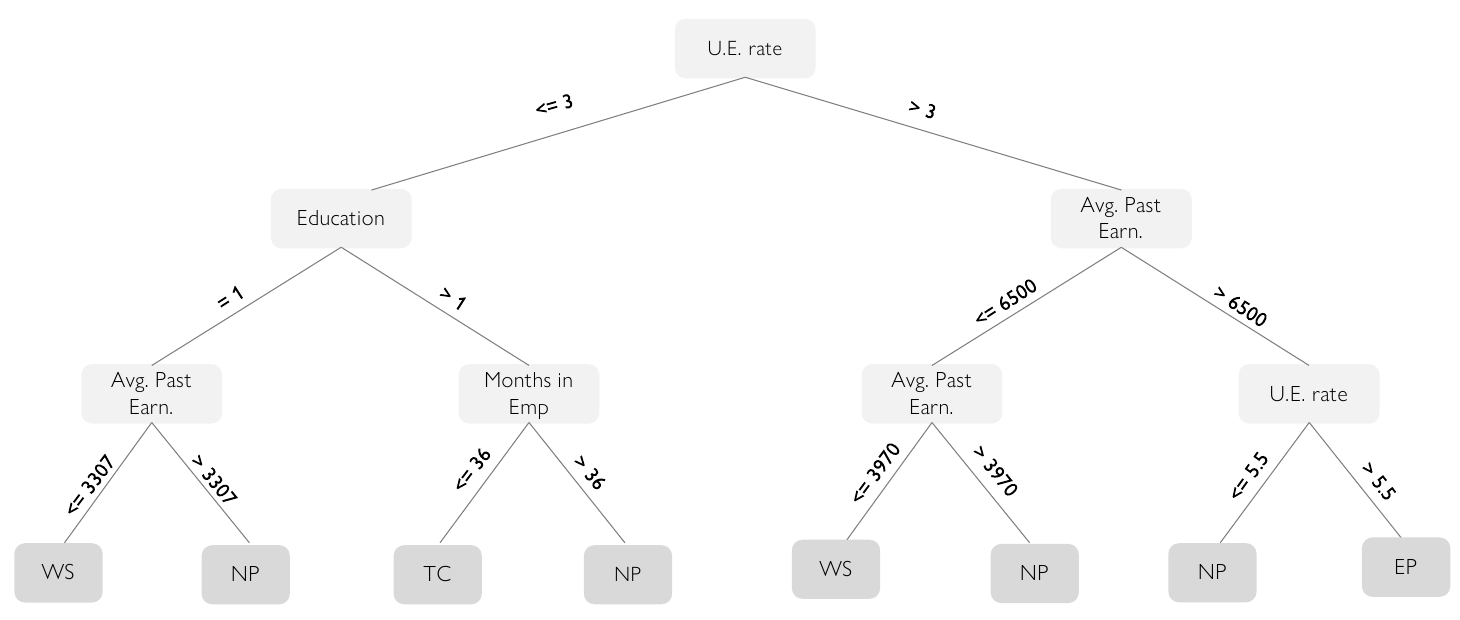}
	\end{minipage}
\caption*{\scriptsize\textit{Note:}  This figure represents a depth-3 policy tree, showing key decision points derived from the dataset. The variables driving the splits are highlighted in light grey rectangles, while the specific values defining these splits are illustrated along the branches. At the terminal nodes, depicted in darker grey, the optimal programme (or absence thereof) is indicated alongside the share of observations. The dataset comprises 9,975 observations for the PERM sample.}
\label{fig:policy_tree_constrained_swiss}
\end{figure}

This section shows the results of the policy trees.
Under the constraint of adhering to the observed assignment shares, we show that simple assignment rules derived from policy trees achieve outcomes that slightly improve the caseworker assignments. Figure \ref{fig:policy_tree_constrained_swiss} shows the assignment rules for the constrained depth-3 shallow tree for the PERM sample. Among the optimal trees, this depth-3 shallow tree not only achieves higher average months in employment compared to other shallow trees, but its simplicity makes it easier to interpret and present.  The shallow tree assigns two groups of individuals to WS. Firstly, those residing in regions with an unemployment rate below 3\%, with only compulsory education and average past earnings below CHF 3,300; secondly, individuals from regions with an unemployment rate above 3\% and average past earnings below CHF 4,000. Interestingly, unemployed individuals residing in regions where the unemployment rate exceeds 5.5\% and who previously earned more than CHF 6,500 are assigned to EP. Finally, individuals living in regions with an unemployment rate below 3\%, with education beyond compulsory schooling, and having 36 months or less of employment in the past five years are assigned to TC. It seems reasonable that individuals with higher levels of education but modest work experience are assigned to such training programmes. All remaining individuals are not assigned to a programme, resulting in zero allocations to BC. We do not present constrained policy trees for the PERM sample because none achieved outcomes as good as the observed assignments.

\begin{figure}[H]
    \captionsetup{font=small}  
    \caption{Depth-3 constrained Policy Tree, months employed in third year}
	\begin{minipage}[t]{\textwidth}
 \includegraphics[width=0.90\textwidth]{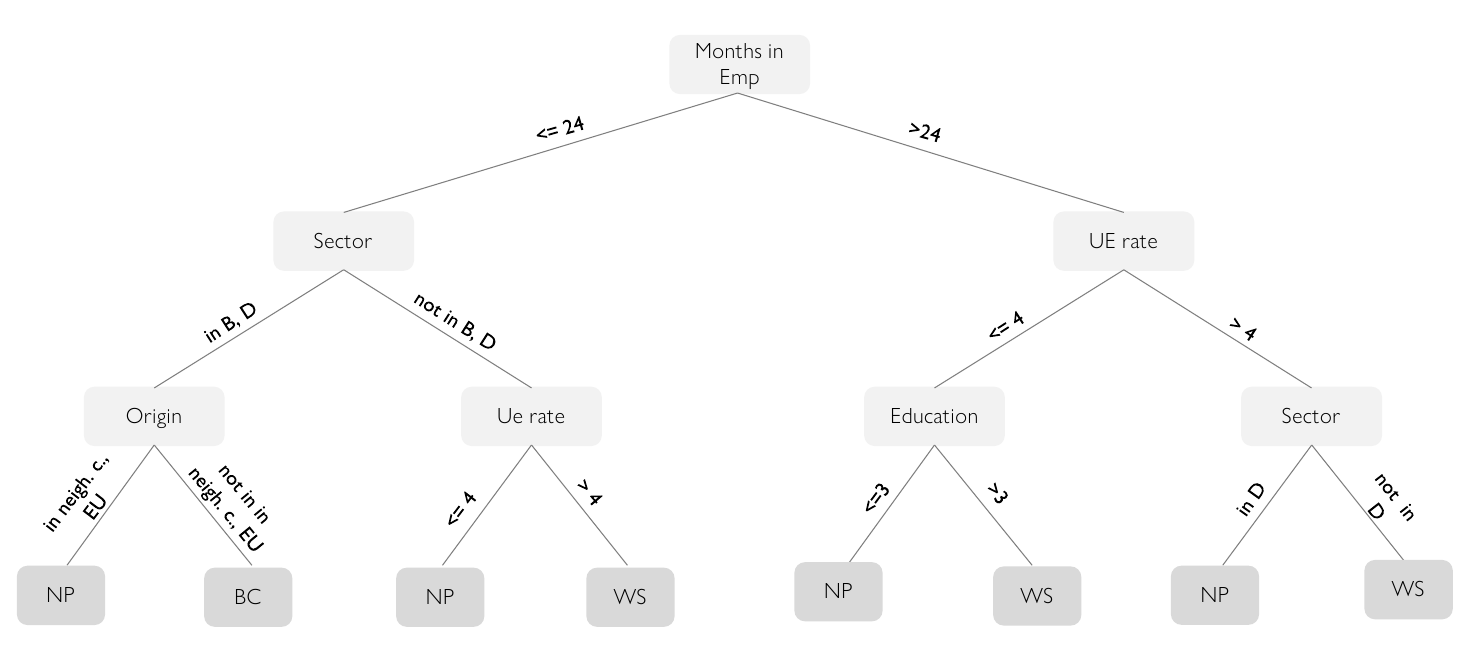}
	\end{minipage}
\caption*{\scriptsize\textit{Note:}  This figure represents a depth-3 policy tree, showing key decision points derived from the dataset. The variables driving the splits are highlighted in light grey rectangles, while the specific values defining these splits are illustrated along the branches. At the terminal nodes, depicted in darker grey, the optimal programme (or absence thereof) is indicated alongside the share of observations. Notably, the dataset comprises 3,020 observations for the TEMP sample. The values for the variable origin are the following: 1 - neighbour country, 2 - EU, 3 - rest of Europe, 4 - rest of the world. Sectors B and D refer to Production in industry and trade and Construction and mining, respectively. }
\label{fig:policy_tree_constrained_temp}
\end{figure}

\subsection{Placebo Tests} \label{Placebo_test}

An additional placebo test selects individuals who became unemployed in 2010 and 2011, estimating the effects of treatment status as assigned in the main analysis. Unlike \citet{cockx2023priority} and the placebo test in Section \ref{sensitivity_analysis}, individuals assigned to any programme during the placebo period are retained.\footnote{The exact design from \citet{cockx2023priority}, which excludes all individuals treated at the time of the placebo, could not be replicated due to an insufficient number of observations for running the \texttt{mcf}.} The outcome of interest is measured one year after the programme start.

Several constraints bound this placebo design. Years closer to the main analysis period are selected to avoid the influence of structural changes in programme composition, while only individuals who became unemployed in 2010 and 2011 are included to ensure a sufficiently wide gap between the outcome measured for the placebo and the unemployment spell in the main analysis, preventing overlap in observed periods. However, the reduced sample size (14,089 individuals) and the overlap of approximately 31\% in WS programme assignment between the main analysis and placebo period could introduce contamination into the analysis.

Table \ref{table:ate_cumemp_pl} (sample: PERM) shows that all average treatment effects (ATEs) are close to zero and statistically insignificant, except for the WS  programme. This exception likely arises from the high proportion of individuals assigned to this programme in the placebo and main analysis. As a result, the placebo test in Section \ref{sensitivity_analysis} excludes individuals assigned to the same programme in both periods, mitigating potential bias. In this test, all estimates remain statistically insignificant. While there may be concerns regarding the comparability of observations assigned to different programmes between the main analysis and placebo, the small sample size of these cases renders their impact negligible. The tables below provide further details on group composition within the placebo sample.

\begin{table}[ht!]
\begin{adjustbox}{width=0.55\columnwidth,  center}
\begin{threeparttable}
    \caption{Placebo Average Treatment Effects for future programmes on 12 cumulative months in employment}
    \centering
    \begin{tabular}{@{}lrrrr@{}}
    \toprule
    \textbf{Programme} & \textbf{Coef.} & \textbf{Std. Err.} &   \textbf{Coef.} & \textbf{Std. Err.}    \\
        \midrule
     Sample: &  \phantom{***} full &  & \phantom{***}  partial*  &   \\
        \midrule
        WS vs NP & -0.22 & (0.10)  ** & -0.01 & (0.12)   \\
        BC vs NP & -0.00 & (0.18) \phantom{**} & 0.15 & (0.18)   \\
        TC vs NP & -0.12 & (0.27) \phantom{**}  & -0.17 & (0.26)   \\
        EP vs NP & 0.40 & (0.24)  \phantom{**}  & 0.27 & (0.20)   \\
        \bottomrule
    \end{tabular}
\begin{tablenotes}
\small
\textit{Note:} Placebo ATEs for Wage Subsidy (WS), Basic Courses (BC), Technical Courses (TC), and Employment Programme (EP). Standard errors are presented in parentheses. *, **, and *** indicate the level of precision for each estimate, denoting if the p-value of a two-sided significance test is below 10\%, 5\%, and 1\%, respectively. *Partial sample refers to the one in which observations with the same treatment, except the NP, in both samples have been dropped.
\end{tablenotes}
\label{table:ate_cumemp_pl}
\end{threeparttable}
\end{adjustbox}
\end{table}

The MCF specification for covariates differs slightly from the main analysis in both placebo tests. Due to the relatively modest sample size, it uses less granular variables for certain covariates. The sector of the last employer is reduced to a single-digit level, and the occupational code of the last job is not included, though control for the individual’s function in the last job is retained. Additionally, cantons are used instead of the more detailed district level for geographic control, while the code for regional employment offices remains. Finally, covariates for unemployment, earnings, and ALMP histories are limited to five years. However, the specifications for earnings, employment, and unemployment status remain consistent for the 36 months preceding programme initiation.
 
\begin{table}[ht]
\begin{adjustbox}{width=0.55\columnwidth,  center}
\begin{threeparttable}
    \caption{Treatment groups and number of individuals for the placebo sample}
    \centering
    \begin{tabular}{@{}lrr@{}} \toprule
    \textbf{Treatment Groups} & \textbf{Number of Individuals} & \textbf{Fraction}\\
        \toprule
            No programme & 6.819& 48.3\% \\
            Temporary Wage Subsidy & 4.234 & 30.3\% \\
            Basic courses & 1.607 & 11.4\% \\
            Employment Programme &  642 & 5.6\% \\
            Training courses & 787 & 4.5\% \\
            \midrule
            Total & 14,089 &  \\
         \bottomrule
    \end{tabular}
\begin{tablenotes}
\textit{Note:} The table shows the treatment groups and the corresponding number of individuals.
\end{tablenotes}
\label{table:programme_placebo}
\end{threeparttable}
\end{adjustbox}
\end{table}

\end{appendices}

\end{document}